\documentclass[a4paper,fontsize=11pt,DIV=13]{scrbook}
\KOMAoptions{BCOR=10mm}

\usepackage{lipsum}
\usepackage[utf8]{inputenc}

\DeclareUnicodeCharacter{0308}{HERE!HERE!}

\usepackage{cite}

\setkomafont{disposition}{\bfseries}

\usepackage{csquotes}
\usepackage{hyperref,bookmark}
\usepackage{newpxtext,newpxmath} 
\usepackage{mathtools,bbm,mathrsfs}
\usepackage{physics,upgreek,tensor,multirow}
\usepackage[T1]{fontenc}
\usepackage[english]{babel}
\usepackage{tikz,tikzpagenodes}

\usepackage{anyfontsize}
\usepackage{graphicx,subcaption}
\usepackage{titlesec}
\usepackage{setspace}
\usepackage{fancyhdr}

\numberwithin{equation}{section}

\setkomafont{dictum}{\normalfont}

\titleformat{\chapter}[display]
{\normalsize \Huge\scshape \color{black}\setstretch{1.0}}
{\hfill{\MakeUppercase{\chaptertitlename}}\hspace{2pt}{\fontsize{60}{70}\selectfont\thechapter}}
{0pt}
{\fontsize{30}{36}\selectfont\rule{\textwidth}{1.5pt}\\[3pt]} 
[\vspace{-18pt}\rule{\textwidth}{1.5pt}]
\titlespacing*{\chapter}{0pt}{0pt}{30pt}

\titleformat{\section}[hang]
{\bfseries\LARGE\color{black}\setstretch{0.9}}
{\LARGE\bfseries\thesection\hspace{5pt}}
{0pt}
{}
[]
\titlespacing*{\section}{0pt}{30pt}{15pt}

\titleformat{\subsection}[hang]
{\bfseries\Large\color{black}\setstretch{0.1}}
{\Large\bfseries\thesubsection\hspace{5pt}}
{0pt}
{}
[]
\titlespacing*{\subsection}{0pt}{25pt}{10pt}

\titleformat{\paragraph}[runin]{\bfseries\large}{}{0pt}{}[]
\titlespacing*{\paragraph}{0pt}{7pt}{7pt}

\fancypagestyle{plain}{%
\fancyhf{}%
\fancyfoot[C]{\thepage}
\fancyhead[C]{}
\renewcommand{\headrulewidth}{0pt}%
\renewcommand{\footrulewidth}{0pt}%
}

\fancypagestyle{fancy}{%
\fancyhead[CE]{\leftmark}
\fancyhead[LE]{\scshape\MakeLowercase\chaptertitlename\hspace{2pt}\large\thechapter}
\fancyhead[RO]{\large\thesection}
\fancyhead[CO]{\rightmark}
\fancyfoot[C]{\thepage}
\renewcommand{\headrulewidth}{0.5pt}
\renewcommand{\footrulewidth}{0pt}
}

\renewcommand*{\dictumauthorformat}[1]{#1}
\renewcommand*{\dictumwidth}{.75\textwidth}

\newcommand{\one}{\mathbbm 1}
\newcommand{\bi}{{\gamma_{\textsc{bi}}}}
\newcommand{\R}{\mathbb{R}}
\newcommand{\C}{\mathbb{C}}

\newcommand{\SUT}{\mathrm{SU}(2)}
\newcommand{\Spin}{\mathrm{Spin}(4)}
\newcommand{\SUO}{\mathrm{SU}(1,1)}
\newcommand{\ISO}{\mathrm{ISO}(2)}
\newcommand{\SL}{\text{SL$(2,\C)$}}
\newcommand{\e}{\mathrm{e}}
\newcommand{\defeq}{\vcentcolon=}
\newcommand{\eqdef}{=\vcentcolon}

\newcommand{\spl}{\mathfrak{sl}\left(2,\C\right)}
\renewcommand{\sumint}[1]{\sum\hspace{-5mm}\int\hspace{-6mm}{\phantom{\int}}_{#1}}

\def\Ds{D_{\textsc s}}
\newcommand{\std}{D}	
\newcommand{\vm}{\vec m}
\newcommand{\vn}{\vec n}
\newcommand{\sfs}{\gamma}
\newcommand{\sfsc}{\gamma_\mathrm{cons}}

\newcommand{\jmin}{{j_{\text{min}}}}
\newcommand{\jmax}{{j_{\text{max}}}}
\newcommand{\Sreg}{S_{\mathrm{R}}}
\newcommand{\GN}{G_{\mathrm{N}}}

\newcommand{\mf}{\phi} 
\newcommand{\mm}{\pi_\phi} 
\newcommand{\pmm}{p_\phi} 
\newcommand{\rf}{\chi} 
\newcommand{\epsp}{\epsilon^\texttt{+}} 
\newcommand{\epsm}{\epsilon^\texttt{-}}
\newcommand{\pip}{\pi_0^\texttt{+}}
\newcommand{\pim}{\pi_0^\texttt{-}}
\newcommand{\tpip}{\tilde{\pi}_0^\texttt{+}}
\newcommand{\tpim}{\tilde{\pi}_0^\texttt{-}}
\newcommand{\slrcw}{\tilde{\sigma}} 
\newcommand{\tlrcw}{\tilde{\tau}} 
\newcommand{\Pl}{\mathrm{Pl}}
\newcommand{\GFT}{\mathrm{GFT}}
\newcommand{\GR}{\mathrm{GR}}

\newcommand{\vbr}{\vb*{\rho}}
\newcommand{\vbk}{\vb*{k}}
\newcommand{\vbn}{\vb*{\nu}}
\newcommand{\vbf}{\vb*{\phi}}
\newcommand{\vbi}{\vb*{i}}
\newcommand{\vbg}{\vb*{g}}
\newcommand{\minus}{\texttt{-}}
\newcommand{\plus}{\texttt{+}}
\newcommand{\zero}{0}
\newcommand{\dloc}{d_{\mathrm{loc}}}
\newcommand{\xiloc}{\xi_{\mathrm{loc}}}
\newcommand{\xinloc}{\xi_{\mathrm{nloc}}}
\newcommand{\Vp}{V_\plus}

\newcommand{\np}{n_\texttt{+}}
\newcommand{\nz}{n_0}
\newcommand{\nm}{n_\texttt{-}}

\newlength{\boxsize}
\renewcommand{\Re}[1]{\mathfrak{Re}\left\{#1\right\}}
\renewcommand{\Im}[1]{\mathfrak{Im}\left\{#1\right\}}

\usetikzlibrary{external}

\definecolor{jred}{rgb}{0.8,0,0}
\definecolor{jgreen}{rgb}{0,0.7,0}
\definecolor{jblue}{rgb}{0,0,0.8}

\usetikzlibrary{arrows,patterns,shapes,positioning,fit}
\tikzstyle{c} =	[coordinate]
\tikzstyle{vc} = [circle, draw=black, line width=1pt, inner sep=0pt, minimum size=2mm]
\tikzstyle{v} =  [circle, draw=black, line width=.2pt, fill=black, inner sep=0pt, minimum size=1.5mm]
\tikzstyle{vb} =[circle, draw=black, line width=.1pt, fill=jred, inner sep=0pt, minimum size=1mm]
\tikzstyle{vh} =[circle, draw=black, line width=.1pt, fill=jblue, inner sep=0pt, minimum size=1.5mm]
\tikzstyle{vs} =[coordinate]
\tikzstyle{e} =	[draw=jred,line width=1pt]
\tikzstyle{eb}= [draw=jgreen,line width=1pt]
\tikzstyle{es}= [dashed, line width=1pt]
\tikzstyle{ev}= [draw=jgreen,line width=1pt]
\tikzstyle{f} = [line width=0.01pt,dotted,fill=blue, fill opacity=.1]
\tikzstyle{cs}= [ellipse, draw=black, line width=.1pt, fill=jred, inner sep=1pt, minimum size=2mm]

\newcommand{\cvfpppp}{
\begin{tikzpicture}[scale=2,x=2ex,y=2ex,baseline={([yshift=-.59ex]current bounding box.center)}] 
\scriptsize{
\node [v]	(1)	at (0,0)	{};
\node [vb, label=above:+]	(b1)	at (-1,1)	{};
\node [vb, label=below:+]	(w1)	at (-1,-1)	{};
\node [vb, label=below:+]	(b2)	at (1,-1)	{};
\node [vb, label=above:+]	(w2)	at (1,1)	{};
\path
\foreach \i/\j in {1/2}{
   	(w\i) edge [ev]   (b\j)
	(b\i) edge [ev]  node [c] {} (w\j)
	}
\foreach \i in {1,2}{
	(w\i)   edge 	[ev]				(b\i)
	(w\i)	edge [ev,bend left=30]	(b\i)
	(w\i)	edge [ev,bend right=30]	(b\i)
	}
\foreach \i in {w1,b1,w2,b2}{
   	(\i) edge [e] (1)
	};
}
\end{tikzpicture}
}

\newcommand{\cvfpppm}{
\begin{tikzpicture}[scale=2,x=2ex,y=2ex,baseline={([yshift=-.59ex]current bounding box.center)}] 
\scriptsize{
\node [v]	(1)	at (0,0)	{};
\node [vb, label=above:\textbf{--}]	(b1)	at (-1,1)	{};
\node [vb, label=below:+]	(w1)	at (-1,-1)	{};
\node [vb, label=below:+]	(b2)	at (1,-1)	{};
\node [vb, label=above:+]	(w2)	at (1,1)	{};
\path
\foreach \i/\j in {1/2}{
   	(w\i) edge [ev]   (b\j)
	(b\i) edge [ev]  node [c] {} (w\j)
	}
\foreach \i in {1,2}{
	(w\i)   edge 	[ev]				(b\i)
	(w\i)	edge [ev,bend left=30]	(b\i)
	(w\i)	edge [ev,bend right=30]	(b\i)
	}
\foreach \i in {w1,b1,w2,b2}{
   	(\i) edge [e] (1)
	};
}
\end{tikzpicture}
}

\newcommand{\cvfppmma}{
\begin{tikzpicture}[scale=2,x=2ex,y=2ex,baseline={([yshift=-.59ex]current bounding box.center)}] 
\scriptsize{
\node [v]	(1)	at (0,0)	{};
\node [vb, label=above:+]	(b1)	at (-1,1)	{};
\node [vb, label=below:+]	(w1)	at (-1,-1)	{};
\node [vb, label=below:\textbf{--}]	(b2)	at (1,-1)	{};
\node [vb, label=above:\textbf{--}]	(w2)	at (1,1)	{};
\path
\foreach \i/\j in {1/2}{
   	(w\i) edge [ev]   (b\j)
	(b\i) edge [ev]  node [c] {} (w\j)
	}
\foreach \i in {1,2}{
	(w\i)   edge 	[ev]				(b\i)
	(w\i)	edge [ev,bend left=30]	(b\i)
	(w\i)	edge [ev,bend right=30]	(b\i)
	}
\foreach \i in {w1,b1,w2,b2}{
   	(\i) edge [e] (1)
	};
}
\end{tikzpicture}
}

\newcommand{\cvfppmmb}{
\begin{tikzpicture}[scale=2,x=2ex,y=2ex,baseline={([yshift=-.59ex]current bounding box.center)}] 
\scriptsize{
\node [v]	(1)	at (0,0)	{};
\node [vb, label=above:\textbf{--}]	(b1)	at (-1,1)	{};
\node [vb, label=below:+]	(w1)	at (-1,-1)	{};
\node [vb, label=below:+]	(b2)	at (1,-1)	{};
\node [vb, label=above:\textbf{--}]	(w2)	at (1,1)	{};
\path
\foreach \i/\j in {1/2}{
   	(w\i) edge [ev]   (b\j)
	(b\i) edge [ev]  node [c] {} (w\j)
	}
\foreach \i in {1,2}{
	(w\i)   edge 	[ev]				(b\i)
	(w\i)	edge [ev,bend left=30]	(b\i)
	(w\i)	edge [ev,bend right=30]	(b\i)
	}
\foreach \i in {w1,b1,w2,b2}{
   	(\i) edge [e] (1)
	};
}
\end{tikzpicture}
}

\newcommand{\cvfppmmc}{
\begin{tikzpicture}[scale=2,x=2ex,y=2ex,baseline={([yshift=-.59ex]current bounding box.center)}] 
\scriptsize{
\node [v]	(1)	at (0,0)	{};
\node [vb, label=above:\textbf{--}]	(b1)	at (-1,1)	{};
\node [vb, label=below:+]	(w1)	at (-1,-1)	{};
\node [vb, label=below:\textbf{--}]	(b2)	at (1,-1)	{};
\node [vb, label=above:+]	(w2)	at (1,1)	{};
\path
\foreach \i/\j in {1/2}{
   	(w\i) edge [ev]   (b\j)
	(b\i) edge [ev]  node [c] {} (w\j)
	}
\foreach \i in {1,2}{
	(w\i)   edge 	[ev]				(b\i)
	(w\i)	edge [ev,bend left=30]	(b\i)
	(w\i)	edge [ev,bend right=30]	(b\i)
	}
\foreach \i in {w1,b1,w2,b2}{
   	(\i) edge [e] (1)
	};
}
\end{tikzpicture}
}

\newcommand{\cvfppmza}{
\begin{tikzpicture}[scale=2,x=2ex,y=2ex,baseline={([yshift=-.59ex]current bounding box.center)}] 
\scriptsize{
\node [v]	(1)	at (0,0)	{};
\node [vb, label=above:+]	(b1)	at (-1,1)	{};
\node [vb, label=below:+]	(w1)	at (-1,-1)	{};
\node [vb, label=below:0]	(b2)	at (1,-1)	{};
\node [vb, label=above:\textbf{--}]	(w2)	at (1,1)	{};
\path
\foreach \i/\j in {1/2}{
   	(w\i) edge [ev]   (b\j)
	(b\i) edge [ev]  node [c] {} (w\j)
	}
\foreach \i in {1,2}{
	(w\i)   edge 	[ev]				(b\i)
	(w\i)	edge [ev,bend left=30]	(b\i)
	(w\i)	edge [ev,bend right=30]	(b\i)
	}
\foreach \i in {w1,b1,w2,b2}{
   	(\i) edge [e] (1)
	};
}
\end{tikzpicture}
}

\newcommand{\cvfppmzb}{
\begin{tikzpicture}[scale=2,x=2ex,y=2ex,baseline={([yshift=-.59ex]current bounding box.center)}] 
\scriptsize{
\node [v]	(1)	at (0,0)	{};
\node [vb, label=above:\textbf{--}]	(b1)	at (-1,1)	{};
\node [vb, label=below:+]	(w1)	at (-1,-1)	{};
\node [vb, label=below:+]	(b2)	at (1,-1)	{};
\node [vb, label=above:0]	(w2)	at (1,1)	{};
\path
\foreach \i/\j in {1/2}{
   	(w\i) edge [ev]   (b\j)
	(b\i) edge [ev]  node [c] {} (w\j)
	}
\foreach \i in {1,2}{
	(w\i)   edge 	[ev]				(b\i)
	(w\i)	edge [ev,bend left=30]	(b\i)
	(w\i)	edge [ev,bend right=30]	(b\i)
	}
\foreach \i in {w1,b1,w2,b2}{
   	(\i) edge [e] (1)
	};
}
\end{tikzpicture}
}

\newcommand{\cvfppmzc}{
\begin{tikzpicture}[scale=2,x=2ex,y=2ex,baseline={([yshift=-.59ex]current bounding box.center)}] 
\scriptsize{
\node [v]	(1)	at (0,0)	{};
\node [vb, label=above:\textbf{--}]	(b1)	at (-1,1)	{};
\node [vb, label=below:+]	(w1)	at (-1,-1)	{};
\node [vb, label=below:0]	(b2)	at (1,-1)	{};
\node [vb, label=above:+]	(w2)	at (1,1)	{};
\path
\foreach \i/\j in {1/2}{
   	(w\i) edge [ev]   (b\j)
	(b\i) edge [ev]  node [c] {} (w\j)
	}
\foreach \i in {1,2}{
	(w\i)   edge 	[ev]				(b\i)
	(w\i)	edge [ev,bend left=30]	(b\i)
	(w\i)	edge [ev,bend right=30]	(b\i)
	}
\foreach \i in {w1,b1,w2,b2}{
   	(\i) edge [e] (1)
	};
}
\end{tikzpicture}
}

\newcommand{\cvftpppp}{
\begin{tikzpicture}[scale=2,x=2ex,y=2ex,baseline={([yshift=-.59ex]current bounding box.center)}] 
\node [v]	(v)	at (0,0) 	{};
\node [vb, label=above:+]	(1)	at (-1,1)	{};
\node [vb, label=below:+]	(2)	at (-1,-1)	{};
\node [vb, label=below:+]	(3)	at (1,-1)	{};
\node [vb, label=above:+]	(4)	at (1,1)	{};
\path
\foreach \i in {1,2,3,4}{
    (\i) edge [e] (v)
    }
\foreach \i/\j in {1/2,2/1,3/4,4/3}{
   (\i)	edge [ev,bend left=15]	(\j)
   (\i)	edge [ev,bend left=40]	(\j)
   };
\end{tikzpicture}
}

\newcommand{\cvftpppm}{
\begin{tikzpicture}[scale=2,x=2ex,y=2ex,baseline={([yshift=-.59ex]current bounding box.center)}] 
\node [v]	(v)	at (0,0) 	{};
\node [vb, label=above:+]	(1)	at (-1,1)	{};
\node [vb, label=below:+]	(2)	at (-1,-1)	{};
\node [vb, label=below:+]	(3)	at (1,-1)	{};
\node [vb, label=above:\textbf{--}]	(4)	at (1,1)	{};
\path
\foreach \i in {1,2,3,4}{
    (\i) edge [e] (v)
    }
\foreach \i/\j in {1/2,2/1,3/4,4/3}{
   (\i)	edge [ev,bend left=15]	(\j)
   (\i)	edge [ev,bend left=40]	(\j)
   };
\end{tikzpicture}
}

\newcommand{\cvftppmma}{
\begin{tikzpicture}[scale=2,x=2ex,y=2ex,baseline={([yshift=-.59ex]current bounding box.center)}] 
\node [v]	(v)	at (0,0) 	{};
\node [vb, label=above:+]	(1)	at (-1,1)	{};
\node [vb, label=below:+]	(2)	at (-1,-1)	{};
\node [vb, label=below:\textbf{--}]	(3)	at (1,-1)	{};
\node [vb, label=above:\textbf{--}]	(4)	at (1,1)	{};
\path
\foreach \i in {1,2,3,4}{
    (\i) edge [e] (v)
    }
\foreach \i/\j in {1/2,2/1,3/4,4/3}{
   (\i)	edge [ev,bend left=15]	(\j)
   (\i)	edge [ev,bend left=40]	(\j)
   };
\end{tikzpicture}
}

\newcommand{\cvftppmmb}{
\begin{tikzpicture}[scale=2,x=2ex,y=2ex,baseline={([yshift=-.59ex]current bounding box.center)}] 
\node [v]	(v)	at (0,0) 	{};
\node [vb, label=above:\textbf{--}]	(1)	at (-1,1)	{};
\node [vb, label=below:+]	(2)	at (-1,-1)	{};
\node [vb, label=below:+]	(3)	at (1,-1)	{};
\node [vb, label=above:\textbf{--}]	(4)	at (1,1)	{};
\path
\foreach \i in {1,2,3,4}{
    (\i) edge [e] (v)
    }
\foreach \i/\j in {1/2,2/1,3/4,4/3}{
   (\i)	edge [ev,bend left=15]	(\j)
   (\i)	edge [ev,bend left=40]	(\j)
   };
\end{tikzpicture}
}

\newcommand{\cvftppmza}{
\begin{tikzpicture}[scale=2,x=2ex,y=2ex,baseline={([yshift=-.59ex]current bounding box.center)}] 
\node [v]	(v)	at (0,0) 	{};
\node [vb, label=above:+]	(1)	at (-1,1)	{};
\node [vb, label=below:+]	(2)	at (-1,-1)	{};
\node [vb, label=below:0]	(3)	at (1,-1)	{};
\node [vb, label=above:\textbf{--}]	(4)	at (1,1)	{};
\path
\foreach \i in {1,2,3,4}{
    (\i) edge [e] (v)
    }
\foreach \i/\j in {1/2,2/1,3/4,4/3}{
   (\i)	edge [ev,bend left=15]	(\j)
   (\i)	edge [ev,bend left=40]	(\j)
   };
\end{tikzpicture}
}
\newcommand{\cvftppmzb}{
\begin{tikzpicture}[scale=2,x=2ex,y=2ex,baseline={([yshift=-.59ex]current bounding box.center)}] 
\node [v]	(v)	at (0,0) 	{};
\node [vb, label=above:0]	(1)	at (-1,1)	{};
\node [vb, label=below:+]	(2)	at (-1,-1)	{};
\node [vb, label=below:+]	(3)	at (1,-1)	{};
\node [vb, label=above:\textbf{--}]	(4)	at (1,1)	{};
\path
\foreach \i in {1,2,3,4}{
    (\i) edge [e] (v)
    }
\foreach \i/\j in {1/2,2/1,3/4,4/3}{
   (\i)	edge [ev,bend left=15]	(\j)
   (\i)	edge [ev,bend left=40]	(\j)
   };
\end{tikzpicture}
}

\newcommand{\cvfnpppp}{
\begin{tikzpicture}[scale=2,x=2ex,y=2ex,baseline={([yshift=-.59ex]current bounding box.center)}] 
\scriptsize{
\node [v]	(1)	at (0,0)	{};
\node [vb, label=above:+]	(b1)	at (-1,1)	{};
\node [vb, label=below:+]	(w1)	at (-1,-1)	{};
\node [vb, label=below:+]	(b2)	at (1,-1)	{};
\node [vb, label=above:+]	(w2)	at (1,1)	{};
\path
\foreach \i/\j in {1/2,2/1}{
   	(w\i) edge [ev,bend left=30]  (b\j)
   	(w\i) edge [ev,bend right=30] (b\j)
	}
\foreach \i in {1,2}{
	(w\i)	edge [ev,bend left=30]  (b\i)
	(w\i)	edge [ev,bend right=30]	(b\i)
	}
\foreach \i in {w1,b1,w2,b2}{
   	(\i) edge [e] (1)
	};
}
\end{tikzpicture}
}

\newcommand{\cvfnpppm}{
\begin{tikzpicture}[scale=2,x=2ex,y=2ex,baseline={([yshift=-.59ex]current bounding box.center)}] 
\scriptsize{
\node [v]	(1)	at (0,0)	{};
\node [vb, label=above:\textbf{--}]	(b1)	at (-1,1)	{};
\node [vb, label=below:+]	(w1)	at (-1,-1)	{};
\node [vb, label=below:+]	(b2)	at (1,-1)	{};
\node [vb, label=above:+]	(w2)	at (1,1)	{};
\path
\foreach \i/\j in {1/2,2/1}{
   	(w\i) edge [ev,bend left=30]  (b\j)
   	(w\i) edge [ev,bend right=30] (b\j)
	}
\foreach \i in {1,2}{
	(w\i)	edge [ev,bend left=30]  (b\i)
	(w\i)	edge [ev,bend right=30]	(b\i)
	}
\foreach \i in {w1,b1,w2,b2}{
   	(\i) edge [e] (1)
	};
}
\end{tikzpicture}
}

\newcommand{\cvfnppmma}{
\begin{tikzpicture}[scale=2,x=2ex,y=2ex,baseline={([yshift=-.59ex]current bounding box.center)}] 
\scriptsize{
\node [v]	(1)	at (0,0)	{};
\node [vb, label=above:+]	(b1)	at (-1,1)	{};
\node [vb, label=below:+]	(w1)	at (-1,-1)	{};
\node [vb, label=below:\textbf{--}]	(b2)	at (1,-1)	{};
\node [vb, label=above:\textbf{--}]	(w2)	at (1,1)	{};
\path
\foreach \i/\j in {1/2,2/1}{
   	(w\i) edge [ev,bend left=30]  (b\j)
   	(w\i) edge [ev,bend right=30] (b\j)
	}
\foreach \i in {1,2}{
	(w\i)	edge [ev,bend left=30]  (b\i)
	(w\i)	edge [ev,bend right=30]	(b\i)
	}
\foreach \i in {w1,b1,w2,b2}{
   	(\i) edge [e] (1)
	};
}
\end{tikzpicture}
}

\newcommand{\cvfnppmmb}{
\begin{tikzpicture}[scale=2,x=2ex,y=2ex,baseline={([yshift=-.59ex]current bounding box.center)}] 
\scriptsize{
\node [v]	(1)	at (0,0)	{};
\node [vb, label=above:\textbf{--}]	(b1)	at (-1,1)	{};
\node [vb, label=below:+]	(w1)	at (-1,-1)	{};
\node [vb, label=below:\textbf{--}]	(b2)	at (1,-1)	{};
\node [vb, label=above:+]	(w2)	at (1,1)	{};
\path
\foreach \i/\j in {1/2,2/1}{
   	(w\i) edge [ev,bend left=30]  (b\j)
   	(w\i) edge [ev,bend right=30] (b\j)
	}
\foreach \i in {1,2}{
	(w\i)	edge [ev,bend left=30]  (b\i)
	(w\i)	edge [ev,bend right=30]	(b\i)
	}
\foreach \i in {w1,b1,w2,b2}{
   	(\i) edge [e] (1)
	};
}
\end{tikzpicture}
}

\newcommand{\cvfnppmza}{
\begin{tikzpicture}[scale=2,x=2ex,y=2ex,baseline={([yshift=-.59ex]current bounding box.center)}] 
\scriptsize{
\node [v]	(1)	at (0,0)	{};
\node [vb, label=above:+]	(b1)	at (-1,1)	{};
\node [vb, label=below:+]	(w1)	at (-1,-1)	{};
\node [vb, label=below:0]	(b2)	at (1,-1)	{};
\node [vb, label=above:\textbf{--}]	(w2)	at (1,1)	{};
\path
\foreach \i/\j in {1/2,2/1}{
   	(w\i) edge [ev,bend left=30]  (b\j)
   	(w\i) edge [ev,bend right=30] (b\j)
	}
\foreach \i in {1,2}{
	(w\i)	edge [ev,bend left=30]  (b\i)
	(w\i)	edge [ev,bend right=30]	(b\i)
	}
\foreach \i in {w1,b1,w2,b2}{
   	(\i) edge [e] (1)
	};
}
\end{tikzpicture}
}

\newcommand{\cvfnppmzb}{
\begin{tikzpicture}[scale=2,x=2ex,y=2ex,baseline={([yshift=-.59ex]current bounding box.center)}] 
\scriptsize{
\node [v]	(1)	at (0,0)	{};
\node [vb, label=above:\textbf{--}]	(b1)	at (-1,1)	{};
\node [vb, label=below:+]	(w1)	at (-1,-1)	{};
\node [vb, label=below:0]	(b2)	at (1,-1)	{};
\node [vb, label=above:+]	(w2)	at (1,1)	{};
\path
\foreach \i/\j in {1/2,2/1}{
   	(w\i) edge [ev,bend left=30]  (b\j)
   	(w\i) edge [ev,bend right=30] (b\j)
	}
\foreach \i in {1,2}{
	(w\i)	edge [ev,bend left=30]  (b\i)
	(w\i)	edge [ev,bend right=30]	(b\i)
	}
\foreach \i in {w1,b1,w2,b2}{
   	(\i) edge [e] (1)
	};
}
\end{tikzpicture}
}

\newcommand{\cvfsppppp}{
\begin{tikzpicture}[scale=2,x=2ex,y=2ex,baseline={([yshift=-.59ex]current bounding box.center)}] 
\scriptsize{
\node [v]	(0)	at (0,0)	{};
\foreach \i in {72,288,360}{
\begin{scope}[rotate=\i]
\node [vb, label=above:+]	(\i)	at (0,1.6)	{};
\end{scope}
}
\foreach \i in {144,216}{
\begin{scope}[rotate=\i]
\node [vb, label=below:+]	(\i)	at (0,1.6)	{};
\end{scope}
}
\path
\foreach \i in {72,144,216,288,360}{
    \foreach \j in {72,144,216,288,360}{
   	    (\i) edge [ev]  (\j)
	    }
    }
\foreach \i in {72,144,216,288,360}{
  	(\i) edge [e] (0)
	};
}
\end{tikzpicture}
}

\newcommand{\cvfspppppnolocal}{
\begin{tikzpicture}[scale=2,x=2ex,y=2ex,baseline={([yshift=-.59ex]current bounding box.center)}] 
\scriptsize{
\foreach \i in {72,288,360}{
\begin{scope}[rotate=\i]
\node [vb, label=above:\scalebox{1.8}{+}]	(\i)	at (0,1.6)	{};
\end{scope}
}
\foreach \i in {144,216}{
\begin{scope}[rotate=\i]
\node [vb, label=below:\scalebox{1.8}{+}]	(\i)	at (0,1.6)	{};
\end{scope}
}
\path
\foreach \i in {72,144,216,288,360}{
    \foreach \j in {72,144,216,288,360}{
   	    (\i) edge [ev]  (\j)
	    }
    };
}
\end{tikzpicture}
}

\newcommand{\cvfsppppm}{
\begin{tikzpicture}[scale=2,x=2ex,y=2ex,baseline={([yshift=-.59ex]current bounding box.center)}] 
\scriptsize{
\node [v]	(0)	at (0,0)	{};
\foreach \i in {72}{
\begin{scope}[rotate=\i]
\node [vb, label=above:\textbf{--}]	(\i)	at (0,1.6)	{};
\end{scope}
}
\foreach \i in {288,360}{
\begin{scope}[rotate=\i]
\node [vb, label=above:+]	(\i)	at (0,1.6)	{};
\end{scope}
}
\foreach \i in {144,216}{
\begin{scope}[rotate=\i]
\node [vb, label=below:+]	(\i)	at (0,1.6)	{};
\end{scope}
}
\path
\foreach \i in {72,144,216,288,360}{
    \foreach \j in {72,144,216,288,360}{
   	    (\i) edge [ev]  (\j)
	    }
    }
\foreach \i in {72,144,216,288,360}{
  	(\i) edge [e] (0)
	};
}
\end{tikzpicture}
}

\newcommand{\cvfspppmm}{
\begin{tikzpicture}[scale=2,x=2ex,y=2ex,baseline={([yshift=-.59ex]current bounding box.center)}] 
\scriptsize{
\node [v]	(0)	at (0,0)	{};
\foreach \i in {288}{
\begin{scope}[rotate=\i]
\node [vb, label=above:+]	(\i)	at (0,1.6)	{};
\end{scope}
}
\foreach \i in {72,360}{
\begin{scope}[rotate=\i]
\node [vb, label=above:\textbf{--}]	(\i)	at (0,1.6)	{};
\end{scope}
}
\foreach \i in {144,216}{
\begin{scope}[rotate=\i]
\node [vb, label=below:+]	(\i)	at (0,1.6)	{};
\end{scope}
}
\path
\foreach \i in {72,144,216,288,360}{
    \foreach \j in {72,144,216,288,360}{
   	    (\i) edge [ev]  (\j)
	    }
    }
\foreach \i in {72,144,216,288,360}{
  	(\i) edge [e] (0)
	};
}
\end{tikzpicture}
}

\newcommand{\cvfspppmz}{
\begin{tikzpicture}[scale=2,x=2ex,y=2ex,baseline={([yshift=-.59ex]current bounding box.center)}] 
\scriptsize{
\node [v]	(0)	at (0,0)	{};
\foreach \i in {288}{
\begin{scope}[rotate=\i]
\node [vb, label=above:+]	(\i)	at (0,1.6)	{};
\end{scope}
}
\foreach \i in {72}{
\begin{scope}[rotate=\i]
\node [vb, label=above:\textbf{--}]	(\i)	at (0,1.6)	{};
\end{scope}
}
\foreach \i in {360}{
\begin{scope}[rotate=\i]
\node [vb, label=above:0]	(\i)	at (0,1.6)	{};
\end{scope}
}
\foreach \i in {144,216}{
\begin{scope}[rotate=\i]
\node [vb, label=below:+]	(\i)	at (0,1.6)	{};
\end{scope}
}
\path
\foreach \i in {72,144,216,288,360}{
    \foreach \j in {72,144,216,288,360}{
   	    (\i) edge [ev]  (\j)
	    }
    }
\foreach \i in {72,144,216,288,360}{
  	(\i) edge [e] (0)
	};
}
\end{tikzpicture}
}

\newcommand{\cvfsppmzz}{
\begin{tikzpicture}[scale=2,x=2ex,y=2ex,baseline={([yshift=-.59ex]current bounding box.center)}] 
\scriptsize{
\node [v]	(0)	at (0,0)	{};
\foreach \i in {288}{
\begin{scope}[rotate=\i]
\node [vb, label=above:0]	(\i)	at (0,1.6)	{};
\end{scope}
}
\foreach \i in {72}{
\begin{scope}[rotate=\i]
\node [vb, label=above:\textbf{--}]	(\i)	at (0,1.6)	{};
\end{scope}
}
\foreach \i in {360}{
\begin{scope}[rotate=\i]
\node [vb, label=above:0]	(\i)	at (0,1.6)	{};
\end{scope}
}
\foreach \i in {144,216}{
\begin{scope}[rotate=\i]
\node [vb, label=below:+]	(\i)	at (0,1.6)	{};
\end{scope}
}
\path
\foreach \i in {72,144,216,288,360}{
    \foreach \j in {72,144,216,288,360}{
   	    (\i) edge [ev]  (\j)
	    }
    }
\foreach \i in {72,144,216,288,360}{
  	(\i) edge [e] (0)
	};
}
\end{tikzpicture}
}

\usepackage{tikzit}

\tikzstyle{Node}=[fill=black, draw=black, shape=circle, scale=0.3px]
\tikzstyle{coh}=[fill=white, draw=black, shape=circle, scale=0.6px, line width=1px]
\tikzstyle{coh_big}=[fill=white, draw=black, shape=circle, scale=0.6px]
\tikzstyle{coh_black}=[fill=black, draw=black, shape=circle, scale=0.6px, line width=1px]
\tikzstyle{coh_blue}=[fill=blue, draw=blue, shape=circle, tikzit fill=blue, scale=0.5px]
\tikzstyle{coh_wb}=[fill=white, draw=blue, shape=circle, scale=0.6px, line width=1px]
\tikzstyle{square  coh}=[fill=black, draw=black, shape=regular polygon, scale=0.4px, line width=1px, regular polygon sides=3]
\tikzstyle{square coh white}=[fill=white, draw=black, shape=regular polygon, scale=0.4px, line width=1px, regular polygon sides=3]
\tikzstyle{orange}=[fill={rgb,255: red,255; green,128; blue,0}, draw=none, shape=circle, scale=0.4px]
\tikzstyle{blue}=[fill=blue, draw=none, shape=circle, scale=0.4px]

\tikzstyle{line}=[-, fill=none, line width=1px]
\tikzstyle{blockline}=[-, fill=black, line width=3px]
\tikzstyle{boost}=[-, fill={rgb,255: red,128; green,128; blue,128}, draw={rgb,255: red,128; green,128; blue,128}, tikzit fill={rgb,255: red,128; green,128; blue,128}, tikzit draw={rgb,255: red,128; green,128; blue,128}, line width=5px]
\tikzstyle{specialsu2}=[-, line width=5px, fill=black, draw={rgb,255: red,0; green,0; blue,189}, tikzit fill=white, tikzit draw=black]
\tikzstyle{boxdash}=[-, line width=5px, fill=black, dash pattern=on 2pt off 2pt]
\tikzstyle{arrow}=[line width=1.1px, ->]
\tikzstyle{arrowdotted}=[line width=1.1px, ->, dash pattern=on 2pt off 1.5pt]
\tikzstyle{dashed}=[-, line width=1px, dash pattern=on 2pt off 1pt, fill=none]
\tikzstyle{thindash}=[-, line width=0.5px, dash pattern=on 1pt off 3pt, draw={rgb,255: red,158; green,158; blue,158}]
\tikzstyle{grayfill}=[-, fill={rgb,255: red,234; green,234; blue,234}, draw=black, line width=1px]
\tikzstyle{thingray}=[-, line width=0.8px, draw={rgb,255: red,158; green,158; blue,158}]
\tikzstyle{arrowgray}=[draw={rgb,255: red,158; green,158; blue,158}, ->, line width=1px]
\tikzstyle{line_blue}=[-, line width=1.5px, draw=blue, tikzit draw=blue]
\tikzstyle{blue_dashed}=[-, draw=blue, tikzit draw=blue, line width=1.5px, dash pattern=on 2pt off 1.5pt]
\tikzstyle{orange_dashed}=[-, draw={rgb,255: red,255; green,128; blue,0}, tikzit draw={rgb,255: red,255; green,128; blue,0}, line width=1.5px, dash pattern=on 2pt off 1.5pt]
\tikzstyle{new edge style 0}=[-]
\tikzstyle{line_orange}=[-, draw={rgb,255: red,255; green,128; blue,0}, tikzit draw={rgb,255: red,255; green,128; blue,0}, line width=1.5px]
\tikzstyle{black fill}=[-, fill=black, draw=none]

\usetikzlibrary{shapes.geometric}
\usepackage{manfnt}
\newcommand*\cube{\mbox{\mancube}}
\renewcommand\bra[1]{{\langle{#1}|}}
\makeatletter
\renewcommand\ket[1]{%
  \@ifnextchar\bra{\k@t{#1}\!}{\k@t{#1}}%
}
\newcommand\k@t[1]{{|{#1}\rangle}}
\makeatother
\usetikzlibrary{patterns}
\usetikzlibrary{matrix,decorations.pathreplacing}
\pgfkeys{tikz/mymatrixenv/.style={decoration={brace},every left delimiter/.style={xshift=8pt},every right delimiter/.style={xshift=-8pt}}}
\pgfkeys{tikz/mymatrix/.style={matrix of math nodes,nodes in empty cells,left delimiter={(},right delimiter={)},inner sep=1pt,outer sep=4pt,column sep=8pt,row sep=8pt,nodes={minimum width=20pt,minimum height=10pt,anchor=center,inner sep=0,outer sep=0}}}
\pgfkeys{tikz/mymatrixbrace/.style={decorate, thick, decoration={raise=3pt}}}

\author{Alexander Florian Jercher}
\title{Effective Lorentzian Geometries from Spin-Foams and Group Field Theories}

\begin{document}
\pagenumbering{roman}
\clearpage
\begin{titlepage}
\areaset[0mm]{210mm}{297mm}

\thispagestyle{empty}
\begin{center}
\begin{minipage}[c]{0.8\textwidth}
    \centering
\vspace*{2cm}
    {\fontsize{30}{36}\scshape Emergent Lorentzian Geometries\\
     from Spin-Foams and\\
      Group Field Theories\\}
    \vspace{150pt}
    {\fontsize{30}{36}\scshape Dissertation\\}
    \vspace{150pt}
    {\Large\scshape zur Erlangung des akademischen Grades Doctor Rerum Naturalium.\\
    Vorgelegt dem Rat der Physikalisch-Astronomischen Fakultät der Friedrich-Schiller-Universität Jena.\\}
    \vspace{20pt}
    {\Large\scshape von\\}
    \vspace{20pt}
    {\LARGE\scshape Alexander Florian Jercher\\}
    {\Large\scshape geboren am 22.06.1998 in München}
\end{minipage}
\end{center}
\end{titlepage}

\clearpage  
{

\thispagestyle{empty}

\begin{center}
\hspace{-50pt}
\begin{minipage}[l]{1.0\textwidth}
    \vspace*{100mm}
    \hspace{40mm}
    \includegraphics[width=0.5\textwidth]{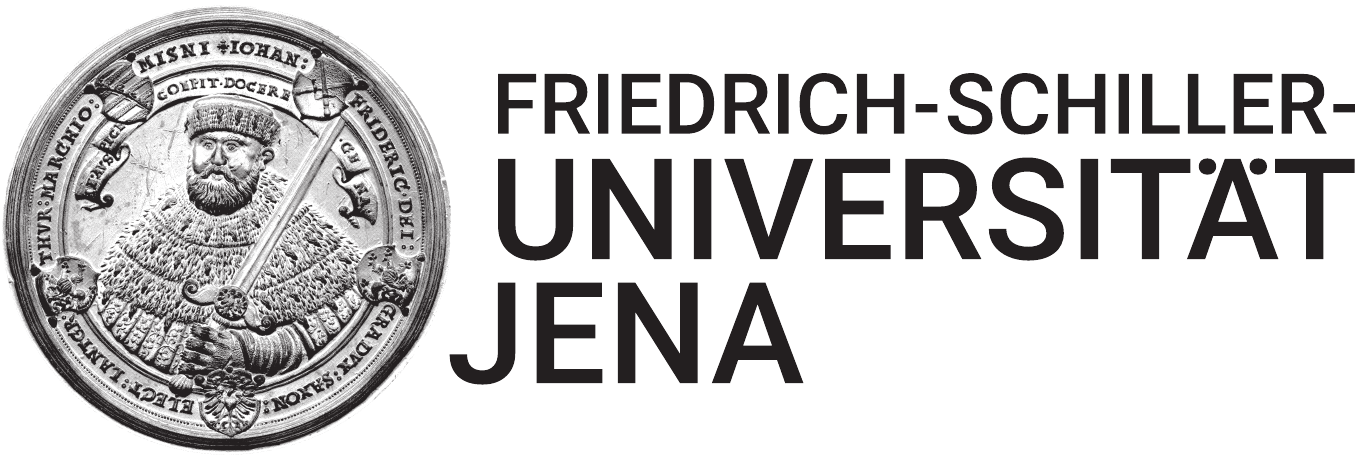}\\[70mm]
    \textsc{Gutachter:}
    \vspace{-10pt}
    \begin{enumerate}
    \itemsep-2mm
    \item Dr.~Sebastian Steinhaus, \textit{Friedrich-Schiller-Universität Jena}
    \item Prof.~Dr.~Edward Wilson-Ewing, \textit{University of New Brunswick} 
    \item Prof.~Dr.~Steffen Gielen, \textit{University of Sheffield}
    \end{enumerate}   
    \textsc{Datum der Disputation: 29.04.2025}
\end{minipage}
\hfill
\end{center}
}

\clearpage
\thispagestyle{empty}
\vspace*{\fill}
\dictum[\normalfont{\textsc{Bertrand Russell~--~ History of western philosophy (1945)}}]{\textit{The changes in the meanings of words are often very instructive. \dots; now I want to speak about the word \enquote{theory.} This was originally an Orphic word, which Cornford interprets as \enquote{passionate sympathetic contemplation.} In this state, he says, \enquote{The spectator is identified with the suffering God, dies in his death, and rises again in his new birth.} For Pythagoras, the \enquote{passionate sympathetic contemplation} was intellectual, and issued in mathematical knowledge. In this way, through Pythagoreanism, \enquote{theory} gradually acquired its modern meaning; but for all who were inspired by Pythagoras it retained an element of ecstatic revelation. To those who have reluctantly learnt a little mathematics in school this may seem strange; but to those who have experienced the intoxicating delight of sudden understanding that mathematics gives, from time to time, to those who love it, the Pythagorean view will seem completely natural even if untrue. It might seem that the empirical philosopher is the slave of his material, but that the pure mathematician, like the musician, is a free creator of his world of ordered beauty.}}

\clearpage
\thispagestyle{empty}
\vspace*{40mm}
\begin{center}
\Large{\textit{To my family.}}
\end{center}

\chapter*{Preface}\label{sec:Preface}
\pagestyle{plain}
\addcontentsline{toc}{chapter}{\textsc{Preface}}

\section*{Abstract}
The emergence of Lorentzian geometries is investigated within the spin-foam, Part~\hyperref[Part I]{I}, and group field theory (GFT), Part~\hyperref[Part II]{II}, approach to quantum gravity.

The spectral dimension of periodic Euclidean spin-foam frusta is studied. At large scales, the spectral dimension is generically four. At lower scales, a non-trivial flow of the spectral dimension is observed, sensitive to quantum effects, curvature induced oscillations and the parameters of the theory. The removal of numerical cutoffs and a thermodynamic limit is discussed, suggesting a phase transition from zero to four large-scale dimensions. 

Lorentzian Regge calculus for (3+1) cosmology, modelled with Lorentzian 4-frusta, coupled to a massless free scalar field is studied. It is shown that causal regularity, solutions to the Regge equations and a continuum limit only exist if the cells connecting neighboring slices are timelike. The dynamics can be expressed relationally only in the small deficit angle limit.

Effective (2+1) spin-foam cosmology with a minimally coupled massive scalar field is investigated. The scalar field mass is shown to ensure convergence of the path integral. The classicality of expectation values is shown to be intimately connected to causal regularity and the path integral measure. 

A causal completion of the Barrett-Crane group field theory model is developed. Its amplitudes are explicitly computed in spin representation using methods from integral geometry.

A Landau-Ginzburg analysis is applied to the complete Barrett-Crane (BC) group field theory model. It is shown that mean-field theory is generically self-consistent, and that timelike faces do not contribute to the critical behavior.

Employing the complete BC model, a physical Lorentzian reference frame is coupled, and scalar cosmological perturbations are extracted from entangled GFT coherent states. The dynamics of these perturbations are shown to agree with classical results up to quantum corrections.

\cleardoublepage
\section*{Kurzfassung}
Es wird die Emergenz Lorentz'scher Geometrien innerhalb der Quantengravitationsansätze von Spinschäumen, Teil~\hyperref[Part I]{I}, und Gruppenfeldtheorien (GFT), Teil~\hyperref[Part II]{II}, untersucht. 

Die spektrale Dimension periodischer Euklidischer Spinschaum-Frusta wird analysiert. Auf großen Skalen ist die spektrale Dimension typischerweise vier. Auf kleineren Skalen wird ein nicht-trivialer Fluss beobachtet, welcher von Quanteneffekten, krümmungsinduzierten Oszillationen und den Parametern der Theorie beeinflusst wird. Die Beseitigung numerischer Schranken und ein thermodynamischer Limes werden besprochen, welcher den Phasenübergang von Dimension null zu vier auf großen Skalen suggeriert.

Lorentz'sches Reggekalkül für (3+1) Kosmologie, modelliert durch Lorentzsche 4-Frusta, gekoppelt an ein freies, masseloses Skalarfeld, wird untersucht. Es wird gezeigt, dass kausale Regularität, Lösungen der Reggegleichungen und ein Kontinuumslimes nur dann bestehen, wenn die Zellen, welche benachbarte Flächen verbinden, zeitartig sind. Die Dynamik kann nur im Limes kleiner Defizitwinkel relational ausgedrückt werden. 

Effektive (2+1) Spinschaumkosmologie mit einem minimal gekoppelten, freien, massiven Skalarfeld wird analysiert. Es wird gezeigt, dass die Skalarfeldmasse die Konvergenz des Pfadintegrals garantiert. Des Weiteren wird aufgezeigt, dass die Klassikalität von Erwartungswerten eng mit kausaler Regularität und dem Pfadintegralmaß zusammenhängt. 

Eine kausale Vervollständigung des Barrett-Crane Gruppenfeldtheoriemodells wird entwickelt. Die definierenden Amplituden werden explizit in der Spindarstellung mithilfe von Integralgeometriemethoden berechnet.

Eine Landau-Ginzburg-Analyse wird auf das vollständige Barrett-Crane (BC) Gruppenfeldtheoriemodell angewandt. Es wird gezeigt, dass Molekularfeldtheorie typischerweise selbstkonsistent ist und dass zeitartige Flächen nicht zum kritischen Verhalten beitragen.

Unter Verwendung des vollständigen BC-Modells wird ein physikalisches, Lorentzsches Bezugssystem gekoppelt und skalare kosmologische Perturbationen von verschränkten kohärenten GFT-Zuständen extrahiert. Es wird gezeigt, dass die Dynamik dieser Perturbationen mit klassischen Resultaten, bis auf Quantenkorrekturen, übereinstimmt.

\clearpage
\section*{Acknowledgements}
First, I would like to thank Sebastian Steinhaus for his close supervision and the continuous support he offered beyond academic matters. I am grateful for our scientific (and often also political) discussions, that he created the incentive to get me into numerics and the freedom he granted me to continue collaborations outside spin-foams. I would like to express my gratitude to Andreas Pithis for introducing me so patiently to the field of quantum gravity, for our fruitful collaborations and for the friendship that arose out of it. I must thank  Daniele Oriti for his intellectual guidance provided in numerous stimulating discussions, and David Symhoven for inspiring me in the very first place to study physics in Munich. I am indebted to my collaborators for what they taught me and for establishing striving yet friendly environments: Roukaya Dekhil, Luca Marchetti, Daniele Oriti, Andreas Pithis, Jos\'{e} Diogo Sim\~{a}o, Sebastian Steinhaus and Johannes Th\"{u}rigen. Stefan Flörchinger, Martin Ammon and especially Holger Gies were always approachable and ready to help, thereby creating this homely atmosphere which characterizes daily life at the TPI Jena. I would like to thank Katrin Kanter and Lisann Schmidt for their attentive and committed help in bureaucratic matters. 

\enquote{People make the place!} is what I have been taught here in Jena. I must thank Jos\'{e} Diogo Sim\~{a}o for easing the beginning of my time in Jena, for his warm friendship and our inspiring conversations on life, philosophy and sometimes spin-foams. I thank Richard Schmieden for his kindness and for making times at the institute so much more fun. I thank Ann-Sophie Schneck for the intellectual connection we share, from physics to religion and everything in between. I am grateful for randomly having met Maike Buck (Team Turbo!) who introduced me to so many beautiful people and places I would have otherwise missed. Office 103 HHW4 offered so much more than just a place to work. Here, we celebrated Friday nights with fancy drinks, the infamous \href{https://chesslemania.com}{Chesslemania} tournament, conversations on existentialism, One Piece, physics, Orgonomy, gym exercises, \dots To the members and honorary members of O103 HHW4, and in particular to Tom Angrick, Leopold Peter and Johannes Schmechel: Thank you! I am grateful to Athanasios Kogios, David Rumler, Dimitrios Gkiatas, Ivan Soler, Jakob Hollweck, Julian Schirrmeister, Kemal D\"{o}ner, Kira Klebesz, Lars Maiwald, Markus Schr\"{o}fl, Marta Picciau, Michael Mandl, Seth Asante and Tobias Bommer for  forest excursions, movie nights, lunch breaks, Quergasse nights and for their overall comradery. I thank the swole patrol, consisting of Christian Schmidt, Markus Schr\"{o}fl and Tim St\"{o}tzel, for gym sessions, ongoing encouragement and their friendship.

I would like to express my deepest gratitude to my family; to my parents, to whom I owe a profound trust in life and whose appreciation of education ultimately led to this work, to my sister for teaching me so many things, and to my little nieces for showing me that the feeling of family overcomes geographical distance. I thank Aurora Capobianco for her honest support and so many intense moments in such a short period of time; we have an exciting future ahead of us! Tobias Dreher and Philipp L\"{a}ngle know the invaluable role they play in my life. In them, I have found kindred spirits.

\vspace*{\fill}
\noindent I would like to thank Andreas Pithis and Sebastian Steinhaus for attentive and insightful remarks on the manuscript. This work was partially funded by the Munich Center of Quantum Science and Technology via the seed fundings Aost 862983-4 and 862933-9, granted to Andreas Pithis, and Aost 862981-8, granted to Jibril Ben Achour, by the Deutsche Forschungsgemeinschaft (DFG, German Research Foundation) under Germany's Excellence Strategy - EXC-2111 - 390814868. This work was partially funded by the DFG under Grant No 422809950 and 406116891 within the Research Training Group RTG 2522/1.

\clearpage
\vspace*{\fill}
\renewcommand*{\raggeddictum}{\raggedright}
\renewcommand*{\raggeddictumauthor}{\raggedright}
\renewcommand*{\raggeddictumtext}{\raggedright}
\dictum[\normalfont{\textsc{Lew Tolstoi -- War and peace (1869)}}]{Lew Tolstoi on the matter of emergence and causality\dots\\
\textit{Many historians say that the French did not win the battle of Borodino because Napoleon had a cold, and that if he had not had a cold the orders he gave before and during the battle would have been still more full of genius and Russia would have been lost and the face of the world have been changed. To historians who believe that Russia was shaped by the will of one man, \dots, 
to say that Russia remained a power because Napoleon had a bad cold on the twenty-fourth of August may seem logical and convincing. If it had depended on Napoleon's will to fight or not to fight the battle of Borodino, and if this or that other arrangement depended on his will, then evidently a cold affecting the manifestation of his will might have saved Russia, and consequently the valet who omitted to bring Napoleon his waterproof boots on the twenty-fourth would have been the savior of Russia. Along that line of thought such a deduction is indubitable, \dots
But to men who do not admit that Russia was formed by the will of one man, \dots
that argument seems not merely untrue and irrational, but contrary to all human reality. To the question of what causes historic events another answer presents itself, namely, that the course of human events is predetermined from on high, depends on the coincidence of the wills of all who take part in the events, and that a Napoleon's influence on the course of these events is purely external and fictitious.}}

\clearpage
\vspace*{\fill}
\renewcommand*{\raggeddictum}{\raggedleft}
\renewcommand*{\raggeddictumauthor}{\raggedleft}
\dictum[\normalfont{\textsc{Fjodor M. Dostojewski -- Aufzeichnungen aus dem Untergrund (1864)}}]{\textit{Sie werden sagen, dass der Mensch sich auch jetzt noch, wenn er auch schon gelernt habe, in manchen Dingen klarer zu sehen als in barbarischen Zeiten, doch noch lange nicht gewöhnt habe, so zu handeln, wie es ihm die Vernunft und die Wissenschaften vorschreiben. Immerhin sind Sie, meine Herrschaften, vollkommen überzeugt, dass er sich bestimmt daran gewöhnen werde, \dots\\
Dann wird die Wissenschaft selbst den Menschen belehren (\dots) und ihm sagen, \dots, dass er selbst nichts anders sei als eine Art Klaviertaste oder Drehorgelstiftchen, und dass auf der Welt außerdem noch Naturgesetze vorhanden wären: so dass alles, was er auch tun mag, nicht durch seinen Wunsch und Willen getan werde, sondern ganz von selbst geschehe, einfach nach den Gesetzen der Natur. \dots Selbstverständlich werden dann alle menschlichen Handlungen nach diesen Gesetzen mathematisch in der Art der Logarithmentafeln bis 10000 berechnet \dots Natürlich kann man nicht garantieren (\dots), dass es dann zum Beispiel nicht furchtbar langweilig sein werde (\dots), dafür wird es aber ungemein Vernünftig zugehen. Aber was denkt man sich schließlich nicht aus Langeweile aus! Es würde mich zum Beispiel nicht im geringsten wundern, wenn sich dann mir nichts dir nichts inmitten der allgemeinen zukünftigen Vernünftigkeit plötzlich ein Gentleman \dots vor uns aufstellte, die Hände in die Seiten stemmte und zu uns allen sagte: \enquote{Nun wie, meine Herrschaften, sollten wir nicht diese ganze Vernünftigkeit mit einem einzigen Fußtritt zertrümmern, damit alle diese verfluchten Logarithmen zum Teufel gehen und wir wieder nach unserem törichten Willen leben können!?}}}

\clearpage
\vspace*{\fill}
\noindent\textit{The compilation of this thesis is solely due to the author. However, Sebastian Steinhaus supported the preparation this thesis in an advisory capacity as a supervisor. Furthermore, the chapters of the main body are based on the following collaborations:
\begin{itemize}
    \item Chapter~\ref{chapter:specdim} is based on~\cite{Jercher:2023rno}, published in \href{https://journals.aps.org/prd/abstract/10.1103/PhysRevD.108.066011}{Phys.~Rev.~D., vol.~108, no.~06, p.~066011, 2023}, \href{https://arxiv.org/abs/2304.13058}{arXiv: 2304.13058 [gr-qc]}, with Sebastian Steinhaus and Johannes Thürigen;
    \item Chapter~\ref{chapter:LRC} is based on~\cite{Jercher:2023csk}, published in \href{https://iopscience.iop.org/article/10.1088/1361-6382/ad37e9}{Class.~Quant.~Grav., vol.~41, no.~10, p.~105008, 2024}, \href{https://arxiv.org/abs/2312.11639}{arXiv: 2312.11639 [gr-qc]}, with Sebastian Steinhaus;
    \item Chapter~\ref{chapter:3d cosmology} is based on~\cite{Jercher:2024kig}, published in \href{https://iopscience.iop.org/article/10.1088/1361-6382/ad9700}{Class.~Quant.~Grav., vol.~42, no.~1, p.~017001, 2025}, \href{https://arxiv.org/abs/2404.16943}{arXiv: 2404.16943 [gr-qc]}, and~\cite{Jercher:2024hlr}, available as a pre-print at \href{https://arxiv.org/abs/2411.08109}{arXiv: 2411.08109 [gr-qc]}, with Jos\'{e} Diogo Sim\~{a}o and Sebastian Steinhaus;
    \item Chapter~\ref{chapter:cBC} is based on~\cite{Jercher:2022mky}, published in \href{https://journals.aps.org/prd/abstract/10.1103/PhysRevD.106.066019}{Phys.~Rev.~D., vol.~106, no.~06, p.~066019, 2022}, \href{https://arxiv.org/abs/2206.15442}{arXiv: 2206.15442 [gr-qc]}, with Daniele Oriti and Andreas Pithis;
    \item Chapter~\ref{chapter:LG} is based on~\cite{Dekhil:2024ssa}, published in \href{https://link.springer.com/article/10.1007/JHEP08(2024)050}{JHEP, vol.~08, p.~050, 2024}, \href{https://arxiv.org/abs/2404.04524}{arXiv: 2404.04524 [hep-th]}, with Roukaya Dekhil, Daniele Oriti and Andreas Pithis, and \cite{Dekhil:2024djp}, published in \href{https://journals.aps.org/prd/abstract/10.1103/PhysRevD.111.026014}{Phys.~Rev.~D., vol.~111, no.~02, p.~026014, 2025}, \href{https://arxiv.org/abs/2404.04524}{arXiv: 2404.04524 [hep-th]}, with Roukaya Dekhil and Andreas Pithis;
    \item Chapter~\ref{chapter:perturbations} is based on~\cite{Jercher:2021bie}, published in \href{https://iopscience.iop.org/article/10.1088/1475-7516/2022/01/050}{JCAP, vol.~01, no.~01, p.~050, 2022}, \href{https://arxiv.org/abs/2112.00091}{arXiv: 2112.00091 [gr-qc]}, with D.~Oriti and A.~G.~A.~Pithis, and~\cite{Jercher:2023nxa}, published in \href{https://journals.aps.org/prd/abstract/10.1103/PhysRevD.109.066021}{Phys.~Rev.~D., vol.~109, no.~06, p.~066021, 2024}, \href{https://arxiv.org/abs/2308.13261}{arXiv: 2308.13261 [gr-qc]}, and~\cite{Jercher:2023kfr}, published in \href{https://iopscience.iop.org/article/10.1088/1361-6382/ad6f67}{Class.~Quant.~Grav., vol.~41, no.~18, 18LT01, 2024} | \href{https://arxiv.org/abs/2310.17549}{arXiv: 2310.17549 [gr-qc]}, with Luca Marchetti and Andreas Pithis.
\end{itemize} 
The corresponding appendices are likewise based on these collaborations. See also the \hyperref[Ehrenwörtliche Erklärung]{Ehrenwörtliche Erklärung}.}

\clearpage
\pagenumbering{arabic}
\setcounter{page}{1}
\tableofcontents
\clearpage
\fancyhead[CE]{\leftmark}
\fancyhead[LE]{\scshape\MakeLowercase\chaptertitlename\hspace{2pt}\large\thechapter}
\fancyhead[RO]{\scshape\MakeLowercase\chaptertitlename\hspace{2pt}\large\thechapter}
\fancyhead[CO]{\leftmark}
\fancyfoot[C]{\thepage}
\renewcommand{\headrulewidth}{0.5pt}
\renewcommand{\footrulewidth}{0pt}
\chapter[\textsc{Introduction}]{Introduction}
\renewcommand*{\dictumwidth}{.5\textwidth}
\dictum[\normalfont{\textsc{Rudolf May}}]{\textit{Das mit dem Gravitationsfeld ist sehr sonderbar. Es ist wie mit dem Heiligenschein der Maria; man sieht ihn nicht, aber er ist vorhanden.}}
\bigskip

\noindent Among the fundamental forces of nature, gravity is the most immediate to our perception. From the very first moment of our lives, we feel its pull - a perpetual force grounding us. It takes us about a year of determined effort to stand up straight, briefly overcoming its grip. This struggle of defying gravity has accompanied humans throughout history, challenging our ingenuity and culminating in feats such as aviation and space travel.

Einstein's theory of general relativity (GR)~\cite{Wald1984} is up to now the most accurate description of gravitational phenomena, captured by Einstein's field equations,
\begin{equation*}
    G_{\mu\nu} = 8\pi \GN T_{\mu\nu}\,,
\end{equation*}
with $\GN$ Newton's constant. It relates spacetime curvature, described by the Einstein tensor $G_{\mu\nu}$, with the matter distribution, captured by the stress-energy tensor $T_{\mu\nu}$. The agreement of predictions from GR with experiments is astonishing. Examples of measurements that demonstrate the accuracy of GR are the direct detection of gravitational waves~\cite{LIGOScientific:2016aoc}, the direct detection of black holes~\cite{EventHorizonTelescope:2019dse} and the timing of double pulsars~\cite{Will:2001mx}.\footnote{For the non-physicist readers inclined towards pragmatism asking what GR is actually useful for: global positioning systems (GPS) only work correctly if effects of special and general relativity are taken into account~\cite{Ashby2003}.}

GR reforms our understanding of space and time. The theory is based on two key principles: the \emph{equivalence principle} and \emph{general covariance}. The equivalence principle states that the laws of physics in an inertial frame subject to gravitational force are the same as in an accelerated frame absent of gravity. The global inertial frames of special relativity are replaced by the local frames of freely falling observes that move along geodesics of a curved spacetime metric. Consequently, gravity is identified with the spacetime metric itself. General covariance, the second principle, implies that the laws of physics are independent of the reference frame we use to describe them. Thereby, the notion of absolute space and time are ultimately deemed unphysical.
General covariance is mathematically realized as the invariance under active diffeomorphisms. Consequently, physical observables are required to be diffeomorphism-invariant so that spatial localization and temporal evolution can only be understood \emph{relationally}~\cite{Rovelli:1990ph,Rovelli:2001bz}, i.e. with respect to other physical degrees of freedom.

\section{Gravity and the quantum}

Despite the remarkable success of GR in describing gravitational phenomena from 10$\upmu$m scales~\cite{Tan2020,Lee:2020zjt} to the size of the Universe~\cite{HawkingBook}, its deterministic classical nature contrasts the characteristics of quantum field theory (QFT)~\cite{Zee} which is the framework successfully incorporating the remaining fundamental forces. Beyond such conceptual incompatibilities, there exist physical phenomena where classical GR fails to provide a valid description. These include black holes~\cite{Hawking1976,Wald:1999vt}, early cosmology~\cite{Ellis2012} and spacetime singularities in general~\cite{HawkingBook}. A heuristic argument on the breakdown of GR near the Planck scale, $l_{\Pl}\sim 10^{-35}\mathrm{m}$, has been put forward by Bronstein~\cite{Bronstein1936a,Bronstein1936b}. He argues that resolving length scales below the Planck length would require energies so high that a black hole forms, thus posing a limit on the smooth spacetime description.

Taken together, there exists theoretical evidence for the need of a theory of \emph{quantum gravity} (QG) which incorporates the principles of quantum mechanics and GR into a unified consistent framework. Before dedicating ourselves to the quest for QG, it must be highlighted that so far, there exists no definite proof for the necessity of such a theory~\cite{Carlip:2008zf,Rovelli:1999hz}. In particular, there is up to now no experiment that rules out the coexistence of classical GR and quantum theory. However, table-top experiments in the near future could unveil new insights on this matter~\cite{Pikovski2012,Bose:2017nin,Marletto:2017kzi,Mehdi:2022oct,Christodoulou:2022mkf}. 

Incorporating gravity into the framework of QFT by splitting the metric into background and perturbation is doomed to fail as a fundamental theory (although it serves as an effective field theory~\cite{Donoghue:2012zc}) as it is perturbatively non-renormalizable~\cite{Goroff1986,vandeVen1992}. The asymptotic safety program~\cite{Weinberg1978,Eichhorn:2020mte} aims at resurrecting this approach by studying the non-perturbative renormalizability of gravity as a QFT. Also, a canonical quantization based on the Arnowitt-Deser-Misner formulation~\cite{Arnowitt:1962hi}, known as Wheeler-de Witt gravity~\cite{DeWitt1967,Wheeler1957}, fails due to obstacles in defining a kinematical Hilbert space and due to a mathematically ill-defined Hamiltonian constraint~\cite{Kiefer2012}.

Driven by the failure of these typical quantization strategies, a plethora of non-perturbative QG approaches emerged, such as string theory~\cite{Blumenhagen:2013fgp}
or the anti-de Sitter/conformal field theory (AdS/CFT) correspondence~\cite{Maldacena:1997re}. Canonical loop quantum gravity (LQG)~\cite{Rovelli:2004tv,Thiemann2007a} is the most prominent approach emphasizing background independence as fundamental principle.
Non-perturbative gravity path integral~\cite{Feynman1942} approaches include quantum Regge calculus~\cite{Williams2009,Hamber2009}, matrix~\cite{DiFrancesco:1993cyw} and tensor models~\cite{Gurau:2016cjo,Gurau:2011xp}, causal sets~\cite{Surya:2019ndm} and causal dynamical triangulations (CDT)~\cite{Loll:2019rdj,Ambjorn:2012jv}. The latter two are particularly noteworthy due to their emphasis on the importance of the causal structure of spacetime which plays a crucial role throughout this thesis. The QG approaches we will focus on for the rest of this work are spin-foam models~\cite{Perez:2012wv} and group field theories (GFTs)~\cite{Freidel:2005jy,Oriti:2012wt} which we introduce briefly in the following.

\section{Spin-foam models}\label{sec:intro to spin-foams}

Spin-foam models aim at turning the formal continuum gravity path integral~\cite{Hawking1978,Misner1957},
\begin{equation}
Z_{\mathrm{QG}} = \int\mathcal{D}\mathrm{g}\,\e^{iS_{\mathrm{EH}}[\mathrm{g}]}\,,
\end{equation}
into a rigorously defined and computable expression. Here, $\mathrm{g}$ is the spacetime metric on a continuum manifold $\mathcal{M}$ which we assume here for simplicity to have no boundary, $\partial\mathcal{M} = \emptyset$ (spin-foam boundary states are discussed at the end of this section).  A huge obstacle to actually define and compute this object is to characterize the integration over geometries, i.e.~over diffeomorphism-invariant metrics~\cite{Perez:2012wv}.

Following~\cite{Baez:1999sr,Perez:2012wv}, the derivation of the spin-foam partition function in (3+1) dimensions is commenced by a series of classically equivalent reformulations of vacuum Einstein-Hilbert gravity. First, one changes to a first-order Palatini-Holst formulation~\cite{Holst:1995pc},
\begin{equation}\label{eq:PH action}
S_{\mathrm{PH}}[e,\omega] = \int_\mathcal{M}\left(*e_I\wedge e_J+\frac{1}{\bi}e_I\wedge e_J\right)\wedge F^{IJ}(\omega)\,,
\end{equation}
which recasts gravity into a local $\SL$ (double cover of $\mathrm{SO}(1,3)$) gauge theory with connection 1-form $\omega$ and curvature 2-form $F$. The tetrad vector fields $\{e_I\}$ form a local Lorentz frame and thus directly realize the equivalence principle. The additional Holst-term (comparable to the $\theta$-term in QCD~\cite{Ashtekar1991}) leaves the vacuum Einstein equations unaltered but induces torsion if fermions are coupled~\cite{Perez:2005pm}. The Barbero-Immirzi parameter $\bi$ is a priori a free parameter and plays an important role in the semi-classical and continuum limit of spin-foams (see e.g.~\cite{Asante:2020qpa}). Suggestions for its value come from black hole entropy calculations~\cite{Perez:2017cmj,Pigozzo:2020zft}. Whether $\bi$ runs under renormalization is an important open question~\cite{Benedetti:2011yb,Charles:2016mjn}.

Spin-foams propose a reformulation of Palatini-Holst gravity as constrained $BF$-theory based on the work of Plebanski~\cite{Peldan:1993hi,Plebanski:1977zz}. Quantization is first performed for unconstrained $BF$-theory which is well understood (however, see~\cite{Bonzom:2012mb} for a discussion on subtleties). Only thereafter, the constraints are imposed on the quantum level. Explicitly, the $BF$-action is given by
\begin{equation}
S_{BF}[B,\omega] = \int_\mathcal{M}B_{IJ}\wedge F^{IJ}(\omega)\,,
\end{equation} 
where $B$ is an $\spl$-valued 2-form. $S_{\mathrm{PH}}$ is obtained upon the imposition of the \emph{simplicity} constraint which enforces $B_{IJ} = *e_I\wedge e_J+\frac{1}{\bi}e_I\wedge e_J$. Formally, the $B$-integration of the partition function of $BF$-theory can be executed, yielding
\begin{equation}\label{eq:Z BF pre}
Z_{BF} = \int\mathcal{D}\omega\mathcal{D}B\,\e^{iS_{BF}} = \int\mathcal{D}\omega\,\delta(F(\omega))\,.
\end{equation}
To regularize the path integral measure $\mathcal{D}\omega$, we introduce a triangulation $\Delta$ (and its dual 2-complex $\Gamma$), consisting of 4-simplices each of which is bounded by five tetrahedra. The connection 1-form $\omega$ is discretized on dual edges $e\in\Gamma$ by $\omega\mapsto g_e = \mathcal{P}\exp \int_e\omega\in\SL$, with $\mathcal{P}$ indicating path-ordering, such that $\mathcal{D}\omega\mapsto \prod_e\dd{g_e}$, where $\dd{g}$ is the Haar measure~\cite{Haar1933}. Curvature is discretized as $F\mapsto g_f = \prod_{e\subset f}g_e$ on dual faces $f\in\Gamma$. The functions $\delta(g_f)$ entering $Z_{BF}$ are decomposed as in Eq.~\eqref{eq:delta on SL2C}, using the Plancherel decomposition of $L^2(\SL)$ into unitary irreducible representations of the principal series $(\rho,\nu)\in\R\times\mathbb{Z}/2$ with $\SL$ Wigner matrices $\vb*{D}^{(\rho,\nu)}$~\cite{Ruehl1970}, details of which are presented in Appendix~\ref{app:Representation Theory of SL2C}. Inserting this decomposition into Eq.~\eqref{eq:Z BF pre}, re-ordering terms and performing the group integrations, one obtains the spin-foam partition function of $BF$-theory,
\begin{equation}\label{eq:Z BF}
Z_{BF}(\Gamma) = \int\dd{\{\rho_f\}}\sum_{\{\nu_f\},\{\iota_e\}}\prod_{f\in\Gamma}\mathcal{A}_f\prod_{e\in\Gamma}\mathcal{A}_{e}\prod_{v\in\Gamma}\mathcal{A}_v\,.
\end{equation}
The $\{\iota_e\}$, referred to as intertwiners, arise from a projection onto the $\SL$-invariant subspace,
\begin{equation}\label{eq:SL2C invariant projection}
    P_e(\{\rho_f\}_{f\supset e}) = \int_\SL \dd{g_e}\bigotimes_{f\supset e}\vb*{D}^{(\rho_f,\nu_f)}(g_e)\bigg{\vert}_{\mathrm{reg}} \eqdef \sum_{\iota_e} \frac{\ket{\iota_e}\bra{\iota_e}}{\vert\vert\iota_e\vert\vert^2}\,,
\end{equation}
where we have introduced by hand an inverse factor $\vert\vert\iota_e\vert\vert^2$ to regularize $P_e$. Note that otherwise, $P_e^2 = \mathrm{vol}(\SL)P_e$, which diverges as $\SL$ is non-compact. The face amplitude $\mathcal{A}_f = \rho_f^2+\nu_f^2$ entering $Z_{BF}$ is given by the Plancherel measure. The edge amplitude carries the normalization of the intertwiner, $\mathcal{A}_e^{-1} = \vert\vert\iota_e\vert\vert^2$. The vertex amplitude $\mathcal{A}_v$ associates a quantum amplitude to the 4-simplex dual to $v$ and forms an essential part of the spin-foam model. For $BF$-theory, it is given as $\mathcal{A}_v = \Tr\left(\vec{\bigotimes}_{e\supset v}\ket{\iota_e}\right)$, i.e.~a non-local contraction of intertwiners according to the combinatorics of a 4-simplex. Disregarding issues of convergence and computability for the moment, Eq.~\eqref{eq:Z BF} defines the partition function of $BF$-theory. Following the prescription above, we are left to discuss the simplicity constraints which are also important for the definition of boundary states when $\partial\Gamma\neq \emptyset$. 

The $B$ field is mapped as $B\mapsto b_t = \int_{t}B$ under discretization, where $t\in\Delta$ a triangle. Then, simplicity is satisfied if at every tetrahedron $\tau\in\Delta$ and every $t\subset\tau$, the equation $X_\tau\cdot (*b_t+\frac{1}{\bi}b_t) = 0$ holds~\cite{Gielen:2010cu}.\footnote{Here, we exclusively consider the \emph{linear} simplicity constraint. For a discussion on why this is advantageous over the earlier quadratic formulation, see~\cite{Gielen:2010cu}.} Simplicity thus requires choosing a vector $X_\tau\in\R^{1,3}$ normal to $\tau$, implying in particular a choice of causal character, i.e. whether  $\tau$ is spacelike, lightlike or timelike. Quantizing these constraints amounts to picking 1) the range of values $\bi$ takes,\, 2) the allowed causal characters of $X_\tau$, and 3) an isomorphism $\beta$ between the space of bivectors $\bigwedge^2\R^{1,3}$ and $\spl$~\cite{Baez:1999tk}. Given $\beta$, the simplicity constraints are translated to conditions on the generators of $\spl$ which in turn yields particular relations for the $\SL$ representations $(\rho,\nu)$, referred to as \emph{simple} representations. Thus, the spin-foam partition function is given by Eq.~\eqref{eq:Z BF} with the sum/integral over representation data restricted to simple representations. 
%
%
The existing spin-foam models in the literature differ by the choices 1) -- 3), the two most prominent of which are briefly listed now. We make the simplicity conditions on the labels $(\rho,\nu)$ explicit in Chapter~\ref{chapter:cBC}.

The Engle-Pereira-Rovelli-Livine (EPRL)~\cite{Engle:2007uq,Engle:2007wy} model is characterized by $0<\bi<\infty$ and the restriction to spacelike tetrahedra.  Due to its close connection to LQG, the EPRL model is currently favored among the various spin-foam models. Its boundary states are given by $\SUT$ spin networks~\cite{Rovelli:2004tv}. However, for actual computations such as an asymptotic analysis of the vertex amplitude~\cite{Barrett:2009gg,Barrett:2009mw}, coherent states~\cite{Livine:2007vk} are often utilized. The Conrady-Hnybida (CH) extension~\cite{Conrady:2010kc,Conrady:2010vx} of the EPRL model alleviates the restriction to spacelike tetrahedra and includes timelike (but not lightlike) tetrahedra and faces. 

Barrett and Crane (BC) formulated a spin-foam model~\cite{Barrett:1999qw,Barrett:1998fp} with $\bi\rightarrow\infty$ and the restriction to spacelike tetrahedra, although indicating already a possible extension to other causal characters. Its boundary states correspond to $\SL$ projected spin networks~\cite{Dupuis:2010jn}. A formulation where every tetrahedron is assumed to be timelike was put forward in~\cite{Perez:2000ep}. In Chapter~\ref{chapter:cBC}, an extension of the BC model to incorporate all types of causal building blocks will be developed.

The spin-foam partition function $Z_{\mathrm{SF}}$ and transition amplitudes will generically depend on the choice of $\Gamma$, corresponding to a truncation of the degrees of freedom. Therefore, defining a discretization independent partition function is indispensable. A conceivable strategy for obtaining discretization independence is to sum over all possible discretizations, which is realized by GFTs~\cite{Freidel:2005jy,Oriti:2012wt}, introduced in the following. 


\section{Group field theories}\label{sec:intro to GFT}

Group field theories are statistical and quantum field theories on $r$ copies of a Lie group $G$, characterized by combinatorially non-local interactions. They can be seen as field theories \emph{of spacetime} rather than standard QFTs on spacetime. One motivation for GFTs is that they generate spin-foam partition functions as the Feynman amplitudes. In this setting, the 2-complex $\Gamma$ is interpreted as a stranded Feynman diagram. The perturbative expansion of a GFT partition function can be written as
\begin{equation}
Z_{\mathrm{GFT}} = \int\mathcal{D}\varphi \,\e^{-S_{\mathrm{GFT}}[\varphi]} = \sum_\Gamma \frac{\lambda^{V_\Gamma}}{\mathrm{sym}(\Gamma)}Z_{\mathrm{SF}}(\Gamma)\,,
\end{equation}
with $Z_{\mathrm{SF}}(\Gamma)$ the corresponding spin-foam partition function. Here, $\varphi: G^r\rightarrow \R$ is the group field, $\lambda$ is the coupling entering the GFT action $S_{\mathrm{GFT}}$, $V_\Gamma$ is the number of vertices of $\Gamma$ and $\mathrm{sym}(\Gamma)$ is a normalization factor. This expansion is to be understood formally, as the sum over 2-complexes is difficult to control. Furthermore, it has been shown in~\cite{DePietri:2000ii,Gurau:2010nd,Gurau:2010mhz} that the generated $\Gamma$ are generically not dual to triangulations and exhibit topological singularities. 

These issues are remedied in \emph{colored tensor models}~\cite{Gurau:2011xp,Gurau:2016cjo} which are higher dimensional generalizations of matrix models~\cite{DiFrancesco:1993cyw}. GFTs can in fact be obtained from tensor models if the tensor indices are promoted to Lie group variables, which has proven to be a fruitful connection. It has been shown for colored tensor models that integrating out all but one color, one obtains so-called tensor-invariant interactions~\cite{Bonzom:2012hw,Gurau:2011tj}. These insights have been imported to GFTs by equipping the interaction term with tensor-invariant combinatorics. This class of theories, termed tensorial GFTs (TGFTs)\footnote{Throughout this work, we use the term GFT for the entire class of field theories defined on $r$ copies of a Lie group irrespective of the precise form of combinatorially non-local interactions.} in~\cite{Carrozza:2013oiy}, provide control over the theory space and enable rigorous renormalization studies~\cite{Carrozza:2024gnh,BenGeloun:2011rc,Carrozza:2012uv,Carrozza:2013wda,BenGeloun:2013vwi}.


As introduced so far, GFTs involve a large class of models characterized by choices of rank $r$, the group $G$, the action $S_{\mathrm{GFT}}$ and the domain and target space of the group fields $\varphi$. In this work, we shall focus on the subclass of GFT models which are viable candidates for (3+1)-dimensional Lorentzian QG. Such GFT models are closely related to the $BF$-theory spin-foam model introduced above, supplemented by simplicity constraints. In general, the 1-particle excitations of $\varphi$ correspond to $(r-1)$-dimensional simplicial building blocks, such that the generated complexes consist of $r$-dimensional cells. Thus, we fix $r=4$ for the remainder. Furthermore, $G=\SL$ is chosen as the double cover of the (3+1) Lorentz group. The Ooguri GFT model~\cite{Ooguri:1992eb} generates the spin-foam amplitudes $Z_{BF}(\Gamma)$ and is defined in terms of the action\footnote{For simplicity of the presentation, we assume here real-valued uncolored fields with a trivial kinetic term and simplicial interactions. Furthermore, we disregard potential divergences due to empty $\SL$ integrations. All these points will be clarified in detail in Chapter~\ref{chapter:cBC}.}
\begin{equation}
S_{\mathrm{GFT}}[\varphi] = \frac{1}{2}\int_{\SL^4}\left[\dd{g}\right]^4\varphi_{1234}\varphi_{1234} + \lambda\int_{\SL^{10}}\left[\dd{g}\right]^{10}\varphi_{1234}\varphi_{4567}\varphi_{7389}\varphi_{9620}\varphi_{0851}\,,
\end{equation}
where $\varphi(g_1,g_2,g_3,g_4) \equiv \varphi_{1234}$. The kinetic and interaction term respectively encode the identification of tetrahedra and the gluing of tetrahedra into 4-simplices. A crucial ingredient of the Ooguri model is the assumption of right invariance of $\varphi$, i.e. $\varphi(g_1,g_2,g_3,g_4) = \varphi(g_1 h,g_2 h,g_3 h,g_4 h)$ for all $h\in\SL$. This symmetry is imposed via group averaging, which yields a (pseudo-)projection onto the $\SL$-invariant subspace and is thus
reminiscent of the projection in Eq.~\eqref{eq:SL2C invariant projection}. 
In spin representation, the details of which are presented in Appendix~\ref{app:Representation Theory of SL2C}, the GFT propagator is identified  with the edge amplitude $\mathcal{A}_e$, and the GFT interaction directly involves the $BF$-vertex amplitude $\mathcal{A}_v$.

The simplicity constraints at the level of the GFT are most straightforwardly formulated for the BC model with spacelike tetrahedra. To that end, the right invariance above is supplemented with $\varphi(g_1u_1,g_2u_2,g_3u_3,g_4u_4) = \varphi(g_1,g_2,g_3,g_4)$ for $u_i\in\SUT$, yielding the model in~\cite{Perez:2000ec}. Due to the non-commutativity of these constraints, an extended formulation using normal vectors has been developed in~\cite{Baratin:2011tx}, generalized to Lorentzian signature in~\cite{Jercher:2021bie} and Chapter~\ref{chapter:cBC}. A model without normal vectors incorporating exclusively timelike tetrahedra, but spacelike and timelike faces, has been put forward in~\cite{Perez:2000ep}. The lack of a quantum geometric GFT model that incorporates all types of causal characters, formulated in terms of normal vectors, motivated the construction of the complete BC model in~\cite{Jercher:2022mky} which is presented in Chapter~\ref{chapter:cBC}. A GFT formulation of the EPRL spin-foam model using $\SL$ as the underlying group is still missing.\footnote{An \enquote{EPRL-like} GFT model has been introduced in~\cite{Oriti:2016qtz}, based on the group $\SUT$ and with the simplicity constraints implicitly defined in the kernel of the vertex term. A GFT model in terms of non-commutative flux variables for Euclidean Plebanski-Holst gravity has been put forward in~\cite{Baratin:2011hp}.}

Despite the apparent direct connection between spin-foam models and GFTs, the two approaches evolved, in practice, into seemingly independent research directions. The difference of perspectives spin-foams and GFTs take on the nature of quantum geometry presents itself most evidently in the way how the recovery of semi-classical smooth spacetimes is conceived of, which we discuss next.

\section{Emergent Lorentzian spacetimes from quantum gravity}

Spin-foam models and GFTs constitute two promising candidates for non-perturbative and background independent QG. Although starting from the continuum Einstein-Hilbert action, spin-foams ultimately prescribe a partition function $Z_{\mathrm{SF}}(\Gamma)$ and therefore reduce to combinatorial $(\Gamma)$ and algebraic (irreducible representations and intertwiners) data. The GFT approach goes even one step further and defines an abstract and rather peculiar field theory on a Lie group generating $Z_{\mathrm{SF}}(\Gamma)$. Clearly, these structures are far from resembling the continuum spacetime metric of classical GR. Thus, even if the models introduced in Secs.~\ref{sec:intro to spin-foams} and~\ref{sec:intro to GFT} are well-defined and computable\footnote{Turning these models into computable theories constitutes in fact a considerable portion of ongoing research.}, reproducing the physics of GR in a semi-classical and continuum limit is \emph{the} most important consistency check. This challenge is in fact shared among all the background independent QG approaches. Bridging the gap between continuum GR and the microscopic theories of quantum spacetime forms the main motivation of the present work. 

Spin-foams and GFTs can take very different perspectives on how semi-classical continuum spacetimes should be recovered. These differences concern in particular the 2-complex $\Gamma$ and how it should be interpreted. We contrast these perspectives in the following. 

One possible way of viewing $\Gamma$ in spin-foams is that it corresponds to a regulator of transition amplitudes, resulting in a truncation of the number of degrees of freedom. Semi-classical continuum physics is envisioned as arising in the semi-classical limit (i.e.~$\hbar\rightarrow 0$) of the quantum theory itself on a sufficiently fine triangulation.\footnote{Along the examples of the discretized particle in 1d and Lamor precession, it is stressed in~\cite{Engle:2021xfs} that the continuum limit must be taken \emph{before} the classical limit as otherwise, accidental constraints arise. 
} 
Encouraged by the results of the Ponzano-Regge model~\cite{Ponzano:1968wi}, the semi-classical limit of spin-foams has been studied extensively on a fixed $\Gamma$ given by a single vertex. For spacelike tetrahedra, the BC\footnote{Note however, that the BC asymptotics are dominated by geometries that only span a 1-dimensional subspace. We pick up this point again in the introduction of Chapter~\ref{chapter:cBC}.} and EPRL vertex amplitudes asymptote to the cosine of the Regge action~\cite{Regge:1961ct} associated to a 4-simplex, see~\cite{Barrett:2002ur,Baez:2002rx} and~\cite{Barrett:2009gg,Barrett:2009mw}, respectively. Although remarkable, these results require further studies as 1)  the crucial extension to larger 2-complexes reveals the so-called flatness problem~\cite{Conrady:2008mk,Dona:2020tvv}, 2) they are plagued by vector geometries~\cite{Dona:2020yao}, 3) the asymptotics of timelike interfaces in the EPRL-CH extension are ill-defined~\cite{Simao:2021qno}. Recent advancements in these directions are achieved via the hybrid algorithm idea~\cite{Asante:2022lnp}, the complex critical points method~\cite{Han:2021kll,Han:2023cen,Han:2024lti} and effective spin-foams~\cite{Asante:2020qpa,Asante:2020iwm,Asante:2021phx,Asante:2021zzh}. Characterizing the behavior of spin-foam amplitudes under the refinement of $\Gamma$ is a pressing issue subject to active research~\cite{Asante:2022dnj}. The results of effective spin-foam models tentatively suggest that taking such a refinement limit needs to be accompanied by a renormalization group flow of $\bi$ to increasingly smaller values. If a fixed point $(\Gamma_*,{\bi}_*)$ of this procedure is found
one has reached the continuum limit of the theory~\cite{Steinhaus:2020lgb}. Pioneering work in this direction has been developed in the context of (decorated) tensor network renormalization~\cite{Dittrich:2014mxa,Dittrich:2011zh,Liu:2013nsa,Cunningham:2020uco,Delcamp:2016yix}.  

Contrary to the preceding, GFTs take an atomistic perspective where $\Gamma$ corresponds to a perturbative Feynman diagram. Semi-classical continuum physics is conceived of as arising from the collective behavior of a quantum many-body system, e.g.~captured by coherent states~\cite{BenAchour:2024gir,Oriti:2024qav}. In analogy, the classical electromagnetic field can be obtained from a coherent state of photons~\cite{Loudon2000}. A curious consequence is that in this picture classical physics does not arise in the typical limit of $\hbar\rightarrow 0$ of the theory itself. In particular, the classical limit of a GFT is not at all obviously related to classical spacetimes. Theoretical hints supporting the assumption of an atomic substructure of spacetime, where the \enquote{atoms of spacetime} correspond to the 1-particle excitations of the group field~\cite{Oriti:2013jga}, come from the finiteness of black hole and Rindler horizon entropy~\cite{Jacobson:2003wv}. In~\cite{Jacobson:1995ab}, the Einstein equations are furthermore reinterpreted as a thermodynamic equation of state. Identifying collective macroscopic quantities associated to continuum spacetime that capture the dynamics of many microscopic degrees of freedom requires some form of coarse-graining. To do so, the field theoretic nature of GFTs allows importing powerful tools from local field theories (see~\cite{Pithis:2018eaq,Marchetti:2020xvf,Benedetti:2014qsa,BenGeloun:2015xrk,Benedetti:2015yaa,BenGeloun:2016rqa,Carrozza:2016vsq,Pithis:2020sxm,Pithis:2020kio,Geloun:2023ray}), such as the functional renormalization group (FRG) methodology~\cite{Berges:2000ew,Delamotte:2007pf,Kopietz:2010zz,Dupuis:2020fhh} and Landau-Ginzburg (LG) mean-field theory~\cite{Zinn-Justin:2002ecy,Zinn-Justin:2007uvz}. In particular, the LG method has been applied recently to the BC GFT model, finding a remarkable robustness of the mean-field approximation~\cite{Marchetti:2022igl,Marchetti:2022nrf}, and suggesting the existence of a condensate phase occupied by a large number of GFT quanta. In Chapter~\ref{chapter:LG} we shall apply the LG methodology to the causal completion of the BC model developed in Chapter~\ref{chapter:cBC}.

Common to both approaches is the need for observables that facilitate a connection to the quantities characterizing the smooth spacetime of classical GR. The spectral dimension provides a notion of an effective dimension on different scales and thus allows checking whether the emergent spacetimes obtained from QG exhibit four dimensions at large scales. We will investigate the spectral dimension of Euclidean spin-foams frusta in Chapter~\ref{chapter:specdim}.

Another promising testing ground for QG is that of cosmology. The large degree of symmetry greatly reduces the complexity of the considered QG model, thus rendering analytical and numerical studies more accessible. Quantum cosmology from QG also holds the promise of shedding light on the Big Bang singularity, potentially replacing it by a quantum Big Bounce~\cite{Brandenberger:2016vhg,Oriti:2016qtz,Agullo:2016tjh}, and leaving observable traces e.g.~in the cosmic microwave background~\cite{Ashtekar:2020gec}. In spin-foams, one possible strategy to identify a cosmological subsector is to truncate the degrees of freedom via a symmetry reduction~\cite{Bianchi:2010zs,Bahr:2017eyi,Han:2024ydv}. Recent studies~\cite{Dittrich:2023rcr} in the context of effective spin-foams unveiled that the causal structure plays an important role for the discrete cosmological path integral. In Chapter~\ref{chapter:LRC}, we develop the classical discrete theory of spatially flat, homogeneous and isotropic cosmology in (3+1) dimensions, forming the foundation of the analysis of the (2+1) quantum cosmological spin-foam model in Chapter~\ref{chapter:3d cosmology}. In GFT on the other hand, coherent (or condensate) states are utilized to explore a hypothetical condensate phase, the existence of which is supported by the LG analyses in~\cite{Marchetti:2020xvf,Marchetti:2022igl} and Chapter~\ref{chapter:LG}. These states capture the collective behavior of many GFT quanta and allow identifying macroscopic cosmological observables~\cite{Gielen:2013naa,Gielen:2014ila,Gielen:2014uga,Oriti:2016ueo,Oriti:2016qtz,Marchetti:2020umh,Marchetti:2020qsq}. Their dynamics are governed by the GFT field equations at mean-field level, in close analogy to the Gross-Pitaevskii~\cite{pitaevskii2016bose} equation for Bose-Einstein condensates. In Chapter~\ref{chapter:perturbations}, this program is advanced by describing scalar cosmological perturbations as emerging from the entanglement of spacelike and timelike tetrahedra within the causally complete BC model, again stressing the importance of faithfully incorporating the causal structure in QG models.
\clearpage
\phantomsection
\addcontentsline{toc}{part}{Part I: Spectral Dimension and Cosmology in Spin-Foams}\label{Part I}
\fancyhead[CE]{\leftmark}
\fancyhead[LE]{\scshape\MakeLowercase\chaptertitlename\hspace{2pt}\large\thechapter}
\fancyhead[RO]{\large\thesection}
\fancyhead[CO]{\rightmark}
\fancyfoot[C]{\thepage}
\renewcommand{\headrulewidth}{0.5pt}
\renewcommand{\footrulewidth}{0pt}
\chapter[\textsc{Curvature Effects in the Spectral Dimension of Spin-Foams}]{Curvature Effects in the Spectral Dimension of Spin-Foams}\label{chapter:specdim}

Observables are a crucial tool for understanding the physical consequences of any QG approach and for bridging the gap to the well-known physics of GR. One such observable which characterizes the global properties of a quantum spacetime is the spectral dimension. It provides a notion of effective dimension at different scales and allows testing whether a quantum spacetime exhibits the observed dimension of $d=4$ at large length scales. This constitutes already a highly non-trivial consistency check that many QG approaches fail. Tensor models~\cite{Gurau:2013cbh} or the non-geometric phases of CDT~\cite{Loll:2019rdj} for instance exhibit a fractal dimension less than 4 despite involving 4-dimensional building blocks. On the other hand, new small-scale effects beyond classical continuum gravity can be examined, such as a dimensional flow to values $0 < d < 4$ at small scales found in many QG approaches~\cite{Carlip:2017eud,Crane:1985ex,Crane:1986yg,Ambjorn:2005qt,Coumbe:2014noa,Lauscher:2005qz,Horava:2009if,Benedetti:2008gu,Alesci:2011cg,Arzano:2014jfa,Steinhaus:2018aav,Eckstein:2020gjd,Reitz:2022dbj,Calcagni:2013dna,Calcagni:2014cza,Eichhorn:2017djq}, possibly leaving observable traces in gravitational wave astronomy~\cite{Calcagni:2019ngc}. In any case, this phenomenon is interesting as it allows comparing conceptually different QG approaches at small scales. 


Determining the spectral dimension $\Ds$ of spin-foams has been attempted in~\cite{Steinhaus:2018aav} in the setting of spin-foam cuboids~\cite{Bahr:2015gxa,Bahr:2016hwc,Allen:2022unb,Ali:2022vhn} within the Euclidean EPRL-Freidel-Krasnov (FK) model~\cite{Engle:2007wy,Freidel:2007py} extended to hypercubic combinatorics~\cite{Kaminski:2009fm}.\footnote{Earlier works attempted to infer the spectral dimension from the LQG area spectrum~\cite{Modesto:2008jz} and were restricted to a single building block~\cite{Magliaro:2009if,Caravelli:2009gk}.} There, a non-trivial flow of $\Ds$ to intermediate values, controlled by a face amplitude parameter $\alpha$, has been observed for $\mathcal{N}$-periodic configurations. The results of~\cite{Steinhaus:2018aav} come with the limitations that 1) only semi-classical amplitudes are being utilized, and 2) spin-foam cuboids are inherently flat. Point 2) implies that the semi-classical amplitudes entering the sum over spins exhibit only a simple scaling behavior. In the presence of curvature, however, spin-foam amplitudes generically oscillate. The main purpose of this chapter is to overcome these two limitations. To that end, we employ spin-foam frusta~\cite{Bahr:2017eyi} which exhibit oscillating amplitudes while still being computationally feasible due to their high degree of symmetry.

\section{Euclidean spin-foam frusta and the spectral dimension}\label{sec:Spin-foam hyperfrusta and the spectral dimension}



\subsection{Euclidean spin-foam frusta in the EPRL-FK model}\label{sec:Euclidean spin-foam frusta in the EPRL-FK model}

In this section, we employ the generalization of the Euclidean EPRL-FK model~\cite{Engle:2007wy,Freidel:2007py} developed in~\cite{Kaminski:2009fm,Oriti:2014yla} with a Barbero-Immirzi parameter $\bi < 1$. The underlying combinatorics are captured by the $2$-complex $\Gamma$ which contains vertices~$v$, edges $e$ and faces $f$ and is assumed here to be hypercubical, i.e. dual to a hypercubical lattice $\mathcal{X}^{(4)}$. This choice is well-suited to introduce the frusta geometries we focus on in this chapter, and allows for a simpler definition of the Laplace operator and $\mathcal{N}$-periodicity discussed below. To $\Gamma$, we associate the partition function
\begin{equation}
Z(\Gamma) = \sum_{\{j_f,\iota_e\}} \prod_f\mathcal{A}_f\prod_e\mathcal{A}_e\prod_v\mathcal{A}_v\,.
\end{equation}
As we assume periodic boundary conditions for $\Gamma$, i.e. $\partial\Gamma = \emptyset$, $Z$ does not depend on any boundary data. The $j_f$ are irreducible $\SUT$ representations associated to the faces $f$, and $\iota_e$ are $\SUT$ intertwiners associated to the edges $e$. $\mathcal{A}_f,\mathcal{A}_e$ and $\mathcal{A}_v$ are the face, edge and vertex amplitudes, respectively, defined in the following.

A distinctive property of the Euclidean EPRL-FK model
is the simplicity constraint which provides an embedding~$Y_{\bi}$ of $\text{SU}(2)$ into $\text{Spin}(4)\cong\SUT\times\SUT$ representations~\cite{Engle:2007wy}. For $\bi < 1$, the explicit relation between $\SUT$ spins~$j$ and $\Spin$ labels $(j^+,j^-)$ is given by
\begin{equation}\label{eq:EPRL condition}
j^{\pm} = \frac{\vert1\pm \bi\vert}{2}j,\qquad\text{with}\qquad j^\pm\overset{!}{\in}\frac{\mathbb{N}}{2}\,.
\end{equation}
For this map to be non-empty, $\bi$ is required to be rational, and we choose for the remainder $\bi = 1/3$. Face, edge and vertex amplitudes are respectively defined as
\begin{equation}\label{eq:quantum amplitudes}
    \mathcal{A}_f^{(\alpha)} = \left((2j_f^+ + 1)(2j_f^- + 1)\right)^{\alpha} \,,\qquad  \mathcal{A}_e = \frac{1}{\vert\vert Y_{\bi}\iota_e\vert\vert^2}\,,\qquad \mathcal{A}_v = \mathrm{Tr}\left(\bigotimes_{e\subset v}Y_{\bi}\ket{\iota_e}\right)\,.
\end{equation}
Constrained $BF$-quantization yields a face amplitude $\mathcal{A}_f^{(\alpha)}$ with $\alpha = 1$. However, modifications of $\mathcal{A}_f$ are proposed in the literature%
\footnote{Following~\cite{Bianchi:2010fj}, the two most common choices in the Euclidean setting are either the dimension of $\Spin$ representations ($\alpha =1 $) or $\SUT$ representation ($\alpha=1/2$).},
the choice of which has direct consequences for the critical behavior of the partition function~\cite{Riello:2013bzw}, as well as the spectral dimension~\cite{Steinhaus:2018aav}. We follow~\cite{Bahr:2015gxa,Bahr:2017eyi,Steinhaus:2018aav} and parametrize this ambiguity by $\alpha\in\R$. $\mathcal{A}_e$ is introduced as a normalization factor of intertwiners. The trace entering the vertex amplitude is understood so that either $Y_{\bi}\ket{\iota_e}$ or $\bra{\iota_e}Y_{\bi}^{\dagger}$ are contracted, depending on the orientation of the edge $e$.

\begin{figure}
    \centering
    \includegraphics[width=0.4\textwidth]{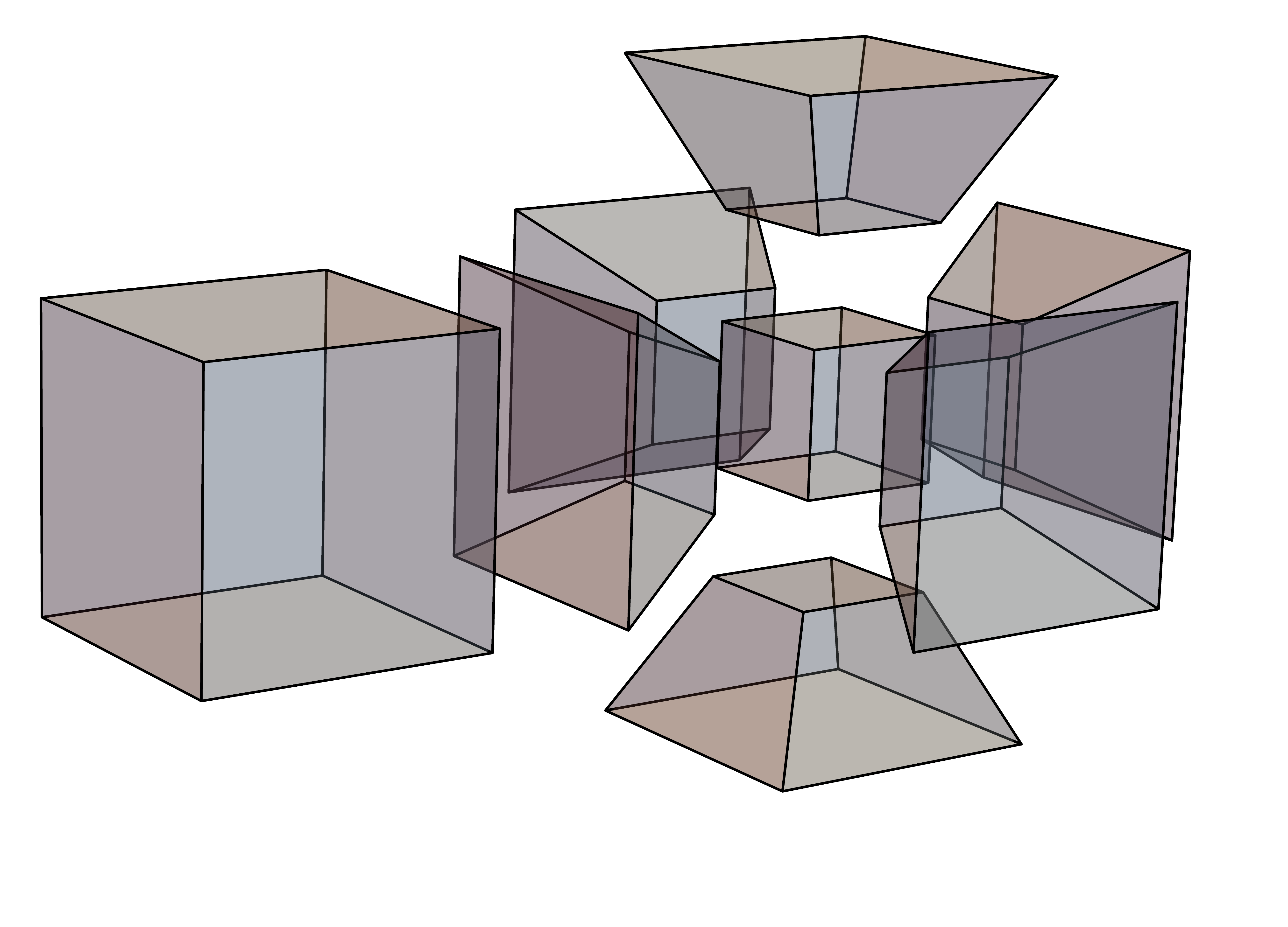}
    \caption{Boundary of a $4$-frustum given by two $3$-cubes and six equal $3$-frusta.}
    \label{fig:4-frustum}
\end{figure}

In the following, we restrict the geometry to Euclidean $4$-frusta (see Fig.~\ref{fig:4-frustum})~\cite{Bahr:2017eyi}, the boundary of which consists of two generically different cubes and six equal $3$-frusta. That is, we impose a $3$-frustum shape on the Livine-Speziale coherent intertwiners~\cite{Livine:2007vk}, as these are peaked on the geometry of $3$-dimensional polyhedra.
Explicitly, 
\begin{equation}\label{eq:intertwiner}
\ket{\iota_{j_1 j_2 k}} = \int\limits_{\SUT}\dd{g}\: g\triangleright \left(\ket{j_1, \hat{e}_3}\otimes\ket{j_2, -\hat{e}_3}\bigotimes_{l=0}^3 \ket{k, \hat{r}_l}\right)\,,
\end{equation}
where the case $j_1 = j_2 = k$ reproduces cubical intertwiners~\cite{Bahr:2015gxa}. Here, \enquote{$\triangleright$} denotes the $\SUT$ action, $\hat{e}_3$ is the unit vector in $\R^3$ along the axis $e_3$ and $\hat{r}_l \defeq \e^{-i\frac{\pi}{4}l\sigma_3}\e^{-i\frac{\phi}{2}\sigma_2}\triangleright \hat{e}_3$ with $l\in\{0,1,2,3\}$. 


Semi-classically, the spins $(j_1, j_2,k)$ correspond to the three areas~\cite{Barrett:2009gg,Barrett:2009mw} of the base and top square and of the bounding trapezoids, respectively, and fully characterize a $3$-frustum.
Another convenient parametrization is given by $j_1,j_2$ and the slope angle $\phi = \cos^{-1}\left(\frac{j_1-j_2}{4k}\right)$, requiring $-1\leq \frac{j_1-j_2}{4k}\leq 1$. Note that for 4-dimensional dihedral angles to be well-defined, the stronger condition $-1/\sqrt{2}\leq \frac{j_1-j_2}{4k}\leq 1/\sqrt{2}$ needs to hold.

Gluing two cubes and six $3$-frusta as indicated above, one obtains a $4$-frustum. The two $3$-cubes lie in separated ``spatial''\footnote{The terms ``spatial'', ``spacelike'', ``temporal'', etc. serve illustrative purposes in the present Euclidean context. Only in the subsequent chapters, we work in an explicitly Lorentzian setting, making these notions precise.} hypersurfaces at the base and top, connected by the six $3$-frusta. A visualization of the unwrapped boundary of a single $4$-frustum is given in Fig.~\ref{fig:4-frustum}.

As a whole complex, the discretization can be understood as a slicing, where the $n$th thick slice is bounded by two spatial hypersurfaces cubulated by $3$-cubes with area $j_n$ and~$j_{n+1}$, respectively, and connected by $3$-frusta with ``spatio-temporal'' areas given by $k_n$. Due to the gluing conditions, the spins $j_n, j_{n+1}$ and $k_n$ are constant throughout a whole slice. Therefore, spin-foam frusta are spatially homogeneous and thus present an ideal setting for spatially flat discrete cosmology~\cite{Bahr:2017eyi}. We make this connection explicit in Chapters~\ref{chapter:LRC} and~\ref{chapter:3d cosmology}.


The quantum amplitudes of Eq.~\eqref{eq:quantum amplitudes} are highly involved functions of the representation labels even in this symmetry restricted setting. To obtain an analytical expression for later purposes, we perform a semi-classical approximation~\cite{Barrett:2009gg,Barrett:2009mw} of the frustum quantum amplitudes~\cite{Bahr:2017eyi}, thereby connecting to classical discrete gravity~\cite{Rovelli:2011eq}. Introducing a uniform rescaling of spins, $j_i \rightarrow \lambda j_i$, with $\lambda\rightarrow\infty$, the quantum face amplitude in~\eqref{eq:quantum amplitudes} is approximated by $\mathcal{A}_{f} = \left[\left(1-\bi^2\right)j_n^2\right]^{\alpha}$. To find the asymptotics of the vertex and the edge amplitude $\mathcal{A}_i$, we notice the factorization property $\mathcal{A}_i = \mathcal{A}^+_i\mathcal{A}^-_i$ with $\mathcal{A}_i^{\pm}$ being the $\SUT$ amplitudes evaluated on the spins $j^{\pm}$. This results from the $Y_{\bi}$-map for $\bi < 1$~\cite{Engle:2007uq} and the fact that $\Spin\cong\SUT\times\SUT$. Applying a stationary phase approximation~\cite{Barrett:2009gg}, one finds~\cite{Bahr:2017eyi} 
%
\begin{equation}\label{eq:Spin4Vamp}
    \mathcal{A}_v 
    = 
    \frac{1}{\pi^7(1-\bi^2)^{21/2}}\left(\frac{\e^{ \frac{i}{\GN}\Sreg }}{-\det H}+\frac{\e^{-\frac{i}{\GN}\Sreg }}{- \det H^{*}}+2\frac{\cos(\frac{ \bi} {\GN} \Sreg-\frac{\Lambda}{\GN}V^{(4)})}{\sqrt{\det H \det H^{*}}}\right)\,,
\end{equation}
where $\det H$ is the Hessian determinant and $\Sreg$ is the boundary Regge action~\cite{Regge:1961ct} of a 4-frustum\footnote{Strictly speaking, $\Sreg$ is the \emph{area} Regge action~\cite{Asante:2018wqy,Barrett:1997tx}. However, in this symmetry restricted setting the map between area and length variables is one-to-one~\cite{Dittrich:2023rcr}. For an introduction to Regge calculus, see~\cite{Hamber2009}.}, both given explicitly in Appendix~\ref{app:Euclidean frusta}. $\GN$ is the gravitational constant, which has been added by hand as in~\cite{Bahr:2018gwf}.
Note that we introduced a cosmological constant $\Lambda$, following the work of~\cite{Han:2011aa} and~\cite{Bahr:2018vv}. Therein, the quantum amplitudes are deformed in a heuristic fashion by a real parameter which, in a semi-classical limit, can be related to a cosmological constant of either sign. It enters with the strictly positive $4$-volume $V^{(4)}$ defined in Eq.~\eqref{eq:4Volume}.
%
%

Repeating the analysis for the $\SUT$ edge amplitude, one obtains
%
%
%
\begin{equation}\label{eq:Spin4Eamp}
\mathcal{A}_e = \pi(1-\bi^2)^{3/2}\frac{k_n\sin^2(\phi)}{2}\Big{(}j_n+j_{n+1}+2k_n\big{(}1-\cos^2(\phi)\big{)}\Big{)}^2\,.
\end{equation}
On a hypercubical lattice, edges and faces are shared by two, respectively four vertices. This allows to define a \textit{dressed} vertex amplitude $\hat{\mathcal{A}}_v \defeq \prod_{f\supset v}\mathcal{A}_f^{1/4}\prod_{e\supset v}\mathcal{A}_e^{1/2}\mathcal{A}_v$. Assuming that a semi-classical approximation is performed at each vertex individually\footnote{Following~\cite{Asante:2022lnp}, it is expected that the stationary phase approximation of amplitudes on extended complexes does not correspond to the product of semi-classical amplitudes. However, since for the former case an analytical formula has not been developed yet, we stick to this simplification here and also in Chapter~\ref{chapter:3d cosmology}.}, the amplitude of whole complex can then be written as the product of dressed vertex amplitudes. Note that the measure factor $\upmu$ of $\hat{\mathcal{A}}_v$, i.e. its scaling part, behaves under uniform rescaling as $\upmu(\lambda j_n,\lambda j_{n+1},\lambda k_n) \sim \lambda^{12\alpha -9}$, agreeing with spin-foam cuboids~\cite{Bahr:2015gxa}. However, in contrast, the entire $\hat{\mathcal{A}}_v$ is \emph{not} a homogeneous function in the spins due to the oscillations of the amplitude.

\paragraph{Quantum amplitudes from extrapolation.} The quantum amplitudes in Eq.~\eqref{eq:quantum amplitudes} are a necessary ingredient to determine expectation values. Despite the symmetry-reduced setting, a contraction of the intertwiners in Eq.~\eqref{eq:intertwiner} is costly due to their higher cuboidal valency which sets numerical limits to the computation of the vertex amplitude already at low spins $j\sim 4$~\cite{Allen:2022unb}. However, as explained later in Sec.~\ref{subsec:1-periodic spectral dimension}, to resolve an interesting non-trivial flow of the spectral dimension many more data points of the quantum amplitude are necessary. 

While resorting to semi-classical amplitudes is a sensible choice~\cite{Steinhaus:2018aav}, the approximation deviates significantly for small spins $j\lesssim 10$~\cite{Allen:2022unb}. 
To find a better approximation at small spins that still agrees with semi-classical amplitudes at large spins, a method to extrapolate quantum amplitudes in the simplest case of a single 4-frustum with $j_1=j_2$ (later referred to as 1-periodic 4-frustum) has been developed in~\cite{Jercher:2023rno}. Such configurations represent a specific subclass of spin-foam cuboids, exhibiting a pure scaling behavior. For given data points, one computes $\varepsilon_i(j,k)\defeq \frac{\abs{\mathcal{A}_i^{\mathrm{sc}} - \mathcal{A}_i^{\mathrm{qu}}}}{\mathcal{A}_i^{\mathrm{sc}}}$, where $\mathcal{A}^{\mathrm{qu}}$ and $\mathcal{A}^{\mathrm{sc}}$ denote the $\SUT$ quantum, respectively the semi-classical amplitudes and $i=v,e$ indicates vertices and edges, respectively. $\varepsilon_i$ is then extrapolated in $k$ and $j$ direction from which $\mathcal{A}^{\mathrm{qu}}$ can be deduced. The numerical results of~\cite{Jercher:2023rno} show that 
the such extrapolated amplitudes deviate less from the quantum amplitudes than the semi-classical amplitudes, in particular for small spins.\footnote{Of course, the whole procedure hinges on the assumption of a single 4-frustum with $j_1=j_2$ and cannot be straightforwardly applied to the more general case of $j_1\neq j_2$ due to curvature-induced oscillations.} The extrapolated $\SUT$ amplitudes evaluated on spins $j^{\pm}$ can be combined to an extrapolated dressed vertex amplitude using the $Y_{\bi}$-map. For the remainder we choose $\bi = \frac{1}{3}$ and discuss the influence of this choice on the spectral dimension in Sec.~\ref{subsec:1-periodic spectral dimension}. The results of~\cite{Jercher:2023rno} show that also for the dressed vertex amplitude, extrapolation provides a better approximation at low spins than the semi-classical amplitude. The extrapolated amplitudes will be explicitly employed in Sec.~\ref{subsec:1-periodic spectral dimension}.

\subsection{Spectral dimension of spin-foam frusta}



The spectral dimension serves as an effective measure of the dimension of a space. Consider a Riemannian manifold $(\mathcal{M},\mathrm{g})$ together with the heat kernel $K(x,x_0;\tau)$ solving the heat equation $\partial_{\tau}K(x,x_0;\tau) = \Delta K(x,x_0;\tau)$, with $\Delta$ the Laplace operator implicitly depending on $\mathrm{g}$~\cite{Carlip:2017eud}. Here, $x,x_0\in\mathcal{M}$ and $\tau$ provides a measure of the size of the probed region, often referred to as diffusion time. In $d$-dimensional flat space, the so-called return probability $P(\tau) = \int_{\mathcal{M}}\dd{x}\sqrt{\mathrm{g}}~  K(x,x;\tau)$ exhibits a scaling $P(\tau)\sim\tau^{-\frac{d}{2}}$. This motivates to extract the classical spectral dimension $\Ds^{\mathrm{cl}}(\tau)$ from the scaling of the return probability by the relation $\Ds^{\mathrm{cl}}(\tau) = -2\frac{\dd \log P(\tau)}{\dd \log\tau}$ for general manifolds. Intuitively, $P(\tau)$ can be understood as the probability of a random walker to return to its starting point, hence the name. Since $P(\tau)$ is a functional of the geometry, one can in principle compute the expectation value of the return probability, $\ev{P(\tau)}$, from the gravity path integral and define the spectral dimension as $\Ds(\tau) \defeq -2\frac{\dd \log \ev{P(\tau)}}{\dd \log\tau}$. Notice, that we do not compute $\langle \Ds^\mathrm{cl}(\tau)\rangle$ but define the quantum spectral dimension as the scaling of $\langle P(\tau)\rangle$. 

To translate these notions to the context of spin-foams, we introduce now a discrete formulation of the Laplace operator, the return probability and the spectral dimension. 
Given a hypercubical lattice, we denote vertices in the dual graph $\Gamma$ by $\vec{n}\in\mathbb{Z}^4$. Interpreting the return probability as the trace over the heat kernel, its discrete form is simply given by $P_{\mathrm{disc}}(\tau) = \sum_{\lambda\in\mathrm{spec}(\Delta)}e^{-\tau\lambda}$~\cite{Calcagni:2012cv,Thurigen:2015uc}, where $\mathrm{spec}(\Delta)$ is the spectrum of the discrete Laplacian. 

Following~\cite{Calcagni:2012cv}, a Laplace operator can be defined on general cellular complexes using methods of discrete exterior calculus~\cite{Desbrun:2005ug}. To that end, one introduces a scalar test field%
\footnote{Note that the test fields do not obey spatial homogeneity or the notion of $\mathcal{N}$-periodicity introduced below.} 
discretized on the complex, 
which we choose to place on dual vertices.\footnote{The scalar field can also be placed on primary vertices as e.g. in~\cite{Ali:2022vhn} and Chapters~\ref{chapter:LRC} and~\ref{chapter:3d cosmology}. Beyond scalar fields, other tensor or $p$-form fields might yield different results for ``generalized'' spectral dimensions~\cite{Reitz:2022dbj}.} The discrete Laplacian is defined by its action on the scalar test field $\phi_{\vn}$~\cite{Thurigen:2015uc},
\begin{equation}\label{eq:defLaplace}
-(\Delta \phi)_{\vn} = -\sum_{\vm\sim\vn}\Delta_{\vn\vm}\left(\phi_{\vn} - \phi_{\vm}\right) = \frac{1}{V^{(4)}_{\vn}}\sum_{\vm\sim\vn}\frac{V^{(3)}_{\vn\vm}}{l^*_{\vn\vm}}\left(\phi_{\vn}-\phi_{\vm}\right)\,,
\end{equation}
where the sum runs over all adjacent vertices. $\Delta_{\vn\vm}$ are the coefficients of the discrete Laplacian, which can be split into a diagonal part and a part proportional to the adjacency matrix of $\Gamma$. $V^{(3)}_{\vn\vm}$ and $l^*_{\vn\vm}$ indicate the $3$-volume, respectively the dual edge length of $(\vn\vm)$. The positive $4$-volume of the frustum dual to $\vn\in\Gamma$ is denoted by $V^{(4)}_{\vn}$ entering also Eq.~\eqref{eq:Spin4Vamp}.

As discussed in~\cite{Thurigen:2015uc}, the definitions of volume, dual edge length and Laplacian are not unique. A construction of the dual 2-complex and a definition of the geometric quantities entering Eq.~\eqref{eq:defLaplace} is given in Appendix~\ref{app:Euclidean frusta}. Note that the spins $j_n,k_n$ enter these definitions as areas which is only valid in a semi-classical regime. However, we assume that the definition $\Delta(j_n,k_n)$ holds for arbitrarily small spins, which can be seen as a continuation of the semi-classical Laplacian and provides \emph{one} possible definition of $\Delta$ in the quantum regime.

\paragraph{$\pmb{\mathcal{N}}$-periodicity.} Evaluating the spectral dimension even in the setting of restricted spin-foams is challenging for the following reasons. The return probability for a given geometric configuration requires full knowledge of the Laplacian spectrum. On a lattice of $L^4$ sites, this amounts to diagonalizing an $L^4\times L^4$ matrix. Furthermore, to compute the expectation value of the return probability, this diagonalization needs to be performed for all configurations, the number of which scales exponentially in $L$. At the same time, $L$ determines the scale $\tau_{\mathrm{comp}}(L)$ at which boundary effects become dominant. To avoid fixed data on the boundary, we henceforth assume periodic boundary conditions, equipping the lattice with a compact toroidal topology. The resulting compactness leads to a fall-off in spectral dimension for $\tau>\tau_{\mathrm{comp}}$. That is because $P(\tau)\rightarrow 1$, which reflects heuristically that a compact space looks point-like from far away~\cite{Thurigen:2015uc}.  Consequently, $L\gg 1$ is required to observe a non-trivial spectral dimension between the smallest lattice scale $\tau\sim j_{\mathrm{min}}$ and $\tau_{\mathrm{comp}}$, which is in conflict with computability. 

To solve these issues, we adopt the assumption of $\mathcal{N}$-periodicity proposed in~\cite{Steinhaus:2018aav}, based on results of~\cite{Sahlmann:2009rk}. The key idea is to assume that the geometry of the spin-foam repeats after $\mathcal{N}$ steps in each direction. As a consequence, the spectrum of the Laplacian is obtained by performing a Fourier transform and diagonalizing an $\mathcal{N}^4\times\mathcal{N}^4$-matrix, rather than diagonalizing a matrix of size $L^4\times L^4$ with $L\gg 1$. The spectrum is then given in terms of four momenta~$p_{\mu}$, which are discrete (continuous) for a finite (infinite) lattice, and which lie in the Brillouin zone.  The return probability is obtained as a sum (integral) over the momenta. Periodicity also reduces the computational complexity at the level of the amplitudes, as the number of spin variables reduces to $2\mathcal{N}$. Although numerically very useful, the assumption of $\mathcal{N}$-periodicity is not physical and needs to be removed in a limit $\mathcal{N}\rightarrow \infty$. We discuss this limit at the end of this chapter.

The $\mathcal{N}$-periodic return probability is given by
\begin{equation}\label{eq:return_prob_N}
P_{\mathcal{N}}(\tau) = \sum_{i=1}^{\mathcal{N}}\prod_{\mu=0}^3\sumint{}\dd{p_\mu}\e^{-\tau\omega_i(\{p_{\mu}\})}\,,
\end{equation}
with $\omega_i(\{p_\mu\})$, $i\in\{1,\dots,\mathcal{N}\}$, the eigenvalues of the Laplacian derived in Appendix~\ref{app:Laplacian spectrum}.\footnote{The $\omega_i$ are analogous to the branches of phonon dispersion relations in solid state physics.} Eq.~\eqref{eq:return_prob_N} contains either a sum over discrete $p_\mu=\frac{2\pi}{L}k_{\mu}$ with $k_\mu\in\mathbb{Z}_L$, or an integral over continuous $p_\mu\in[-\pi,\pi]$ in the Brillouin zone.\footnote{For $L\rightarrow\infty$, the integrals are evaluated numerically using the \href{https://github.com/giordano/Cuba.jl}{\texttt{Cuba}}-package in \href{https://arxiv.org/pdf/Julia}{\texttt{Julia}}. Higher-dimensional integrations are possible but more costly and issues of convergence are more likely to arise.} Notice, that the eigenvalues $\omega_i(\{p_{\mu}\})$ depend on the geometry of the entire lattice, turning the return probability into a highly non-local quantity. 

The expectation value of the return probability for a finite $\mathcal{N}$-periodic lattice in this setting is
\begin{equation}\label{eq:return_prob_exp pre}
 \langle P_{\mathcal{N}}(\tau) \rangle = \frac{1}{Z} \sum_{\{j_i,k_i\}}\left(\prod_{n=1}^{\mathcal{N}}\hat{\mathcal{A}}(j_n,j_{n+1},k_n)^{L^3}\right)^{L/\mathcal{N}}P_{\mathcal{N}}(\tau,\{j_i,k_i\})\,,
\end{equation}
where $j_{\min} \leq j_i,k_i \leq j_{\max}$ with lower and upper cutoffs $\jmin,\jmax$. A priori, $\hat{\mathcal{A}}$ denotes the full quantum amplitude. However, as we argue in Sec.~\ref{subsec:1-periodic spectral dimension}, the semi-classical dressed vertex amplitude already captures the behavior of the spectral dimension sufficiently for $\tau\gtrsim 10^2$. 

As discussed previously, $L$ is required to be large to resolve a non-trivial spectral dimension between the scales $\tau\sim j_{\mathrm{min}}$ and $\tau_{\mathrm{comp}}$. Consequently, the amplitudes enter~\eqref{eq:return_prob_exp pre} with large powers, requiring the utilization of arbitrary precision floating point numbers which is costly in memory and computation time. To circumvent this issue, we truncate the total number of amplitudes by assuming that the amplitudes of a single $\mathcal{N}$-cell sufficiently capture the relevant information of the whole spin-foam. Within this approximation, $\ev{P_{\mathcal{N}}(\tau)}$ is finally given by
\begin{equation} \label{eq:return_prob_exp}
    \langle P_{\mathcal{N}}(\tau) \rangle =  \frac{1}{Z} \sum_{\{j_i,k_i\}}\prod_{n=1}^{\mathcal{N}} \hat{\mathcal{A}}(j_n,j_{n+1},k_n)^{\mathcal{N}^3} \; P_{\mathcal{N}}(\tau,\{j_i,k_i\})\,,
\end{equation}
which can be computed efficiently by a tensor contraction of the spin indices (rather than with \texttt{for}-loops) of the amplitude and the return probability. This holds the advantage that $P_{\mathcal{N}}(\tau)$ needs to be computed only once for all configurations. It can then be contracted with the fast computed amplitudes $\hat{\mathcal{A}}$ for different parameter values $(\alpha,\GN,\bi,\Lambda)$. 

To summarize, we compute $\ev{P(\tau)}$ by summing up the return probability for all possible frusta geometries, weighted by the amplitudes $\hat{\mathcal{A}}$ for various diffusion times $\tau$. From this expectation value, we derive the spectral dimension  as $D^{\mathcal{N}}_\textsc{s} = - 2 \frac{\dd{\log\ev{P_{\mathcal{N}}(\tau)}}}{\dd{\log\tau}}$.

\section{Spectral dimension from spin-foam frusta}\label{sec:results}

In this section, 
the spectral dimension is computed numerically for $1$-periodic frusta using quantum amplitudes and semi-classical amplitudes with non-vanishing cosmological constant. Thereafter, the $2$-periodic spectral dimension is being investigated. 
We close this section with an analytical estimate of the spectral dimension.

\subsection{Spectral dimension of 1-periodic frusta}\label{subsec:1-periodic spectral dimension}

Restricting to $1$-periodic frusta, the only branch of the Laplacian spectrum separates into momentum components, $\sum_\mu\omega^{(\mu)}(p_{\mu})$, and thus,
%
the momentum integrals entering $P(\tau)$ factorize which is numerically advantageous.
%
%

\begin{figure}
    \centering
    \begin{subfigure}{0.45\textwidth}
    \includegraphics[width=\linewidth]{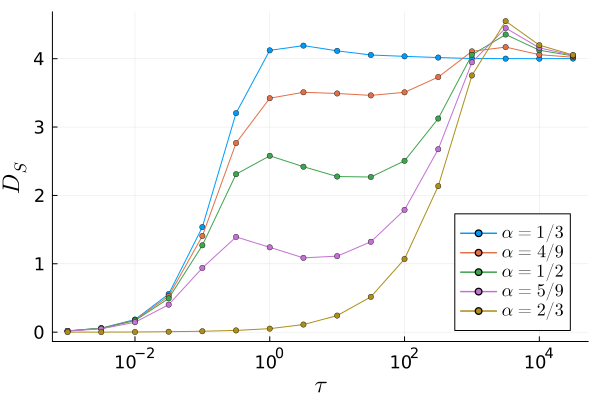}
    \end{subfigure}\hspace{0.05\textwidth}
    \begin{subfigure}{0.45\textwidth}
    \includegraphics[width=\linewidth]{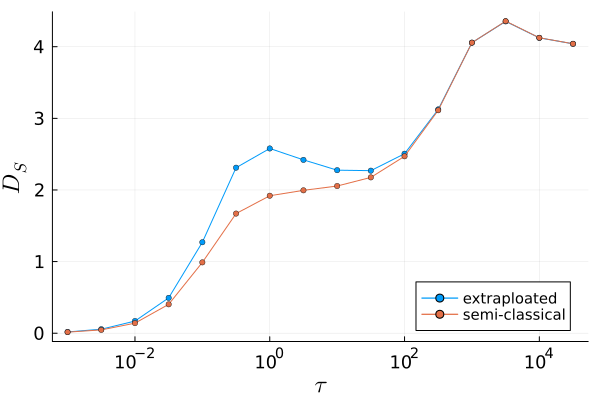}
    \end{subfigure}
    \caption{Left: expectation value of the spectral dimension for various values of $\alpha$ computed with extrapolated dressed vertex amplitudes. Right: comparison of $\Ds$ computed from extrapolated and semi-classical amplitudes at $\alpha = 0.5$. Both sets of data are computed with cutoffs $j_{\mathrm{min}} = \frac{1}{2}$ and $j_{\mathrm{max}} = 7500$.}
    \label{fig:specdim_quantum}
\end{figure}

Employing the extrapolated amplitudes introduced at the end of Sec.~\ref{sec:Euclidean spin-foam frusta in the EPRL-FK model}, we compute the expectation value of the return probability via Eq.~\eqref{eq:return_prob_exp}. 
%
%
The numerical results for 
the spectral dimension are presented in Fig.~\ref{fig:specdim_quantum} for different values of $\alpha$. 
Probing spacetime at scales below the lowest lattice scale, $\tau\ll j_{\mathrm{min}}$, $\Ds$ is zero. Above the largest scale, i.e. $\tau\gg j_{\mathrm{max}}$, every classical configuration exhibits a spectral dimension of four and hence the quantum spectral dimension is four as well. Similar to the findings of~\cite{Steinhaus:2018aav}, we observe a non-trivial dimensional flow between $0$ and $4$ for $\alpha$ in a certain interval $[\alpha_{\mathrm{min}},\alpha_{\mathrm{max}}]$. 
We discuss briefly a selection of influence factors of the spectral dimension in the following paragraphs.

\paragraph{$\pmb{\alpha}$-parameter.} The most salient factor driving the spectral dimension is $\alpha$. As the left panel of Fig.~\ref{fig:specdim_quantum} shows, only a small range of $\alpha\in[\alpha_{\mathrm{min}},\alpha_{\mathrm{max}}]$ leads to an intermediate spectral dimension $\Ds<\std$. For any $\alpha < \alpha_{\mathrm{min}}$, the spectral dimension attains the value $\std=4$ without any non-trivial behavior before. 
Similarly, $\alpha >\alpha_{\mathrm{max}}$ suppresses an intermediate dimension $\Ds\neq 0$ and only at scales $\tau\gtrsim j_{\mathrm{max}}$, the value $\Ds=4$ is obtained. 
In Sec.~\ref{subsec:Analytical estimate of the spectral dimension}, we support the statements on the role of $\alpha$ with analytical considerations and present an estimate of the interval $[\alpha_{\mathrm{min}},\alpha_{\mathrm{max}}]$.

\paragraph{Cutoffs.} An intermediate spectral dimension between $0$ and $4$ is only resolved if $\frac{j_{\mathrm{max}}}{j_\mathrm{min}}\gg 1$. 
While $\jmin > 0$ can be motivated from LQG\footnote{Excluding $j=0$ is an additional condition that does not follow from constrained $BF$-quantization. Note that the action of the Laplace operator in Eq.~\eqref{eq:defLaplace} is ill-defined for $j=0$ as the 4-volume vanishes in this case.}, $\jmax<\infty$ is introduced for numerical purposes\footnote{Alternatively, a cutoff $j_{\mathrm{max}}\sim\pi/(\Lambda l_{\mathrm{P}}^2)$ could be introduced by employing the quantum deformation $\SUT_q$~\cite{Turaev:1992hq,Major:1995yz,Han:2010pz,Fairbairn:2010cp,Dittrich:2018dvs,Dupuis:2020ndx,Dupuis:2013lka} with $q$ a root of unity related to a cosmological constant $\Lambda > 0$. However, inserting the observed value of $\Lambda$, this $\jmax$ is much larger than what can be numerically implemented.}. 
To recover a physical interpretation of our results, we therefore need to consider the limit $j_{\mathrm{max}}\rightarrow\infty$. 
In this limit, the intermediate regime of $\Ds$ extends to infinite $\tau$.

\paragraph{Barbero-Immirzi parameter.} For $1$-periodic frusta, $\bi$ controls the spacing of allowed $\SUT$ spins according to the $Y_\bi$-map. Thus, changing the value of $\bi$ results in a rescaling of the allowed spins, which can be absorbed into $\tau$.
In contrast, for $\mathcal{N}>1$, $\bi$ controls the relative phase of the oscillations in the semi-classical vertex amplitude of Eq.~\eqref{eq:Spin4Vamp} and is therefore expected to have a non-trivial effect on $\Ds$.


\paragraph{Semi-classical amplitudes.} 
A direct comparison of $\Ds$ computed from extrapolated and semi-classical amplitudes is presented in the right panel of Fig.~\ref{fig:specdim_quantum} for $\alpha = 0.5$. With semi-classical amplitudes, the spectral dimension is constant in the intermediate regime. 
%
%
Considering the analytical explanations of~\cite{Steinhaus:2018aav}, the deviation between the two curves is a consequence of the different \emph{effective scaling} $\sfs$, defined as
\begin{equation}\label{eq:effective scaling}
    \sfs \defeq -\frac{\dd{\log\hat{\mathcal{A}}}}{\dd{\log\lambda}}\,,
\end{equation}
where $\lambda$ is the parameter of a uniform rescaling of the spins. Numerical results in~\cite{Jercher:2023rno} show that for $\alpha > 0.24$ the scaling $\gamma$ of extrapolated amplitudes is larger than the semi-classical value $\gamma_{\textsc{const}} = 9-12\alpha$, implying a larger value $\Ds$.
Also, since the effective scaling of the extrapolated amplitudes is non-constant, there is a non-constant flow of $\Ds$ to the semi-classical constant value at larger scales.

The different behavior of $\Ds$ due to the different amplitudes appears in the regime $10^{-2} < \tau < 10^2$ and is of quantitative nature. Although providing an increasingly bad approximation at low spins, this suggests that the semi-classical amplitude is sufficient for extracting the spectral dimension on large scales. In particular, there is agreement with the quantum amplitude results for scales $\tau > 10^2$, even in the limit of infinite upper cutoff. Therefore, we are going to employ semi-classical amplitudes for the rest of this work, offering the following two advantages: 1) a cosmological constant can be straightforwardly included via an ad hoc deformation of the amplitudes~\cite{Han:2011aa,Bahr:2018vv}, and 2) semi-classical amplitudes allow studying the spectral dimension at higher periodicities with $\mathcal{N}>1$.

\subsection{Cosmological constant}\label{subsec:Cosmological constant} 

A way to add a cosmological constant $\Lambda$ to the simplicial Euclidean EPRL-FK model was introduced in~\cite{Han:2011aa} and generalized to arbitrary $4$-dimensional polyhedra in~\cite{Bahr:2018vv}. 
In essence, the vertex amplitudes of the model are deformed while keeping the boundary Hilbert space fixed. 
Relating the deformation parameter with $\Lambda$, the asymptotic vertex amplitude yields the Regge action with a cosmological constant as in Eq.~\eqref{eq:Spin4Vamp}. The introduction of a $\Lambda$ term allows considering oscillations even when $\mathcal{N} = 1$. In this setting, the sign of the cosmological constant is irrelevant for the amplitudes, as Eq.~\eqref{eq:cc oscillations} below shows. These oscillations are of a particular type in comparison to cases of non-vanishing Regge curvature, as 
they are of simple cosine shape and enter with the $4$-volume which scales quadratically in the spins. 
Despite its simple form, the $\Lambda$ term allows to get a first glimpse of the effects of oscillating amplitudes on the spectral dimension.

\begin{figure}
    \centering
    \begin{subfigure}{0.47\textwidth}
    \includegraphics[width=\linewidth]{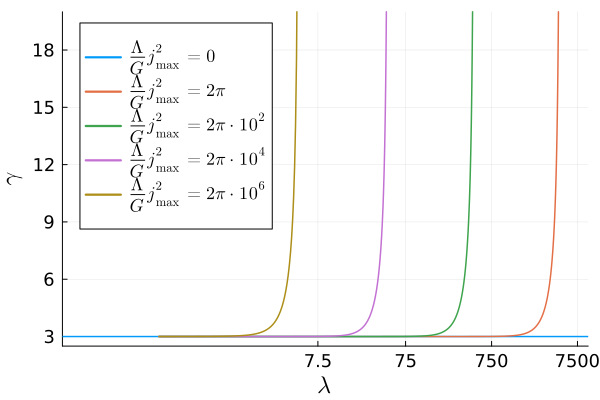}
    \end{subfigure}\hspace{0.05\textwidth}
    \begin{subfigure}{0.45\textwidth}
    \includegraphics[width=\linewidth]{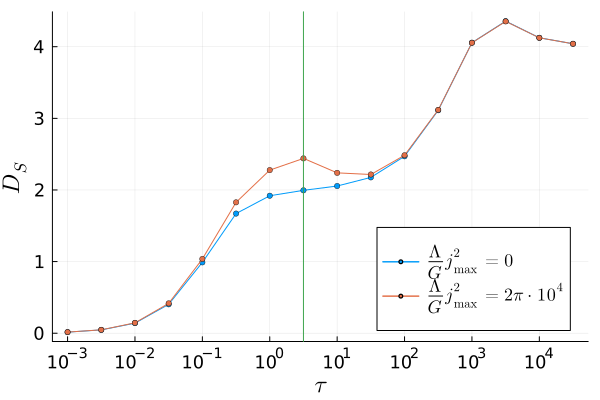}
    \end{subfigure}
    \caption{Left: effective scaling $\sfs$ of the $\Lambda$ amplitude as a function of the scaling parameter $\lambda < \lambda_0^{(\Lambda)}$. At $\lambda = \lambda_0^{(\Lambda)}$, the amplitude completes its first oscillation. Right: expectation value of $\Ds$ with vanishing and non-vanishing $\Lambda$. Both plots are generated for $\alpha = 0.5$.}
    \label{fig:specdim1_Lambda}
\end{figure}

Writing $\hat{\mathcal{A}} = \upmu\,\mathcal{C}$, with scaling part $\upmu$, the oscillating part $\mathcal{C}$ is given by
\begin{equation}\label{eq:cc oscillations}
\mathcal{C}(j,k) = \cos\left(\frac{\Lambda}{\GN} jk\right)-\frac{\mathfrak{Re}\{\det H\}}{\vert \det H\vert}\,,
\end{equation}
where $\det H$ is the Hessian determinant, see Eq.~\eqref{eq:determH}, and $\mathfrak{Re}\{\det H\}/\abs{\det H}$ is a scale-invariant function of $(j,k)$. For a given upper cutoff $j_{\mathrm{max}}$, the amplitudes are not altered if $\frac{\Lambda}{\GN}\ll \frac{1}{j_{\mathrm{max}}^2}$. However, for $\frac{\Lambda}{\GN}\gtrsim \frac{1}{j_{\mathrm{max}}^2}$ effects of non-vanishing $\Lambda$ become relevant as Fig.~\ref{fig:specdim1_Lambda} shows. 

The strongest deviation from the $\Lambda = 0$ case is localized around the scale at which the first oscillation takes place, defined by $\lambda_0^{(\Lambda)} \defeq \left(\frac{\GN}{\Lambda}\arccos\frac{\mathfrak{Re}\{\det H\}}{\vert \det H\vert}\right)^{1/2}$, and marked by a green vertical line in the right panel of Fig.~\ref{fig:specdim1_Lambda}. If this scale is in the regime where $0 < \Ds <4$, the spectral dimension is larger than for $\Lambda = 0$. To visualize this, the left panel of Fig.~\ref{fig:specdim1_Lambda} displays the effective scaling $\sfs$, defined in Eq.~\eqref{eq:effective scaling}, of the amplitude for all $\lambda < \lambda_0^{(\Lambda)}$. For $\lambda > \lambda_0^{(\Lambda)}$, the scaling oscillates rapidly, which however does not affect the spectral dimension. 
As before, a value $\sfs>9-12\alpha$ implies a larger spectral dimension which is the case for all $\Lambda\neq 0$. 

The results presented here are corroborated by the integrable toy model computations of~\cite{Jercher:2023rno}. Therein, the expectation value of the return probability is computed analytically on a 1$d$ lattice for an oscillating measure with constant shift, reminiscent of Eq.~\eqref{eq:cc oscillations}. Indeed, one finds that the oscillating cosine term yields a local maximum in $\Ds$ when its argument is $\approx 1$, while the constant term yields the intermediate plateau of $\Ds$. The final flow of $\Ds$ is then a superposition of these two effects, leading precisely to the shape observed in Fig.~\ref{fig:specdim1_Lambda}.

\subsection{Spectral dimension of 2-periodic frusta}\label{subsec:2-periodic spectral dimension}


In this section, we employ 2-periodic semi-classical amplitudes to study the spectral dimension. The $2\times 2$ Laplace operator in momentum space and its eigenvalues are presented in Appendix~\ref{app:Laplacian spectrum}. 
%
%
%
%
Compared to $\mathcal{N}=1$, the expression of the $2$-periodic return probability is more involved, in particular because the momentum integrals entering $P_2(\tau)$
%
%
do not factorize. Thus, full $4$-dimensional momentum integration is required, leading to larger numerical computation times. The expectation value of $P_2(\tau)$ is computed according to Eq.~\eqref{eq:return_prob_exp} with cutoffs chosen as $j_{\mathrm{min}}=\frac{1}{2}$ and $j_{\mathrm{max}}=201$. Since for $\mathcal{N}=2$, the $3$-cubes are not restricted to equal size, genuine frustum geometries with non-vanishing Regge curvature arise, exhibiting oscillations even for $\Lambda = 0$, which we assume here if not stated otherwise. 

\paragraph{Large $\pmb{\GN}$.} First, the limiting case $\GN\rightarrow\infty$ is considered for which $\Sreg\rightarrow 0$. We have checked that setting $\GN =10^{10}$ effectively captures this limit as larger $\GN$ do not change the results. The numerical results detailed in~\cite{Jercher:2023rno} show that there exists an intermediate flow of the spectral dimension for $\alpha\in[\alpha_{\mathrm{min}},\alpha_{\mathrm{max}}]$ with $\alpha_{\mathrm{min}}\approx 0.68$ and $\alpha_{\mathrm{max}}\approx 0.7$. Compared to the $1$-periodic case, the size of this interval of $\alpha$ is smaller which is in accordance with the findings of~\cite{Steinhaus:2018aav}. 
Following the arguments of Sec.~\ref{subsec:Analytical estimate of the spectral dimension}, this is expected from the higher powers of the amplitudes.

\begin{figure}
    \centering
    \begin{subfigure}{0.45\textwidth}
    \includegraphics[width=\linewidth]{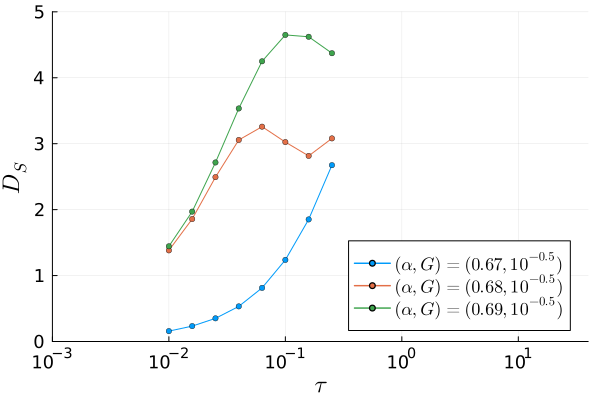}
    \end{subfigure}\hspace{0.05\textwidth}
    \begin{subfigure}{0.45\textwidth}
    \includegraphics[width=\linewidth]{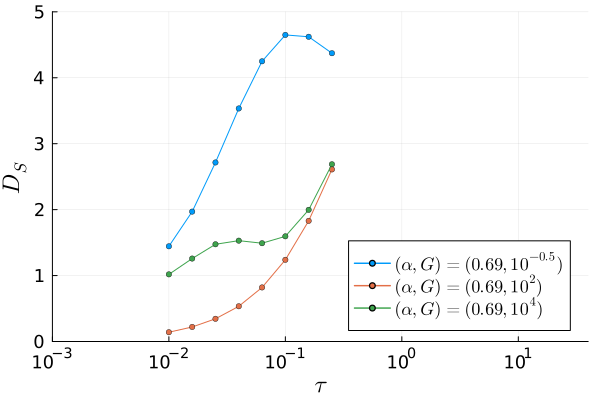}
    \end{subfigure}
    \caption{2-periodic $\Ds$ for varying $\alpha$ (left) and Newton's constant $\GN$ (right). Parameters are set to $\bi = 1/3$, $\Lambda = 0$, $\jmin =1/2$ and $\jmax=201$.}
    \label{fig:specdim2}
\end{figure}

\paragraph{Regge curvature oscillations.} Oscillations induced from Regge curvature  are expected to become relevant for $\GN$ being comparable to $\Sreg$. In this case, the numerical results of $\Ds$ are depicted in Fig.~\ref{fig:specdim2} for varying $\alpha$ (left panel) and varying $\GN$ (right panel). We find that the flow of $\Ds$ is highly sensitive to the value of $\GN$. In comparison to the $\GN\rightarrow\infty$ case, the Regge oscillations induce either a positive, negative or negligible correction, depending intricately on the value of $\GN$.
In contrast to previous cases, the flow of the $2$-periodic spectral dimension is not straightforwardly understood by considering the effective scaling $\sfs$ of the amplitudes.

\paragraph{Intermediate regime.} By computing the spectral dimension for various $\GN$ and a wide range of $\alpha$, we observe that the $\alpha$-interval $[\alpha_{\min},\alpha_{\max}]$, for which an intermediate spectral dimension exists, depends on $\GN$. For instance, at $\GN = 10^{-0.5}$, we find $\alpha_{\mathrm{min}} \approx 0.67$ and $\alpha_{\mathrm{max}}\approx 0.69$, while at $\GN = 10^2$ we obtain $\alpha_{\mathrm{min}}\approx 0.69$ and $\alpha_{\mathrm{max}} \approx 0.71$. Furthermore, in contrast to previous cases, the intermediate values of $\Ds$ do not depend approximately linearly on $\alpha$. 

\paragraph{$\pmb{\Lambda\neq 0}$.} Following Eq.~\eqref{eq:Spin4Vamp}, $\Lambda\neq 0$ introduces a phase shift to the cosine term containing the parameter $\bi$. The numerical results of~\cite{Jercher:2023rno} show that effects of $\Lambda\neq 0$ become important only for $\Lambda$ not much smaller than $\GN$. In this case, $\Lambda$ significantly affects the flow of $\Ds$. In comparison to the $1$-periodic case, we observe that 1) high frequency oscillations due to $\Lambda\gg 1$ are not negligible, 2) the region of $\tau$'s, where $\Lambda$ leads to a deviation in the flow of $\Ds$, is not as clearly localized as previously, 3) the direction in which $\Ds$ is altered by the presence of $\Lambda$, so to lower or larger values than for $\Lambda = 0$, is obscured. Similar to the above, one finds that the range $[\alpha_{\mathrm{min}},\alpha_{\mathrm{max}}]$, for which an intermediate dimension exists, depends on the value of $\Lambda$.

\subsection{An analytical estimate and the thermodynamic limit}\label{subsec:Analytical estimate of the spectral dimension}

An analytical estimate of the spectral dimension is summarized in this section, a derivation of which is given in Appendix~\ref{app:Derivation of analytical estimate}. First, define $r^2 \defeq \frac{1}{N}\sum j_f^2$, with $N = 2\mathcal{N}$, as a ``radial'' coordinate and $\Omega$ as the  angular part in the space of configurations $j_f$. Then, the amplitudes utilized in this work can be written as $\mathcal{A}_v = r^{-\sfsc}\mathcal{C}_v(r,\Omega)$, with a scaling part and a \emph{correction} term $\mathcal{C}_v$ capturing quantum amplitude effects and oscillations. Following~\cite{Sahlmann:2009rk,Calcagni:2014cza,Steinhaus:2018aav} and assuming that the Laplace operator obeys $\Delta(j_f)\sim\frac{1}{r}\Delta$ with $\Delta$ the Laplace matrix on the equilateral hypercubical lattice, it has been shown in~\cite{Jercher:2023rno} that the spectral dimension takes the form
\begin{equation}\label{eq:approximation D}
    \Ds = 2(\sfsc V - N)-2\sum_v\frac{\int\dd{\Omega}\dd{r}\;r^{N-1}\frac{r}{\mathcal{C}_v}\frac{\partial\mathcal{C}_v}{\partial r}\prod_{v'}\mathcal{A}_{v'}\Tr\left(\e^{\frac{\tau\Delta}{r}}\right)}{\int\dd{\Omega}\dd{r}\;r^{N-1}\prod_{v'}\mathcal{A}_{v'}\Tr\left(\e^{\frac{\tau\Delta}{r}}\right)}-2\frac{\partial I}{\langle P\rangle Z}\,.
\end{equation}
Here, $V = \mathcal{N}^4$ is the number of vertices, $\partial I$ is a boundary term defined in Eq.~\eqref{eq:bdr term} and the spin summation has been replaced by an integration, justified for $\jmax/\jmin\gg 1$. The essential step to arrive at this expression is to exchange $\Tr\left(\frac{\tau\Delta}{r}\e^{\frac{\tau\Delta}{r}}\right)$, entering $\Ds$ and arising from the log-log derivative with respect to $\tau$, with $-r\frac{\partial}{\partial r}\Tr\left(\e^{\frac{\tau\Delta}{r}}\right)$. An ensuing integration by parts yields the first two terms complemented by the boundary term $\partial I$. 

For purely scaling amplitudes, $\mathcal{C}_v = \text{const.}$, and negligible boundary term, $\partial I = 0$, the above yields $\Ds^{\alpha} = 2\left((9-12\alpha)V-N\right)$ for the intermediate spectral dimension, consistent with~\cite{Steinhaus:2018aav}.

For $\alpha\gtrsim 0.24$, extrapolated 1-periodic quantum amplitudes exhibit an effective scaling larger than the semi-classical value which implies that $-\frac{r}{\mathcal{C}_v}\frac{\partial\mathcal{C}_v}{\partial r} > 0$, and hence, that the spectral dimension is corrected to a larger value. This is exactly what we observed in Fig.~\ref{fig:specdim_quantum}. 

$1$-periodic frusta with $\Lambda\neq 0$ exhibit the correction term in Eq.~\eqref{eq:cc oscillations} which hits many zeros, such that the effective scaling diverges at these points. Consequently, Eq.~\eqref{eq:approximation D} is only valid on a restricted domain. Still, for $r_0^{(\Lambda)}$ the first zero of $\mathcal{C}_v$, the effective scaling of $\mathcal{C}_v$ is larger than zero in the interval $[j_{\mathrm{min}},r_0^{(\Lambda)}]$ thus suggesting a positive correction to $\Ds$, observed also numerically in Sec.~\ref{subsec:Cosmological constant}. We conjecture that in the remaining integration range, the rapid oscillations of the scaling behavior average out, but it is currently not in reach to substantiate this statement further.

$2$-periodic frusta exhibit a highly non-trivial correction term $\mathcal{C}_v(r,\Omega)$ with the integral in Eq.~\eqref{eq:approximation D} having a highly restricted domain of validity. Due to these obstacles we can only draw an estimate of $\Ds$ for large values of $\GN$ where $\mathcal{C}_v$ is approximately constant. In this case, $\Ds = 32(9-12\alpha)-8$, suggesting that an intermediate spectral dimension can be observed for $\alpha\in[0.72,0.73]$. The results of Sec.~\ref{subsec:2-periodic spectral dimension} are in partial accordance with this prediction. Qualitatively, we find a flow to intermediate values of $\Ds$ controlled solely by $\alpha$. Also, the window $[\alpha_{\min},\alpha_{\max}]$ for such a regime is smaller compared to the $1$-periodic case. However, the quantitative predictions do not agree with the numerical results. In particular, $\alpha_{\mathrm{min}},\alpha_{\mathrm{max}}$ and $\Ds$ of the analytical derivation appear to be shifted with respect to the numerical values. We suspect the assumption of a scaling Laplacian for $\mathcal{N}>1$ as the main source of this discrepancy.

\paragraph{Thermodynamic limit.} The numerical computations of this chapter hinge on the upper cutoff $\jmax$ and most importantly on the assumption of $\mathcal{N}$-periodicity, both of which ultimately have to be removed. In the limit $\jmax\rightarrow\infty$, the intermediate regime of the spectral dimension extends to arbitrarily large scales $\tau$. Taking the limit $\mathcal{N}\rightarrow\infty$ thereafter, the results of~\cite{Steinhaus:2018aav} and this chapter suggest that the interval $[\alpha_{\mathrm{min}},\alpha_{\mathrm{max}}]$ for which such an intermediate regime exists shrinks to a point $\alpha_*$. Since the number of degrees of freedom, $2\mathcal{N}$, and the combinatorial length, $L = \mathcal{N}$, are taken to infinity while keeping their ratio fixed, $\mathcal{N}\rightarrow\infty$ corresponds to a thermodynamic limit. As noted already in~\cite{Steinhaus:2018aav}, the point $\alpha_*$ is reminiscent of a phase transition as the large-scale spectral dimension changes discontinuously from $0$ to $4$.  Following the results of Sec.~\ref{subsec:2-periodic spectral dimension}, the interval $[\alpha_{\min},\alpha_{\max}]$ depends on $\GN$ and $\Lambda$, suggesting that in this case $\alpha_*$ is a function of the parameters $(\GN,\Lambda)$ (If $\bi$ is not fixed, it is expected to have a similar effect.). Consequently, $\alpha_*(\GN,\bi,\Lambda)$ defines a non-trivial embedded surface in the parameter space which, tentatively speaking, marks the critical surface of a phase transition. Future investigations will hopefully allow to further quantify $\alpha_*$ at large $\mathcal{N}$ such that its value can be compared with the renormalization group results of~\cite{Bahr:2016hwc,Bahr:2017kr,Bahr:2018gwf,Steinhaus:2020lgb}.\\

\begin{center}
    \textit{Summary.}
\end{center}

\noindent The overarching insight of this chapter is that $\mathcal{N}$-periodic spin-foam frusta exhibit a large-scale spectral dimension of $4$ and a non-trivial flow at lower scales, due to the superposition of geometries, controlled by the face amplitude parameter $\alpha$. In comparison to semi-classical computations at $\Lambda = 0$, the 1-periodic spectral dimension is subject to additive corrections when employing extrapolated quantum amplitudes or semi-classical amplitudes with $\Lambda\neq 0$. Both effects can be understood qualitatively from the effective scaling $\sfs$. $2$-periodic amplitudes with an intricate oscillatory behavior lead to a flow of the spectral dimension which sensitively depends on the full set of parameters $(\alpha, \GN,\bi,\Lambda)$. Based on the analytical and numerical results, we conjecture the existence of a critical $\alpha_*$ in the limit $\mathcal{N}\rightarrow\infty$ at which the large-scale spectral dimension changes discontinuously from $0$ to $4$ and which defines a critical surface in the parameter space $(\GN,\bi,\Lambda)$. In contrast to tensor models~\cite{Gurau:2013cbh} and the non-geometric phases of CDT~\cite{Loll:2019rdj} with fractal dimensions, our results suggest that spin-foam frusta exhibit a phase with $\Ds = 4$, thus satisfying an important consistency check. A comparison to the spectral dimension flow of CDT~\cite{Ambjorn:2005qt,Reitz:2022dbj} requires tuning to $\alpha_*$, constituting an intriguing future research direction. 

\begin{center}
    \textit{Closing remarks.}
\end{center}

\paragraph{Lifting the restriction to frusta.} Spin-foam frusta with their inherent high degree of symmetry present a strong restriction of the quantum geometry compared to the general case. If all restrictions on the geometry are lifted, it is advantageous to directly work on triangulations, where semi-classical amplitudes are well studied~\cite{Conrady:2008mk,Barrett:2009gg,Barrett:2009mw,Kaminski:2017eew,Liu:2018gfc,Simao:2021qno,Han:2021bln} and more numerical methods are available~\cite{Dona:2019dkf,Gozzini:2021kbt,Asante:2020qpa,Han:2021kll,Asante:2022lnp,Asante:2024eft,Steinhaus:2024qov}. In this case, new challenges are posed by 1) defining $\mathcal{N}$-periodicity, 2) dealing with vector geometries that have no well-defined 4-volume~\cite{Barrett:2009gg}, and 3) constructing the Laplace operator on quantum geometries with non-matching $4$-simplices.

\paragraph{Monte Carlo methods.} Monte Carlo (MC) methods serve to compute expectation values without exponentially scaling with the number of variables of the system. The generically complex spin-foam amplitudes do not define a probability distribution to be used for MC methods. A way around this might be to define Markov Chain MC on Lefschetz thimbles of the spin-foam partition function~\cite{Han:2020npv}, where the integration contour is changed such that the imaginary part of the system is constant and thus non-oscillatory. Alternatively, a potential strategy might be to propose a uniform probability distribution as in~\cite{Dona:2023myv} or an importance sampling as in~\cite{Steinhaus:2024qov}.

\paragraph{Lorentzian signature.} The studies of the spectral dimension presented here assumed a Euclidean signature. It will be interesting to see how these concepts generalize to the Lorentzian setting, e.g. as done in causal set theory~\cite{Eichhorn:2013ova,Carlip:2015mra}. As a first step towards this direction, classical and quantum dynamics of Lorentzian frusta are studied in the next two chapters.
\chapter[\textsc{Spatially Flat Cosmology in Lorentzian Regge Calculus}]{Spatially Flat Cosmology in \\Lorentzian Regge Calculus}\label{chapter:LRC}

Quantum cosmology with its high degree of symmetry presents an ideal setting for performing explicit analytical and numerical computations within QG. It allows testing whether a prescribed semi-classical and/or continuum limit yields the continuum Friedmann dynamics of GR~\cite{Ellis2012} at late times and what kind of quantum effects, such as a Big Bounce~\cite{Brandenberger:2016vhg,Oriti:2016qtz,Agullo:2016tjh}, arise at early times. Advancing beyond the sector of spatially homogeneous geometries, quantum cosmology holds the promise for establishing a direct bridge between QG and cosmological observations~\cite{Ashtekar:2021kfp,Ashtekar:2020gec}.

One way to identify a cosmological subsector in spin-foams is to symmetry reduce to those quantum geometric variables which capture the desired spatially homogeneous and isotropic dynamics. Significant steps towards this goal were achieved in~\cite{Bianchi:2010zs,Bianchi:2011ym,Rennert:2013pfa,Sarno:2018ses,Gozzini:2019nbo,Frisoni:2022urv,Frisoni:2023lvb} within the full EPRL spin-foam model.  Most recently~\cite{Han:2024ydv}, the complex critical point method~\cite{Han:2021kll,Han:2023cen,Han:2024lti} has been applied to investigate cosmological transition amplitudes in the EPRL-CH model. 

Effective spin-foams~\cite{Asante:2021zzh,Asante:2020iwm,Asante:2020qpa} have been applied to the cosmological setting in~\cite{Dittrich:2023rcr,Dittrich:2021gww,Asante:2021phx} extending earlier works of quantum Regge cosmology~\cite{CorreiadaSilva:1999cg,CorreiadaSilva:1999es,Hartle:1985wr,Hartle:1986up,Liu:2015gpa}. Therein, equilateral triangulations of the $3$-sphere describing positively curved spatial slices, i.e.~spatially spherical geometries, have been frequently considered. Already in these simple models, intriguing physical features have been revealed. In particular, causally irregular configurations in the form of more or less than two light cones located at spacelike subsimplices have been detected, some of which can be interpreted as spatial topology change. Such configurations are typically obtained if the subcells connecting the spatial $3$-spheres are spacelike.  


In this chapter we lay the foundation for setting up an effective spin-foam model of spatially flat cosmology. To that end, we utilize 4-frusta as in the previous section but in Lorentzian signature, and study their classical dynamics within Lorentzian Regge calculus. A particularly important feature of Lorentzian 4-frusta is that the boundary 3-frusta can be either spacelike or timelike, bearing non-trivial consequences on the causal structure of the discrete spacetime. To render the dynamics non-trivial we consider as matter content of the universe a massless scalar field coupled to the Regge action, following the suggestion by Hamber~\cite{Hamber:1993gn}. Thus, it can also be investigated whether the scalar field can be used to deparametrize the theory, expressing the cosmological dynamics relationally~\cite{Rovelli:1990ph,Hoehn:2019fsy}.

\section{Kinematics}\label{sec:Spatially flat cosmology in Regge calculus}

Regge calculus~\cite{Regge:1961ct} is a discrete gravitational theory on simplicial manifolds $\Delta$ with the length of edges $\{l_e\}$ dynamical, encountered in a similar form already in the previous chapter. In a Lorentzian setting, a simplicial manifold consists of intrinsically flat Lorentzian $4$-simplices $\sigma\in\Delta$. Curvature is located at triangles $t\in\Delta$ and captured by deficit angles $\delta_t(\{l_e\})$, lying in the space orthogonal to $t$. For $t$ spacelike (timelike), this space is isomorphic to $\R^{1,1}$ ($\R^2$).\footnote{For the remainder of this chapter, we neglect the presence of lightlike edges, triangles and tetrahedra. In fact, the Regge action in $d$ dimensions is insensitive to the causal character of $(d-k)$-dimensional subcells for $k>2$ including edges in the $4$-dimensional case. Furthermore, since the volumes of $(d-2)$ cells enter the Regge action linearly, contributions from lightlike hinges (with vanishing volume) are zero. Finally, $(d-1)$-dimensional null cells require the definition of angles between lightlike vectors for which we refer the reader to~\cite{Sorkin:2019llw}. In this work, we choose boundary data such that $(d-1)$-dimensional cells are not null.} Lorentzian deficit angles are defined as $\delta_t^{\mathrm{L},\pm} = \mp i 2\pi  -\sum_{\sigma\supset t}\psi_{\sigma,t}^{\mathrm{L},\pm}$, following the conventions of~\cite{Sorkin:2019llw,Asante:2021zzh}, where the $\psi_{\sigma,t}$ are dihedral angles. \enquote{$\pm$} indicates a choice of sign which becomes important for causally irregular configurations~\cite{Sorkin:2019llw,Asante:2021zzh,Asante:2021phx} detailed in Sec.~\ref{sec:causal regularity}. The bulk action of Lorentzian Regge calculus is given by~\cite{Asante:2021zzh,Sorkin:2019llw,Regge:1961ct}
\begin{equation}\label{eq:general Regge action}
\Sreg[\{l_e\}] = \sum_{t\in\Delta:\text{ tl}}\abs{A_t(\{l_e\})}\delta_t^{\mathrm{E}}(\{l_e\})+\sum_{t\in\Delta:\text{ sl}}\abs{A_t(\{l_e\})}\delta_t^{\mathrm{L},\pm}(\{l_e\})\,,
\end{equation}
with $\abs{A_t}$ the absolute value of the triangle area. In the presence of a boundary, $\partial\Delta\neq\emptyset$, $\Sreg$ attains boundary terms such that overall additivity is ensured. The equations of motion are given by $\pdv{\Sreg}{l_e} = 0$ supplemented by the Schl\"{a}fli identity, $\sum_t \abs{A_t}\pdv{\delta_t}{l_e} = 0$. 

Regge calculus can be formulated in terms of different variables~\cite{Barrett:1994nn,Dittrich:2008va}, with area variables~\cite{Asante:2018wqy,Makela:2000ej,Barrett:1997tx} playing a particularly important role here due to their connection to LQG and spin-foams. The relation between area and length variables is in general not invertible, and area variables need to be further constrained to yield the dynamical equations of length Regge calculus~\cite{Barrett:1997tx,Dittrich:2008va,Asante:2021zzh}. However, in the symmetry reduced setting introduced below, the relation of area and length is globally invertible and the constraints mentioned before trivialize. 

\subsection{Lorentzian 4-frusta}\label{sec:Lorentzian frustum}

As argued in~\cite{Bahr:2017eyi} and the previous chapter, $4$-dimensional frusta (see again Fig.~\ref{fig:4-frustum}) are ideal to capture spatially flat, homogeneous and isotropic discrete geometries. In contrast to Chapter~\ref{chapter:specdim} however, we consider here \emph{Lorentzian} $4$-frusta, where the two $3$-cubes are spacelike and where the six boundary $3$-frusta can be either spacelike or timelike. Its entire geometry is captured by the length of the $3$-cube edges, $l_0$ and $l_1$, and its height $H$, defined as the distance between the midpoints of the two $3$-cubes. The general notions of Regge calculus introduced above are straightforwardly adapted to this setting.  

Gluing $4$-frusta along boundary $3$-frusta in spatial direction and along boundary $3$-cubes in temporal direction, one obtains an extended cosmological spacetime of hypercubical combinatorics $\mathcal{X}^{(4)}_\mathcal{V}$ with cubulated slices labelled by $n\in\{0,...,\mathcal{V}\}$. We say that a $4$-frustum lies in a \enquote{slab} between slices $n$ and $n+1$.

\begin{figure}
    \centering
    \includegraphics[width=0.7\textwidth]{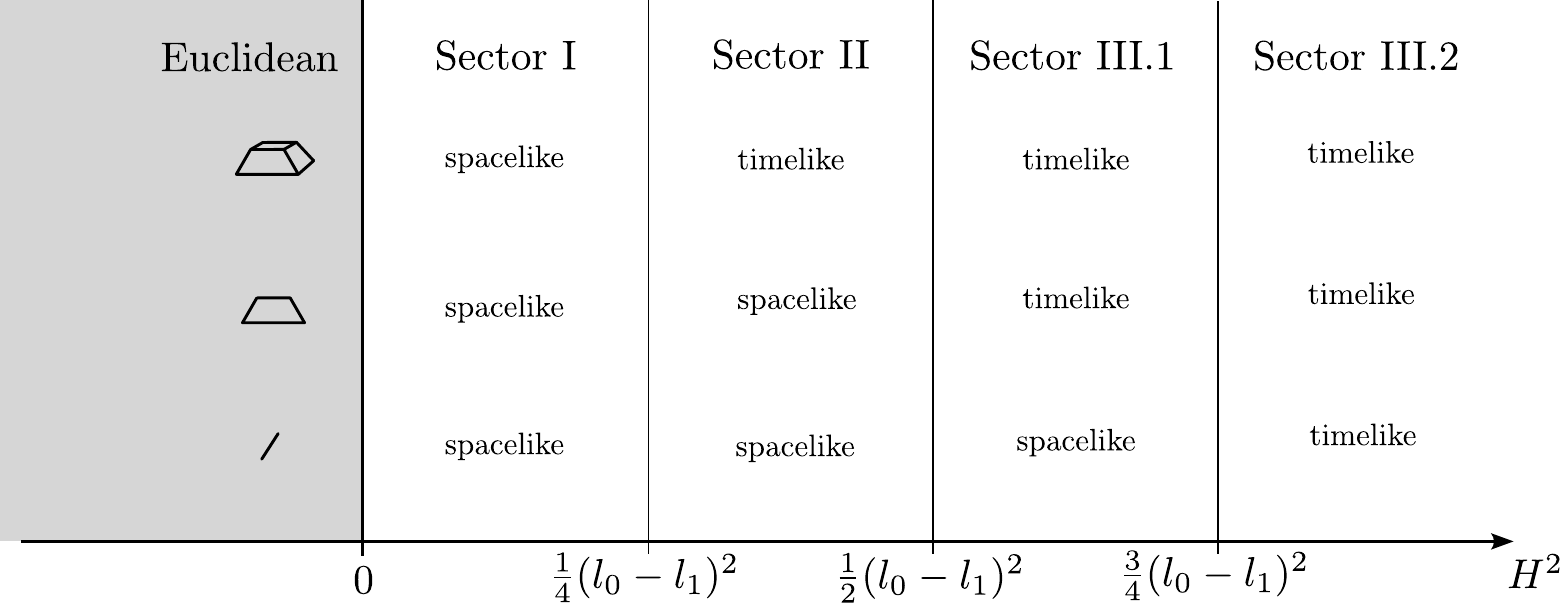}
    \caption{Sectors, determined by the relation of height $H$ and difference in spacelike edge length $(l_0-l_1)$, in which the subcells of Lorentzian $4$-frusta take different signature.}
    \label{fig:regions}
\end{figure}

The geometry of $4$-frusta as a function of $(l_n,l_{n+1},H_n)$ is detailed in Appendix~\ref{app:Lorentzian frustum}. Most importantly, the squared volume of subcells connecting neighboring slices, i.e. 3-frusta, trapezoids and edges, carries information on the causal structure. If it is positive (negative), the corresponding subcell is timelike (spacelike). In the symmetry reduced setting, the ratio $\frac{l_n-l_{n+1}}{H_n}$ fully characterizes the causal character of subcells suggesting separating the $4$-frusta configurations into three sectors summarized in Fig.~\ref{fig:regions}. Embeddability of $4$-frusta into $\R^{1,3}$ requires the height to be real. Otherwise, the configuration corresponds to a Euclidean $4$-frustum, employed in Chapter~\ref{chapter:specdim}.  As argued below, $H_n$ is naturally associated to the lapse function $N$ which, in the simple context of homogeneous cosmology, relates the Lorentzian and Euclidean sector via an analytical continuation, commonly referred to as Wick rotation.

Dihedral and deficit angles are a crucial ingredient for the Lorentzian Regge action. Here, there are two types of dihedral angles, associated either to spacelike squares or to trapezoids connecting slices. Crucially, the explicit expressions of these angles depend on the causal characters of subcells which are given in Appendix~\ref{app:Lorentzian angles} as a function of $(l_n,l_{n+1},H_n)$.

\subsection{Lorentzian Regge action for spatially flat cosmology}\label{sec:Lorentzian Regge action for spatially flat cosmology}

The underlying combinatorics of the model are hypercubical, denoted by $\mathcal{X}^{(4)}_{\mathcal{V}}$, such that every square or trapezoid is shared by four $4$-frusta. As a result, the Euclidean and Lorentzian exterior boundary deficit angles for a single frustum are defined as $\delta^{\mathrm{E}} = \frac{\pi}{2} - \psi^{\mathrm{E}}$ and $\delta^{\mathrm{L},\pm} = \mp i\frac{\pi}{2} - \psi^{\mathrm{L},\pm}$, respectively, where $\psi$ are the corresponding dihedral angles. Since the single $4$-frustum consists of six squares in slice $n$, twelve trapezoids and six squares in slice $n+1$, spatial homogeneity then implies that the Regge action is given by
\begin{equation}\label{eq:slice Regge action}
\begin{aligned}
S_{\mathrm{R}}^{(n)} = 6l_n^2\left(\mp i\frac{\pi}{2}-\varphi_{nn+1}\right)+6l_{n+1}^2\left(\mp i\frac{\pi}{2}-\varphi_{n+1n}\right)+ 12\abs{k_n}\left[\Theta_{\mathrm{sl}}\left(\mp i\frac{\pi}{2}-\theta_n^{\mathrm{L}}\right) + \Theta_{\mathrm{tl}}\left(\frac{\pi}{2} - \theta_n^{\mathrm{E}}\right)\right]\,.
\end{aligned}
\end{equation}
Here, $\varphi_{nn+1},\varphi_{n+1n}$ are the dihedral angles located at squares in the $n$th, respectively $(n+1)$th slice and $\theta_n$ is the dihedral angle located at trapezoids. $\Theta_{\mathrm{sl,tl}}=0,1$ is a toggle for the trapezoid being spacelike or timelike.

By construction, the $4$-frustum action is additive such that on a discretization $\mathcal{X}^{(4)}_{\mathcal{V}}$ with $L^3$ vertices in spatial direction and $\mathcal{V}+1$ spatial slices, the action is given by $S_{\mathrm{R}} = L^3\sum_{n=0}^{\mathcal{V}-1}S_{\mathrm{R}}^{(n)}$. Additivity of the action per slab implies that the equations of motion for a given geometric quantity, being either $l_n$ or $H_n$, will only depend on neighboring geometric labels. We discuss the consequences of this slicing structure for the dynamics of the model in more detail in Sec.~\ref{sec:Regge equations and deparametrisation}. 

The Lorentzian Regge action is related to the Euclidean theory discussed previously by a Wick rotation in the height variable $H_n$.  Analogous to continuum studies~\cite{Feldbrugge:2017kzv}, the Wick rotation is performed by complexifying the lapse function, represented in the discrete setting by the squared height $H_n^2\rightarrow R_n^2\e^{i\alpha}$, where $\alpha$ is the rotation angle and $R_n^2$ is the modulus. This procedure has been carried out rigorously first in~\cite{Asante:2021phx}. As a consequence of the Wick rotation, the area of trapezoids as well as the dihedral angles are complexified, denoted by $k_n(\alpha),\theta^\pm_n(\alpha)$ and $\varphi^\pm_{nn+1}(\alpha)$. The resulting Regge action can be analytically continued, the details of which crucially depend on the structure of branch cuts and thus on the causal sector under consideration. Remarkably, the Lorentzian Regge action in Eq.~\eqref{eq:slice Regge action} constructed in an ad hoc manner is directly connected to the Euclidean action derived in~\cite{Bahr:2017eyi} from the semi-classical limit of an underlying quantum geometric model. Although this result suggests that there may exist a symmetry restricted Lorentzian spin-foam model whose semi-classical limit leads to the Lorentzian Regge action of spatially flat cosmology, the ultimate answer to this question remains open, at least in (3+1) dimensions. That is because the semi-classical limit of full Lorentzian spin-foam models, being the EPRL-CH model~\cite{Conrady:2010vx,Conrady:2010kc} or the complete Barrett-Crane model (see Chapter~\ref{chapter:cBC}), is not well understood for timelike interfaces and subject to active research, see~\cite{Simao:2021qno}. In Chapter~\ref{chapter:3d cosmology}, a similar scenario will be studied in (2+1) dimensions where differences of Lorentzian Regge calculus and spin-foam asymptotics can be shown to emerge.

\subsection{Causal (ir)regularity}\label{sec:causal regularity}

The Regge action can attain complex values when the imaginary parts of the dihedral angles at a spacelike face do not sum up to $\mp i2\pi$. This indicates the presence of causal irregularities in the form of a degenerate light cone structure, two examples of which are given by the trouser singularity and the yarmulke singularity associated to spatial topology change~\cite{Horowitz:1990qb,Louko:1995jw,Dowker:1999wu}. Choosing \enquote{$+$} in the convention of Lorentzian angles, the amplitude $\e^{i\Sreg}$ is exponentially suppressed for trouser and exponentially enhanced for yarmulke configurations, vice versa for \enquote{$-$}~\cite{Asante:2021zzh}.  

Causal irregularities can be located at spacelike subcells of different dimension. In (3+1) dimensions, causal regularity conditions are therefore referred to as hinge, edge and vertex causality, respectively~\cite{Asante:2021phx}. In the following, we summarize the results of~\cite{Jercher:2023csk} studying these three types of causality conditions for the present model.

\paragraph{Hinge causality.} Hinge causality violations are characterized by a non-vanishing imaginary part of the Lorentzian deficit angle. Each summand of $\mp i\frac{\pi}{2}$ for a dihedral angle counts one light cone crossing. Thus, there are exactly four light rays at a hinge if the imaginary parts of the dihedral angles sum up to $\mp i2\pi$. At spatial squares, causality violations occur if the 3-frusta are spacelike. At trapezoids, causality violations occur if the trapezoid is spacelike. Overall, the hinge causally regular configurations are found in Sector III (see Fig.~\ref{fig:regions}) where trapezoids and $3$-frusta are timelike. The Regge action in this regime is given by
\begin{equation}\label{eq:regular Regge action}
\begin{aligned}
S_{\textsc{iii}}^{(n)} & = 6(l_n^2-l_{n+1}^2)\sinh^{-1}\left(\frac{l_{n+1}-l_n}{\sqrt{4H_n^2-(l_n-l_{n+1})^2}}\right)
& + 12\abs{k_n}\left[\frac{\pi}{2}-\cos^{-1}\left(\frac{(l_n-l_{n+1})^2}{4H_n^2-(l_n-l_{n+1})^2}\right)\right]\,.
\end{aligned}
\end{equation}  
Hinge causality is not sensitive to the signature of the edges connecting the $n$th and $(n+1)$th slice, and therefore, both Sector III.1 and III.2 are considered causally regular at the hinges.  

\paragraph{Edge causality.} The causal structure at a spacelike edge $e$ is considered regular if the intersection $P_e\cap\Sigma$ of the orthogonal space of $e$ containing its midpoint and the union $\Sigma = \cup_{\sigma\supset e}\sigma$ of $4$-cells sharing $e$, contains exactly two light cones. The two types of spacelike edges here are 1) those contained in $3$-cubes, and 2) those connecting neighboring slices if $H_n^2<\frac{3}{4}(l_n-l_{n+1})^2$ is met. Following~\cite{Jercher:2023csk} and referring to the details given therein, one finds that the causal structure at edges of type 1) is regular if and only if the trapezoids are timelike, i.e. if $H_n^2 > \frac{1}{2}(l_n-l_{n+1})^2$ holds. On the other hand, the intersection $P_e\cap\Sigma$ for edges of type 2) contains no light cones at all and thus, edge causality is violated everywhere except in Sector III.2.

\paragraph{Vertex causality.} At any vertex $v$ of the cellular complex, consider the union of all $4$-dimensional building blocks that contain that vertex. Then, the causal structure is said to be regular if the intersection of the light cone at $v$ with the boundary of the union are two disconnected spheres~\cite{Asante:2021phx}. Following~\cite{Jercher:2023csk}, this form of causality is also violated everywhere except in Sector III.2.

The results on causality violations in this simplified setting suggest that higher dimensional causality conditions imply lower-dimensional ones, e.g. vertex causality implies hinge causality. It is conceivable that this arises due to the projections onto orthogonal subspaces of $k$-dimensional cells (e.g. edges, $k=1$ or hinges, $k=2$). The relation of different causality violations in a more general setting, however, remains an interesting question for future research.

\section{A minimally coupled massless scalar field} \label{sec:Minimally coupled scalar field}

Coupling a massless scalar field serves the purpose of rendering the dynamics non-trivial in the absence of spatial curvature and a cosmological constant. Furthermore, it allows investigating whether this scalar field can be used to deparametrize the theory, i.e. to express the cosmological dynamics relationally~\cite{Rovelli:1990ph,Hoehn:2019fsy,Rovelli:2001bz,Dittrich:2005kc,Goeller:2022rsx,Giesel:2012rb} with the field acting as a relational clock~\cite{Domagala:2010bm,Giesel:2012rb}. In spatially flat continuum cosmology, such a field is strictly monotonic in coordinate time and thus ideal to deparametrize the cosmological system. Examples of quantum cosmology approaches where this strategy is routinely followed are loop quantum cosmology~\cite{Bojowald:2008wn,Banerjee:2011qu,Ashtekar:2021kfp}, Wheeler-de Witt cosmology~\cite{Kiefer2012} and GFT condensate cosmology (see Chapter~\ref{chapter:perturbations}).

Starting from a continuum perspective a scalar field is discretized by placing it either on primary or dual vertices of $\mathcal{X}^{(4)}_{\mathcal{V}}$~\cite{Desbrun:2005ug}. It is for our purposes advantageous to discretize the scalar field on primary vertices, $\phi(t)\rightarrow \phi_n$, therefore attaining the index of the containing slice. 

\subsection{Discrete scalar field action}

Adapting the results of~\cite{Hamber:2009wl,Hamber:1993gn} to Lorentzian $4$-frusta, the action of a minimally coupled massless free scalar field on a single slab is defined as
\begin{equation}\label{eq:slice scalar field action}
S_\phi^{(n)} = w_n(\phi_{n+1}-\phi_n)^2\,,\qquad w_n \defeq \frac{(l_n+l_{n+1})^3}{16H_n}\,.
\end{equation}
Clearly, this action exhibits translation ($\phi\mapsto \phi+\text{const.})$ and reflection ($\phi\mapsto -\phi$) symmetry and is quadratic in the fields, in analogy to the continuum. Here, $w_n(l_n,l_{n+1},H_n)$ is a geometrical coefficient that plays the role of the continuum factor $\frac{a^3}{2N}$ in the discrete and yields the correct continuum limit of the coupled gravity and matter system, as we show in Sec.~\ref{sec:Linearisation, deparametrisation and continuum limit}.
%
On an extended lattice $\mathcal{X}^{(4)}_{\mathcal{V}}$ with $L^3$ spatial vertices and $\mathcal{V}+1$ spatial slices, the total action for $\phi$ is given by $S_\phi = L^3\sum_{n = 0}^{\mathcal{V}-1}w_n(\phi_n-\phi_{n+1})^2$, showing the same slicing structure as the Regge action, thus simplifying the equations of motion. The total action of the coupled gravity and matter system is finally given by
\begin{equation}\label{eq:total action}
S_{\mathrm{tot}} = L^3\sum_{n = 0}^{\mathcal{V}-1}S_{\mathrm{tot}}^{(n)} = L^3\sum_{n = 0}^{\mathcal{V}-1}\left(\frac{1}{8\pi \GN}S_{\mathrm{R}}^{(n)}+\lambda S_\phi^{(n)}\right)\,.
\end{equation}
We introduced the factor of Newton's coupling $\GN$ and a parameter $\lambda$ in the sum of the two actions in order to 1) account for the correct dimensions of the two terms, and 2) have a parameter that controls the strength and sign of the matter term. 

\subsection{Equations of motion}

Variation of the scalar field action with respect to $\phi_n$ yields
\begin{equation}\label{eq:discrete scalar field equation}
\phi_n(w_{n-1}+w_n)-(\phi_{n-1}w_{n-1}+\phi_{n+1}w_n) = 0,\qquad n \in \{1,...,\mathcal{V}-1\}\,.
\end{equation}
This system of equations can be solved recursively for a single scalar field $\phi_n$ as a function of the boundary data $(\phi_0,\phi_\mathcal{V})$ and the geometric quantities $\{l_n,H_n\}$. Doing so, one obtains 
\begin{equation}\label{eq:phi_n sol}
\phi_n = \frac{\phi_\mathcal{V}\mathcal{W}_0^{n-1}+\phi_0\mathcal{W}_n^{\mathcal{V}-1}}{\mathcal{W}_0^{\mathcal{V}-1}}\,,
\end{equation} 
where we introduced the symbols $\mathcal{W}_{n_1}^{n_2} \defeq \sum_{m =n_1}^{n_2}\frac{1}{w_m}$, which explicitly depend on the geometric quantities $\{H_n,l_n\}$. Clearly, scalar field solutions for boundary conditions $\phi_0 = \phi_\mathcal{V}$ are constant, i.e. $\phi_n = \phi_0$ for all $n\in\{0,...,\mathcal{V}\}$. To check for monotonicity, it is instructive to re-express the scalar field solutions as $\phi_{n+1}-\phi_n = \frac{1}{w_n\mathcal{W}_0^{\mathcal{V}-1}}(\phi_\mathcal{V}-\phi_0)$. Since each $w_m$ is positive, the scalar field evolves monotonically, the sign of which depends on the sign of $\phi_\mathcal{V}-\phi_0$, i.e. $\mathrm{sgn}\left(\phi_{n+1}-\phi_n\right) = \mathrm{sgn}\left(\phi_\mathcal{V}-\phi_0\right)$. For the later purpose of deparametrizing the system with respect to the scalar field, this is one of the crucial properties.

\subsection{Continuum time limit}\label{sec:scalar field cont limit}

The simple form of the scalar field equations allows for a straightforward definition of a map from discrete to continuum variables, serving as a consistency check of the previous construction. $H_n$ is associated to a finite proper time difference and is therefore identified as $H_n\rightarrow \dd{\tau} = N\dd{t}$ in a continuum time limit. Following~\cite{Bahr:2017eyi}, we propose
\begin{align}
l_n\rightarrow a(t)\,,&\qquad l_{n\pm 1}\rightarrow a(t) \pm \dot{a}(t)\dd{t} +\frac{1}{2}\left(\ddot{a}(t)-\frac{\dot{N}}{N}\dot{a}(t)\right)\dd{t}^2,\label{eq:s cont limit}\\[7pt]
\phi_n \rightarrow \phi(t)\,,&\qquad \phi_{n\pm 1} \rightarrow \phi(t) \pm \dot{\phi}(t)\dd{t}+\frac{1}{2}\left(\ddot{\phi}(t)-\frac{\dot{N}}{N}\dot{\phi}(t)\right)\dd{t}^2,\label{eq:phi cont limit}
\end{align}
where $a(t)$ is the scale factor, $\phi(t)$ the continuum scalar field and dot denotes a derivative with respect to time $t$. Indeed, applying this limit to Eq.~\eqref{eq:discrete scalar field equation}, one finds at lowest order in $\dd{t}$ the familiar equation $\dv{}{t}\left(\frac{a^3}{N}\dot{\phi}\right) = 0$, corresponding to the continuum scalar field equation. 

\section{Regge equations and deparametrization}\label{sec:Regge equations and deparametrisation} 

Given a discretization $\mathcal{X}^{(4)}_{\mathcal{V}}$ of $\mathcal{V}$ slabs and thus $\mathcal{V}+1$ spatial slices, the coupled system exhibits in total $2\mathcal{V}-1$ geometric bulk variables, being $\mathcal{V}-1$ spatial edge lengths and $\mathcal{V}$ height variables. The Regge equations are obtained by imposing $\frac{\partial S_{\mathrm{tot}}}{\partial H_n}\overset{!}{=} 0$, and $\frac{\partial S_{\mathrm{tot}}}{\partial l_n}\overset{!}{=} 0$, forming a coupled system of $2\mathcal{V}-1$ transcendental equations.\footnote{Strictly speaking, the Regge equations are obtained via a variation with respect to the edge length. The height $H_n$ is not such an edge length but the Jacobian between length and height variables is invertible for non-null edges.} In the following, we restrict to causally regular configurations, and only briefly discuss hinge causality violations at the end of this chapter.

\paragraph{Vacuum Regge equations.} To gain a first grasp of the transcendental Regge equations, consider the vacuum case by setting $\lambda = 0$. One finds that $\partial\Sreg/\partial H_n = 0$ holds exclusively for flat geometries, i.e. when $l_n = l_{n+1}$ for all $n$. In this case, all the deficit angles vanish for arbitrary values of $H_n$. Thus, the total Regge action vanishes and becomes independent of the height variables. Note that the Regge equations with respect to the $\{l_n\}$ are similarly satisfied. Overall, imposing the equations of motion on every slice, we obtain an extended globally flat cubulation, the height variables of which become arbitrary.

From these considerations, we learn that 1) the only vacuum solution is the flat solution, consistent with the dynamics of the continuum, where the scale factor $a$ is constant in spatially flat vacuum, and 2) for flat configurations, the height variables $\{H_n\}$ become arbitrary. The emergence of such a symmetry is common for flat solutions in 4d Regge calculus~\cite{Rocek:1981ama,Rocek:1982tj} and signifies the restoration of diffeomorphism symmetry which has been broken initially due to the discretization~\cite{Dittrich:2008pw,Bahr:2009qc}. Due to the spatially homogeneous setting, this symmetry amounts to moving spacelike slices (instead of individual vertices), see also the discussion in~\cite{Bahr:2015gxa}. This is analogous to the arbitrariness of the lapse function in the continuum.  

\subsection{Regge equations with matter} 

For $\mathcal{V}=1$, the only dynamical variable is $H_0$. There exist solutions of the equation $\partial S_{\mathrm{tot}}/\partial H_0$ for non-trivial boundary data $l_0\neq l_1$ and $\phi_0\neq \phi_1$, provided that $\lambda >0$ and 
\begin{equation}\label{eq:ineq for existence of sols}
\left(\frac{2}{l_0+l_1}\right)^2\left(\frac{l_1-l_0}{\phi_1-\phi_0}\right)^2 < \lambda\frac{4\pi\GN}{3}\,.
\end{equation}
From this inequality, we extract that large changes of spatial edge length need to be accompanied by large scalar field differences such that solutions are allowed. Note that configurations violating this inequality only admit Euclidean solutions.

For $\mathcal{V}+1$ slices with $\mathcal{V} > 1$, non-trivial equations for bulk spatial length $\{l_n\}$ and scalar field values $\{\phi_n\}$ must be satisfied. As for the case of $\mathcal{V}=1$, the equations for the height variables $\{H_n\}$ exhibit solutions, provided that the inequality of Eq.~\eqref{eq:ineq for existence of sols} holds on every slice. Notice that these conditions now also affect bulk spatial lengths and scalar field values. That is, the equations for the $\{H_n\}$ constrain the equations of bulk spatial edge lengths and scalar fields.\footnote{It was argued in~\cite{Brewin1999,Bahr:2017eyi} that the equation $\partial S/\partial H=0$ corresponds to an analogue of the Hamiltonian constraint.}

Following the results of~\cite{Jercher:2023csk}, the equation $\partial S_{\mathrm{tot}}/\partial l_n$ can be re-expressed  as $\frac{l_{n+1}-l_n}{H_n} = \frac{l_n-l_{n-1}}{H_{n-1}}$, the form of which is reminiscent of the scalar field equation~\eqref{eq:discrete scalar field equation}. It implies that relative to the $4$-height, the differences of spatial edge lengths within a $4$-frustum remain constant along the entire discrete spacetime. Introducing $\mathcal{H}_{m_1}^{m_2} \defeq \sum_{m=m_1}^{m_2}H_m$ and solving for $l_n$, we find
\begin{equation}
    l_n = \frac{l_\mathcal{V}\mathcal{H}_0^{n-1}+l_0\mathcal{H}_n^{\mathcal{V}-1}}{\mathcal{H}_0^{\mathcal{V}-1}}\,.
\end{equation}
With these formulas, the differences of neighboring spatial length, say on slice $n+1$ and $n$ is expressed as $\frac{l_{n+1}-l_n}{H_n} = \frac{l_\mathcal{V}-l_0}{\mathcal{H}_0^{\mathcal{V}-1}}$, that is, the length difference of spacelike edges relative to the height of a given $4$-frustum is the same as the length difference of the boundary spacelike edges, $(l_\mathcal{V}-l_0)$, relative to the total height $\mathcal{H}_0^{\mathcal{V}-1}$. Importantly, from this equation it follows that the spatial edge lengths evolve monotonically, with the sign determined by the relation of $l_\mathcal{V}$ and $l_0$. Consequently, the boundary data selects either the contracting or expanding branch of solutions.

\subsection{Linearization, deparametrization and continuum limit}\label{sec:Linearisation, deparametrisation and continuum limit}

We consider in the following an expansion of the Regge equations around small deficit angles to connect to the spatially flat Friedmann equations and to obtain analytically feasible equations. Deficit angles are a function of $\eta_n \defeq \frac{l_{n+1}-l_n}{H_n}$ and expanding in small $\eta_n$ yields an expansion in small deficit angles with the flat case given by $\eta_n = 0$.

An expansion of the equation $\partial S/\partial H_n = 0$ around $\eta_n = 0$ yields
\begin{equation}\label{eq:1st CRE expansion}
6(l_n+l_{n+1})\left(\frac{\eta_n^2}{4}+\frac{\eta_n^4}{8}+\mathcal{O}(\eta_n^6)\right)-\lambda 8\pi\GN\frac{(l_n+l_{n+1})^3}{16 H_n^2}(\phi_{n+1}-\phi_n)^2 = 0\,.
\end{equation}
As a consistency check, we notice that cutting the equation at vanishing order of $\eta_n$ yields the flat solution where $l_n = l_{n+1}$ and $\phi_n = \phi_{n+1}$ for all slices $n$.

At first non-vanishing order, $\eta_n$ enters Eq.~\eqref{eq:1st CRE expansion} quadratically and the dependence on the height variable $H_n$ drops out, yielding
\begin{equation}\label{eq:discrete relational Friedmann}
\left(\frac{2}{l_n+l_{n+1}}\right)^2\left(\frac{l_{n+1}-l_n}{\phi_{n+1}-\phi_n}\right)^2 = \lambda\frac{4\pi\GN}{3},
\end{equation}
which corresponds exactly to the limiting case of the inequality~\eqref{eq:ineq for existence of sols}. This equation is \textit{relational} in the sense that it only involves the spatial edge length and the scalar field values, independent of the height of the frusta. In fact, this equation leaves the $H_n$ undetermined, similar to the vacuum case. 

Schematically, the continuum limit corresponds to infinitesimally small but many time steps, at each of which the deficit angles are small~\cite{Bahr:2017eyi,Brewin1999}. Therefore, it is at the lowest non-vanishing order of the expansion in $\eta_n$, where one can define a continuum time limit. Furthermore, the height of a $4$-frustum is mapped to $H_n\rightarrow \dd{\tau} = N\dd{t}$. Spatial edge lengths $\{l_n\}$ and scalar field values $\{\phi_n\}$ are mapped according to Eqs.~\eqref{eq:s cont limit} and~\eqref{eq:phi cont limit}, respectively. Following this prescription, one indeed finds at vanishing order of $\dd{t}$ the continuum Friedmann equations $\left(\frac{\dot{a}}{a}\right)^2 = \lambda\frac{4\pi \GN}{3}\dot{\phi}^2$. Thus, Lorentzian 4-frusta correctly capture the dynamics of spatially flat cosmology in the discrete. Note that Eq.~\eqref{eq:discrete relational Friedmann} is already given in relational form such that one can alternatively define the mapping $l_n\rightarrow a(\phi)$ and $l_{n\pm 1}\rightarrow a(\phi)\pm a'(\phi)\dd{\phi}+\frac{1}{2}a''(\phi)\dd{\phi}^2$, yielding the relational continuum Friedmann equations, $\left(\frac{a'}{a}\right)^2 = \lambda\frac{4\pi\GN}{3}$, where prime denotes $\phi$-derivatives.

Including higher orders of the parameter $\eta_n$ to the full system of equations increases in general the degree of the resulting polynomial equations. In this case, the equation $\partial S_{\mathrm{tot}}/\partial H_n$ becomes explicitly dependent on $H_n$.  It has been furthermore observed in~\cite{Jercher:2023csk} that 1) the inequality in Eq.~\eqref{eq:ineq for existence of sols} is required to hold at any order, and 2) there exists only a single non-perturbative solution for the height variable.\\

\begin{center}
    \textit{Summary.}
\end{center}

\noindent The main result of this chapter was to show that Lorentzian 4-frusta capture the discrete classical dynamics of spatially flat, homogeneous and isotropic geometries, displaying intriguing effects of the causal structure. Moreover, going beyond previous works~\cite{Dittrich:2021gww,Bahr:2017eyi,Dittrich:2023rcr}, dynamical matter in the form of a minimally coupled massless free scalar field was studied. We showed explicitly that the Friedmann equations emerge in the continuum limit of the Regge equations. Existence of the continuum limit, solutions to the Regge equations and the possibility of deparametrizing the system with respect to the scalar field are intimately connected to the causal regularity of the discrete spacetime. This is only given in the sector of timelike 3-frusta and trapezoids, highlighting the importance of other than spacelike building blocks for the recovery of Lorentzian continuum geometries. 

\begin{center}
    \textit{Closing remarks.}
\end{center}

\paragraph{Dynamics of causality violations.} The continuum limit as defined above required $\eta_n\ll 1$ and can therefore not be applied to causally irregular configurations with $\eta_n > \sqrt{2}$. In the present setting, causality violations are thus an explicit feature of discreteness. Moreover, such configurations do not exhibit classical solutions if the geometric variables are assumed to be real. Complexified geometries have been studied in the context of the Euclidean gravity path integral~\cite{Louko:1995jw,Gibbons:1976ue,Gibbons:1978ac,Witten:2021nzp}, continuum Lorentzian quantum cosmology~\cite{Feldbrugge:2017kzv}, semi-classical analyses of Lorentzian spin-foams~\cite{Han:2020npv,Han:2023cen,Han:2021kll} as well as Lorentzian effective spin-foams~\cite{Asante:2021phx,Asante:2021zzh}. However, the results of~\cite{Asante:2021phx} and~\cite{Jercher:2023csk} suggest that even for complexified variables, the complex Regge action does not exhibit saddle points in the causally irregular regime.

\paragraph{Relational dynamics.} Expanding the Regge equations around small deficit angles, the system deparametrizes and the evolution of spatial geometric data is described in a relational sense with respect to the scalar field. However, this property is broken by taking higher orders into account. Thus, introducing a discretization can in principle obstruct a straightforward deparametrization of the system which is instead possible in the corresponding continuum system. Note, however, that by explicitly solving the equation of the height variable, the spatial edge length may still be described relationally as a function of the scalar field.

\paragraph{Timelike building blocks.} The entirely timelike sector of the theory appears preferred both from causal regularity and the dynamics. This casts doubts on the viability to describe cosmology in such a symmetry reduced setting with entirely spacelike 3$d$ building blocks, as prescribed by the Lorentzian EPRL model. This argument is corroborated by the results of Chapters~\ref{chapter:3d cosmology} and~\ref{chapter:perturbations}. Of course, this is not conclusive so far, as the symmetry reduction used here is highly restrictive.

\paragraph{Quantum cosmology.} The insights on the kinematics and dynamics of the coupled gravity and matter system developed in this chapter serve as a foundation for investigations of the symmetry restricted path integral in the spirit of effective spin-foams~\cite{Asante:2020qpa,Asante:2020iwm,Asante:2021zzh,Dittrich:2023rcr} as conducted in the next chapter.

\chapter[\textsc{(2+1) Lorentzian Quantum Cosmology from Spin-Foams}]{(2+1) Lorentzian Quantum \\ Cosmology from Spin-Foams}\label{chapter:3d cosmology}

On the basis of the previous chapter, we want to construct now an effective model of spin-foam quantum cosmology. The importance of timelike building blocks suggests identifying a cosmological subsector within causally extended Lorentzian spin-foam models, such as the EPRL-CH model or the complete BC model, and studying their semi-classical behavior. For the latter, introduced in Chapter~\ref{chapter:cBC}, such an analysis is still missing. For the EPRL-CH model, a stationary phase approximation has been successfully applied for spacelike interfaces between tetrahedra of arbitrary causal character~\cite{Barrett:2009mw,Kaminski:2017eew}.
However, for timelike interfaces, a closed asymptotic formula remains unknown.
That is because 1) the critical points are non-isolated~\cite{Liu:2018gfc,Simao:2021qno}\footnote{In~\cite{Liu:2018gfc}, this is circumvented by picking a single critical point by hand. A geometric reconstruction is then achieved also for 4-simplices containing timelike faces. The lack of a closed asymptotic expression remains.} ,\, 2) $\SUO$ coherent states in the continuous series either require an asymptotic approximation~\cite{Liu:2018gfc} or a regularization~\cite{Simao:2024don}, and 3) the integrand exhibits a branch cut at the critical points~\cite{Simao:2021qno}. The results of Chapter~\ref{chapter:LRC} show that, unfortunately, it is precisely the timelike interfaces which are relevant to identify a causally regular subsector of Lorentzian spin-foams.

In order to address problems 1) and 2), a coherent state model has been developed in the simpler setting of (2+1)-dimensional Lorentzian spin-foams~\cite{Simao:2024don}. In contrast to previous works on the Lorentzian Ponzano-Regge model~\cite{Freidel:2000uq,Freidel:2002hx,Davids:1998bp,Garcia-Islas:2003ges}, this model explicitly incorporates the full set of causal configurations. By introducing a regularization of the $\SUO$ coherent states in the continuous series and supplementing them with an ad hoc Gaussian constraint that ensures the correct gluing condition, a closed semi-classical formula can be attained for every causal configuration. 
Investigations in~\cite{Jercher:2024kig} have shown that for faces either all timelike or all spacelike, the typical cosine-like asymptotics is recovered, while in every other case only a single factor of the exponentiated Regge action is obtained.


When neglecting perturbations, the homogeneous continuum metric of cosmology effectively reduces to a one-dimensional field, being therefore insensitive to the spatial dimension (except for numerical factors entering the dynamical equations). 
Furthermore, propagating degrees of freedom only become relevant when including perturbations. Therefore, the (2+1)-dimensional coherent model proposed in~\cite{Simao:2024don} is well-suited as a first feasible model for investigating the homogeneous sector of quantum cosmology as long as an explicit asymptotic expression for (3+1)-dimensional EPRL-CH model for all causal configurations is still to be developed. 

\section{An effective amplitude for (2+1) quantum cosmology}

Obtaining a numerically computable model from the (2+1) model in~\cite{Simao:2024don} that effectively captures the dynamics of spatially flat, homogeneous and isotropic cosmology involves several steps of modifications and simplifications. These are summarized by 1) the factorization of the amplitude associated to an extended cubical lattice $\mathcal{X}^{(3)}$ into a product of coherent vertex amplitudes,\, 2) the replacement of the quantum amplitudes by their semi-classical approximation, 3) the assumption of toroidal spatial topology, and 4) the restriction of boundary data to that of Lorentzian 3-frusta. The detailed steps of this derivation form an integral part of the work in~\cite{Jercher:2024hlr} and can be seen as complementary to the construction of the semi-classical model employed in Chapter~\ref{chapter:specdim}. Here, we use the results of this construction and refer to Appendix~\ref{app:derivation of the effective model} for details. 

\subsection{Lorentzian 3-frusta}\label{sec:Lorentzian 3-frusta}

We consider a lattice $\mathcal{X}^{(3)}_{\mathcal{V}}$ of $\mathcal{V}$ cuboidal 3-cells, organized in a linear chain along the temporal direction. The cosmological principle is implemented by demanding the 2-cells to be squares, characterized by a single spatial edge length. Consecutive squares at different time steps are connected by four edges termed \emph{struts}, which are identified amongst themselves to impose spatial toroidal topology. As a result of this construction, spacetime is reduced to 3-dimensional Lorentzian frusta. Their geometry is fully characterized by the spatial edge lengths $l_0,l_1$ of squares in different slices and the strut length $m$.\footnote{The notation of edge lengths here and in Chapter~\ref{chapter:LRC} is the same, despite working here (2+1) dimensions. Note that the dimensionality affects the functional dependence of geometrical quantities, such as the height, on $(l_0,l_1,m)$.} These lengths correspond to absolute values, i.e. $l_0,l_1,m > 0$. While the spatial edges are always spacelike by assumption, the struts are allowed to be either spacelike or timelike. Note that in contrast to Chapter~\ref{chapter:LRC}, we do not substitute the strut length with the 3-frustum height to keep the presentation close to Regge calculus. Still, the construction outlined here is in close analogy to that of Chapters~\ref{chapter:specdim} and~\ref{chapter:LRC}. 

The signed squared volume (denoted in boldface) of Lorentzian 3-frusta and its subcells is given in Appendix~\ref{app:Identifying boundary data with Lorentzian 3-frusta} as a function of $(l_0,l_1,m)$. Following the discussion of Sec.~\ref{sec:Lorentzian frustum}, the different causal characters of trapezoids and struts are conveniently captured by three sectors of the theory, similar to Fig.~\ref{fig:regions}. In Sector I, defined by $-\frac{(l_0-l_1)^2}{2} < \vb*{m}^2 < -\frac{(l_0-l_1)^2}{4}$, struts and trapezoids are spacelike. In Sector II, $-\frac{(l_0-l_1)^2}{4} < \vb*{m}^2 <0$, with a spacelike strut and a timelike trapezoid. At $\vb*{m}^2 = -\frac{(l_0-l_1)^2}{4}$, the trapezoid is lightlike. In Sector III, $\vb*{m}^2 > 0$, and the trapezoid and struts are timelike. This range also captures the Lorentzian cuboid with $l_0 = l_1$.

\subsection{Asymptotic vertex amplitude and measure factors}

As in effective spin-foams~\cite{Asante:2021phx,Dittrich:2023rcr}, the domain of length variables is imported from the spectrum of the Casimir operator in terms of the boundary spins in the full spin-foam model. Following the discussion of Appendix~\ref{app:Identifying boundary data with Lorentzian 3-frusta}, this amounts to spacelike edges having a continuous spectrum $[\frac{1}{2},\infty)$ and timelike edges having a discrete spectrum $\mathbb{N}/2$. The length gap of spacelike edge lengths arises from the gap in the spectrum of the $\SUO$ Casimir in the continuous series.  

Given a discretization $\mathcal{X}^{(3)}_{\mathcal{V}}$ and boundary data $(l_0,l_{\mathcal{V}})$ assigned to initial and final slice, the gravitational part  of the effective amplitude is given by (see Appendix~\ref{app:derivation of the effective model} for a derivation)
\begin{equation}
\mathcal{A}^{\mathrm{grav}}(\mathcal{X}_\mathcal{V}^{(3)},l_0,l_{\mathcal{V}}) = \prod_{n=0}^{\mathcal{V}-1} \mathcal{A}_v^{\mathrm{asy}}(l_n,l_{n+1},m_n)\mathcal{A}_f(m_n)\prod_{n'=1}^{\mathcal{V}-1}\mathcal{A}_f(l_{n'})\,,
\end{equation}
where $\mathcal{A}_f$ is a face amplitude\footnote{For a discussion of different choices of face and edge amplitudes, see Appendix~\ref{app:Spin-foam amplitude simplifications}.} 
evaluating to $\mathcal{A}_f = l\tanh(\pi l)$ for spacelike, and $\mathcal{A}_f = 2m-1$ for timelike edges. The amplitude $\mathcal{A}_v^{\mathrm{asy}}$ is obtained via a stationary phase approximation~\cite{Simao:2024don} of the quantum vertex amplitude $\mathcal{A}_v$, and (neglecting constant pre-factors) given by 
\begin{equation}
    \mathcal{A}_{v}^{\mathrm{asy}} =\left(\upmu_{\one}(l_0,l_1,m)\e^{-i\,\mathfrak{Re}\{S_{\mathrm{R}}\}}+\Theta\,\upmu_{\vartheta}(l_0,l_1,m)\e^{i\,\mathfrak{Re}\{S_{\mathrm{R}}\}}\right) \,.
\end{equation}
In comparison to the previous chapters, we set $\GN = 1$ here and leave an analysis of the effect of $\GN$ to future research. The number of critical points depends on the causal characters of the boundary data~\cite{Jercher:2024kig}, encoded here in $\Theta$: if the quadrilaterals are either all timelike or all spacelike, then $\Theta=1$. Thus, $\Theta=1$ in Sector I and $\Theta=0$ in II and III. The phases $\Re{\Sreg}$ contain the Lorentzian Regge action, given explicitly in Eq.~\eqref{eq:S} for the three sectors. Importantly, only the real part of $\Sreg$ enters $\mathcal{A}_v^{\mathrm{asy}}$ which bears important consequences as we show later. 

The functions $\upmu_{\one,\vartheta}$ constitute measure factors arising from the two critical points of the asymptotics and consist of the inverse square root of the Hessian determinant and factors of $\vartheta$. These $\vartheta$ are defined as exponentials of the Lorentzian dihedral angles and originate from the ad hoc Gaussians introduced in~\cite{Simao:2024don} to ensure well-defined asymptotics of the model. For the critical point associated to the identity solution, $g_a=\one$, a closed expression for the Hessian determinant as a function of $(l_0,l_1,m) $ can be computed, given in Eq.~\eqref{eq:det H 1}. At the non-identity solution, $g_a\neq \one$, $\det H_\vartheta$ takes a more involved form, and as a result, its functional form can only be defined implicitly, given in Eq.~\eqref{eq:det H vartheta}. Both Hessian determinants satisfy the correct scaling behavior $\det H(\lambda l_0,\lambda l_1,\lambda m)= \lambda^{15}\det H(l_0,l_1,m)$ consistent with the stationary phase approximation conducted in Appendix.~\ref{app:The semi-classical limit of the vertex}. It has been checked numerically in~\cite{Jercher:2024hlr} that the measure factors are finite even for lightlike trapezoids with $m = (l_0-l_1)^2/4$.

Deriving a measure from spin-foam asymptotics is a strategy that has been employed in Chapter~\ref{chapter:specdim} and in symmetry reduced spin-foam models~\cite{Bahr:2015gxa,Bahr:2016hwc,Bahr:2017klw,Steinhaus:2018aav,Bahr:2018gwf,Allen:2022unb,Bahr:2017eyi}. Therein, the measure plays an important role for the behavior of the coarse-graining flow and, as we have seen, the spectral dimension. Although asymptotics provide a reasonable motivation for these measure factors, different choices are in principle conceivable. For instance in effective spin-foams, $\upmu = 1$ is a common choice~\cite{Asante:2021zzh,Asante:2020qpa,Asante:2020iwm}. Another approach, followed in~\cite{Dittrich:2023rcr,Asante:2021phx}, is to derive a measure from continuum quantum cosmology~\cite{Feldbrugge:2017kzv}. Discretization independence has been also used as a guiding principle choosing the measure, see~\cite{Dittrich:2011vz,Borissova:2024pfq,Borissova:2024txs,Dittrich:2014rha}. 

\subsection{Minimally coupled massive scalar field}\label{sec:Minimally coupled massive scalar field}

We consider here the minimal coupling of a massive scalar field to induce non-trivial dynamics. Whether this scalar field can serve as a relational clock will be discussed below. In a strict sense, matter should be included by coupling classical $BF$-theory, already fully describing gravity in (2+1) dimensions, to the relevant matter fields, subsequently proceeding with a spin-foam quantization to arrive at a spin-foam model of gravitational and matter degrees of freedom. Such an approach is however difficult to implement, having only been attained in the context of 3-dimensional Riemannian gravity coupled to fermions \cite{Fairbairn:2006dn} or Yang-Mills \cite{Speziale:2007mt}. Alternative approaches include justifying reasonable Ansätze for the partition function of the coupled system \cite{Mikovic:2001xi, Bianchi:2010bn, Han:2011as},  matter as topological defects~\cite{Freidel:2005bb,Freidel:2005me,Livine:2024iyk}, modifying spin-foam amplitudes in analogy with LQG coupled to scalar fields \cite{Kisielowski:2018oiv}, or by conceiving of each spin-foam history as providing a geometry where the discrete matter action is defined \cite{Oriti:2002bn, Mikovic:2002uq, Ali:2022vhn, Jercher:2023rno}. 
Here we follow the latter perspective by adding factors of $\e^{iS_\phi}$ to the amplitude $\mathcal{A}^{\mathrm{grav}}$, with $S_\phi$ the massive scalar field action on frusta geometries. While such an approach is evidently limited, the hope is that it is sufficient to describe gravity-matter interactions effectively~\cite{Ali:2022vhn}.

The spatially homogeneous minimally coupled massive scalar field is discretized on primary vertices of $\mathcal{X}^{(3)}_{\mathcal{V}}$, i.e. $\phi(t)\rightarrow\phi_n$. 
We consider a massive scalar field since 1) it is more general than the massless case and 2) the mass acts as a regulator for the partition function as discussed in Secs.~\ref{sec:freezing oscillations} and~\ref{sec:mass term regularization}. Its action discretized on a single 3-frustum is given by
\begin{equation}\label{eq:S_phi}
S_\phi = w_0(l_0,l_1,{m}_0)(\phi_0-\phi_1)^2-M_0(l_0,l_1,{m}_0)(\phi_0^2+\phi_1^2)\,,
\end{equation}
with $w_n(l_n,l_{n+1},m_n) \defeq (l_n+l_{n+1})^2/(8\sqrt{(l_n-l_{n+1})^2/2+\vb*{m}_n^2})$ characterizing the kinetic term and $M_n(l_n,l_{n+1},m_n) \defeq \mu^2V(l_n,l_{n+1},\vb*{m}_n)/4$ defining the mass term. Here, $\mu$ is the scalar field mass and $V(l_0,l_1,\vb*{m}_0)$ is the 3-volume, given in Eq.~\eqref{eq:3-volume}. 

%
%

The vertex amplitude $\mathcal{A}_v^\phi$ for the coupled system of geometry and matter for a single 3-frustum is obtained via the replacement
\begin{equation}
\mathcal{A}_v^{\mathrm{asy}}(l_0,l_1,m)\mapsto \mathcal{A}_v^\phi(l_0,l_1,m;\phi_0,\phi_1) = \left(\upmu_\one\e^{-i(\Re{\Sreg}+S_\phi)}+\Theta\,\upmu_\vartheta\e^{i(\Re{\Sreg}+S_\phi)}\right)\,.
\end{equation}
This defines the final amplitude $\hat{\mathcal{A}}$ for the boundary data being complemented with additional scalar field data, $(l_0,\phi_0,l_{\mathcal{V}},\phi_{\mathcal{V}})$. We demonstrate below that the scalar field does render the dynamics of the effective model non-trivial in that its solutions are non-stationary.

In the present setting we explicitly consider the scalar field to be massive. In the continuum picture, a finite mass $\mu\neq 0$ spoils the global monotonicity property, such that there is no global inversion of scalar field values in time. Recent works in the context of Hamiltonian quantum mechanics and quantum cosmology~\cite{Bojowald:2021uqo,Martinez:2023fsd} show however that a massive scalar field still defines a relational clock globally, so long as the clock values are supplemented with a cycle count - much in the same way time is read from a watch. We expect follow-up investigations to require a careful analysis of this notion of clock time.

\subsection{Effective cosmological partition function}\label{sec:Effective cosmological partition function}

The partition function for a discretization $\mathcal{X}_\mathcal{V}$ (dropping the index \enquote{$(3)$}) and boundary data $\Phi = (l_0,\phi_0,l_{\mathcal{V}},\phi_\mathcal{V})$ is given by
\begin{equation}\label{eq:Z tot}
\begin{aligned}
    & Z_{\mathcal{X}_{\mathcal{V}}}(\Phi) =\sumint{\{l,m,\phi\}} \prod_{n=0}^{\mathcal{V}-1}\hat{\mathcal{A}}(l_n,l_{n+1},m_n,\phi_n,\phi_{n+1})\,.
\end{aligned}
\end{equation}
The sum/integration over all bulk variables includes an unbounded sum over timelike strut lengths in Sector III and bounded integrations over spacelike strut lengths in Sectors I and II. The partition function can be used to define expectation values of functions $\mathcal{O}(\{l,m,\phi\})$ of the bulk variables by injecting $\mathcal{O}$ into Eq.~\eqref{eq:Z tot} and dividing by $Z$.

In analogy to Chapter~\ref{chapter:LRC}, for real variables classical solutions and their correct continuum limit are found exclusively in Sector III. In contrast, configurations of Sectors I and II are off-shell and exhibit causal irregularities as discussed in the section hereafter. Restricting the partition function to causally regular configurations therefore amounts to a restriction to Sector III. We denote the restricted amplitude, partition function and expectation values with an index \enquote{\textsc{III}}. 

In terms of the \textit{general boundary formulation} introduced by Oeckl~\cite{Oeckl:2003vu,Oeckl:2005bv,Oeckl:2006rs,Oeckl:2011qd}, the partition function $Z_{\mathcal{X}_{\mathcal{V}}}$ can be formally understood as an amplitude map from the space of boundary states to the complex numbers. Assume $\Phi = \Phi_0\cup\Phi_{\mathcal{V}}$ splitting into data on initial and final slice. Then, $Z_{\mathcal{X}_{\mathcal{V}}}$ enters the \textit{transition probability} $\mathbb{P}(\Phi_0\rightarrow\Phi_{\mathcal{V}}) \defeq \abs{Z_{\mathcal{X}_{\mathcal{V}}}(\Phi)}^2/\int_{\Phi'}\abs{Z_{\mathcal{X}_{\mathcal{V}}}(\Phi')}^2$. Even for a diverging denominator, the ratio of probabilities might still be a meaningful quantity. Following~\cite{Oeckl:2011qd}, also expectation values can be defined within the general boundary formulation. In the present setting with $\Phi = \Phi_0\cup\Phi_{\mathcal{V}}$, the definition of~\cite{Oeckl:2011qd} in fact reduces to our definition above. Since $\hat{\mathcal{A}}$ is complex, also the expectation values of real observables are generically complex. In~\cite{Dittrich:2023rcr}, the authors consider this as representing the quantum nature of the model. Our computations will reveal non-vanishing imaginary parts of expectation values as well, albeit tending to a constant half imaginary unit. Understanding these effects further is obstructed by the fact that the effective model only provides amplitudes and the partition function but no a priori notion of a boundary Hilbert space. Future research in this direction could show whether the imaginary part arises from operator ordering ambiguities and would furthermore allow defining hermitian operators with real eigenvalues.

\subsection{Semi-classical limit and causality violations}\label{sec:causality violations}

The effective cosmological vertex amplitude derived in Appendix~\ref{app:derivation of the effective model} shows a particularity on which we elaborate in this section: in the semi-classical limit of the spin-foam vertex only the real part of the Lorentzian Regge action is recovered. This has important consequences for semi-classical configurations with an irregular light cone structure, as we detail now.

Following the discussion on causality violations in Sec.~\ref{sec:causal regularity}, we observe that the Lorentzian Regge actions in Eq.~\eqref{eq:S} attain imaginary values in Sectors I and II. These configurations are therefore considered as hinge causality violating. In effective spin-foams~\cite{Dittrich:2023rcr,Asante:2021zzh}, these imaginary parts play a crucial role as they provide a physical mechanism for suppressing causality violating configurations instead of ad hoc excluding such configurations from the partition function. However, the asymptotic vertex amplitude $\mathcal{A}_v^{\mathrm{asy}}$ only contains Regge exponentials $\e^{\pm i\mathfrak{Re}\{\Sreg\}}$. Consequently, the mechanism of exponential suppression of causality violations does not figure in the effective cosmological amplitude constructed here. 

The mismatch of Lorentzian Regge calculus and the effective model derived from spin-foam asymptotics could arise because 1) we work in the particular setting of three dimensions, 2) performing the stationary phase approximation at each vertex individually, which is a particularly strong assumption since causal regularity is defined for the gluing of multiple building blocks, or 3) it reflects an inherent property of Lorentzian spin-foams and is not merely an artifact of the simplifications performed here.  Tentatively, this would imply that for causality violations, full spin-foam amplitudes are not related to the amplitudes of Lorentzian effective spin-foams via an asymptotic limit. Future investigations will hopefully give a definite answer to this question.  

Without the exponential suppression of causality violations, such configurations can be dealt with by excluding them by hand and thus only considering $Z_{\textsc{iii}}$. However, beyond the symmetry restricted setting here, an ad hoc exclusion of causality violations might be computationally unfeasible~\cite{Asante:2021phx}. Including these configurations, there are two potential suppression mechanisms besides the one from effective spin-foams, being destructive interference due to an absence of classical solutions in Sectors I and II, or the measure factors $\upmu_{\one,\vartheta}$. We explicitly investigate this question in Sec.~\ref{sec:including I and II}.

\section{Numerical evaluation: strut in the bulk}\label{sec:Numerical evaluation: Strut in the bulk}

In this section, we numerically evaluate the effective partition function for a single 3-frustum, $\mathcal{V}=1$, with one bulk strut. In Secs.~\ref{sec:freezing oscillations} and~\ref{sec:mass term regularization}, we restrict to the causally regular Sector III and discuss the inclusion of causality violations in Sec.~\ref{sec:including I and II}. The unbounded sum in Sector III is evaluated using convergence acceleration techniques~\cite{Wynn1956,Weniger2003}, which have been applied to effective spin-foams in~\cite{Dittrich:2023rcr} and which are summarized in Appendix~\ref{app:Wynn}.

\subsection{Freezing oscillations}\label{sec:freezing oscillations}

The partition function for a single 3-frustum contains in particular an unbounded sum over the timelike strut length $m\in\mathbb{N}/2$ in Sector III. For fixed boundary data $\Phi = (l_0,\phi_0,l_1,\phi_1)$ and in the case of vanishing scalar field mass, $\mu = 0$, the Regge action scales as $\Sreg\sim 1/m$ for large strut length, and thus $\e^{\pm i\Sreg}\rightarrow 1$ for $m\rightarrow \infty$. This asymptotic freezing of oscillations obstructs the sum in III to converge. We remark that this is not an effect of merely considering a single strut length as bulk variable but also occurs for $\mathcal{V}>1$.

Even if the spin-foam integrand exhibits saddle points, expectation values of the strut length diverge for $\mu = 0$. That is because the action gets effectively stationary in the limit $m\rightarrow \infty$. Notice in particular that while an upper cutoff $m_{\mathrm{max}}$ regularizes $Z$, the result is cutoff dependent and $m_{\max}$ cannot be removed. A similar argument applies to continuum quantum cosmology~\cite{Feldbrugge:2017kzv} and effective spin-foams~\cite{Dittrich:2023rcr} if considered in the spatially flat case without cosmological constant. The issue of freezing oscillations is absent in~\cite{Feldbrugge:2017kzv,Dittrich:2023rcr} precisely due to spatial curvature, $k=1$, and a cosmological constant, $\Lambda > 0$. We demonstrate next that a non-zero scalar field mass acts as a regulator of the partition function rendering finite expectation values. 

\subsection{Strut length expectation values in Sector III}\label{sec:mass term regularization}

To regularize the divergent partition function in Sector III, we work in the remainder with a non-vanishing scalar field mass, $\mu\neq 0$. Furthermore, we restrict here to causally regular configurations, i.e. to Sector III, indicated by \enquote{$\ev{\cdot}_{\textsc{iii}}$}. These expectation values are compared with classical solutions $m_{\mathrm{cl}}$, computed from the Regge equation $\partial(S_{\textsc{iii}}+S_\phi)/\partial m = 0$. 

The expectation value of $m$ can be computed via the effective cosmological partition function in Sector III for a variety of boundary data and mass parameters. Summarizing the results of~\cite{Jercher:2024hlr}, one finds that $\mathfrak{Re}\{\ev{m}_{\textsc{iii}}\}$ generically agrees with the classical solutions $m_{\mathrm{cl}}$. Deviations between these quantities arise mainly due to the discrete spectrum of timelike length, $m\in\mathbb{N}/2$, which can lead to an insufficient resolution of the saddle point at $m_{\mathrm{cl}}$, as already discussed in~\cite{Dittrich:2023rcr}.
Another numerical challenge are highly oscillatory summands combined with small amplitudes which can be potentially resolved by utilizing arbitrary precision arithmetic. The imaginary part of $\ev{m}_{\textsc{iii}}$ is approximated by $-\frac{1}{2}$ for a range of boundary data. Its existence has been deemed a quantum effect in~\cite{Dittrich:2023rcr} and is to be expected from the perspective of the general boundary formulation, discussed above. Whether real expectation values can be obtained by considering different boundary states or observables is an intriguing question for future research. 

\begin{figure}
    \centering
    \begin{subfigure}{0.45\textwidth}
    \includegraphics[width=\linewidth]{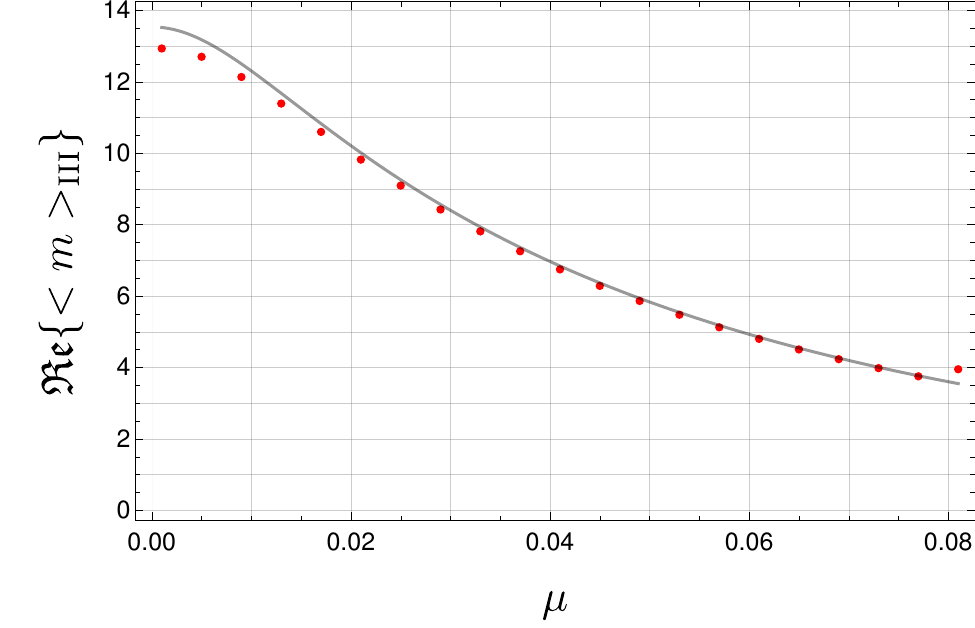}
    \end{subfigure}\hspace{0.05\textwidth}
    \begin{subfigure}{0.45\textwidth}
    \includegraphics[width=\linewidth]{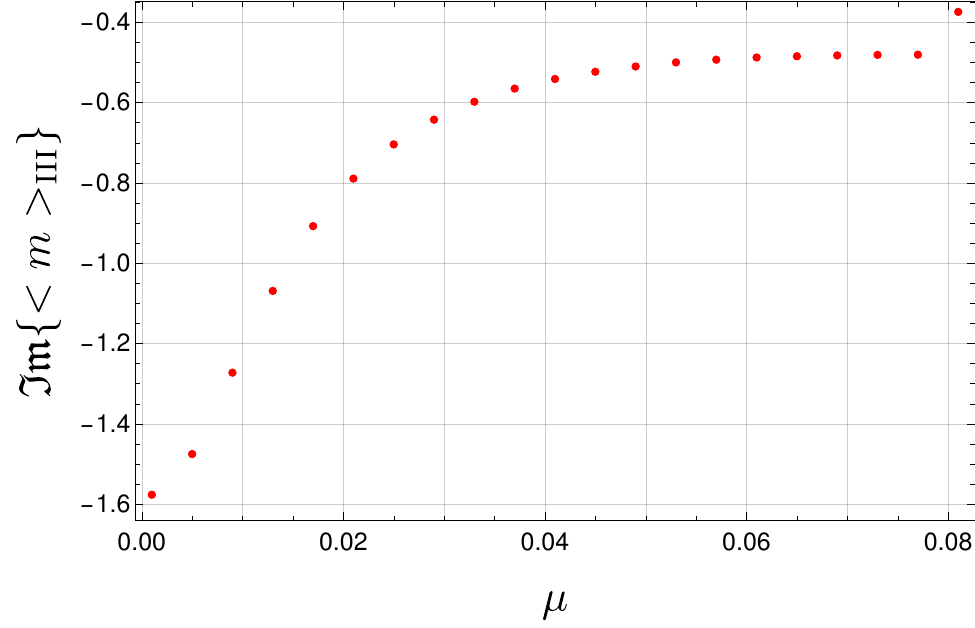}
    \end{subfigure}
    \caption{Left: real part of $\ev{m}_{\textsc{iii}}$ (red dots) and the classical solutions $m_{\mathrm{cl}}$ (gray graph). Right: the imaginary part of the strut length expectation value.  Both quantities are computed for varying scalar field mass $\mu = 10^{-3}(1+4n)$ with $n\in\{0,\dots,20\}$. Boundary data is fixed to $l_0= 10, l_1 = 30$, $\phi_0=2$ and $\phi_1=4$.}
    \label{fig:m expval mu}
\end{figure}

\paragraph{Mass dependence.} As an example, let us detail the mass dependence of the strut length expectation value in comparison to the classical solution $m_{\mathrm{cl}}$ for fixed boundary data $(l_0,\phi_0,l_1,\phi_1)$. The classical solution is a continuous function of the scalar field mass in particular at the point $\mu=0$, i.e. $\lim_{\mu\rightarrow 0}m_{\mathrm{cl}}(\mu) = m_{\mathrm{cl}}(0)$. Also in the quantum theory, the strut length expectation value is given as a continuous function of the mass parameter in the regime $\mu>0$ as plotted in the left panel of Fig.~\ref{fig:m expval mu}. The real part of $\langle m\rangle_{\textsc{iii}}$ shows good agreement with the classical solutions for small mass values. $\Im{\ev{m}_{\textsc{iii}}}$ tends towards $-1/2$, representing a generic feature of $\ev{m}_{\textsc{iii}}$. For larger masses, the dependence of $\mathfrak{Re}\{\expval{m}_{\textsc{iii}}\}$ on $\mu$ follows an inverse square law, similar to what has been obtained in~\cite{Ali:2022vhn}. Remarkably, given the results of Sec.~\ref{sec:freezing oscillations}, the strut length expectation value is \emph{discontinuous} in the mass at $\mu=0$, i.e. $\lim_{\mu\rightarrow 0^+}\ev{m}_{\textsc{iii}}(\mu)\neq\ev{m}_{\textsc{iii}}(\mu=0)$. This is to be expected since the introduction of a non-zero mass $\mu$ guarantees oscillations linear in the summation variable. In the continuum, the mass term explicitly breaks the lapse independence of the action, similar to the breaking of gauge symmetry in Proca theory and massive gravity~\cite{Veltman1970}. The results observed here serve as a discrete analogon of such phenomenona.

\subsection{Strut length expectation value in Sectors I and II}\label{sec:including I and II}

The full effective partition function contains not only a sum over the strut length but also over its causal character. Therefore, a complete evaluation of expectation values requires the inclusion of Sectors I and II. For the amplitude $\hat{\mathcal{A}}^{\textsc{ii}} = \mu_\one\e^{-i(S_\textsc{ii}+S_\phi)}$, a closed expression exists and thus, the bounded integral $Z_{\textsc{ii}}$ can be evaluated straightforwardly using numerical integration techniques.\footnote{We utilize the \href{https://github.com/giordano/Cuba.jl}{\texttt{Cuba}}-package in \href{https://arxiv.org/pdf/Julia}{\texttt{Julia}} or the \textsc{NIntegrate}-function in \textsc{Mathematica}.} The evaluation of the partition function in Sector I is more challenging since an analytical formula of the Hessian determinant at $\vartheta\neq 1$ is not at hand. In principle, the numerical integration algorithms can also be applied to $\hat{\mathcal{A}}^{\textsc{i}}$ without an explicit formula of $\det H_\vartheta$. In this case the determinant $\det H_\vartheta$ needs to be computed for every sampling point of the integrand. Unfortunately, many samples and integrand evaluations are required for convergence due to rapid oscillations. As a result, convergent numerical integration requires an unfeasible amount of computation time. To surpass these obstacles, we interpolate the Hessian determinant between discrete points and use this function then for the numerical integration. 

The influence of Sectors I and II is quantified by first computing $\langle m^2\rangle = -\langle m^2\rangle_{\textsc{i}}-\langle m^2\rangle_{\textsc{ii}}+\langle m^2\rangle_{\textsc{iii}}$, taking into account the causal character of the strut. The deviation between $\langle m^2\rangle$ and $\langle m^2\rangle_{\textsc{iii}}$ is then measured by $\Delta \defeq \abs{\mathfrak{Re}\{\langle m^2\rangle_{\textsc{iii}}\}-\mathfrak{Re}\{\langle m^2\rangle\}}/{\mathfrak{Re}\{\langle m^2\rangle_{\textsc{iii}}\}}$. 

\begin{figure}
    \centering
    \begin{subfigure}{0.45\textwidth}
    \includegraphics[width=\linewidth]{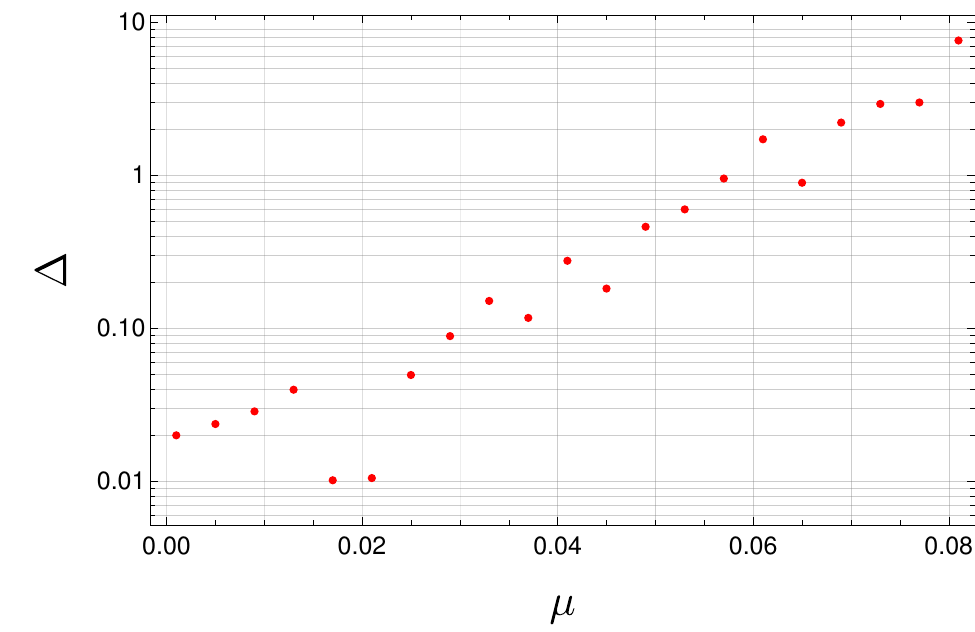}
    \end{subfigure}\hspace{0.05\textwidth}
    \begin{subfigure}{0.45\textwidth}
    \includegraphics[width=\linewidth]{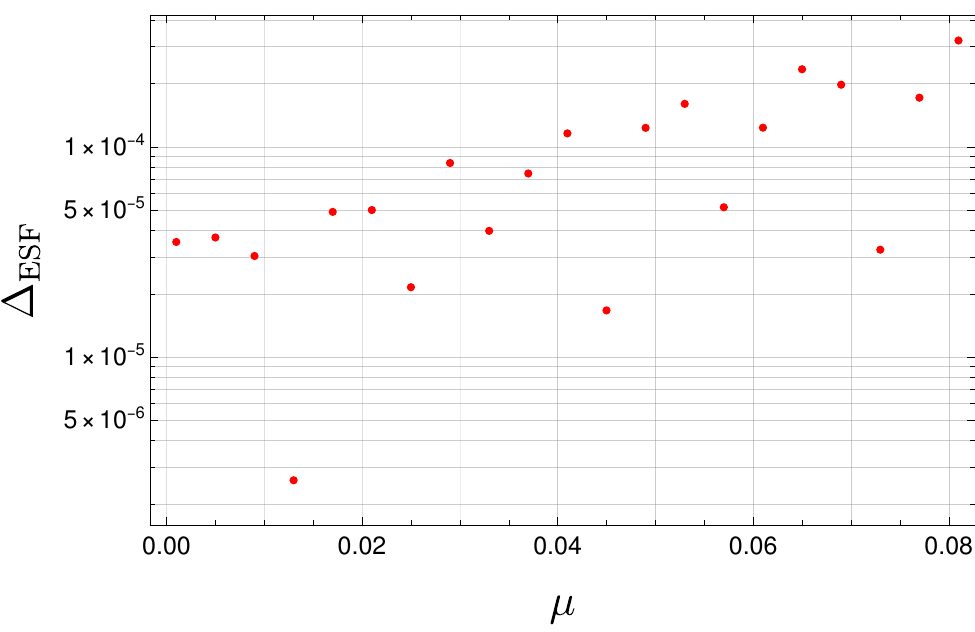}
    \end{subfigure}
    \caption{Relative deviation $\Delta$ of the squared strut length expectation values when including Sectors I and II, computed from the effective model (left) and effective spin-foams (right). Scalar field mass is given by $\mu = 10^{-3}(1+4n)$ with $n\in\{0,\dots,20\}$ and boundary data is fixed to $l_0=10$, $l_1 =30$, $\phi_0=2$ and $\phi_1 = 4$.}
    \label{fig:IIIvstot}
\end{figure}

Explicit numerical investigation shows that only in a regime of small scalar field mass $\mu$, the full partition function with spacelike and timelike struts can be satisfactorily approximated with that of Sector III. Outside this regime, the deviations $\Delta$ are substantial, see Fig.~\ref{fig:IIIvstot} for an exemplary plot. This behavior is sourced by a lack of an exponential suppression which can be traced back to the failure of the asymptotics to reproduce complex Lorentzian deficit angles. Although a suppression mechanism for causality violations is missing for the effective cosmological model constructed here, we emphasize that in the present setting one can consistently exclude these configurations as done in Secs.~\ref{sec:mass term regularization} and ~\ref{sec:1slice}.\footnote{Alternatively, the amplitudes could be complemented to include the exponentially suppressing terms, amounting to a modification of face amplitudes. The results depicted in Fig.~\ref{fig:IIIvstot} suggest that this is indeed promising.} Thus, the effective partition function restricted to Sector III still provides a viable model for quantum cosmology.  

\paragraph{Comparison to effective spin-foams.} Effective spin-foams~\cite{Asante:2021phx,Dittrich:2023rcr,Asante:2020qpa,Asante:2020iwm} assume as vertex amplitude the exponentiated Lorentzian Regge, including its imaginary part. Originally applied to spatially spherical cosmologies with a cosmological constant $\Lambda>0$~\cite{Asante:2021phx,Dittrich:2023rcr}, effective spin-foams can also be applied to the present setting. Adapting the measure proposed in~\cite{Dittrich:2023rcr,Asante:2021phx} to this scenario, expectation values of $m$ are computed straightforwardly. One observes that 1) the real part of $\ev{m}^{\textsc{esf}}_{\textsc{iii}}$ generically agrees with the classical solutions and is finite for any $\mu>0$ while it diverges at $\mu = 0$ for the same reasons explained above, 2) the deviation between classical solutions and expectation values is generically smaller than the one obtained above, arising from the different measure terms, 3) the imaginary part of $\ev{m}^{\textsc{esf}}_{\textsc{iii}}$ is non-constant and shows oscillatory behavior, and 4) the influence of Sectors I and II is negligible, rooted in the exponential suppression that arises from the imaginary parts of the deficit angles. For a comparison of $\Delta$ from effective spin-foams and the present model, see Fig.~\ref{fig:IIIvstot}.  

\section{Outlook: spacelike bulk slices}\label{sec:1slice}

In this section we give an outlook on the evaluation of the effective partition function for a discretization consisting of $\mathcal{V} = 2$ cubes containing a spacelike slice in the bulk. The dynamical variables are given by two strut lengths, one spatial edge length and one scalar field value, $(m_0,l_1,\phi_1,m_1)$. We restrict in this section to causally regular configurations and compute the partition function and expectation values only with respect to the amplitudes of Sector III.

For comparison, we compute the solutions to the classical equations of motion. To that end, the results of Chapter~\ref{chapter:LRC} can in part be transferred to the present setting, with the difference being that the volume of hinges entering the Regge action are lengths rather than areas, and the scalar field is considered here to be massive. The scalar field equation can be solved analytically in terms of $(m_0,l_1,m_1)$ such that the remaining dynamical equations are given by $\partial S_{\mathrm{tot}}/\partial x = 0$ with $x\in\{m_0,l_1,m_1\}$ and $S_{\mathrm{tot}}$ the gravity + matter action on the discretization $\mathcal{X}_{2}$. Although forming an intricate set of coupled transcendental equations, they can be explicitly solved for given boundary data using the \textsc{FindRoot} method of \textsc{Mathematica}.


\subsection{Evaluation of the partition function}\label{sec:Evaluation of the partition function}

The key object for computing expectation values is the partition function which, restricted to timelike strut lengths, is explicitly given by
\begin{equation}\label{eq:Z1slice}
Z_{\mathcal{X}_2}(\Phi) = \sum_{m_0,m_1\in\mathbb{N}/2}\int_{1/2}^{\infty}\dd{l_1}\int_{\R}\dd{\phi_1} \mathcal{A}_f(l_1)\hat{\mathcal{A}}^{\textsc{iii}}(l_0,l_1,m_0,\phi_0,\phi_1)\hat{\mathcal{A}}^{\textsc{iii}}(l_1,l_2,m_1,\phi_1,\phi_2)\,,
\end{equation}
for boundary data $\Phi = (l_0,\phi_0,l_2,\phi_2)$. It is evaluated in three steps: 1) analytical scalar field integration,\, 2) $l_1$-integration for sufficiently many values of $(m_0,m_1)$,\, 3) summation over $(m_0,m_1)$ using Wynn's algorithm for multiple variables (see Appendix~\ref{app:Wynn}). As the analysis of~\cite{Jercher:2024hlr} shows, steps 1) and 3) are straightforward to execute while performing step 2) uncovers an intricate interplay of the path integral measure and semi-classical physics. We elaborate on the latter point in the following while referring to~\cite{Jercher:2024hlr} for details on steps 1) and 2). 

\begin{figure}
    \centering
    \begin{subfigure}{0.45\textwidth}
    \includegraphics[width=\linewidth]{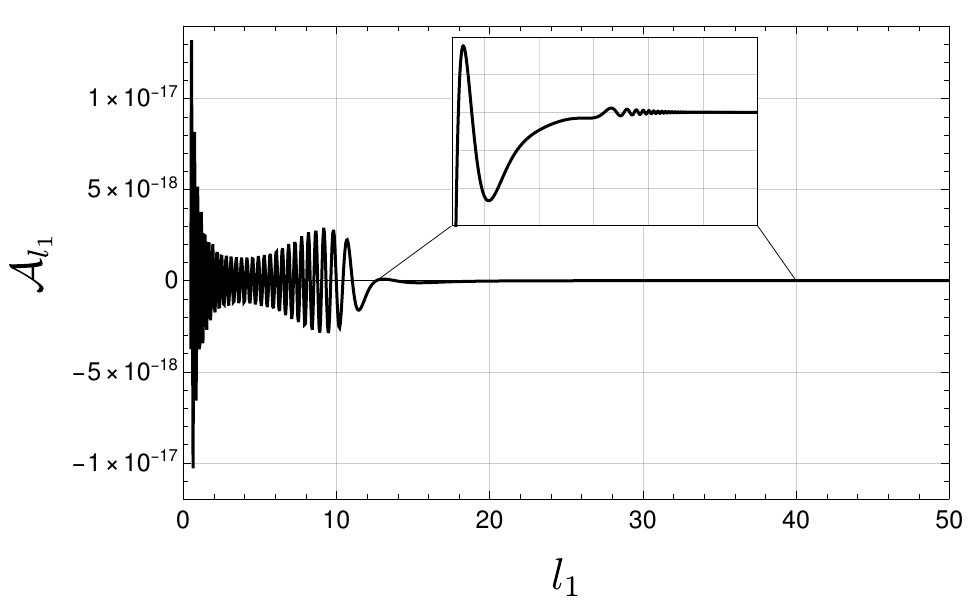}
    \end{subfigure}\hspace{0.05\textwidth}
    \begin{subfigure}{0.45\textwidth}
    \includegraphics[width=\linewidth]{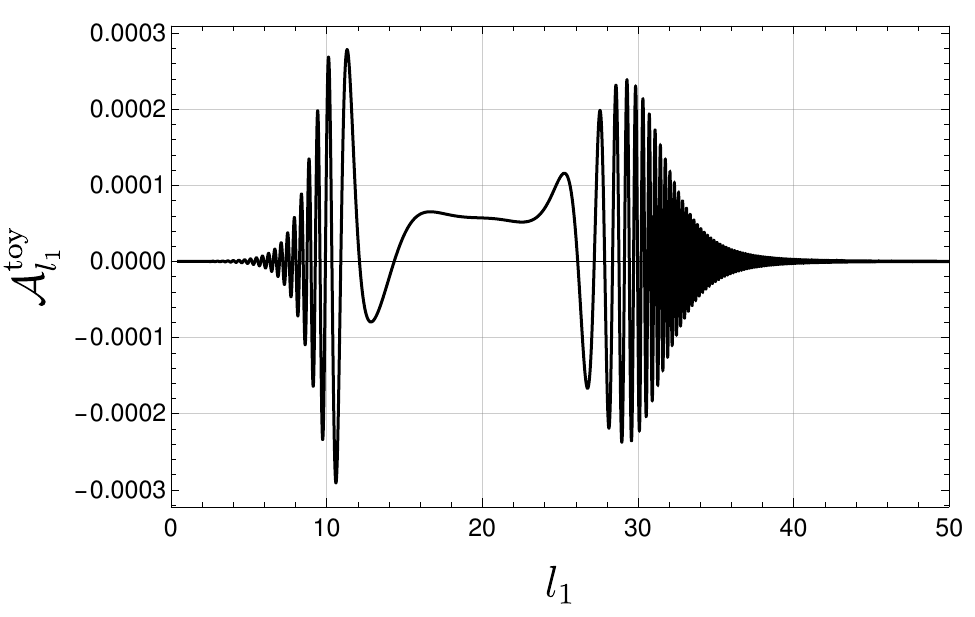}
    \end{subfigure}
    \caption{$l_1$-integrand $\mathcal{A}_{l_1}$ of the effective model (left) and the toy model (right), for fixed bulk strut length $m_0 = 2$ and $m_1 = 3$. Both graphics are generated for boundary data $l_0=10$, $l_2=30$, $\phi_0=2$, $\phi_2=4$ and a scalar field mass $\mu = 0.05$.}
    \label{fig:l1 integrand}
\end{figure}

Performing the $l_1$-integration for fixed $(m_0,m_1)$ yields an \emph{effective} amplitude $\mathcal{A}_{\mathrm{eff}}(m_0,m_1;\Phi) \defeq \int\dd{l_1}\mathcal{A}_{l_1}$. Doing so for many values of $(m_0,m_1)$, $\mathcal{A}_{\mathrm{eff}}$ can be stored as a matrix of size $2m_{\max}\times 2m_{\max}$. 
The integrand, denoted as $\mathcal{A}_{l_1}$, diverges for $l_1\rightarrow 0$ as $l_1^{-1}$ which follows from the measure $\upmu_{\one}$. However, due to the assumed length gap,  $l\in[\frac{1}{2},\infty)$, $\mathcal{A}_{l_1}$ is finite in the region of integration. $\mathcal{A}_{l_1}$ is decaying as $l_1^{-33/2}$ for large values of $l_1$ which is advantageous for convergence of the integration, but suppresses the region of the classical solution which spoils the resolution of this saddle point, see the left panel of Fig.~\ref{fig:l1 integrand}. Numerical evaluation shows that the effective amplitude $\mathcal{A}_{\mathrm{eff}}$ does not display a saddle point in the $(m_0,m_1)$ space at the classical solution $(m_0^{\mathrm{cl}},m_1^{\mathrm{cl}})$. These are generic features that do not depend on the choice of boundary data.

\paragraph{Results.}  A summary of the expectation values of bulk variables for fixed boundary data $\Phi$ is given in Tab.~\ref{tab:exp vals}. Comparing the real part of expectation values to the classical solutions we find strong deviations for all observables. Thus, the expectation values computed with $Z_{\mathcal{X}_2}$ do not reproduce classical results despite the presence of saddle points. This is a direct consequence of the effective measure which suppresses the region of the $l_1$-integration where the saddle point $l_1^{\mathrm{cl}}$ is located. In particular, one obtains similar results for different mass parameters and boundary data. This demonstrates an intricate interplay between the path integral measure and semi-classical behavior. We remind the reader that the measure utilized here is given by a product of measures obtained from a stationary phase approximation of the single vertex amplitude. Whether the measure obtained from a stationary phase approximation of the amplitude on an extended complex recovers semi-classical behavior remains as an intriguing open question. 

\begin{table}[]
    \centering
    \begin{tabular}{ | c | c | c | c | c |}
    \hline
    cl. solutions & $2.20$ & $3.04$ & $17.83$ & $3.15$\\
    \hline
    \hline
    $\quad \text{amplitude}\quad$   &   $\quad \ev{m_0}\quad$   &  $\quad\ev{m_1}\quad$ & $\quad \ev{l_1}\quad$ & $\quad\ev{\phi_1}\quad$ \\
    \hline
    \rule{0pt}{4pt} $Z_{\mathcal{X}_2}$  &   $22.05+ 5.06i$  &   $12.76-1.00 i$ & $2.56-1.83i$ & $1.43-2.13i$   \\[3pt]
    \rule{0pt}{4pt} $Z_{\mathcal{X}_2}^{\mathrm{toy}}$   &   $3.82+0.27i$  &   $3.11 -0.64i$  & $21.26+1.32i$ & $3.27+0.14i$ \\[3pt]
    \hline
    \end{tabular}
    \caption{Upper part: solutions of the classical equations of motion. Lower part: expectation values of the same variables for the effective partition function $Z_{\mathcal{X}_2}$ and the toy model $Z_{\mathcal{X}_2}^{\mathrm{toy}}$ of~\cite{Jercher:2024hlr}. Boundary values are fixed to $l_0=10$, $l_2=30$, $\phi_0=2$, $\phi_2=4$ and $\mu = 0.05$.}
    \label{tab:exp vals}
\end{table}

To substantiate the relation between measure and semi-classical physics, a toy measure $\upmu_{\mathrm{toy}}$ has been constructed in~\cite{Jercher:2024hlr} which  does not suppress the saddle point of the $l_1$-integration, see the right panel of Fig.~\ref{fig:l1 integrand}. For the same fixed boundary data, expectation values are presented in Tab.~\ref{tab:exp vals}. We find that $\mathfrak{Re}\{\ev{m_1}_{\mathrm{toy}}\}$ and $\mathfrak{Re}\{\{\phi_1\}_{\mathrm{toy}}\}$ are close to the classical solution. $\mathfrak{Re}\{\ev{m_0}_{\mathrm{toy}}\}$ deviates from $m_0^{\mathrm{cl}}$, which is a result of the discrete spectrum $m_0\in\mathbb{N}/2$. $\mathfrak{Re}\{\ev{l_1}\}_{\mathrm{toy}}$ is slightly larger than $l_1^{\mathrm{cl}}$ due to a plateau of the total action around $l_1^{\mathrm{cl}}$.  The toy example shows that semi-classical physics can be obtained from an effective partition function on an extended cellular complex by following the recipe outlined above. Crucially, we identified the measure of the $l_1$-integration as the essential factor for the success or failure of this strategy.\\

\begin{center}
    \textit{Summary.}
\end{center}

\noindent As the overarching result of this chapter, we demonstrated that the effective cosmological spin-foam model, coupled to a massive scalar field, constitutes a viable and computationally feasible model for quantum cosmology. We have shown that in the spatially flat case with $\Lambda = 0$, a non-vanishing scalar field mass ensures convergence of the partition function due to the otherwise asymptotic stationary Regge action. Semi-classicality of expectation values hinges on two factors: the measure of the effective path integral and causal regularity. Thereby, our results emphasize again the significance of timelike building blocks. By extending the analysis towards spatial bulk slices at the end of this chapter, our results represent a significant advancement in effective spin-foam cosmology~\cite{Dittrich:2023rcr}, facilitating future studies on physically intriguing scenarios such as a quantum bounce. 

\begin{center}
    \textit{Closing remarks.}
\end{center}

\paragraph{Causality violations in spin-foams.} Effective spin-foam models~\cite{Asante:2021zzh,Asante:2020iwm,Asante:2020qpa} have proven numerically feasible and suitable to study physically interesting scenarios, see e.g.~\cite{Dittrich:2023rcr,Asante:2021phx,Dittrich:2024awu,Borissova:2024pfq,Borissova:2024txs}. These models are however proposed ad hoc, leaving it as an open question if and how they can be obtained from fundamental spin-foams. In the cosmological setting, this question motivated the investigation of the present model. Tentatively, our results suggest that while causality violating configurations generically appear in the semi-classical limit, their exponential suppression does not emerge.
Of course, our argument is limited to the (2+1)-dimensional symmetry reduced setting and the assumption of factorizing semi-classical vertex amplitudes. Thus, spin-foams may still possess a mechanism for suppressing causality violations. This constitutes an intriguing new research direction, which can be explored via complex critical points~\cite{Han:2021kll,Han:2023cen} or via numerical methods~\cite{Dona:2019dkf,Gozzini:2021kbt,Dona:2022dxs,Dona:2023myv,Steinhaus:2024qov,Asante:2024eft}.

\paragraph{Outlook.} The results of Sec.~\ref{sec:1slice} present a promising outlook for future investigations on cosmological partition functions with spatial bulk slices. Despite the restriction to causal regularity and the simplification of the measure, $Z_{\mathcal{X}_2}^{\mathrm{toy}}$ can be utilized to study physically and conceptually interesting questions. These include a bouncing scenario, where initial and final spatial edges on the boundary are equal but boundary scalar field values evolve. Conceptually interesting is also the role of the scalar field as a relational clock. In particular, $Z_{\mathcal{X}_2}^{\mathrm{toy}}$ offers the possibility to study the influence of the scalar field mass and under which conditions fluctuations dominate.

\begin{center}
    $***$
\end{center}

\noindent This closes Part~\hyperref[Part I]{I} on the spectral dimension and cosmology from spin-foams. Subject of the next part is the emergence of effective Lorentzian geometries from group field theories. 

\clearpage
\phantomsection
\addcontentsline{toc}{part}{Part II: Effective Geometries from Group Field Theories}\label{Part II}
\thispagestyle{fancy}
\chapter[\textsc{The Complete Barrett-Crane Model and its Causal Structure}]{The Complete Barrett-Crane \\Model and its Causal Structure}\label{chapter:cBC}

Given that the causal structure of spacetime is a key ingredient of GR, one expects a theory of QG to address the role of causality, either by encoding it directly into the quantum theory or by demonstrating why and how it arises in a classical and/or continuum limit. The relevance of this point is highlighted also by the results of the previous two chapters.

In their most common formulations, GFTs and spin-foam models restrict to exclusively spacelike building blocks. An alternative to the Barrett-Crane GFT model with spacelike tetrahedra~\cite{Perez:2000ec} has been proposed in~\cite{Perez:2000ep}, involving only timelike tetrahedra but featuring spacelike and timelike faces. In spin-foams, the Conrady-Hnybida extension~\cite{Conrady:2010kc,Conrady:2010vx} of the EPRL spin-foam model includes spacelike \textit{and} timelike but not lightlike tetrahedra. An explicit GFT formulation for the EPRL model and its CH extension is still missing.

The main objective of this chapter is to develop a GFT and spin-foam model that transparently incorporates building blocks of every causal character. In contrast to the EPRL model, the Lorentzian Barrett-Crane model~\cite{Barrett:1999qw} exhibits an explicit GFT formulation~\cite{Perez:2000ec}, extended in~\cite{Jercher:2021bie} by a timelike normal vector variable. As we demonstrate here, this formulation is ideal for a straightforward generalization to include timelike, lightlike and spacelike tetrahedra.

Several criticisms have been raised towards the BC model, listed in the following with a corresponding response, based on the discussion of~\cite{Baratin:2011tx}:

\textit{The BC model is employing the \enquote{wrong} boundary states~\cite{Alesci:2007tx}:} The findings of~\cite{Alesci:2007tx} in the context of the graviton propagator reflect a mismatch between boundary states in LQG and the BC model. However, the BC model was not intended to make contact with LQG. In particular, while LQG results from a quantization of Palatini-Holst gravity with Barbero-Immirzi $\bi$, the BC model originates from Palatini gravity with $\bi\rightarrow \infty$. Thus, the mismatch should not be surprising or upsetting, and has per se no bearing on the validity of the BC model. 

\textit{Degenerate geometries in the asymptotics~\cite{Barrett:2002ur,Freidel:2002mj}:} In these asymptotic analyses degenerate geometries are shown to dominate in a semi-classical limit. Although being a serious issue, the argument is limited to a single $4$-simplex and thus its implications for the emergent continuum geometry are unclear. Results suggesting that the BC model may still yield a viable continuum limit are given in the context of effective spin-foams~\cite{Dittrich:2021kzs} and GFT condensate cosmology~\cite{Jercher:2021bie}. Furthermore, note that in analogous computations for the coherent EPRL model such configurations can in principle occur and dominate as well, but are excluded by choice of boundary data. For the BC model $\SUT$ coherent states cannot be constructed by definition. However, future investigations on $\SL$ coherent states could show that degenerate configurations can also be excluded for the BC model. 


\textit{The BC constraints are imposed \enquote{too strongly}~\cite{Engle:2007uq}:} Following~\cite{Baratin:2011tx,Baratin:2011hp}, the linear simplicity constraints are first class for $\bi\rightarrow\infty$ and thus should be imposed strongly. This is realized via a projector, rigorously defined in an extended formulation utilizing normal vectors~\cite{Baratin:2011tx,Jercher:2021bie}.

We conclude that the BC model has not been ruled out conclusively and is therefore worth of any further attention. In contrast to the EPRL model, it is simpler in structure and exhibits an explicit GFT formulation. Furthermore, intriguing results concerning its effective continuum gravitational description have been obtained: 1) Landau-Ginzburg analyses of the spacelike~\cite{Marchetti:2022igl,Marchetti:2022nrf} and causally complete (Chapter~\ref{chapter:LG}) BC model demonstrate the validity of mean-field theory and thereby suggest the existence of a condensate phase with a continuum geometric interpretation. 2) This phase is explored in GFT condensate cosmology~\cite{Gielen:2013kla,Gielen:2013naa,Oriti:2016qtz,Gielen:2016dss,Oriti:2016acw,Pithis:2019tvp} where key properties such as  emergent Friedmann dynamics and a quantum bounce can be recovered in the BC GFT model as well~\cite{Jercher:2021bie}. 3) GR-like perturbations can be extracted from the BC model with spacelike and timelike tetrahedra~\cite{Jercher:2023nxa,Jercher:2023kfr}, as detailed in Chapter~\ref{chapter:perturbations}.  

\section{Causally complete Barrett-Crane group field theory model}\label{sec:Extended Barrett-Crane Group Field Theory Model}

\subsection{Definition of the complete model}\label{subsec:Definition of the Complete Model}

In~\cite{Jercher:2021bie}, the Lorentzian BC model restricted to spacelike tetrahedra was extended by an additional variable $X_\texttt{+}$ lying in the 3-hyperboloid $\mathrm{H}_\texttt{+}$, interpreted as the timelike normal vector of tetrahedra. This auxiliary variable allows for a covariant and commuting imposition of the geometricity constraints introduced below, see also~\cite{Baratin:2011tx}. The key idea of the complete BC model is to include group fields with spacelike ($X_\texttt{-}\in\mathrm{H}_{\texttt{-}}$) and lightlike ($X_0\in\mathrm{H}_0$) normal vectors\footnote{As discussed later on in Sec.~\ref{subsec:Spacetime Orientation}, the choice of upper or lower part of the two-sheeted hyperboloid $\mathrm{H}_\texttt{+}$ and the light cone $\mathrm{H}_0$ is irrelevant for the construction of the model.} to remove the restriction of the causal characters of tetrahedra and its subcells.

The defining ingredients are the three group fields $\varphi_\alpha:\SL^4\times\mathrm{H}_\alpha\rightarrow\mathbb{K}$ with $\alpha\in\{\texttt{+},0,\texttt{-}\}$ indicating the signature of the normal vector, $\mathrm{H}_\alpha$ being the homogeneous space of $\SL$ containing the normal vector $X_\alpha$ and $\mathbb{K} = \R,\C$.\footnote{The Feynman graphs of colored tensor models are bipartite if the tensors are complex-valued.  It has been shown in~\cite{Caravelli:2010nh} that this property is related to the orientability of the dual pseudo-manifold. In Chapter~\ref{chapter:LG}, $\varphi$ is assumed to be real-valued while here and in Chapter~\ref{chapter:perturbations}, $\varphi$ is assumed to be complex-valued.} The fields $\varphi_\alpha$ satisfy closure and simplicity
\begin{align}
    \varphi_\alpha(\vbg,X_{\alpha}) &= \varphi_\alpha(\vbg h^{-1},h\cdot X_{\alpha})\,,\quad\forall h\in\SL\,,\label{eq:extended closure}\\
    \varphi_\alpha(\vbg,X_{\alpha}) &= \varphi_\alpha(\vbg \vb*{u},X_{\alpha})\,,\quad\qquad\: \forall \vb*{u}\in \text{U}_{X_{\alpha}}^4\,,\label{eq:simplicity}
\end{align}
collectively referred to as \emph{geometricity}. Here, $\vbg\in\SL^4$, $\mathrm{U}_{X_\alpha}$ denotes the stabilizer subgroup of $\SL$ with respect to $X_\alpha$ and \enquote{$\cdot$} denotes the action of $\SL$ on $\mathrm{H}_\alpha$. Note that $\mathrm{U}_{X_\alpha}\cong \mathrm{U}^{(\alpha)}$ given by $\SUT$, $\ISO$ and $\SUO$ for $\alpha$ respectively \enquote{$\texttt{+}$}, \enquote{$0$} and \enquote{$\texttt{-}$}. For an introduction to these notions, see Appendix~\ref{app:Homogeneous spaces}. 

Changing from group to Lie algebra variables via the non-commutative Fourier transform~\cite{Baratin:2010wi,Guedes:2013vi,Oriti:2018bwr}, it is demonstrated in~\cite{Jercher:2022mky} that Eq.~\eqref{eq:extended closure} corresponds to the closure of bivectors, $\sum_i B_i = 0$, irrespective of $\alpha$, and Eq.~\eqref{eq:simplicity} imposes the linear simplicity constraint $X_\alpha\cdot(*B) = 0$.

The dynamics of the $\varphi_\alpha$ are governed by the GFT action $S = \mathfrak{K} + \mathfrak{V}$ with kinetic term $\mathfrak{K}$,
\begin{equation}\label{eq:kinetic action}
    \mathfrak{K}
    =
    \sum_{\alpha}\int\left[\dd{g}\right]^8\int\dd{X_{\alpha}}\bar{\varphi}(\vbg,X_{\alpha})\mathcal{K}_{\alpha}(\vbg;\vbg')\varphi(\vbg',X_{\alpha})\,,
\end{equation}
where a bar denotes complex conjugation. Geometrically, the $\mathcal{K}_\alpha$ encode the gluing of tetrahedra chosen here to be diagonal in $\alpha$, i.e.~only tetrahedra of the same signature are glued together. Normal vectors of tetrahedra are identified which is why $\varphi$ and $\bar{\varphi}$ are evaluated at the same $X_\alpha$. The kinetic kernels $\mathcal{K}_\alpha$ remain in most general form here, and we refer to Chapters~\ref{chapter:LG} and~\ref{chapter:perturbations} for a discussion of explicit choices. Due to geometricity, the kinetic term $\mathfrak{K}$ and the vertex term $\mathfrak{V}$, introduced below, exhibit a divergence in the form of an empty $\SL$ integration, not carrying any physical information. For the remainder, we implicitly assume a straightforward regularization of such trivial redundancies, see also~\cite{Jercher:2021bie} for a discussion.


GFT models are most commonly introduced with simplicial interactions but can be straightforwardly generalized to include also tensor-invariant interactions which will be explicitly employed in Chapter~\ref{chapter:LG}. Irrespective of the causal character of tetrahedra, the combinatorics of such general interactions are captured by a \emph{vertex graph} $\upgamma$~\cite{Oriti:2014yla,Marchetti:2020xvf,Marchetti:2022igl}. Its vertices and edges respectively represent the group fields and its group arguments. The connectivity of vertices encodes the non-local gluing of tetrahedra along the faces. In the causal completion here, the notion of a vertex graph is enriched by associating a causal character to the tetrahedra, thus captured by a \textit{causal vertex graph} $\upgamma_\textsc{c}$. In the most general case, the interaction term is thus written as 
\begin{equation}\label{eq:general vertex}
\mathfrak{V}= \sum_{\upgamma_{\textsc{c}}}\lambda_{\upgamma_{\textsc{c}}}\int\dd{\vb*{X}}\Tr_{\upgamma_{\textsc{c}}}\left[\varphi_\texttt{+}^{\np}\varphi_0^{\nz}\varphi_\texttt{-}^{\nm}\right]+\mathrm{c.c.}\,.
\end{equation}
Here, the $\lambda_{\upgamma_{\textsc{c}}}$ are couplings and $n_\alpha$ is the number of spacelike, lightlike and timelike tetrahedra, respectively (encoded in $\upgamma_{\textsc{c}}$). The trace $\Tr_{\upgamma_{\textsc{c}}}$ encodes the pairwise contraction of group elements according to $\upgamma_{\textsc{c}}$.  Note that the normal vectors are integrated over separately for every field entering the interaction suggesting their interpretation as auxiliary variables~\cite{Jercher:2021bie,Jercher:2022mky}. The corresponding integration measure is denoted for short by $\dd{\vb*{X}}$. To illustrate the definition above, consider as an example $\upgamma_{\textsc{c}} = \raisebox{2pt}{\scalebox{0.3}{\cvfspppppnolocal}}$, associated to a $4$-simplex with five spacelike tetrahedra. Writing $\varphi_{1234}(X_{\alpha}) \equiv \varphi(g_1,g_2,g_3,g_4,X_{\alpha})$, the interaction is then given by 
\begin{equation}\label{eq:simplicial interactions}
    \mathfrak{V}\left(\scalebox{0.5}{\cvfspppppnolocal}\right) = \int\left[\dd{g}\right]^{10}\left[\dd{X_\texttt{+}}\right]^5\varphi_{1234}(X^1_\texttt{+})\varphi_{4567}(X^2_\texttt{+})\varphi_{7389}(X^3_\texttt{+})\varphi_{9620}(X^4_\texttt{+})\varphi_{0851}(X^5_\texttt{+})+\text{c.c.}\,.
\end{equation}
For completeness, let us close this section with two extensions of the model, being coloring and the coupling of a scalar field.  

Following~\cite{Gurau:2010nd}, coloring GFTs with simplicial interactions ensures that the GFT Feynman diagrams are dual to topological pseudo-manifolds.\footnote{As pointed out in~\cite{Gurau:2010ba} in the context of colored tensor models, coloring allows for a $1/N$ expansion, with $N$ the index range of the tensors, which is a key ingredient for renormalization~\cite{Gurau:2010ba,Gurau:2011aq,Gurau:2011xq,Bonzom:2012hw,Gurau:2013pca,Gurau:2012vk}.} Such a coloring consists of an extension $\varphi(\vbg,X_{\alpha})\rightarrow \varphi^i(\vbg,X_{\alpha})$ with labels $i \in\{0,\dots,4\}$. The colored kinetic term, $\mathfrak{K}_{\mathrm{col}} = \sum_i\mathfrak{K}_i$, ensures that only tetrahedra of the same color $i$ are being glued. The vertex $\mathfrak{V}_{\mathrm{col}}$ exhibits simplicial combinatorics with every color $i$ appearing exactly once, e.g. Eq.~\eqref{eq:simplicial interactions} is schematically generalized to $\mathfrak{V}_{\mathrm{col}} = \int\varphi_{1234}^0\varphi_{4567}^1\varphi_{7389}^2\varphi_{9620}^3\varphi_{0851}^4+\text{c.c. }$. For more involved assignments of causal characters, summing over permutations of colors is required as to not couple colors and causal characters~\cite{Jercher:2022mky}. For instance a given color $\hat{\imath}$ should not be associated with a preferred causal character $\hat{\alpha}$. In this way, the combinatorial properties of generated Feynman diagrams governed by the colors are separated from the causal characters associated to the dual building blocks. 

A massless free scalar field $\phi$ is minimally coupled to the GFT\footnote{The $\mf$ are coupled so that the GFT Feynman amplitudes correspond to simplicial gravity path integrals with minimally coupled massless free scalar fields~\cite{Oriti:2016qtz,Li:2017uao,Gielen:2018fqv}.} by extending the group field domain by the scalar field value~\cite{Oriti:2016qtz,Li:2017uao,Gielen:2018fqv}, $\varphi(\vbg,X_\alpha)\rightarrow \varphi(\vbg,\phi,X_\alpha)$. The kernels $\mathcal{K}_\alpha$ are extended to $\mathcal{K}_\alpha(\vbg,\vbg') \rightarrow \mathcal{K}_\alpha(\vbg,\vbg',(\phi_v-\phi_w)^2)$, respecting the translation and reflection invariance of the classical action of $\phi$. They encode information about the propagation of $\phi$ between neighboring $4$-simplices, denoted $v$ and $w$. Note that the interaction $\mathfrak{V}$ is local in $\phi$~\cite{Li:2017uao,Jercher:2021bie}, i.e. the group fields in $\mathfrak{V}$ are evaluated at the same scalar field value $\phi$.

\subsection{The spin representation}\label{subsec:Spin Representation of the Group Field and its Action}

The spin representation allows expressing a GFT explicitly as a spin-foam model and is an expedient step for the computations in Chapters~\ref{chapter:LG} and~\ref{chapter:perturbations}. A detailed derivation of the spin representation for the complete BC model based on~\cite{Jercher:2021bie,Jercher:2022mky} is presented in Appendix~\ref{app:Projection onto invariant subspaces}. Let $D_\alpha \defeq \SL^4\times\mathrm{H}_\alpha/\sim$ be the domain of the field with \enquote{$/\sim$} encoding the quotient structure due to closure and simplicity in Eqs.~\eqref{eq:extended closure} and~\eqref{eq:simplicity}, respectively. A field $\varphi_\alpha\in L^2(D_\alpha)$ is then expanded in terms of unitary irreducible $\SL$ representations $(\rho,\nu)\in\R\times\mathbb{Z}/2$ as
\begin{equation}\label{eq:spin representation of group field}
\varphi(\vbg,X_{\alpha})
=
\left[\prod_{c=1}^4 \sum_{\nu_c}\int\dd{\rho_c}(\rho_c^2+\nu_c^2)\varpi^{(\rho_c,\nu_c)}_\alpha\sum_{j_c m_c l_c n_c}\right]\varphi^{\vbr\vbn,\alpha}_{\vb*{j} \vb*{m}}\prod_c D^{(\rho_c,\nu_c)}_{j_c m_c l_c n_c}(g_c g_{X_{\alpha}})\bar{\mathcal{I}}^{(\rho_c,\nu_c),\alpha}_{l_c n_c},
\end{equation}
where $(\rho_c^2+\nu_c^2)$ is the Plancherel measure, $\varphi^{\vbr\vbn,\alpha}_{\vb*{j} \vb*{m}}$ are the expansion coefficients and $D^{(\rho,\nu)}_{jmln}$ are $\SL$ Wigner matrix elements in the canonical basis, see also Appendix~\ref{app:Representation Theory of SL2C}. Throughout this and the following chapters, $c\in\{1,\dots, 4\}$ is an index labelling the four group elements $\vbg\in\SL^4$ or the corresponding $\SL$ representations. 

The $\varpi^{(\rho,\nu)}_\alpha$ impose simplicity onto the representations $(\rho,\nu)$ arising from a projection onto $\mathrm{U}^{(\alpha)}$-invariant subspaces. Following~\cite{VilenkinBook,Barrett:1999qw}, $\varpi_\alpha^{(\rho,\nu)} = \delta_{\nu,0}$ for timelike ($\alpha=\texttt{+}$) and lightlike ($
\alpha=0$) normal vectors, and $\varpi_\texttt{-}^{(\rho,\nu)} = \delta_{\nu,0} + \delta(\rho)\chi_{\nu}$ for spacelike normals, where $\chi_\nu$ is a characteristic function imposing $\nu\in2\mathbb{Z}\setminus\{0\}$. The $\mathcal{I}^{(\rho,\nu),\alpha}_{jm}$ are the canonical basis coefficients of $\mathrm{U}^{(\alpha)}$-invariant vectors with $\mathcal{I}^{(\rho,0),\alpha}_{jm}$ for $\alpha\in\{\texttt{+},0\}$, and $\mathcal{I}^{(\rho,0),\texttt{-}}_{jm}$ and $\mathcal{I}^{(0,\nu),\texttt{-}}_{jm}$, respectively. For $\alpha = \texttt{+}$, the well-known results of~\cite{Perez:2000ec,Perez:2000ep,Barrett:1999qw} are reproduced by noticing that $\mathcal{I}^{(\rho,0)}_{jm} = \delta_{j,0}\delta_{m,0}$. In Sec.~\ref{sec:Explicit Expressions for the Vertex Amplitudes} the $\mathcal{I}$ are endowed with a quantitative meaning, given their rather formal introduction here.

Following~\cite{Barrett:1999qw,Perez:2000ep}, the \emph{simple} representations $(\rho,0)$ and $(0,\nu)$ are associated to spacelike and timelike faces, respectively, and relate to the squared area via the first $\SL$ Casimir with eigenvalues $A^2\sim -\rho^2+\nu^2-1$. Clearly, $A^2 < 0$ for $\nu = 0$ with a continuous spectrum and an area gap, and $A^2 > 0$ for $\rho = 0$ and $\nu^2>1$ with a discrete spectrum and an area gap.\footnote{Similarly, in Lorentzian (2+1) spin-foam models~\cite{Simao:2024don,Davids:1998bp,Garcia-Islas:2003ges,Freidel:2000uq}, spacelike (timelike) edges exhibit a continuous (discrete) spectrum. Following~\cite{Alexandrov:2005ar}, an analogy of continuous space and discrete time can be found in the 't Hooft model of a point particle in (2+1)-dimensional quantum gravity~\cite{Matschull:1997du,tHooft:1993jwb}. In contrast, the EPRL-CH model~\cite{Conrady:2010kc,Conrady:2010vx} proposes discrete spectra for both types of faces, where the spacelike spectrum is controlled by $\bi$.} The projection $\varpi^{(\rho,\nu)}_\alpha$ can be understood geometrically as well: for timelike normal vectors, $(\rho,0)$ is singled out, which is consistent with the fact that spacelike tetrahedra consist exclusively of spacelike faces. In contrast, a tetrahedron orthogonal to a spacelike vector can contain an arbitrary combination of spacelike and timelike faces, and thus, representations are either given by $(\rho,0)$ or $(0,\nu)$. Simplicity implies $A^2\neq 0$, interpreted as the exclusion of lightlike faces labelled by $(\pm i,0)$ and $(0,\pm 1)$. Remarkably, this is in contrast to classical Lorentzian geometry, where lightlike and timelike tetrahedra can contain lightlike faces. Importantly, when applying a Landau-Ginzburg analysis to the BC GFT model as presented in~\cite{Marchetti:2022igl,Marchetti:2020xvf,Dekhil:2024ssa} and Chapter~\ref{chapter:LG}, representations with $(\pm i,0)$ defined in the sense of hyperfunctions~\cite{Ruehl1970} play a crucial role.\footnote{Intriguingly, the zero modes of $\SUO$, which play a similar role to the representations $(\pm i,0)$ here, are important for unitarity in (2+1)-dimensional spin-foams coupled to matter~\cite{Livine:2024iyk}.}

Inserting Eq.~\eqref{eq:spin representation of group field} into the GFT action, one extracts the spin representation of the kinetic kernel, $\mathcal{K}^{\vbr\vbn,\vbr'\vbn'}_{\alpha}$, defining the inverse edge amplitude of the corresponding spin-foam model~\cite{Jercher:2022mky}. The vertex term directly characterizes the vertex amplitude $\mathcal{A}_v$. Given an interaction with tetrahedra labelled by $a,b$, sharing faces $(ab)$ which are either spacelike or timelike, $\mathcal{A}_v$ writes~\cite{Jercher:2022mky}
\begin{equation}\label{eq:generalized vertex integral form}
    \mathcal{A}_v
    =
    \int\dd{\vb*{X}} \prod_{(ab)\text{ sl}}K_{\alpha_a\alpha_b}^{(\rho_{ab},0)}(X_a,X_b)\prod_{(ab)\text{ tl}}K_{\alpha_a\alpha_b}^{(0,\nu_{ab})}(X_a,X_b)\,,
\end{equation}
where the $\alpha_a$ encode the causal characters of tetrahedra, determining the domain of the normal vectors $X_a\in\mathrm{H}_{\alpha_a}$. The kernels $K$ are the defining ingredient of $\mathcal{A}_v$, given by
\begin{equation}\label{eq:definition of K}
    K^{(\rho_{ab},\nu_{ab})}_{\alpha_a\alpha_b}(X_a,X_b)
    \defeq
    \sum_{jmln}\mathcal{I}^{(\rho_{ab},\nu_{ab}),\alpha_a}_{jm}D^{(\rho_{ab},\nu_{ab})}_{jmln}(g_{X_a}^{-1}g_{X_b})\bar{\mathcal{I}}^{(\rho_{ab},\nu_{ab}),\alpha_b}_{ln}\,,
\end{equation}
effectively depending on normal vectors $X_a\in\mathrm{H}_{\alpha_a}$, for which the $g_{X_a}$ are representatives.\footnote{Note that in Eq.~\eqref{eq:spin representation of group field}, every Wigner $D$ is accompanied by one invariant vector $\mathcal{I}$. By factorizing the matrix $D$ in Eq.~\eqref{eq:definition of K} with respect to the two group elements, one finds the same structure.} The $K$ are defined in analogy to~\cite{Barrett:1999qw,Perez:2000ep} and play a similar role as the state pairings entering $\SUT$ $BF$-theory~\cite{Dona:2017dvf}, the (2+1) model of~\cite{Simao:2024don} (see Eq.~\eqref{eq:SU11 general vertex}) and the EPRL-CH model~\cite{Livine:2007vk,Barrett:2009gg,Barrett:2009mw,Conrady:2010vx,Conrady:2010kc}. Crucially, however, the $\mathrm{U}^{(\alpha)}$ invariance of the BC kernels obstructs the definition of widely used $\mathrm{U}^{(\alpha)}$ coherent states in other models. Goal of the following section is to provide explicit expressions of these functions using the methods developed in Appendix~\ref{app:Harmonic analysis on homogeneous spaces}. 

\section{Computing the kernels of the vertex amplitudes}\label{sec:Explicit Expressions for the Vertex Amplitudes}

The vertex amplitude $\mathcal{A}_v$ is the essential object to explicitly define the complete BC model. As discussed in the previous section, $\mathcal{A}_v$ is given in terms of the kernels $K_{\alpha_1\alpha_2}(X_1,X_2)$ introduced in Eq.~\eqref{eq:definition of K}. The computations of these kernels are simplified by noticing the symmetry relation $K_{\alpha_1\alpha_2}(X_1,X_2) = \overline{K_{\alpha_2\alpha_1}(X_2,X_1)}$. Consequently, there are in total six independent types of kernels, given by $K_{\texttt{++}},K_{\texttt{-}\texttt{-}},K_{00},K_{\texttt{+}0},K_{\texttt{+-}}$ and $K_{\texttt{-}0}$. Another important property is that the $K$ are invariant under simultaneous action of $\SL$, i.e. $K_{\alpha_1\alpha_2}(X_1,X_2) = K_{\alpha_1\alpha_2}(h\cdot X_1,h\cdot X_2)$ for all $h\in\SL$. Consequently, the $K_{\alpha_1\alpha_2}(X_1,X_2)$ effectively depend on the Minkowski product of $X_1$ and $X_2$, for which a convenient parametrization can be chosen without loss of generality. 


\subsection{Kernels of non-mixed type}\label{subsec:Kernels of Non-Mixed Type}

Following~\cite{Barrett:1999qw}, the kernels $K_{\alpha\alpha}^{(\rho,0)}$ and $K_{\alpha\alpha}^{(0,\nu)}$ define a projection onto the $(\rho,0)$, respectively the $(0,\nu)$ component of functions $f\in L^2(\mathrm{H}_\alpha)$.  That is
\begin{equation}
f^{(\rho,0)}(X) \defeq \int_{\mathrm{H}_\alpha}\dd{Y}\overline{K^{(\rho,0)}_{\alpha\alpha}(Y,X)}f(Y)\,,
\end{equation}
similarly for $f^{(0,\nu)}(X)$. Upon integration over $\rho$ (respectively summation over $\nu$) one re-obtains $f(X)$. Alternatively, the kernels $K_{\alpha\alpha}$ of non-mixed type define the $\delta$-function on $\mathrm{H}_\alpha$, the expression of which can be identified with an expansion of the $\delta$-function in terms of the Gel'fand transform, developed in~\cite{VilenkinBook}. Consequently, an explicit expression for $K_{\alpha\alpha}$ can be extracted. We present here the results of the derivation given in Appendix~\ref{app:Harmonic analysis on homogeneous spaces}.

For $\alpha = \texttt{+}$, the kernel $K_{\texttt{++}}$ takes the integral form given in Eq.~\eqref{eq:D_++ before evaluation}. Choosing a parametrization $X = (1,0,0,0)$, $Y = (\cosh(\eta),0,0,\sinh(\eta))$ with $\eta\in\R$, the kernel $K_{\texttt{++}}^{(\rho,0)}$ evaluates to
\begin{equation}
K_{\texttt{++}}^{(\rho,0)}(\eta) = \frac{\sin(\rho\eta)}{\rho\sinh(\eta)}\,,
\end{equation}
agreeing with the results obtained in~\cite{Barrett:1999qw,Perez:2000ec}. Note that this function is regular in $\rho$ and $\eta$ and appears frequently in the literature on $\SL$ representations~\cite{Ruehl1970}.

For $\alpha = \texttt{-}$, the kernel $K_{\texttt{-}\texttt{-}}$ comes with two components, $K_{\texttt{-}\texttt{-}}^{(\rho,0)}$ and $K_{\texttt{-}\texttt{-}}^{(0,\nu)}$ associated to spacelike and timelike faces, respectively. The integral forms of $K_{\texttt{-}\texttt{-}}^{(\rho,0)}$ and $K_{\texttt{-}\texttt{-}}^{(0,\nu)}$ are given in Eqs.~\eqref{eq:D-- rho} and~\eqref{eq:D-- nu}, respectively. Then, one finds~\cite{Perez:2000ep}
\begin{equation}
\begin{aligned}
 K^{(\rho,0)}_{\texttt{-}\texttt{-}}(\eta,\hat{\vb*{r}})
=
\int\limits_0^{2\pi}\frac{\dd{\phi}}{4\pi}\int\limits_{-1}^{1}\dd{t}\abs{\sinh(\eta)-\cosh(\eta)\left(\sqrt{1-t^2}\sqrt{1-r_z^2}\sin(\phi)+tr_z\right)}^{-i\rho-1}\abs{t}^{i\rho-1}\,,
\end{aligned}
\end{equation}
using the parametrization $X = (0,0,0,1)$, $Y = (\sinh(\eta),\cosh(\eta)\hat{\vb*{r}})$ with $\hat{\vb*{r}}\in S^2$. For the restriction $r_z=\pm 1$,  the integral above readily simplifies to $K^{\rho}_{\texttt{-}\texttt{-}}(\eta,\pm 1) = \frac{\sin(\rho\eta)}{\rho\sinh(\eta)}$, therefore agreeing with $K_{\texttt{++}}$~\cite{Perez:2000ep}. The component associated to timelike faces evaluates to~\cite{Perez:2000ep}
\begin{equation}
\begin{aligned}
K^{(\nu,0)}_{\texttt{-}\texttt{-}}(\eta,\hat{\vb*{r}}) = \frac{32e^{i2\nu\Theta(\eta,\hat{\vb*{r}})}}{\abs{\nu}\sin(\Theta)}\,,\qquad\text{ for }\qquad0\leq\Theta\leq\frac{\pi}{2}\,,
\end{aligned}
\end{equation}
and vanishes otherwise. Here, $\cos(\Theta(\eta,\hat{\vb*{r}})) = \abs{\cosh(\eta)r_z}$. As noted in~\cite{Perez:2000ep}, the real part of $K^{\nu}_{\texttt{-}\texttt{-}}$ diverges for $\eta = 0$ and $r_z = \pm 1$, which corresponds to the special case where $X$ and $Y$ are equal, and is regular otherwise. Whether this isolated point of divergence  presents an issue for the well-definedness of the vertex amplitude is left as an open question to future research.    

The computation of the lightlike kernel $K_{00}$ is novel, but follows exactly the same lines as for the cases above. The $\delta$-function on $\mathrm{H}_0$ in Eq.~\eqref{eq:delta function on cone} is written in a parametrization $X\in\mathrm{H}_0$ as $X = \lambda\xi$ with $\xi = (1,\hat{\vb*{r}}(\theta,\phi))$, $\hat{\vb*{r}}\in S^2$ and $\theta\in [0,\pi),\phi\in [0,2\pi)$. Topologically, the light cone is given as $S^2\times [0,\infty)$, with the sphere at the origin $S^2\times \{0\}$ identified to a point. On this space, $\lambda\in\R^+$ linearly parametrizes the non-compact direction, while $\hat{\vb*{r}}$ parametrizes $S^2$. Setting $Y = (1,0,0,1)$ and using Eq.~\eqref{eq:delta function on cone}, $K_{00}$ evaluates to
\begin{equation}
    K_{00}^{(\rho,0)}(\lambda,\theta) = \frac{\delta(\theta)}{\sin(\theta)}\lambda^{i\rho-1}\,,
\end{equation}
where the term $\frac{\delta(\theta)}{\sin(\theta)}$ arises from a $\delta$-function on $S^2$, which acts regularly upon integration. In particular, $\sin(\theta)$ in the denominator is canceled when the measure $\dd{\Omega}$ on $S^2$ is considered. 

\subsection{Kernels of mixed type}\label{subsec:Kernels of Mixed Type}

Extending the arguments of~\cite{Barrett:1999qw} to the case of mixed signatures, $\alpha_1\neq\alpha_2$, the kernels $K_{\alpha_1\alpha_2}$ define a projection of functions $f\in L^2(\mathrm{H}_{\alpha_2})$ onto the $(\rho,0)$, respectively the $(0,\nu)$ components of $L^2(\mathrm{H}_{\alpha_1})$. Since the spaces $L^2(\mathrm{H}_{\texttt{+}})$ and $L^2(\mathrm{H}_0)$ only decompose into $(\rho,0)$ representations, it follows immediately that $K^{(0,\nu)}_{\alpha_1\alpha_2} = 0$ for $\alpha_1\neq\alpha_2$. This reflects on the level of $\SL$ representations that classically, timelike tetrahedra can only be glued to spacelike or lightlike tetrahedra along spacelike faces. To obtain the kernels $K_{\alpha_1\alpha_2}^{(\rho,0)}$ one expands a function $f\in L^2(\mathrm{H}_{\alpha_1})$ via its Gel'fand transform $F$, given in Appendix~\ref{app:Harmonic analysis on homogeneous spaces}, only keeping the $\rho$-component. Disregarding the $\nu$-component precisely corresponds to the fact that $K^{(0,\nu)}_{\alpha_1\alpha_2} = 0$ for $\alpha_1\neq\alpha_2$. Then, one inserts for the function $F$ the inverse expression on $\mathrm{H}_{\alpha_2}$. The integrand of the resulting $\mathrm{H}_{\alpha_2}$-integration is then identified with the kernel $\overline{K_{\alpha_1\alpha_2}^{(\rho,\nu)}(X_1,Y_2)}$.

Applying the Gel'fand expansion given in Eq.~\eqref{eq:inverse Gel'fand transform on 3-hyperboloid} to a function $f$ on $\mathrm{H}_{\texttt{+}}$,
\begin{equation}
f(X_+) = \int\dd{\rho}\rho^2\int_{S^2}\dd{\Omega}F(\xi;\rho)(X^{\mu}_+\xi_{\mu})^{-i\rho-1}\,,
\end{equation}
and inserting the $\rho$-component of the Gel'fand transform $F(\xi;\rho)$ on $\mathrm{H}_\texttt{-}$, given in Eq.~\eqref{eq:Gel'fand transform on H21 with rho}, one obtains
\begin{equation}
f(X_+)
=
\int\dd{\rho}\rho^2\int\dd{Y_-}\int_{S^2}\dd{\Omega}(X_+^{\mu}\xi_{\mu})^{-i\rho-1}\abs{Y^{\nu}_{-}\xi_{\nu}}^{i\rho-1}f(Y_-)\,.
\end{equation}
From this equation, we extract the mixed kernel $K_{\texttt{+-}}^{(\rho,0)}$ which, in the parametrization $X_+ = (1,0,0,0)$, $Y_- = (\sinh(\eta),\cosh(\eta)\hat{\vb{r}})$ with $\eta\in\R,\hat{\vb*{r}}\in S^2$, evaluates to
\begin{equation}
K_{+-}^{(\rho,0)}(\eta,\hat{\vb*{r}})
=
\int\frac{\dd{\phi}}{4\pi}\int\dd{t}\abs{\sinh(\eta)-\cosh(\eta)\left(\sqrt{1-t^2}\sqrt{1-r_z^2}\sin(\phi)+tr_z\right)}^{-i\rho-1}\,.
\end{equation}
We can gain some further intuition by considering $r_z = \pm 1$, which is of course a restriction of the general case where $r_z\in [-1,1]$, yielding $K_{+-}^{\rho}(\eta,\pm 1) = \frac{i\cos(\rho\eta)}{\rho\cosh(\eta)}$. This expression is of a similar structure as $K^{\rho}_{++}(\eta)$ and $K^{\rho}_{--}(\eta,\pm 1)$, and is regular at $\rho = 0$ if considered under an integral of $\rho$, due to the Plancherel measure.

$K_{0\texttt{-}}$ is similarly derived by considering the Gel'fand expansion of a function $f$ on $\mathrm{H}_0$, given in Eq.~\eqref{eq:inverse Gel'fand transform on cone}, and inserting for $F(\xi;\rho)$ the $\rho$-component of the Gel'fand transform on $\mathrm{H}_{\texttt{-}}$, defined in Eq.~\eqref{eq:Gel'fand transform on H21 with rho}. For $X_0 = \lambda\xi\in\mathrm{H}_0$ and $Y_\texttt{-}\in\mathrm{H}_{\texttt{-}}$, this procedure yields
\begin{equation}
f(\lambda\xi)
=
\int\dd{\rho}\rho^2\int\dd{Y_-}\abs{Y_-^{\mu}\xi_{\mu}}^{-i\rho-1}\lambda^{-i\rho-1}f(Y_-)\,,
\end{equation}
from which we extract $K_{0\texttt{-}}^{(\rho,0)}(X_0,Y_\texttt{-}) = \abs{X^{\mu}Y_{\mu}}^{i\rho-1}$. In the parametrization with $Y_\texttt{-} = (0,0,0,1)$, this simplifies further to $K^{(\rho,0)}_{0\texttt{-}}(\lambda,\theta) = \abs{\lambda\cos(\theta)}^{i\rho-1}$, which is a regular function for all $\lambda\in\R^+$, $\rho\in\R$ and $\theta\in[0,2\pi)\setminus\{\frac{\pi}{2},\frac{3\pi}{2}\}$. Similar to the ($\texttt{+-}$) case, the discrete part of the Gel'fand transform on $\mathrm{H}_{\texttt{-}}$ is projected out, leaving only terms with $\nu = 0$. This again reflects on the level of quantum amplitudes the condition that a lightlike and timelike tetrahedron can only be glued along a spacelike face (as lightlike faces are excluded).

Proceeding with the mixed case of a timelike and a lightlike normal vector, $X_{\texttt{+}}\in\mathrm{H}_\texttt{+},Y_0\in\mathrm{H}_0$, we write down the inverse Gel'fand transform of a function on $\mathrm{H}_\texttt{+}$ according to Eq.~\eqref{eq:inverse Gel'fand transform on H21} and insert for $F(\xi;\rho)$ the Gel'fand transform for functions on the light cone $\mathrm{H}_0$, given in Eq.~\eqref{eq:Gel'fand transform on cone}, leading to
\begin{equation}
f(X_\texttt{+})
=
\int\dd{\rho}\rho^2\int\dd{Y_0}\left(X^{\mu}_+Y_{0{\mu}}\right)^{-i\rho-1},
\end{equation}
from which  the kernel $K_{\texttt{+}0}^{(\rho,0)}(X_\texttt{+},Y_0) = (X^{\mu}_+Y_{0\mu})^{i\rho-1}$ is extracted. Choosing $X_\texttt{+} = (1,0,0,0)$ and $Y_0 = \lambda\xi$, $K_{\texttt{+}0}^{(\rho,0)}$ is further simplified to $K_{+0}^{(\rho,0)}(\lambda) = \lambda^{i\rho-1}$. Clearly, this kernel is regular for all values of $\lambda\in\R^+$ and $\rho\in\R$. In comparison to the mixed cases above, spacelike and lightlike tetrahedra allow for spacelike faces only, and so no $\nu$-components are projected out.

With the computations of all the kernels $K_{\alpha_1\alpha_2}$ achieved, vertex amplitudes of any causal character assignment can be computed as a convolution of these kernels according to Equation~\eqref{eq:generalized vertex integral form}.

\subsection{Spacetime orientation}\label{subsec:Spacetime Orientation}

In addition to the complete set of causal building block, incorporating a notion of time orientability is required to cover every aspect of causality. Heuristically, this should allow distinguishing between past and future, thus inducing a causal ordering. Following~\cite{Livine:2002rh}, the BC model restricted to spacelike tetrahedra does not exhibit a time orientation, reflected by the invariance under time reversal $T = \mathrm{diag}(-1,1,1,1)$, i.e. $K_{\texttt{++}}(T\cdot X,T\cdot Y) = K_{\texttt{++}}(X,Y)$ for all $X,Y\in\mathrm{H}_{\texttt{+}}$. In fact, as detailed in~\cite{Jercher:2022mky}, all the kernels $K_{\alpha_1\alpha_2}$ are invariant under $T$, as well as parity $P=\mathrm{diag}(1,-1,-1,-1)$, and spacetime reversal $PT$. Thus, the $K_{\alpha_1\alpha_2}$ exhibit a symmetry under the larger group $\text{O}(1,3)$. In summary, the complete model defined by the kernels $K_{\alpha_1\alpha_2}$ does neither incorporate causality in the sense of time orientability nor is it sensitive to space and spacetime orientation.

A lack of oriented amplitudes is a typical feature of spin-foam models.
For appropriately chosen boundary data, vertex amplitudes
generically asymptote to $\mathcal{A}_v\sim e^{iS_{\mathrm{R}}}+e^{-iS_{\mathrm{R}}}\sim\cos(S_{\mathrm{R}})$, referred to as the \enquote{cosine problem}~\cite{Barrett:2009mw}, with $S_{\mathrm{R}}$ the Regge action.
%
The proposals for oriented models in~\cite{Livine:2002rh,Bianchi:2021ric} are characterized by an \emph{a posteriori} restriction of the quantum amplitudes, such that the semi-classic limit yields only one Regge exponential (see also~\cite{Engle:2015mra,Engle:2015zqa}). Developing an $\text{O}(1,3)$ GFT and spin-foam model for Lorentzian QG, chosen to produce orientation-dependent amplitudes, therefore constitutes a compelling research direction. In the (2+1) coherent model~\cite{Simao:2024don} employed previously, the boundary states associated to edges pre-select an orientation~\cite{Jercher:2024kig}. Therefore, another interesting question is whether a (3+1) model with boundary states corresponding to edges would come equipped with a notion of orientation.\\

\begin{center}
    \textit{Summary.}
\end{center}

\noindent The causally complete BC model developed in this chapter constitutes the first GFT and spin-foam model which includes the full set of causal building blocks, being spacelike, lightlike and timelike tetrahedra. The kernels defining the vertex amplitude of the model are equipped with explicit expressions for any combination of causal character, using methods of integral geometry~\cite{VilenkinBook}. The importance of this causal completion lies in the variety of applications it offers. For instance, it allows the comparison to other QG models that incorporate a wider class of causal building blocks such as the EPRL-CH spin-foam model or CDT. We briefly elaborate on these connections subsequently. Most importantly, the complete BC GFT model allows studying the impact of an extended set of causal building blocks on the phase structure, as well as the extraction of cosmological perturbations from the entanglement between spacelike and timelike tetrahedra. These two avenues shall be explored in the next two chapters. 

\begin{center}
    \textit{Closing remarks.}
\end{center}

\paragraph{Relation to the EPRL-CH model.} Evidently, the EPRL-CH and the complete BC models are quantizations of different classical theories. The EPRL-CH model is based on first-order Palatini-Holst gravity including the $\bi$ parameter. Besides the Poisson structure~\cite{Baez:1999tk}, also the EPRL-CH simplicity constraint differs, given by $X\cdot(*B+\frac{1}{\bi}B)=0$. For $\bi<\infty$, this constraint is second class and therefore has to be imposed weakly in the EPRL-CH spin-foam model~\cite{Engle:2007wy,Conrady:2010kc,Conrady:2010vx,Baratin:2011hp}. On $\SL$ representations $(\rho,\nu)$, simplicity acts as follows: for a timelike normal vector, $\rho = \bi \nu$ and $\nu = j\in\mathbb{N}/2$ with $j$ an $\SUT$ representation. For a spacelike normal vector and a spacelike face, $\rho=\bi\nu$ and $\nu = -k\in\mathbb{N}/2$ with $k$ an $\SUO$ representation in the discrete series. In both cases, the squared area spectrum is $A^2\sim -\bi^2 j(j+1)$ and thus differs from the BC model by the presence of $\bi$ which controls the gap in the spectrum as well as the discreteness. For a spacelike normal and a timelike face, $\nu = -\bi\rho$ and $\rho = -\sqrt{s^2+1/4}$ with $s$ an $\SUO$ representation in the continuous series. The area spectrum is given by $A^2\sim \bi^2(s^2+1/4) = \nu^2$ and is thus similar to that of the complete BC model. Lastly, we remark that the EPRL-CH model does not treat configurations with lightlike normal vectors and that an explicit GFT formulation of the EPRL model and its CH extension is still to be developed.

\paragraph{A CDT-like model.} The variety of causal configurations entering the complete BC model allows approaching CDT~\cite{Ambjorn:2012jv,Loll:2019rdj} within the complete BC model at increasing degrees of proximity. Following~\cite{Jercher:2022mky} and referring to the details given therein, this consists of 1) a restriction to the interactions $\mathfrak{V}_{(4,1)} = \lambda_{(4,1)}\int\dd{\vb*{X}}\Tr\left[\varphi_{\texttt{+}}\varphi_{\texttt{-}}^4\right] + \text{c.c.}$ and $\mathfrak{V}_{(3,2)} = \lambda_{(3,2)}\int\dd{\vb*{X}}\Tr\left[\varphi_{\texttt{-}}^5\right]+ \text{c.c.}$ mimicking the $(4,1)$ and $(3,2)$ simplices of CDT, 2) a restriction of the causal character of faces, 3) fixing and relating the representations in analogy to the length of CDT. The resulting model, explicitly given in~\cite{Jercher:2022mky}, defines a causal tensor model with the tensors being the group fields in spin representation and the $\SL$ magnetic indices $(jm)$ defining the corresponding tensor indices. In lower dimensions, multi-matrix models~\cite{Ambjorn:2001br,Benedetti:2008hc,Eichhorn:2020sla} have already proven to efficiently encode causality. To complete the definition of the causal tensor model, two steps are remaining: 4) coloring the tensors to ensure that only non-singular simplicial complexes are generated (this is not guaranteed by preceding restrictions) and 5) enforcing a foliation constraint via a modification (dual-weighting~\cite{Kazakov:1995ae,Benedetti:2011nn}) of the kinetic term to prevent spatial topology change~\cite{Horowitz:1990qb,Louko:1995jw,Dowker:1999wu}, following the (1+1) causal matrix model of~\cite{Benedetti:2008hc}. It is an intriguing question for future research whether in (3+1) dimensions causal regularity as defined in Sec.~\ref{sec:causal regularity} requires constraints on bubbles of the colored graphs and what kind of dual-weighting can achieve that.

  
\paragraph{Asymptotic analysis.} An important future avenue is to extend the analysis of the asymptotics and the perturbative finiteness for the BC model with timelike normal, given respectively in~\cite{Barrett:2002ur,Baez:2002rx} and~\cite{Crane:2001as,Crane:2001qk}, to the complete model. In particular, it would be interesting to investigate whether there exists a finite and closed expression of the vertex amplitude including timelike interfaces which remains unknown for the EPRL-CH model, see~\cite{Simao:2024don,Simao:2021qno}.  

\paragraph{Timelike and lightlike boundaries.} The complete model allows considering lightlike as well as timelike boundaries. The most prominent example including the latter is anti-de Sitter space~\cite{kroon2016conformal}, which is of enormous theoretical and physical interest. Lightlike boundaries on the other hand are quintessential to describe cosmological and black hole horizons. In the enlarged setting of the complete model, it would be possible to revisit the observations of~\cite{Oriti:2018qty} on black holes as GFT condensates in terms of lightlike hypersurfaces, foliated into spheres. Studies in this direction could strengthen the area law results of~\cite{Oriti:2018qty} and offer a way to enforce more detailed horizon conditions, with the horizon being understood as a lightlike boundary.
\chapter[\textsc{Landau-Ginzburg Analysis of the Complete Barrett-Crane Model}]{Landau-Ginzburg Analysis of the Complete Barrett-Crane Model}\label{chapter:LG}

Coarse-graining methods are expected to be a crucial tool to describe the emergence of smooth spacetime geometries from an underlying fundamental QG theory~\cite{BenAchour:2024gir,Asante:2022dnj,Oriti:2013jga}. The Kadanoff-Wilson formulation of the renormalization group (RG) implements a coarse-graining operation by progressively eliminating short-scale fluctuations towards the infrared (IR) and thus allows investigating how a physical theory evolves along scales~\cite{Wilson:1983xri}. This procedure allows searching for RG fixed points, charting the phase diagram of the considered model,  studying phase transitions and determining critical exponents~\cite{Zinn-Justin:2002ecy,Zinn-Justin:2007uvz}. A potent realization of this idea is provided by the functional renormalization group (FRG) methodology~\cite{Delamotte:2007pf,Kopietz:2010zz,Dupuis:2020fhh} which, beyond standard local field theories, has been applied to matrix and tensor models~\cite{Sfondrini:2010zm,Eichhorn:2013isa,Eichhorn:2014xaa,BenGeloun:2016tmc,Eichhorn:2017xhy,Eichhorn:2018phj,Eichhorn:2019hsa,Castro:2020dzt,Eichhorn:2020sla} and in particular to GFTs~\cite{Benedetti:2014qsa,BenGeloun:2015xrk,BenGeloun:2016rqa,Benedetti:2015yaa,Carrozza:2016tih,Carrozza:2016vsq,Carrozza:2017vkz,BenGeloun:2018ekd,Pithis:2020sxm,Pithis:2020kio,Baloitcha:2020lha,Geloun:2023ray,Juliano:2024rgu,Carrozza:2024gnh}, making use of their field-theoretic description.

Circumventing the heavy machinery of FRG, a simpler, yet efficient coarse-graining approach is offered by Landau-Ginzburg (LG) mean-field theory which was originally introduced to study phase transitions in local field-theoretic descriptions of lattice systems~\cite{Kopietz:2010zz,zinn2021quantum}. It provides a coarse account of the phase structure of a theory, captured by the mean-field, acting as an order parameter. To check self-consistency of this approximation, one has to verify that fluctuations remain small in the vicinity of the phase transition, known as the Levanyuk-Ginzburg criterion~\cite{levanyuk1959contribution,ginzburg1961some}. Although being a rather simple method, the application of LG theory to GFTs is highly non-trivial mainly due to their combinatorial non-local interactions. Still, considerable progress in this direction has been achieved~\cite{Pithis:2018eaq,Marchetti:2020xvf,Marchetti:2022igl,Marchetti:2022nrf}. In particular, in~\cite{Marchetti:2022igl,Marchetti:2022nrf}, LG theory has been applied to the BC model restricted to spacelike tetrahedra with simplicial and tensor-invariant interactions~\cite{Carrozza:2013oiy,Carrozza:2016vsq}. This analysis showed that due to the hyperbolic geometry of the Lorentz group, mean-field theory provides a self-consistent description of phase transitions towards non-vanishing vacua. This is interesting since such non-perturbative vacua are typically highly excited by GFT quanta which makes a compelling case for a continuum geometric approximation. These results also motivate the extraction of effective cosmological dynamics from GFT condensates as will be presented in the next chapter.

Here, we apply the LG method to the complete BC model introduced in Chapter~\ref{chapter:cBC}. This analysis will elucidate the impact of the enlarged set of configurations, including spacelike and timelike faces, onto the critical behavior and the phase structure at mean-field level.

\paragraph{General strategy.} Before heading into details, let us briefly outline the general strategy pursued. We employ the complete BC model with real-valued group fields $\varphi_\alpha$ minimally coupled to $\dloc$ massless free scalar fields $\vbf\in\R^{\dloc}$ as described in Sec.~\ref{sec:Extended Barrett-Crane Group Field Theory Model}. A priori, no assumptions are posed on the action $S = \mathfrak{K}+\mathfrak{V}$. The LG analysis is conducted in the following steps: 1) The equations of motion, $\delta S/\delta\varphi_\alpha = 0$, are solved for constant field configurations $\varphi^{\mathrm{m}}_\alpha$, constituting the mean-field, or equivalently the order parameter of the system. 2) Fluctuations $\delta\varphi_\alpha$ around the mean-field are studied via $\varphi_\alpha(\vbg,\vb*{\phi},X_\alpha) = \varphi^{\mathrm{m}}_\alpha+\delta\varphi_\alpha(\vbg,\vb*{\phi},X_\alpha)$, effectively described by a Gaussian field theory as higher orders in $\delta\varphi$ are neglected. The correlation function, defined as the 2-point function of fluctuations, allows defining local and non-local correlations via
\begin{equation*}
C_{\alpha\beta}(\vb*{\phi}) \defeq \int\dd{\vb*{g}}C_{\alpha\beta}(\vb*{g},\vb*{\phi})\,,\qquad C_{\alpha\beta}(\vb*{g}) \defeq \int\dd{\vb*{\phi}}C_{\alpha\beta}(\vb*{g},\vb*{\phi})\,.
\end{equation*}
Given an asymptotic exponential decay of $C_{\alpha\beta}(\vbf)$ and $C_{\alpha\beta}(\vbg)$, local ($\xiloc$) and non-local ($\xinloc$) correlation lengths are extracted. 3) Validity of the mean-field ansatz is given if the fluctuations are small compared to the mean-field. This is quantified by the Ginzburg-$Q_{\alpha\beta}$ which will be defined in Eq.~\eqref{eq:definition Q}. If the Ginzburg-Levanyuk criterion~\cite{levanyuk1959contribution,ginzburg1961some}, $Q_{\alpha\beta}\ll 1$ close to criticality, holds, the mean-field approach constitutes a consistent approximation of the phase transition towards a non-perturbative vacuum state. 

\section{A single interaction of arbitrary causal structure}

In this section, we explicitly perform the LG analysis for the complete BC GFT model with a single interaction of double-trace melonic, quartic melonic, necklace or simplicial type and an arbitrary combination of spacelike, lightlike, or timelike tetrahedra (see Fig.~\ref{fig:causal vertex graph} below). Following the detailed discussion of~\cite{Dekhil:2024djp}, a summary of the extension to other types of interactions is summarized at the end of this chapter.

The kinetic term $\mathfrak{K}$ employed in this section is defined by the kinetic kernel
\begin{equation}\label{eq:kinetic kernel}
    \mathcal{K}_{\alpha}(\vbg,\vbf;\vbg',\vbf') = \mu_{\alpha}\delta(\vbg^{-1}\vbg') - Z^\phi_\alpha(\vbg^{-1}\vbg')\Delta_{\phi} - Z^g_\alpha(\abs{\vbf-\vbf'})\sum_{c=1}^4 \Delta_c\,,
\end{equation}
which consists of the following ingredients. \textit{Mass parameters:} the $\mu_\alpha$ play the role of masses which, in a spin-foam picture, lead to a simple multiplicative factor $\mu_\alpha^{-1}$ of the edge amplitudes. For coupled scalar fields, non-unity masses can be further motivated from the zeroth order of a derivative expansion of the kinetic kernel~\cite{Li:2017uao,Oriti:2016qtz,Gielen:2018fqv}. Here, the $\mu_\alpha$ serve as control parameters that allow to separate the phases $\mu_\alpha > 0$ and $\mu_\alpha < 0$. In a thermodynamic system with temperature $T$, the masses $\mu_\alpha$ would be expressed as $\mu_\alpha\sim (T - T_c)$ where $T_c$ is the critical temperature at which the phase transition occurs, see also~\cite{Goldenfeld:1992qy}.\footnote{Note however that this merely serves as an analogy, in particular because the considered GFT particles are particles \textit{of} spacetime. If even possible, relating the masses $\mu_\alpha$ to a temperature \textit{in} spacetime goes beyond the scope of this work.}  \textit{Laplace operators:} the Laplace operators introduce a notion of scale on the geometric and matter domains. $\Delta_{\phi}$ is the Laplace operator acting on the scalar fields $\vbf\in\R^{\dloc}$ and arises from a derivative expansion, similar to the mass terms, but at second order. $\Delta_c$ is the Laplace operator on $\SL$, typically motivated from studies of radiative corrections~\cite{BenGeloun:2011jnm,BenGeloun:2011rc,BenGeloun:2013mgx}, introducing in particular a notion of scale which is crucial to define an RG scheme. \textit{Weights of Laplacians:} we anticipate at this point that one has to generalize the pre-factors of the Laplacians to functions $Z_\alpha^\phi$ and $Z_\alpha^g$ for obtaining well-behaved correlation functions with an asymptotic exponential fall-off and no oscillations or exponential divergences (see the discussions in Secs.~\ref{sec:Local correlation function} and~\ref{sec:Non-local correlation function}.). The factors $Z_\alpha^\phi$ are also known to encode non-trivial features of the minimal coupling~\cite{Oriti:2016qtz,Li:2017uao,Gielen:2018fqv}. The reciprocity of the matter-gravity coupling suggests in turn the factors $Z_\alpha^g$. Note the resemblance of these functions with wave function renormalizations~\cite{WipfRG} which, within a full-fledged RG treatment, would be required for the consistency of the flow equations. Thus, we expect the $Z_\alpha^{\phi,g}$ to emerge in such a procedure, however leaving a deeper analysis of this matter to future investigations. To sustain the symmetries of the kinetic kernel, the particular dependence of $Z_\alpha^\phi$ and $Z_\alpha^g$ on $\Tr(\vbg^{-1}\vbg')$, respectively $\abs{\vbf-\vbf'}$ is assumed.

We consider in this section a real-valued group field with a single interaction term characterized by a fixed but arbitrary causal vertex graph $\upgamma_{\textsc{c}}$ with $n_\upgamma = \sum_\alpha n_\alpha$ tetrahedra, $\np$ being spacelike, $n_0$ being lightlike and $\nm$ being timelike. That is, we employ Eq.~\eqref{eq:general vertex} for a particular $\upgamma_{\textsc{c}}$ supplemented with a local scalar field integration. Examples of $\upgamma_{\textsc{c}}$ are given in Fig.~\ref{fig:causal vertex graph}. The degree $(\np,\nz,\nm)$ determines the symmetries of the model. In particular, if $n_\alpha$ is even, the model exhibits a $\mathbb{Z}_2$ symmetry, $\varphi_\alpha\mapsto -\varphi_\alpha$. For the remainder, we pose no assumptions on the $n_\alpha$ and discuss this point again at the end of this chapter.

\begin{figure}
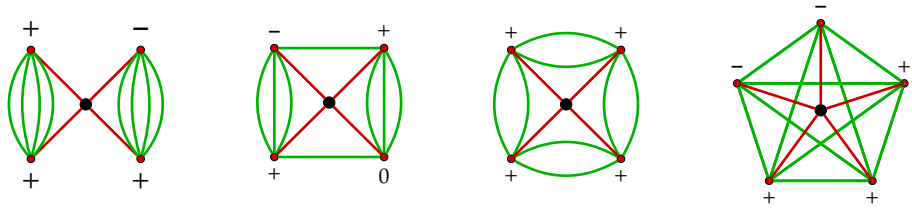

    \centering
\cvftpppm~~~~~~~~~\cvfppmzc~~~~~~~~~\cvfnpppp~~~~~~~~~\cvfspppmm
    \caption{Diagrammatic representation of four exemplary causal vertex graphs $\upgamma_{\textsc{c}}$. Red vertices represent the fields $\varphi_\alpha(\vbg,\vbf)$ (tetrahedra) and green half-edges indicate pairwise convolution of the non-local variables $\vbg$ (faces). The variables $\vbf$ enter $\mathfrak{V}$ locally, indicated by the red edges. From left to right: double-trace melon with three spacelike and one timelike tetrahedron, quartic melon with two spacelike, one lightlike and one timelike tetrahedron, quartic necklace with four spacelike tetrahedra, $4$-simplex with three spacelike and two timelike tetrahedra.}
    \label{fig:causal vertex graph}
\end{figure}

\subsection{Mean-field equations and linearization}\label{sec:Mean-field equations and linearization}

Evaluating the equations of motion on uniform\footnote{Since $\SL$ is non-compact, constant field configurations do not live in an $L^2$-space, requiring the extension to the space of so-called hyperfunctions~\cite{Ruehl1970,hormander2015analysis}. Thereby, one does not rely on a regularization via Wick rotation as in~\cite{Marchetti:2022igl}, keeping the causal structure encoded in spacelike, lightlike and timelike building blocks clearly visible.}, i.e.~constant, field configurations yields
\begin{equation}
    (\varphi^{\mathrm{m}}_\alpha)^{n_\alpha-2}V_\alpha^{n_\alpha-1} = -\frac{\mu_\alpha}{\lambda n_\alpha}\Vp^{4-2n_\upgamma}\prod_{\beta\neq\alpha}(\varphi^{\mathrm{m}}_\beta V_\beta)^{-n_\beta}\,,
\end{equation}
for $\mu_{\alpha} < 0$, where the $V_\alpha$ are divergent volume factors of empty $\SL$ or $\mathrm{H}_\alpha$ integrations, regularized by a cutoff in the non-compact direction. The scaling behavior of the $V_\alpha$ in the cutoff is discussed in Appendix~\ref{app:Empty integrals} and will be made explicit in Sec.~\ref{sec:arbitrary but fixed Q}. The mean-field equations are solved by $\varphi^{\mathrm{m}}_\alpha = 0$ for $\mu_\alpha > 0$ and
\begin{equation}\label{eq:mf solutions}
\varphi^\mathrm{m}_\alpha = V_\plus^{-2}\left(-\frac{\mu_\alpha}{n_\alpha\lambda}\right)^{\frac{2+n_\alpha-n_\upgamma}{2(n_\upgamma-2)}}V_\alpha^{-\frac{n_\alpha}{2(n_\upgamma-2)}-\frac{1}{2}}\prod_{\beta\neq \alpha}\left(-\frac{\mu_\beta}{n_\beta\lambda}\right)^{\frac{n_\beta}{2(n_\upgamma-2)}}V_\beta^{-\frac{n_\beta}{2(n_\upgamma-2)}}\,,
\end{equation}
for $\mu_\alpha < 0$, thus depending on the masses $\mu_\alpha$, the coupling $\lambda$, the $n_\alpha$ characterizing $\mathfrak{V}$ and volume factors. Clearly, the $\varphi^{\mathrm{m}}_\alpha$ act as order parameters, distinguishing between the phases of vanishing and non-vanishing vacuum expectation value. This is analogous to a thermodynamic magnetic system where the order parameter corresponds to the magnetization which is vanishing above the critical temperature and non-vanishing below. 

Introducing fluctuations $\delta\varphi_\alpha$ around the mean-fields and linearizing the equations of motion, one obtains
\begin{equation}
0 = \sum_\beta\int\dd{\vb*{g}'}\dd{\vb*{\phi}'}\dd{X_\beta'}\left[\delta_{\alpha\beta}\mathcal{K}_\alpha(\vb*{g},\vb*{\phi};\vb*{g}',\vb*{\phi}')\delta(X_\alpha,X_\beta')+F_{\alpha\beta}(\vb*{g},\vb*{\phi};\vb*{g}',\vb*{\phi}')\right]\delta\varphi_\beta(\vb*{g}',\vb*{\phi}',X_\beta')\,.
\end{equation}
The kinetic contribution involves a $\delta$-function imposed on the normal vectors $X_\alpha$ and $X_\beta'$, while the Hessian function $F_{\alpha\beta}$ is independent of those arguments. In spin representation, this leads to BC intertwiners (see Eq.~\eqref{eq:definition of generalized BC intertwiners}) appearing in the Hessian term which are however absent in the kinetic term. To symmetrize this imbalance, which otherwise obstructs further evaluation, and to eliminate the auxiliary normal vector variable from the equations of motion, we perform an additional $X_\alpha$-integration. This adds a BC intertwiner to the kinetic term in spin representation as well as an additional volume factor to the Hessian~\cite{Marchetti:2022igl}. As a result, the effective equations of motion for the fluctuating fields take the form
\begin{equation}
0 = \sum_\beta\int\dd{\vb*{g}'}\dd{\vb*{\phi}'}G_{\alpha\beta}(\vb*{g},\vb*{\phi};\vb*{g}',\vb*{\phi}')\delta\varphi_\beta(\vb*{g}',\vb*{\phi}')\,,
\end{equation}
with an effective kinetic kernel $G_{\alpha\beta}$. Note that the dynamical equations can be derived from an effective quadratic action, $S_{\mathrm{eff}}[\delta\varphi_\alpha]$. The Hessian contribution $F_{\alpha\beta}$ to the effective kinetic kernel is given by 
$V_\alpha F_{\alpha\beta}(\vb*{g},\vb*{\phi};\vb*{g}',\vb*{\phi}')  = -\sqrt{\frac{\mu_\alpha\mu_\beta}{n_\alpha n_\beta}}\chi_{\alpha\beta}\delta(\vb*{\phi}-\vb*{\phi}')$, where we used that the ratios $V_\alpha/V_\beta$ converge to unity after regularization, as shown in Appendix~\ref{app:Empty integrals}. 

The matrix $\chi_{\alpha\beta}(\vb*{g},\vb*{g}')$ is the generalization of the function $\mathcal{X}(\vb*{g},\vb*{g}')$ introduced in~\cite{Marchetti:2020xvf,Marchetti:2022nrf}. In the causally extended setting considered here, it plays an essential role in capturing the interplay of combinatorial non-localities and the different causal characters of the tetrahedra in spite of the projection to uniform field configurations. It is appropriately regularized and leads to Kronecker-delta-like symbols $\delta_{\rho,i}$ in spin representation, the details of which are explained in Appendix~\ref{sec:Regularization of Dirac delta function}. An exhaustive list of this matrix for double-trace melonic, quartic melonic, necklace and simplicial interactions is given in Appendix~\ref{sec:Explicit expressions for chi}. As an example, two spacelike and two timelike tetrahedra can be glued in three different ways according to quartic melonic combinatorics, all of which lead to different expressions for $\chi_{\alpha\beta}$. Thus, $\chi_{\alpha\beta}$ crucially depends on the details of the causal vertex graph $\upgamma_{\textsc{c}}$ governing the interaction. 

\subsection{Correlation functions in spin representation}\label{sec:Correlation functions in spin representation}

The correlation function is obtained by inverting the effective kinetic kernel, which is commonly done in Fourier space for both, matter $(\vbf)$ and geometric $(\vbg)$ variables. The local variables $\vbf\in\R^{\dloc}$ are transformed via the standard Fourier transform on $\R^{\dloc}$. The geometric variables are instead expanded in spin representation, introduced in Sec.~\ref{subsec:Spin Representation of the Group Field and its Action} with further details given in Appendix~\ref{sec:Aspects of SL2C and its Representation Theory}. In contrast to previous studies~\cite{Marchetti:2022igl,Marchetti:2022nrf}, however, this is more involved here due to the interplay of the causal characters of tetrahedra and faces. Intuitively, while any two tetrahedra can be correlated via spacelike $(\rho)$ faces, only two timelike tetrahedra ($\alpha = \beta = \texttt{-}$) can be correlated via timelike faces ($\nu$). Following this intuition and the rigorous derivation of Appendix~\ref{app:A derivation of the correlation function} to which we refer to for further details, the spin representation of the effective action $S_{\mathrm{eff}}[\delta\varphi_\alpha]$ is given by
\begin{equation}
\begin{aligned}
S_{\mathrm{eff}} = \frac{1}{2}&\sum_{\alpha,\beta}\int\prod_{c=1}^4\dd{\rho_c}\rho_c^2\:\delta\varphi^{\vbr,\alpha}_{\vb*{j}\vb*{m}}(\vbk)B^{\vbr,\alpha}_{\vb*{l}\vb*{n}}G_{\alpha\beta}^{\vbr}(\vbk)\delta\varphi^{\vbr,\beta}_{\vb*{j}\vb*{m}}(\vbk)B^{\vbr,\alpha}_{\vb*{l}\vb*{n}}+\\[7pt]+\frac{1}{2}\sum_{t=1}^4\sum_{(c_1,...,c_t)}&\prod_{u=1}^t\sum_{\nu_{c_u}}\nu_{c_u}^2\int\prod_{v=t+1}^4\dd{\rho_{c_v}}\rho_{c_v}^2\delta\varphi_{\vb*{j}\vb*{m}}^{(\vbr\vbn)_t,\minus}(\vbk)B^{(\vbr\vbn)_t,\minus}_{\vb*{l}\vb*{n}}G_{\minus\minus}^{(\vbr\vbn)_t}\left(\vbk\right)\delta\varphi_{\vb*{j}\vb*{m}}^{(\vbr\vbn)_t,\minus}(\vbk)B^{(\vbr\vbn)_t,\minus}_{\vb*{l}\vb*{n}}\,,
\end{aligned}
\end{equation}
where the $B^{\vbr\vbn,\alpha}_{\vb*{l}\vb*{n}}$ are generalized BC intertwiners defined in Eq.~\eqref{eq:definition of generalized BC intertwiners}, and a sum over repeated magnetic indices is understood. The sum over $(c_1,...,c_t)$ is performed such that the $t$ timelike and $4-t$ spacelike labels are distributed equally across the four possible entries and  $(\vbr\vbn)_t = \nu_{c_1}...\nu_{c_t}\rho_{c_{t+1}}...\rho_{c_4}$. Following the geometric interpretation of Chapter~\ref{subsec:Spin Representation of the Group Field and its Action}, the splitting of the action reflects the fact that spacelike faces can be shared between two tetrahedra of any signature (first term), whereas timelike faces can only be shared between two timelike tetrahedra (second term). Consequently, it is helpful for the remainder to consider the two terms in the effective action separately. In particular, the inversion of $G_{\alpha\beta}$ is performed for each case individually.

If all faces are spacelike then $G_{\alpha\beta}^{\vbr}$ is matrix-valued, and thus the correlation function is obtained as the matrix inverse, i.e. $\sum_\gamma G^{\vbr}_{\alpha\gamma}(\vbk)C^{\vbr}_{\gamma\beta}(\vbk) = \delta_{\alpha\beta}$. As a consequence, the correlation matrix $C_{\alpha\beta}$ contains an inverse factor of the determinant of $G_{\alpha\beta}$, turning it into a rational function in the variables $\vbr$ and $\vbk$. In the presence of at least one timelike face, $G_{\minus\minus}^{(\vbr\vbn)_t}(\vbk)$ is the only component and thus a scalar. Its inverse is simply the multiplicative inverse 
\begin{equation}
C_{\minus\minus}^{(\vbr\vbn)_t}(\vbk) = \frac{1}{Z_\minus^\phi((\vbr\vbn)_t)\vbk^2+\frac{Z_-^g(\vbk)}{a^2}\sum_c\mathrm{Cas}_{1,c}+b^\minus}\,.
\end{equation}
Here $a$ is the skirt radius of $\mathrm{H}_\plus$ and $b^{\minus} \defeq \mu(1-\chi^{(\vbr\vbn)_t})$ is the \emph{effective mass} depending on the labels $(\vbr\vbn)_t$. Note that $\chi^{(\vbr\vbn)_t}$ is scalar-valued and of the same form as $\mathcal{X}$ in~\cite{Marchetti:2022igl,Marchetti:2020xvf}.

The direct space correlator is obtained by performing the inverse of the Fourier transformation for which a detailed derivation is given in Appendix~\ref{app:A derivation of the correlation function}. Following Eq.~\eqref{eq:spinrepCab}, all but the $(\minus\minus)$-component contain contributions of spacelike faces only. That is, the Fourier components are given by $C_{\alpha\beta}^{\vbr}(\vbk)$. The matrix element $C_{\texttt{-}\texttt{-}}(\vbg,\vbf)$ on the other hand contains $C_{\minus\minus}^{\vbr}(\vbk)$ as well as the contributions from timelike faces, $C_{\minus\minus}^{(\vbr\vbn)_t}(\vbk)$. An explicit formula is given in Eq.~\eqref{eq:spinrepC--}.


\subsection{Local correlation function}\label{sec:Local correlation function}

The local correlation function $C_{\alpha\beta}(\vbf)$ is obtained by integrating out the geometric variables $\vbg$, yielding a set of projections onto the trivial presentation with the label $\rho = i$. Notice in particular that the timelike labels $\nu$ are constrained to vanish, $\nu = 0$, which is in conflict with the simplicity condition that $\nu\in \{2,4,6,...\}$. Thus, the contributions to the local correlation of $C_{\minus\minus}$ with $t>0$ timelike faces vanish, leaving
\begin{equation}\label{eq:loc corr 1}
C_{\alpha\beta}(\vbf) = \int\frac{\dd{\vbk}}{(2\pi)^{\dloc}}\e^{i\vbk\cdot\vbf}C_{\alpha\beta}^{\vbi}(\vbk)\,,
\end{equation}
where $\vbi$ denotes the four labels $\vbr$ being evaluated on the trivial representation $\rho_c = i$. Essential for the qualitative behavior of the correlation function is the matrix $\chi_{\alpha\beta}(\vbr)$ which, evaluated on $\vbr = \vb*{i}$, takes the values $\chi_{\alpha\beta}^{\vbi}= n_\alpha(n_\alpha-1)$ for $\alpha=\beta$, and $\chi_{\alpha\beta}^{\vbi}=n_\alpha n_\beta$ for $\alpha\neq \beta$. In particular, the combinatorial details of the interaction are integrated out and the functions $\chi_{\alpha\beta}^{\vbi}$ depend only on the numbers $\np,\nz$ and $\nm$, equivalent to a local theory with multiple fields.

Since $C^{\vbi}_{\alpha\beta}(\vbk)$ is obtained as the matrix inverse of $G^{\vbi}_{\alpha\beta}(\vbk)$, it contains an inverse factor of the determinant of $G_{\alpha\beta}$. As a result, $C^{\vbi}_{\alpha\beta}(\vbk)$ is a rational function in $\vbk^2$. To evaluate the integral for the local correlation function in Eq.~\eqref{eq:loc corr 1} explicitly, it is therefore expedient to perform a partial fraction decomposition, i.e. we write
\begin{equation}\label{eq:local correlation fraction decomp}
C_{\alpha\beta}^{\vbi}(\vbk) = \sum_{m=1}^3\frac{\varsigma^m_{\alpha\beta}}{\vbk^2+b^m_{\vbi}}\,.
\end{equation}
The $b_{\vbi}^m$ are interpreted as \textit{effective masses} evaluated on four trivial representations and are involved functions of the masses $\mu_\alpha$ and the $Z_\alpha^\phi(\vbi)$, implicitly depending on the form of $\chi^{\vbi}_{\alpha\beta}$. 
The coefficients $\varsigma_{\alpha\beta}^m\in\C$ are constants of the partial fraction decomposition, independent of $\mu_\alpha$ and $Z^\phi_\alpha$. Notice that this decomposition depends on the components of the correlator and therefore carries indices $\alpha,\beta$. Since the local interaction is point-wise, all tetrahedra are correlated with one another, and thus, $\varsigma^m_{\alpha\beta}\neq 0$ for all $\alpha,\beta$.

Following~\cite{Marchetti:2022igl,Dekhil:2024djp}, the integrations of Eq.~\eqref{eq:loc corr 1} can be performed explicitly, yielding
\begin{equation}\label{eq:local correlator K sum}
C_{\alpha\beta}(r) = \frac{1}{(2\pi)^{\frac{d}{2}}r^{d-2}}\sum_{m}\varsigma_{\alpha\beta}^m\left(\sqrt{b_{\vbi}^m}r\right)^{\frac{d-2}{2}}K_{\frac{d-2}{2}}\left(\sqrt{b_{\vbi}^m}r\right)\,,
\end{equation}
where $d\equiv \dloc$, $r\equiv \abs{\vbf}$ and $K_n$ are second type modified Bessel functions~\cite{GradshteynBook}. The asymptotic behavior of this function in the limit $r\gg 1$, which can be understood as the limit of large relational distances on the space $\R^{\dloc}$, is crucially determined by the details of $b_{\vbi}^m$. If non-zero, the $b_{\vbi}^m$ satisfy $b_{\vbi}^m(\mu\mu_{\plus},\mu\mu_0,\mu\mu_{\minus}) = \mu\: b_{\vbi}^m(\mu_{\plus},\mu_0,\mu_{\minus})$ for all  $\mu\in\R$, i.e.~they are homogeneous functions of the $\mu_\alpha$. In particular, in the limit $\mu_\alpha\rightarrow 0$, the effective masses $b_{\vbi}^m$ go to zero as well. Some of the $b_{\vbi}^m$ can in principle vanish, which is entirely determined by the matrix $\chi_{\alpha\beta}^{\vbi}$.\footnote{While a vanishing effective mass can be observed generically in tensorial field theories~\cite{Dekhil:2024ssa}, we emphasize that the vanishing of $b_{\vbi}^m$ is because one considers a multi-field theory. In particular, the regularization scheme suggested in~\cite{Dekhil:2024ssa} is not required here to compute the Ginzburg-$Q$ later on.} In those cases, the corresponding contribution to the correlator decays as a power law, scaling as $C\sim r^{-d+2}$. The remaining $b_{\vbi}^m$ are either positive, negative, or even complex, sensitively depending on the parameter values $Z_\alpha^\phi(\vbi)$. We have checked numerically, that there exists a range of these $Z_\alpha^\phi(\vbi)$ for which the $b^m$ are real and positive, and we restrict the theory to this parameter range for the remainder of this work. Using the asymptotic properties of the Bessel function $K_n$~\cite{GradshteynBook}, the local correlation function behaves asymptotically as
\begin{equation}
C_{\alpha\beta}(r)\underset{r\gg1}{\longrightarrow}\frac{1}{r^{\frac{d-1}{2}}}\tilde{\varsigma}^{m_*}_{\alpha\beta}\exp\left(-\sqrt{b_{\vbi}^{m_*}}r\right)\,,
\end{equation}
where the index $m_* = \arg\min_m(b_{\vbi}^m)$. Clearly, the positive effective masses yield an exponential suppression. Notice that this is only the case for certain values of the $Z_\alpha^\phi(\vbi)$. Outside this range, the local correlator can exhibit oscillatory and/or exponentially decaying behavior.

From the asymptotic behavior of $C_{\alpha\beta}$, the correlation length $\xiloc \defeq (b_{\vbi}^{m_*})^{-1/2}$ is extracted which, in a homogeneous limit $\mu\rightarrow 0$, scales as $\xiloc\rightarrow\mu^{-1/2}$. This is the typical mean-field theory result, where the critical exponent of the correlation length, usually denoted as $\nu_\mathrm{crit}$, is given by $\nu_{\mathrm{crit}} = 1/2$. In an analogous magnetic system, the correlation length would then be characterized by the scaling $\xi\sim\abs{T-T_c}^{-1/2}$, diverging as $T$ approaches the critical temperature, $T\rightarrow T_c$. The scaling of the correlation function at criticality is $C_{\alpha\beta}(r)\sim r^{-d+2}$ and is thus also consistent with standard mean-field theory results. We expect that the scaling of the correlation function in a full RG treatment is modified by an anomalous dimension.

As a consistency check, the case of just one signature, e.g. all tetrahedra spacelike, can be re-obtained by setting $Z^\phi_\zero = Z^\phi_\minus = 0$ and demanding that $Z^\phi_\plus > 0$. Then, the correlation function and correlation length of Refs.~\cite{Marchetti:2022igl,Marchetti:2022nrf} are reproduced.


\subsection{Non-local correlation function}\label{sec:Non-local correlation function}

Integrating out the scalar field dependence of $C_{\alpha\beta}(\vbg,\vbf)$ leads to evaluating the corresponding Fourier components on vanishing momenta, $\vbk = 0$. 
%
For further analysis it is useful to expand $C_{\alpha\beta}(\vbg)$ in terms of zero modes as performed in~\cite{Marchetti:2020xvf,Marchetti:2022igl}. That is, because zero modes 1) arise from the projection onto constant field configurations as in Sec.~\ref{sec:Mean-field equations and linearization}, 2) are necessary to evaluate the $\chi_{\alpha\beta}$ matrix entering the effective masses and containing projections onto zero modes, and 3) can be shown to arise from a Wick rotation~\cite{Dona:2021ldn} of $\SL$ to $\Spin$ and back to regularize divergent volume factors\footnote{Note that in this work, no such Wick rotation to $\Spin$ for regularization is being performed in order to keep aspects of the Lorentz group and the causal structure clearly visible. The results of~\cite{Marchetti:2022igl} show that this is indeed a valid strategy.}, as shown in~\cite{Marchetti:2022igl}. Since zero modes are a priori not part of the spin decomposition of $L^2$-functions on $\SL$, we extend for this purpose the correlation function to 
\begin{equation}
C^{\mathrm{ext}}_{\alpha\beta}(\vbg) = \sum_{s=s_0}^4V_\plus^{-s}\sum_{(c_1,...,c_s)}\int\prod_{c=c_1}^{c_s}\dd{g_{c}}C_{\alpha\beta}(\vbg)\equiv \sum_{s=s_0}^4V_\plus^{-s}\sum_{(c_1,...,c_s)}C_{\alpha\beta}^s(\vbg_{4-s})\,,
\end{equation}
where $s$ labels the number of zero modes and $(c_1,\dots c_s)$ denotes the constant group arguments, i.e.~those arguments which contain zero modes in spin representation.\footnote{Volume factors have been included for regularization and can be derived from a de-compactification from $\Spin$ to $\SL$, as shown in~\cite{Marchetti:2022igl}.} We restrict to $s\geq s_0$ zero modes for which the matrix $\chi_{\alpha\beta}$ is non-vanishing. That is because for $s < s_0$, the mass corrections vanish, generically leading to long-range correlations that are present irrespective of the phase transition~\cite{Marchetti:2022igl,Marchetti:2022nrf,Marchetti:2020xvf}. This justifies their exclusion in the analysis of the critical behavior. The number $s_0$ depends on the combinatorics being double-trace melonic ($s_0 = 0$),  quartic melonic ($s_0 = 1$), necklace ($s_0 =2$) or simplicial ($s_0 = 3$)~\cite{Marchetti:2020xvf}. Due to the projection onto $s$ trivial representations, the \textit{residual correlation function} $C_{\alpha\beta}^s$ only depends on the $4-s$ remaining group variables, $\vbg_{4-s} = (g_{c_{s+1}},\dots,g_{c_4})$.

\paragraph{Contributions with spacelike faces only.}

In the case where all faces are spacelike, the contribution to the correlation function $C_{\alpha\beta}^s(\vbg_{4-s})$ is given in terms of an integral
\begin{equation}\label{eq:nonlocal integrand all sl}
\eval{C^s_{\alpha\beta}(\vbg_{4-s})}_{\mathrm{sl}} = \int\prod_{c=c_{s+1}}^{c_4}\dd{\rho_{c}}\rho_{c}^2D^{(\rho_{c},0)}_{j_{c}m_{c}j_{c}m_{c}}(g_{c})C_{\alpha\beta}^{s,\vbr_{4-s}}(\vb*{0})\,.
\end{equation}
Notably, the matrix $\chi_{\alpha\beta}^{c_1\dots c_s}$ entering ${C}^s$ is evaluated on $s$ zero modes and therefore takes a constant value, depending on the details of $\upgamma_\textsc{c}$, see Appendix~\ref{sec:Explicit expressions for chi} for explicit expressions. 

To extract the asymptotic behavior for large distances on the group manifold, one performs a Cartan decomposition in Eq.~\eqref{eq:nonlocal integrand all sl} of group elements $g_c$ into $\SUT$ elements and a boost matrix, as detailed in Eq.~\eqref{eq:Cartan decomp g}. This induces a decomposition of the $\SL$ Wigner matrices as prescribed by Eq.~\eqref{eq:Cartan decomp D}. The resulting $\SUT$ Wigner matrices are independent of $\rho_{c_1},...,\rho_{c_{4-s}}$ and can therefore be factorized from the integral in Eq.~\eqref{eq:nonlocal integrand all sl}. Following~\cite{Dekhil:2024djp} this leaves a correlation function in the boost parameters $\eta\in\R^+$, given by
\begin{equation}\label{eq:C(eta) integral}
 C^s_{\alpha\beta}\left(\vb*{\eta}_{4-s}\right)_{\vb*{j}_{4-s}\vb*{m}_{4-s}} = \int\prod_{c=c_{s+1}}^{c_4}\dd{\rho_{c}}\rho_{c}^2\: d_{j_cj_cm_c}^{(\rho_c,0)}\left(\frac{\eta_c}{a}\right)C_{\alpha\beta}^{s,\vbr_{4-s}}\,,
\end{equation}
with $d^{(\rho,\nu)}_{jlm}(\eta/a)$ the \emph{reduced} Wigner matrix and $a$ the skirt radius of $\mathrm{H}_\texttt{+}$. In the following, we suppress the dependence of the factorized  $C^s_{\alpha\beta}(\vb*{\eta}_{4-s})$ on the magnetic indices.

Following~\cite{Marchetti:2022igl}, the integrals in Eq.~\eqref{eq:C(eta) integral} can be evaluated by performing a contour integration for one of the variables, say $\rho_{c_1}$, and performing a stationary phase approximation for the remaining $3-s$ variables $\rho_{c_2},...,\rho_{c_{4-s}}$. To that end, it is advantageous to carry out a partial fraction decomposition of $C_{\alpha\beta}^{s,\vbr_{4-s}}(\vb*{0})$, yielding
\begin{equation}\label{eq:C(eta) integral after pfd}
 C^s_{\alpha\beta}\left(\vb*{\eta}_{4-s}\right) = \sum_{m=1}^3 \int\prod_{c=c_{s+1}}^{c_4}\dd{\rho_{c}}\rho_{c}^2\:d_{j_cj_cm_c}^{(\rho_c,0)}\left(\frac{\eta_c}{a}\right)\frac{\varrho_{\alpha\beta}^m}{\frac{1}{a^2}\sum_u (\rho_{c_u}^2+1)+b^m_{c_1\dots c_s}}\,.
\end{equation}
The $b_{c_1\dots c_s}^m$ are \textit{effective masses} evaluated on $s$ zero modes, being intricate functions of $\mu_{\alpha}$ and $Z_\alpha^g(\vb*{0})$, and implicitly depending on the matrix $\chi_{\alpha\beta}^{c_1\dots c_s}$. These masses determine the pole structure of the integrand above. 
%
The coefficients $\varrho_{\alpha\beta}^m\in\C$ arise from the partial fraction decomposition, independent of the parameters $\mu_\alpha$ and $Z^g_\alpha$, and explicitly depend on $\alpha,\beta$. In particular, if tetrahedra of causal character $\bar{\alpha}$ and $\bar{\beta}$ are uncorrelated, then $\varrho_{\bar{\alpha}\bar{\beta}}= 0$.

For the asymptotics of $C^s_{\alpha\beta}(\vb*{\eta}_{4-s})$ in large boost parameters $\eta/a\gg 1$, two properties of the $b_{c_1\dots c_s}^m$ are determining: 1) if non-zero, the $b_{c_1\dots c_s}^m$ are homogeneous functions of the parameters $\mu_\alpha$, i.e. $b_{c_1\dots c_s}^m(\mu\mu_{\plus},\mu\mu_0,\mu\mu_{\minus}) = \mu\: b_{c_1\dots c_s}^m(\mu_{\plus},\mu_0,\mu_{\minus})$ for all $\mu\in\R$. In particular, in the limit $\mu_\alpha\rightarrow 0$, the effective masses scale to zero. 2) the effective mass $b_{c_1\dots c_s}^{\bar{m}}$ for one of the $m$, say $\bar{m}$,  will generically vanish. This occurrence has been highlighted in~\cite{Dekhil:2024ssa} in the context of general tensorial field theories. In particular, it was shown therein that the LG method can be applied also to this case by introducing a regularization via a small parameter $\epsilon>0$, i.e. $b^{\bar{m}}_{c_1\dots c_s} = \epsilon f(\{\mu_\alpha\})$ with $f$ any positive homogeneous function of the $\mu_\alpha$. The computation of the correlation function and the Ginzburg-$Q$ parameter can then be carried through with the limit $\epsilon\rightarrow 0$ taken at the end of the calculation. The sign of the non-vanishing $b_{c_1\dots c_s}^m$ determines whether the correlation function exhibits an exponential decay or an exponential divergence. This behavior has already been noted in~\cite{Marchetti:2022igl} where the parameters of the theory have been restricted to ensure exponential decay. Here, we proceed analogously and restrict for the remainder the $Z_\alpha^g(\vb*{0})$ such that $b_{c_1\dots c_s}^m>0$. It has been verified numerically that this set of $Z_\alpha^g(\vb*{0})$ is uncountably infinite, but its precise characterization remains open. Outside this parameter range the correlation function exhibits an exponential divergence, potentially marking the breakdown of the mean-field description. It is expected that a full RG treatment in the future will shed light on these matters.

Following~\cite{Marchetti:2022igl} and summarizing the detailed steps in~\cite{Dekhil:2024djp}, the asymptotics of Eq.~\eqref{eq:C(eta) integral after pfd} are obtained by 1) utilizing the asymptotic behavior of the residual $\SL$ Wigner matrices, $d^{(\rho,0)}_{jlm}\left(\frac{\eta}{a}\right)\rightarrow c_{\rho}(j,l,m)\e^{-\frac{\eta}{a}(1-i\rho)}$, 2) applying the residue theorem for the $\rho_{c_{s+1}}$-contour integral, 3) performing a stationary phase approximation for the remaining integrations, 4) re-scaling $C^s_{\alpha\beta}(\vb*{\eta}_{4-s})$ by the Jacobian determinant on $\mathrm{H}_\plus^{4-s}$ to $\tilde{C}^s_{\alpha\beta}(\vb*{\eta}_{4-s})$, and 5) expanding in terms of small values of $a^2b^m_{c_1\dots c_s}$. In the \enquote{isotropic} limit where $\eta_{c_u} = \eta$ for all $u\in\{s+1,...,4\}$, these steps yield altogether an asymptotic behavior in large boosts, $\eta/a\gg 1$, given by~\cite{Dekhil:2024djp}
\begin{equation}\label{eq:asymptotics of Cs}
\tilde{C}^s_{\alpha\beta}(\vb*{\eta}_{4-s})\:\underset{\eta/a\gg 1}{\longrightarrow}\:\exp\left(-\frac{1}{2}a b^{m_*}_{c_1\dots c_s}\eta\right)\,,
\end{equation}
where $m_* = \arg \min_m(b^m_{c_1\dots c_s})$. The non-local correlation length is thus given by $\xinloc \defeq 2/(a b^{m_*}_{c_1\dots c_s})$, scaling as $\xinloc\rightarrow(a\mu)^{-1}$ in the homogeneous limit $\mu_\alpha\rightarrow 0$, like in~\cite{Marchetti:2022igl,Marchetti:2022nrf}. Crucially, $\xinloc$ is a correlation length on the space of geometries rather than a correlation length on spacetime. As the $\eta$ parametrize the boost part of $\SL$ holonomies, $\xinloc\rightarrow\infty$ renders non-negligible correlations between tetrahedra characterized by very different half-holonomies. Determining whether and in what manner the length scale $\xinloc$ translates to a length scale on an emergent spacetime is left to future research.   

In retrospect, including the functions $Z_\alpha^g$ turned out essential for obtaining a regime of an exponentially decaying correlation function. For $Z_\alpha^g$ lying outside the chosen regime above, the correlation function contains a mixture of exponentially decaying and diverging terms. Exponentially diverging correlation functions indicate long-range correlations insensitive to the limit $\mu_\alpha\rightarrow 0$ which we therefore exclude for studying phase transitions at $\mu_\alpha = 0$. This exponential behavior, diverging or converging, is characteristic of the underlying hyperbolic geometry of $\SL$. We note that these results are not surprising as they would be obtained also in a local multi-field theory on the two-sheeted hyperboloid $\mathrm{H}_\plus$.


\paragraph{Contributions with timelike faces.}

To complete the analysis of the non-local correlation function, we study the component $C^s_{\minus\minus}$ if at least one of the labels is associated with a timelike face. For $t>0$ labels $\nu_{c_1},...,\nu_{c_t}$ this results in evaluating the following expression
\begin{equation}\label{eq:nonlocal integrand tl}
\eval{C^s_{\minus\minus}(\vb*{\eta}_{4-s})}_{\mathrm{tl}} = \prod_{c=c_{s+1}}^{c_{s+t}}\sum_{\nu_{c}}\nu_{c}^2\prod_{c'=s+t+1}^{4}\int\dd{\rho_{c'}}\rho_{c'}^2\frac{d^{(0,\nu_{c})}_{j_{c}j_{c}m_{c}}\left(\frac{\eta_c}{a}\right)d^{(\rho_{c'},0)}_{j_{c'}j_{c'}m_{c'}}\left(\frac{\eta_{c'}}{a}\right)}{\frac{1}{a^2}Z_\minus^g(\vb*{0})\left(\sum_{c'}\rho_{c'}^2-\sum_c\nu_{c}^2+(4-s)\right)+b^{\minus}_{c_1\dots c_s}}\,,
\end{equation}
where $b^{\minus}_{c_1\dots c_s} $ is the effective mass evaluated on $s$ zero modes with $t$ timelike labels. For further evaluation, we employ the asymptotic form of the reduced Wigner matrices restricted to timelike labels, given by $d^{(0,\nu)}_{jlm}\left(\frac{\eta}{a}\right)\:{\rightarrow}\:\e^{-\frac{\eta}{a}}$, and derived in Appendix~\ref{sec:asymptotics of d}. This scaling is in fact independent of $\nu$, bearing crucial consequences for the asymptotics of the correlation function.

For $t=4-s$ timelike labels, the asymptotics of the correlation function for large $\eta/a$ fully decouple from the labels $\nu_{c_1},...,\nu_{c_{4-s}}$, i.e. $\eval{C_{\minus\minus}^s(\vb*{\eta}_{4-s})}_{t=4-s}\sim \e^{-(4-s)\frac{\eta}{a}}$. This scaling is readily insensitive to the effective mass $b^{\minus}_{c_1\dots c_s}$ and thus to $\mu_\alpha$, which applies also to the re-scaled correlation function $\tilde{C}^s_{\alpha\beta}$. We conclude that due to this independence, the contribution with $4-s$ timelike labels does not affect the critical behavior.

If one or more of the $4-s$ remaining labels are spacelike, one can apply the strategies of the previous paragraph. That is, one performs a contour integration for one of the spacelike labels by applying the residue theorem. The remaining $\rho$-integrals are then performed via a stationary phase approximation, yielding 
\begin{equation}
\eval{C_{\minus\minus}^s(\vb*{\eta}_{4-s})}_{t<4-s} \:\underset{\eta/a\gg 1}{\longrightarrow}\:\sum_{\nu_{c_{s+1}}...\nu_{c_{s+t}}}\exp\left(\frac{\eta}{a}\left(-(4-s)+i(4-s-t)\bar{\rho}\right)\right)\,,
\end{equation}
where $\bar{\rho}$ are the stationary points of the $\rho$-integrals depending on the $\nu_{c_u}$, detailed in~\cite{Dekhil:2024djp}. As $\mu_\alpha\rightarrow 0$, the function $\bar{\rho}$ remains finite and real for any $\nu_{c_u}$. Thus, the asymptotics of the correlation function $C^s_{\minus\minus}$ or its re-scaled form $\tilde{C}^s_{\minus\minus}$ remain unaffected in the critical region $\mu_\alpha\rightarrow 0$. In particular, the correlation length is in this case independent of $\mu_\alpha$. This suggests that contributions containing at least one timelike face do not affect the critical behavior which is instead driven by contributions from spacelike labels $\rho$.

Summarizing, there exists a regime of an exponentially decaying non-local correlation function close to criticality which is driven by contributions from spacelike faces. The extracted correlation length scales as $\xinloc \sim (a\mu)^{-1}$ in a homogeneous limit $\mu_\alpha\rightarrow 0$. Timelike faces do not contribute to this exponential decay and therefore do not drive the critical behavior of the system. These results offer an elegant geometric explanation: The group $\SL$ is topologically given as $\SL\cong_\mathrm{top}\mathrm{H}_\plus\times S^3$, with $S^3\cong \SUT$ the $3$-sphere. Continuous labels $\rho$ of spacelike faces are associated with the non-compact hyperbolic boost part, while for discrete timelike faces, the conjugate variables have only a compact domain and are associated with the rotational part. Following~\cite{Benedetti:2014gja} as well as the results of~\cite{Pithis:2020sxm,Marchetti:2020xvf,Marchetti:2022igl,Marchetti:2022nrf}, phase transitions require a non-compact domain of the fields, which agrees with what we have found here. 

\subsection{Ginzburg-$Q$}\label{sec:arbitrary but fixed Q}

The mean-field approach provides a self-consistent description of phase transitions if, in the limit $\xi\rightarrow\infty$, the fluctuations averaged over the domain set by the correlation lengths, $\langle\delta\varphi_\alpha\delta\varphi_\beta\rangle_{\Omega_\xi}$, are much smaller than the averaged mean-field $\langle\varphi_\alpha^{\mathrm{m}}\varphi_\beta^{\mathrm{m}}\rangle_{\Omega_\xi}$. This is quantified by the Ginzburg-$Q_{\alpha\beta}$,
\begin{equation}\label{eq:definition Q}
    Q_{\alpha\beta} \defeq \frac{\langle\delta\varphi_\alpha\delta\varphi_\beta\rangle_{\Omega_\xi}}{\langle\varphi_\alpha^{\mathrm{m}}\varphi_\beta^{\mathrm{m}}\rangle_{\Omega_\xi}} = \frac{\int_{\Omega_\xi}\dd{\vb*{g}}\dd{\vb*{\phi}}C^{\mathrm{ext}}_{\alpha\beta}(\vb*{g},\vb*{\phi})}{\int_{\Omega_\xi}\dd{\vb*{g}}\dd{\vb*{\phi}}\varphi^{\mathrm{m}}_\alpha\varphi^{\mathrm{m}}_\beta} \,,
\end{equation}
where the 2-point function of fluctuations has been identified with the correlation function $C_{\alpha\beta}$. Validity of mean-field theory is given if $Q_{\alpha\beta}\ll 1$ in the limit $\mu_\alpha\rightarrow 0$~\cite{levanyuk1959contribution,ginzburg1961some}. The integration range $\Omega_{\xi} \defeq \SL^4_{\xi}\times [-\xiloc,\xiloc]^{\dloc}$ is determined by the local and non-local correlation lengths, where the non-compact part of $\SL_{\xi}$ is restricted to $\eta\in [0,\xinloc]$. In the following, we identify volume factors of different signatures $V_\alpha\equiv V$ justified by the arguments in Appendix~\ref{app:Empty integrals}. Furthermore, the $\SL$ volume factors are regulated by cutoffs $L$ and $\xinloc$, and are differentiated as $V_L$ and $V_\xi$, respectively. For large values of $L$ and $\xi$, the volume factors scale as $V_L\sim\e^{2L/a}$ and $V_\xi\sim\e^{2\xinloc/a}$.

First, we compute the denominator of $Q_{\alpha\beta}$ in Eq.~\eqref{eq:definition Q}. Since the $\varphi^{\mathrm{m}}_\alpha$ are constant, the integration yields four volume factors of $V_\xi$ and $\dloc$ volume factors of $\R$, similarly cut off by the local correlation length $\xiloc$. Furthermore, from Eq.~\eqref{eq:mf solutions}, we extract
\begin{equation}
\varphi^{\mathrm{m}}_\alpha\varphi^{\mathrm{m}}_\beta =  V_L^{-4}V_L^{-2\frac{n_\upgamma-1}{n_\upgamma-2}}\lambda^{-\frac{2}{n_\upgamma-2}}M^\upgamma_{\alpha\beta}(\mu_\plus,\mu_\zero,\mu_\minus)\,,
\end{equation}
where $M^\upgamma_{\alpha\beta}$ solely depends on the $n_\alpha$ and $\mu_\alpha$. The relevant property of  $M^\upgamma_{\alpha\beta}$ is its scaling in a homogeneous limit of the masses $\mu_\alpha$, that is $M^\upgamma_{\alpha\beta}(\mu\mu_\plus,\mu\mu_\zero,\mu\mu_\minus) = \mu^{\frac{2}{n_\upgamma-2}}M^\upgamma_{\alpha\beta}(\mu_\plus,\mu_\zero,\mu_\minus)$. Combined with the empty integrations, the denominator of $Q_{\alpha\beta}$ scales as
\begin{equation}\label{eq:denominator scaling}
\int_{\Omega_\xi}\dd{\vbg}\dd{\vbf}\varphi^{\mathrm{m}}_{\alpha}\varphi^{\mathrm{m}}_\beta \:\sim\:(a\xinloc)^{\frac{\dloc}{2}}\left(\frac{V_\xi}{V_L}\right)^4 V_L^{-2\frac{n_\upgamma-1}{n_\upgamma-2}}\lambda^{-\frac{2}{n_\upgamma-2}}\mu^{\frac{2}{n_\upgamma-2}}\,,
\end{equation}
in the homogeneous limit $\mu_\alpha\rightarrow 0$. Since $M_{\alpha\beta}$ is a non-zero function, we neglect its matrix structure which would simply yield different constant proportionality factors. 

The numerator of $Q_{\alpha\beta}$ includes the extended correlation function, containing the expansion in zero modes. Since the $Z_\alpha^\phi(\vbi)$ are chosen such that the local correlation function exhibits an exponential suppression, the local integration domain can be extended to all of $\R^{\dloc}$. This yields
\begin{equation}
\int_{\SL_\xi^4}\dd{\vbg}C_{\alpha\beta}^{\mathrm{ext}}(\vbg) = \sum_{s=s_0}^4 \left(\frac{V_\xi}{V_L}\right)^s\sum_{(c_1\dots c_s)} \int\dd{\vb*{g}_{4-s}}C_{\alpha\beta}^s(\vbg_{4-s})\,,
\end{equation}
where we notice that the contributions from the timelike faces with labels $(0,\nu)$ vanish due to the projection onto the trivial representation $(i,0)$. This is another indication that timelike faces do not affect correlations near criticality. From the remaining contributions, we obtain
\begin{equation}\label{eq:numerator scaling}
\int\dd{\vb*{g}_{4-s}}C_{\alpha\beta}^s(\vbg_{4-s})\: \sim \:\sum_m \frac{\varrho^m_{\alpha\beta}}{b^m_{c_1\dots c_s}}\:\longrightarrow\: \frac{\varrho^{m_*}_{\alpha\beta}}{b_{c_1\dots c_s}^{m_*}}\,,    
\end{equation}
where $m_*$ marks the dominant contribution with the smallest effective mass, $b_{c_1\dots c_s}^{m_*}$. Depending on the specifics of $\upgamma_\textsc{c}$, we notice that this term is possibly vanishing and thus regulated as prescribed in Sec.~\ref{sec:Non-local correlation function}. We keep the constants $\varrho^m_{\alpha\beta}$ as they encode the matrix structure of the correlation function, capturing also potentially vanishing entries. 

Combining numerator and denominator, the Ginzburg-$Q_{\alpha\beta}$ scales as
\begin{equation}
Q_{\alpha\beta}\:\sim\: \varrho^{m_*}_{\alpha\beta}\lambda^{\frac{2}{n_\upgamma-2}}V_L^{2\frac{n_\upgamma-1}{n_\upgamma-2}}(a\xinloc)^{-\frac{\dloc}{2}+\frac{n_\upgamma}{n_\upgamma-2}}\sum_{s=s_0}^4\left(\frac{V_\xi}{V_L}\right)^{-(4-s)}\,.
\end{equation}
Taking the limit of large group volume first, $L\rightarrow \infty$, we notice that the $s_0$-term of the zero mode sum dominates. After absorbing volume factors and the skirt radius into the coupling, $\bar{\lambda} \defeq V_L^{2(n_\upgamma-1)+(4-s_0)(n_\upgamma-2)} a^{\frac{n_\upgamma}{2}-\frac{\dloc(n_\upgamma-2)}{4}}\lambda$, which is consistent with~\cite{Marchetti:2022igl}, the scaling of $Q_{\alpha\beta}$ is finally given by
\begin{equation}\label{eq:Q arbitrary interaction}
Q_{\alpha\beta}\sim\varrho^{m_*}_{\alpha\beta}\bar{\lambda}^{\frac{2}{n_\upgamma-2}}(\xinloc)^{-\frac{\dloc}{2}+\frac{n_\upgamma}{n_\upgamma-2}}\e^{-2(4-s_0)\frac{\xinloc}{a}}\,.
\end{equation}
In the limit $\xinloc\rightarrow \infty$, the non-zero entries of $Q_{\alpha\beta}$ approach zero, proving the validity of the mean-field approach. Remarkably, the exponential suppression is present irrespective of the details of the underlying interactions which only affect its overall strength via $s_0$. The matricial factor $\varrho_{\alpha\beta}^{m_*}$ encodes the interplay of combinatorial non-locality with the causal character of the tetrahedra, i.e. it encodes whether and how tetrahedra of different signatures are correlated.

The exponential suppression is the determining factor for the asymptotic behavior of $Q_{\alpha\beta}$ and arises from the hyperbolic structure of $\SL$. The same result has been found previously~\cite{Marchetti:2022igl,Marchetti:2022nrf}, where all tetrahedra were assumed to be spacelike. Our results go beyond this restricted setting as we include tetrahedra of arbitrary signature. As a secondary result, we find that timelike faces do not contribute to the critical behavior of the theory as they characterize the rotational, and thus compact, part of $\SL$. This is apparent from the non-local correlation function. In the same vein, contributions to the Ginzburg-$Q$ emanating from timelike faces vanish as a result of the projection onto zero modes. 

The possibility of a vanishing mass has been discussed at length in~\cite{Dekhil:2024ssa}. Transferring these results to the present setting, we notice that the asymptotic behavior of $Q_{\alpha\beta}$ does not change if the $b_{c_1\dots c_s}^m$ vanish. This can be seen by explicitly using the $\epsilon$-regularization suggested in Sec.~\ref{sec:Non-local correlation function}. Furthermore, due to the presence of three signatures, there is in fact a set of three effective mass parameters $\{b_{c_1\dots c_s}^m\}$, not all of which are vanishing for the combinatorics considered here. 


\section{Other interactions}\label{sec:other interactions}

The model analyzed in this chapter assumed a single interaction term of arbitrary causal structure, clearly posing a restriction of the theory space. In the following, we briefly summarize two possible extensions. For a more extensive discussion, see~\cite{Dekhil:2024djp}.

\paragraph{Multiple interactions.} For multiple interactions with the same degree $(\np,\nz,\nm)$ but different combinatorics $\upgamma_\textsc{c}$, the mean-field equations are still solvable. Solutions to these are obtained by replacing $\lambda$ in Eq.~\eqref{eq:mf solutions} with $\sum_{\upgamma_{\textsc{c}}}\lambda_{\upgamma_{\textsc{c}}}$ and the Hessian matrix $F_{\alpha\beta}$ is obtained by replacing every entry $\chi_{\alpha\beta}$ with $\sum_{\upgamma_{\textsc{c}}}\tilde{\lambda}_\gamma\chi_{\alpha\beta}^{\upgamma_{\textsc{c}}}$, where $\tilde{\lambda}_{\upgamma_{\textsc{c}}} \defeq \lambda_{\upgamma_{\textsc{c}}}/(\sum_{\upgamma_\textsc{c}'}\lambda_{\upgamma_\textsc{c}'})$ and $\chi_{\alpha\beta}^{\upgamma_{\textsc{c}}}$ captures the specifics of $\upgamma_{\textsc{c}}$. These modifications affect the pole structure of the correlation functions and thus the parameter ranges of $Z_\alpha^\phi(\vbi)$ and $Z_\alpha^g(\vb*{0})$ that lead to positive effective masses. Besides these differences, the non-vanishing components of $Q_{\alpha\beta}$ are still subject to exponential suppression, and thus, mean-field theory is valid also for multiple interactions of the same degree.


\paragraph{Colored simplex.} As discussed in Sec.~\ref{sec:Extended Barrett-Crane Group Field Theory Model}, given simplicial interactios, coloring the group fields guarantees that only topologically well-behaved complexes are  generated by the GFT. The methods developed here enable applying the LG analysis also to a colored simplicial model restricted to spacelike tetrahedra as defined in Sec.~\ref{subsec:Definition of the Complete Model}. Note that this model is symmetric under simultaneous reflection of two or four fields, yielding a set of $\mathbb{Z}_2$ symmetries. The correlators and $Q$ then carry \emph{color indices} $i,j\in\{0,\dots,4\}$ just as they did carry causal character indices $\alpha,\beta$ above for fields $\varphi_\alpha$. Following~\cite{Dekhil:2024djp}, an exponentially decaying local correlator requires the coloring of the Laplacian weights $Z^\phi$, suggesting a non-trivial interplay of colors and matter coupling. Otherwise, the results of this chapter transfer to the colored case, and the parameter $Q_{ij}$ exhibits an exponential suppression for $\mu\rightarrow 0$. In the context of tensor models, integrating out all but one color yields tensor-invariant interactions~\cite{Bonzom:2012hw,Gurau:2011tj} which exhibit a $\mathbb{Z}_2$ symmetry. Extending the arguments of~\cite{Bonzom:2012hw,Gurau:2011tj} to the present case and relating the symmetries before and after integrating out colors constitutes an intriguing task for future research.\\

\begin{center}
    \textit{Summary.}
\end{center}

\noindent The central result of this chapter is that also in the causally extended setting, mean-field theory generically serves as a viable approximation. This is induced by the hyperbolic structure of $\SL$, yielding an exponential suppression of fluctuations as measured by $Q_{\alpha\beta}$. The critical behavior is entirely driven by representations $(\rho,0)$ associated to spacelike faces while timelike faces do not contribute as they characterize the rotational and thus compact part of $\SL$. Our results suggest the existence of a non-trivial vacuum occupied by many GFT quanta, thus forming a condensate. Exploring the physics of this phase via GFT coherent states at mean-field level~\cite{Gielen:2013kla,Gielen:2013naa,Gielen:2016dss,Oriti:2016acw,Pithis:2019tvp,Jercher:2021bie,Oriti:2021oux,Marchetti:2021gcv} forms the motivation of the next chapter. Therein, effective scalar cosmological perturbations are extracted from a GFT with timelike and spacelike tetrahedra.

\begin{center}
    \textit{Closing remarks.}
\end{center}

\paragraph{Restricted parameters.} Our results are limited in that we restricted to a regime of parameters where local and non-local correlation functions are exponentially decaying. Beyond that, correlations in GFTs can exhibit exotic behavior such as oscillations or exponential divergences, potentially marking the breakdown of the mean-field approximation. A source for this behavior could also lie in the fact that the non-local correlations are defined on the space of geometries, where intuition from local field theories might not be applicable. Examining these regimes further might require employing methods beyond mean-field such as the FRG, see e.g.~\cite{Geloun:2023ray}.

\paragraph{Symmetry breaking.} LG theory was originally developed to study second order phase transitions in phenomenological models exhibiting spontaneous symmetry breaking (SSB)~\cite{Goldenfeld:1992qy}. In this work, we took a different perspective by studying models motivated from QG which do not necessarily exhibit SSB. This does not imply the absence of phase transitions but determining the order of transition requires arguments beyond mean-field.  Still, in the parameter range where $Q_{\alpha\beta}$ is exponentially suppressed, the mean-field approximation is valid in the regime $\mu\rightarrow 0$. 

\paragraph{Geometry of vacua.} A deeper understanding of the emergent geometries encoded in the vacuum $\varphi^{\mathrm{m}}_\alpha$ is highly desirable. The main challenge to make progress on this front lies in the identification of suitable operators and observables which would allow extracting for instance the Hausdorff and spectral dimensions of such states, thus connecting to the work of Chapter~\ref{chapter:specdim}.

\paragraph{Universality with EPRL-CH?} It is an important task to apply the LG analysis to a yet to be developed GFT formulation of the EPRL-CH model. Since this model is also based on $\SL$, we expect that much of the work presented here can be carried over. Tentatively, results from area Regge calculus~\cite{Dittrich:2021kzs} and GFT condensate cosmology~\cite{Jercher:2021bie} already suggest that the EPRL and the BC model could lie in the same universality class. This conjecture would be strongly supported if a LG analysis yields the same scaling exponents near criticality.

\chapter[\textsc{Perturbations from Quantum Gravitational Entanglement}]{Scalar Cosmological \\ Perturbations from Quantum \\ Gravitational Entanglement}\label{chapter:perturbations}

Spatially flat and homogeneous cosmological dynamics have been successfully recovered from the mean-field dynamics of GFT coherent states~\cite{Gielen:2013kla,Gielen:2013naa,Oriti:2016qtz}, while replacing at early relational times the initial Big Bang singularity of classical cosmology by a Big Bounce~\cite{Oriti:2016ueo,Oriti:2016qtz,Marchetti:2020umh}. GFT coherent states present the simplest form of coarse-graining since a single function, the condensate wavefunction, captures the behavior of an infinite superposition of quantum geometrical building blocks~\cite{Gielen:2013naa}. This condensate wavefunction acts as a mean-field, or order parameter, allowing to explore a hypothetical condensate phase of the GFT. Existence of such a phase has been supported by LG analyses of Lorentzian GFTs as conducted in~\cite{Marchetti:2022igl,Marchetti:2022nrf,Dekhil:2024ssa,Dekhil:2024djp} and the previous chapter. A special class of coherent states, called coherent peaked states (CPSs)~\cite{Marchetti:2020qsq,Marchetti:2020umh,Marchetti:2021gcv}, allows furthermore for an effective localization of observables in relational space and time in the absence of a background spacetime manifold.

Evidently, modelling our Universe with a spatially homogeneous geometry and matter distribution is valid only on the largest scales. To understand structure formation or the anisotropies of the cosmic microwave background~\cite{Planck:2019evm}, including small inhomogeneities, i.e. small perturbations around homogeneity, of geometry and matter is crucial. For a QG theory, it is therefore an important task to recover cosmological perturbations from the fundamental theory. This does not only provide a test of the QG theory which is far more rigorous than for the homogeneous sector, but also offers connecting QG predictions with cosmological observations~\cite{Ashtekar:2021kfp}. In GFTs, this challenge has been tackled first in~\cite{Gielen:2017eco,Gielen:2018xph,Gerhardt:2018byq} and further advanced in~\cite{Marchetti:2021gcv}, where the classical perturbation equations of GR have been recovered in the super-horizon limit of large perturbation wavelengths. However, deviations from classical results occur there for non-negligible wave vectors, rendering the effective equations incompatible with the observationally successful perturbation theory of classical GR. Since time and space derivatives are not distinguished in the effective equations (contrary to GR), we argue here that the deviations arise from an insufficient coupling between the physical reference frame and the underlying causal structure.

In this chapter, we advance the GFT condensate cosmology program by deriving cosmological perturbation equations from GFT which agree with classical results at sub-Planckian scales and are subject to quantum corrections in the trans-Planckian regime. Within our framework, inhomogeneities of cosmological observables emerge from the entanglement between the quantum geometric degrees of freedom of the model. Thus, we also provide a concrete realization of the general expectation that non-trivial geometries are associated with quantum gravitational entanglement~\cite{Giddings:2005id,Ryu:2006bv,VanRaamsdonk:2010pw,Bianchi:2012ev,Maldacena:2013xja,Giddings:2015lla,Cao:2016mst,Donnelly:2016auv,swingle2018spacetime,colafrancheschi2022emergent,Bianchi:2023avf}. Our constructions are facilitated by the rich causal structure of the complete BC GFT model introduced in Chapter~\ref{chapter:cBC}.


\section{Fock space and a physical Lorentzian reference frame}\label{sec:Fock space and a physical Lorentzian reference frame}

The model employed in this chapter is the causally complete BC model introduced previously, restricted to spacelike and timelike tetrahedra. Closure and simplicity constraints~\cite{Jercher:2022mky} are imposed on the fields $\varphi_{\pm}$ as in Eqs.~\eqref{eq:extended closure} and~\eqref{eq:simplicity}, respectively. We pose no assumptions on the vertex term $\mathfrak{V}$ of the action in Eq.~\eqref{eq:general vertex} as it will be neglected, justified by the arguments given later on. The kinetic kernels $\mathcal{K}_{\pm}$ will be specified momentarily.

\subsection{Fock space}

In analogy to local QFTs, the Hilbert space of GFTs corresponds to a Fock space\footnote{The structure of diffeomorphism-invariant states in LQG and its relation to the GFT Fock space is studied in~\cite{Sahlmann:2023plc}.} with the 1-particle Hilbert space being that of a quantum tetrahedron~\cite{Gielen:2013naa,Oriti:2013aqa}. That is, the 1-particle excitations of $\varphi_\pm$ correspond to tetrahedra. Promoting the group fields $\varphi_{\pm}$ to annihilation operators
requires an extension of the Fock space to incorporate spacelike and timelike 1-particle Hilbert spaces. The individual Fock space sectors are defined as $\mathcal{F}_\pm \defeq \bigoplus_N \mathrm{sym}(\mathcal{H}_\pm^{(1)}\otimes...\otimes \mathcal{H}_\pm^{(N)})$, where the 1-particle Hilbert spaces for spacelike and timelike tetrahedra are respectively given by $\mathcal{H}_\pm \defeq L^2(D_\pm)$, with the domain $D_\pm$ defined in Sec.~\ref{subsec:Spin Representation of the Group Field and its Action}. The total Fock space $\mathcal{F}$ of the theory is constructed as the tensor product of $\mathcal{F}_+$ and $\mathcal{F}_-$, i.e.
\begin{equation}
\mathcal{F}\defeq \mathcal{F}_+\otimes\mathcal{F}_- = \bigoplus_{N_{\text{tot}}}^{\infty}\bigoplus_{N+M = N_{\text{tot}}}\mathrm{sym}\left(\mathcal{H}_+^{\otimes N}\right)\otimes\mathrm{sym}\left(\mathcal{H}_-^{\otimes M}\right)\,.
\end{equation}
Creation, $\hat{\varphi}_{\pm}^\dagger$, and annihilation, $\hat{\varphi}_\pm$, operators on $\mathcal{F}$ are defined in terms of the creation and annihilation operators of the respective sectors. Imposing bosonic commutation rules, the operators satisfy the algebra $\comm{\hat{\varphi}_{\pm}}{\hat{\varphi}_{\pm}^{\dagger}} = \one_{\pm}$ and $\comm{\hat{\varphi}_{\pm}}{\hat{\varphi}_{\pm}} = \comm{\hat{\varphi}_{\pm}^{\dagger}}{\hat{\varphi}_{\pm}^{\dagger}} = 0$, where $\one_\pm$ is the identity on $\mathcal{F}_{\pm}$ respecting closure and simplicity constraints. By construction, operators of different sectors mutually commute, i.e. $\comm{\varphi_{\pm}}{\varphi_{\mp}^{\dagger}} = \comm{\hat{\varphi}_{\pm}}{\hat{\varphi}_{\mp}} = \comm{\hat{\varphi}_{\pm}^{\dagger}}{\hat{\varphi}_{\mp}^{\dagger}} = 0$. The vacuum state $\ket{\emptyset}$ of the total Fock space is defined as $\ket{\emptyset} = \ket{\emptyset}_\texttt{+}\otimes\ket{\emptyset}_\texttt{-}$, with $\ket{\emptyset}_\pm$ being the vacua of the individual sectors. Note that these vacua correspond to \enquote{states of no space}~\cite{Oriti:2013aqa}.

Operators acting on $\mathcal{F}$ are in general defined as convolutions of kernels with creation and annihilation operators, see~\cite{Oriti:2016qtz,Oriti:2013aqa,Marchetti:2020umh}.
Here, we are particularly interested in one- and two-body operators, such as the number operators $\hat{N}_{\pm} = \hat{\varphi}_\pm^\dagger\cdot\hat{\varphi}_\pm$ and the spatial $3$-volume operator $\hat{V} = \hat{\varphi}^\dagger_\texttt{+}\cdot V\cdot\hat{\varphi}_\texttt{+}$, where \enquote{$\cdot$} denotes an integration over the full GFT field domain. In spin representation, the kernel of $\hat{V}$ scales as  $V\sim\rho^{3/2}$~\cite{Oriti:2016qtz,Marchetti:2020umh,Jercher:2021bie} for isotropic representation labels $(\rho_c \equiv \rho)$, and is given in analogy to the eigenvalues of the LQG volume operator~\cite{Rovelli:1994ge,Barbieri:1997ks,Brunnemann:2004xi,Ding:2009jq}. A two-body operator $\hat{O}_{\alpha\beta}$ is generally given by $\hat{O}_{\alpha\beta} = (\hat{\varphi}^{\dagger}_\alpha\times\hat{\varphi}^{\dagger}_\beta)\cdot O$, where $\times$ is either operator multiplication ($\alpha=\beta$) or a tensor product ($\alpha\neq \beta$). Note that $\hat{O}_{\alpha\beta}$ does not factorize in general, thus creating an entangled state when acting on a product state in $\mathcal{F}$. 

\subsection{A physical Lorentzian reference frame}

To define an effective relational localization~\cite{Rovelli:1990ph,Hoehn:2019fsy,Rovelli:2001bz,Dittrich:2005kc,Goeller:2022rsx} of observables in space and time, we follow~\cite{Marchetti:2021gcv} and implement a physical Lorentzian reference frame composed of four minimally coupled massless free (MCMF) scalar fields, $\rf^\mu$, $\mu\in\{0,1,2,3\}$, serving as the dynamical clock $(\chi^0)$ and rods $(\chi^i$, with $i\in\{1,2,3\})$~\cite{Domagala:2010bm,Giesel:2012rb}. An additional {MCMF} \textit{matter} scalar field $\mf$ is coupled which is assumed to dominate the field content of the emergent cosmology. 

The scalar fields $\rf^\mu$ and $\mf$ are coupled to the GFT as prescribed in Sec.~\ref{sec:Extended Barrett-Crane Group Field Theory Model}. In particular, the two kinetic kernels $\mathcal{K}_\pm$ are extended to $\mathcal{K}_\pm(\vbg,\vbg') \rightarrow \mathcal{K}_\pm(\vbg,\vbg',(\rf_v^\mu-\rf_w^\mu)^2,(\mf_v-\mf_w)^2)$. Also, the commutation relations introduced above naturally extend to $\comm{\hat{\varphi}_\pm(\rf^\mu_v,\mf_v)}{\hat{\varphi}^{\dagger}_\pm(\rf^\mu_w,\mf_w)}=\one_\pm\delta^{(4)}(\rf^\mu_v-\rf_w')\delta(\mf_v-\mf_w')$. Operators now include an integration over the full domain including the scalar field values, where we refer to~\cite{Marchetti:2020umh,Jercher:2021bie} for further details.

In GR, perturbation equations distinguish between derivatives with respect to time (clock) and space (rods), as can be seen explicitly from Eq.~\eqref{eq:classical relative perturbed volume equation}. In harmonic coordinates, this difference is reflected by a relative weight of $a^4$ between space and time derivatives with $a$ the scale factor, later to be related with the expectation value of the 3-volume operator. This behavior could not be reproduced in the effective equations of~\cite{Marchetti:2021gcv}, which are derived from a GFT restricted to spacelike tetrahedra, and constitutes the source of the mismatch with GR for non-vanishing perturbation momenta $k> 0$. In the remainder, we show that by carefully coupling the frame to the underlying causal structure, this mismatch is alleviated. 

\begin{figure}
    \centering
    \includegraphics[width=0.5\textwidth]{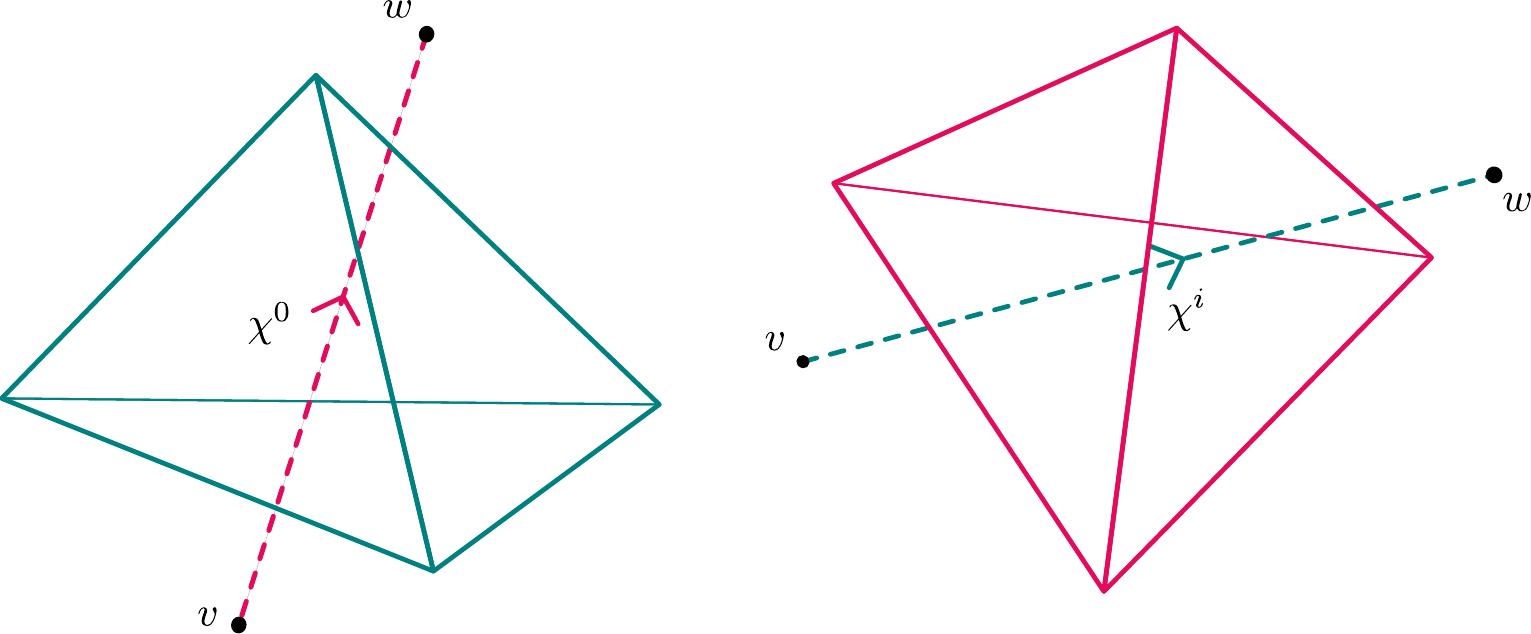}
    \caption{Left: clock $\chi^0$ propagating across spacelike tetrahedron (timelike dual edge). Right: rods $\chi^i$ propagating across timelike tetrahedron (spacelike dual edge). Propagation of matter data is encoded in the kinetic kernels $\mathcal{K}_{\pm}$, hence the restrictions in Eq.~\eqref{eq:kineticrestriction}.}
    \label{fig:spacetime_tetrahedra}
\end{figure}

In the continuum, clock and rods are distinguished by the signature of their gradient, i.e. $\mathrm{g}(\partial\chi^0,\partial\chi^0) > 0$ and $\mathrm{g}(\partial\chi^i,\partial\chi^i) < 0$, where $\mathrm{g}$ is the metric. Note that 1) these conditions are not implied by the Klein-Gordon equation, but constitute an additional physical requirement and 2) despite the point-particle intuition, a massless scalar field does not necessarily have a lightlike gradient. Introducing a discretization and placing the fields on dual vertices $v,w$, the continuum condition is imposed strongly\footnote{\enquote{Strongly} refers here to the fact the spatial (temporal) components of $\partial\rf^0$ ($\partial\rf^i$) are set to zero, which is the strongest assumption to guarantee that $\mathrm{g}(\partial\rf^0,\partial\rf^0) > 0$ ($\mathrm{g}(\partial\rf^i,\partial\rf^i) < 0$).} by requiring that $\chi^0_v-\chi^0_{w} = 0$ for the dual edge $(vw)$ spacelike and $\chi^i_v-\chi^i_{w} = 0$ for the dual edge $(vw)$ timelike. Consequently, the clock propagates along timelike dual edges and rods propagate along spacelike dual edges, as depicted in Fig.~\ref{fig:spacetime_tetrahedra}.

Importantly, the manifestly causal nature of the minimally extended BC GFT model allows us, for the first time, to consistently implement the Lorentzian properties of the physical frame at the QG level. As just discussed, at a classical discrete geometric level, a clock propagates along timelike dual edges, while rods propagate along spacelike dual edges. These conditions can be imposed strongly\footnote{\enquote{Strongly} refers here to the fact that the dependence of $\mathcal{K}_{\texttt{+}}$ ($\mathcal{K}_{\texttt{-}}$) on $\rf^i$ ($\rf^0$) is dropped. Conceivably, this condition could be imposed weaker via a Gaussian.} at the quantum gravity level by requiring that
\begin{equation}\label{eq:kineticrestriction}
    \mathcal{K}_\texttt{+}= \mathcal{K}_\texttt{+}\bigl(\vbg,\vbg'; (\rf^0_v-\rf^{0}_w)^2\bigr)\,,\qquad \mathcal{K}_\texttt{-}= \mathcal{K}_\texttt{-}\bigl(\vbg,\vbg';\abs{\vb*{\rf}_v-\vb*{\rf}_w}^2\bigr)\,.
\end{equation}
Note that no such restriction is assumed for $\phi$, which is suppressed in the notation above.

\section{Entangled coherent peaked states}

Building on a series of previous results~\cite{Oriti:2016qtz,Marchetti:2020umh,Jercher:2021bie}, we suggest that the dynamics of scalar cosmological perturbations can be extracted from perturbed coherent GFT states. The underlying assumption here is that the GFT atoms are in a condensate phase, the existence of which is supported by the results of the previous chapter. It is then assumed that the physics of this phase can be studied via the proposed perturbed coherent states, which appear well-suited since: 1) They allow to systematically implement a mean-field approximation, and thus the simplest form of coarse-graining of the QG theory~\cite{Oriti:2021oux}, with the condensate wavefunction acting as the order parameter. 2) It has been shown \cite{Marchetti:2020umh,Pithis:2016cxg,Calcinari:2023sax} that coherent states lead to small relative fluctuations of geometric and matter field operators, which is essential for a semi-classical interpretation of the dynamics.  Furthermore, in virtue of their simple collective behavior, these states offer a transparent way of connecting macroscopic quantities to microscopic ones, as they only carry few parameters. More precisely, we consider states of the form
\begin{equation}\label{eqn:perturbedstates}
\ket{\Delta;x^0,\vb*{x}} \defeq \mathcal{N}_\Delta e^{\hat{\sigma}\otimes\one_\texttt{-}+\one_\texttt{+}\otimes\hat{\tau}+\hat{\delta\Phi}\otimes\one_\texttt{-}+\hat{\delta\Psi}+\one_\texttt{+}\otimes\hat{\delta\Xi}}\ket{\emptyset}\,,
\end{equation}
where $\mathcal{N}_\Delta$ is a normalization factor and $\ket{\emptyset} = \ket{\emptyset}_\texttt{+}\otimes\ket{\emptyset}_\texttt{-}$ is the $\mathcal{F}$-vacuum. As detailed below, all the operators appearing contain only creation operators such that there is no choice of operator ordering. The one-body operators
\begin{equation}
\hat{\sigma}\equiv\sigma_{\vb*{\xi}_\texttt{+};x^0,p_\phi}\cdot \hat{\varphi}^\dagger_\texttt{+}\,,\qquad 
\hat{\tau}\equiv \tau_{\vb*{\xi}_\texttt{-};x^0,\vb*{x},p_\phi}\cdot\hat{\varphi}^\dagger_\texttt{-}\,,
\end{equation}
generate a background condensate state in the extended Fock space of the form $\ket{\sigma;x^0,p_\phi}\otimes\ket{\tau;x^0,\vb*{x},p_\phi} = \mathcal{N}_\sigma\mathcal{N}_\tau\e^{\hat{\sigma}\otimes\one+\one\otimes\hat{\tau}}\ket{\emptyset}$, with $\ket{\sigma;x^0,p_\phi}$ and $\ket{\tau;x^0,\vb*{x},p_\phi}$ representing spacelike and timelike condensates, respectively. The spacelike $\sigma_{\vb*{\xi}_\texttt{+};x^0,p_\phi}$ and timelike $\tau_{\vb*{\xi}_\texttt{-};x^0,\vb*{x},p_\phi}$ condensate wavefunctions are localized around $\chi^0=x^0$ and $(\rf^0,\vb*{\rf}) = (x^0,\vb*{x})$, respectively. In practice, this is achieved by assuming that they factorize into a peaking function, assumed here to be a Gaussian~\cite{Marchetti:2021gcv}, and a reduced condensate wavefunction, $\tilde{\sigma}$ and $\tilde{\tau}$.\footnote{The peaking properties of the states and the kinetic coupling in Eq.~\eqref{eq:kineticrestriction} manifestly reflect our choice of physical frame. A different choice of relational frame would affect both these aspects of our construction.} The states are furthermore localized in $\phi$-Fourier space ($\phi\rightarrow\pi_\phi$) around an arbitrary scalar field momentum $p_\phi$ which will be identified with the background conjugate momentum of the scalar field below. The peaking properties of the states are collectively represented by $\vb*{\xi}_\pm$, see~\cite{Jercher:2023nxa} for more details.

Requiring $\tilde{\sigma}$ ($\tilde{\tau}$) to contain only gauge-invariant data, the spacelike (timelike) condensate wavefunction can be seen as a distribution of geometric and matter data on a $3$--surface (resp.\ $(2+1)$--surface) localized at, i.e. peaked around, relational time $x^0$ (relational point $(x^0,\vb*{x}))$. Therefore, averages of operators on such relationally localized states can be seen as effective relational observables~\cite{Marchetti:2020umh}. Relational homogeneity of the background structures is then imposed by assuming that $\tilde{\sigma}$ and $\tilde{\tau}$ only depend on the clock variable $\rf^0$. Furthermore, we follow standard protocol~\cite{Oriti:2016qtz,Marchetti:2020umh} and impose isotropy on the condensate wavefunctions by requiring that $\tilde{\sigma}$ and $\tilde{\tau}$ only depend on a single spacelike\footnote{This is a non-trivial requirement for $\tilde{\tau}$, which carries in principle spacelike ($\rho$) and timelike ($\nu$) representations~\cite{Jercher:2022mky}.} representation label $\rho\in\mathbb{R}$. Under these assumptions, we have $\tilde{\sigma}=\tilde{\sigma}_\rho(\rf^0,\pi_\phi)$ and $\tilde{\tau}=\tilde{\tau}_\rho(\rf^0,\pi_\phi)$.

Inhomogeneities in Eq.~\eqref{eqn:perturbedstates} are encoded in the operators $\hat{\delta\Phi}\otimes\one_\texttt{-},\hat{\delta\Psi},$ and $\one_\texttt{+}\otimes\hat{\delta\Xi}$ which are in general $m$-body operators, with integer $m>1$. 
For the remainder, we choose $m=2$, anticipating that, under our assumptions, this is sufficient to capture the physics of scalar, isotropic and slightly inhomogeneous cosmological perturbations.\footnote{Higher $m$-body operators would offer more involved forms of entanglement and could be required for describing phenomenologically richer systems (e.g.\ involving anisotropies).
}
Explicitly, the three $2$-body operators are defined as further
\begin{equation}\label{eqn:entanglementkernels}
    \hat{\delta\Phi} \defeq \hat{\varphi}^\dagger_\texttt{+}\hat{\varphi}^\dagger_\texttt{+}\cdot\delta\Phi\,,\qquad
    \hat{\delta\Psi} \defeq \hat{\varphi}^\dagger_\texttt{+}\cdot\delta\Psi\cdot\hat{\varphi}^\dagger_\texttt{-}\,,\qquad
    \hat{\delta\Xi} \defeq \delta\Xi\cdot\hat{\varphi}^\dagger_\texttt{-}\hat{\varphi}^\dagger_\texttt{-}\,.
\end{equation}
The kernels $\delta\Phi$, $\delta\Psi$ and $\delta\Xi$ are in general bi-local, i.e. their domain is given by two copies of the respective domains of $\varphi_\pm$. As these kernels are assumed to not factorize, the operators above produce non-factorized states, and are therefore the source of the entanglement we refer to in this chapter. More precisely, $\hat{\delta\Phi}$ ($\hat{\delta\Xi}$) leads to an entanglement within the spacelike (timelike) sector, while $\hat{\delta\Psi}$ leads to entanglement between the two sectors. Imposing isotropy\footnote{This condition, although restrictive, is compatible with the isotropic geometric observables we will study.} and gauge invariance, the kernels take the form $\delta\Psi=\delta\Psi_{\rho\rho'}(\rf^\mu_v,\rf^\mu_w,\phi_v,\phi_w)$, similarly for $\delta\Phi$ and $\delta\Xi$.

The relational dynamics of the kernels and the condensate wavefunctions are derived via a mean-field approximation of the full quantum dynamics, i.e.
\begin{equation}\label{eq:lowest SDE}
\expval**{\fdv{S[\hat{\varphi}_\pm,\hat{\varphi_\pm}^{\dagger}]}{\hat{\varphi}(\vbg,x^\mu,\mf,X_\pm)}}{\Delta;x^0,\vb*{x}}= 0\,.
\end{equation}
Following~\cite{Oriti:2016qtz}, the mean-field approximation is only valid in a mesoscopic regime where the expectation value of the GFT particles, or equivalently the modulus of the condensate wavefunctions, is large enough to allow for a continuum geometric interpretation but not too large as otherwise, deviations from mean-field become dominant. According to~\cite{Oriti:2016qtz}, the existence and extent of this mesoscopic regime is linked to the strength of the interaction term $\mathfrak{V}$. Thus, validity of Eq.~\eqref{eq:lowest SDE} is connected to negligible interactions, which we assume for the remainder.\footnote{For phenomenological explorations of interacting condensates, see~\cite{Oriti:2021rvm,deCesare:2016rsf}} Finally, since we are interested in small inhomogeneities, we study Eq.~\eqref{eq:lowest SDE} perturbatively by working with linearized states
\begin{equation}\label{eq:linearized states}
\ket{\Delta;x^0,\vb*{x}} \approx \mathcal{N}_\Delta \left(\one+\hat{\delta\Phi}+\hat{\delta\Psi}+\hat{\delta\Xi}\right)e^{\hat{\sigma}\otimes\one_\texttt{-}+\one_\texttt{+}\otimes\hat{\tau}}\ket{\emptyset}\,.
\end{equation}
Neglecting interactions and simultaneously working perturbatively collapses the whole set of Schwinger-Dyson equations to the lowest order equations~\eqref{eq:lowest SDE}, see~\cite{Jercher:2023nxa}. As we will discuss shortly, this will result in a dynamical freedom at the level of perturbations, i.e.~one of the functions remains indeterminate by the equations. 

\subsection{Dynamics of entangled coherent peaked states}

We now proceed with the perturbative study of equations \eqref{eq:lowest SDE}, based on the derivations of Appendix~\ref{app:Derivation condensate dynamics}. At the background level, and in the limit of negligible interactions, equations for the spacelike and timelike sectors decouple. Following from the peaking in the scalar field momentum $\pi_\phi$ and in the frame variables, the equations of motion for the reduced condensate wavefunctions $\slrcw$ and $\tlrcw$ become second order differential equations in relational time. We consider for the remainder the limit of large modulus of these wavefunctions, associated with late times and classical behavior~\cite{Oriti:2016qtz,Pithis:2016cxg,Marchetti:2020qsq}, and where the condensate is generically dominated by a single representation label $\rho_o$~\cite{Gielen:2016uft,Jercher:2021bie}. We assume the same dominance for the kernels in Eq.~\eqref{eqn:entanglementkernels}, and we will suppress any explicit dependence of functions on representation labels.\footnote{For multiple representation labels dominant, say $\rho_1,\rho_2$, one could in principle also study the entanglement between these modes. However, doing so for a model restricted to spacelike tetrahedra, one cannot straightforwardly reproduce the results of this chapter. That is because of the particular interplay between the two-sector coherent states and expectation values of spacelike operators such as the 3-volume.} In this limit, solutions to the equations of motion are given by
\begin{equation}\label{eq:bkgcondensates}
\slrcw(x^0,\pmm) = \slrcw_0\e^{(\mu_\texttt{+} + i\tpip)x^0}\,,\qquad \tlrcw(x^0,\pmm) = \tlrcw_0\e^{(\mu_\texttt{-} + i\tpim)x^0}\,,
\end{equation}
where $\mu_\pm=\mu_\pm(p_\phi)$ are functions of the peaking parameters, defined in Appendix~\ref{app:Derivation condensate dynamics}. The background solution of $\slrcw$ agrees with previous works~\cite{Oriti:2016qtz,Marchetti:2020umh}.

At first order in perturbations, we obtain two differential equations for the three kernels $\delta\Psi$, $\delta\Phi$ and $\delta \Xi$. This dynamical freedom cannot be reduced beyond mean-field as long as interactions are negligible and the above kernels are small. However, it can be completely fixed by requiring a low-energy agreement with classical physics, as will be seen below. We make an ansatz
\begin{equation}\label{eq:relation of dPsi and dPhi}
\delta\Phi(\rf^\mu,\mm) = \mathrm{f}(\rf^\mu)\delta\Psi(\rf^\mu,\mm)\,,  
\end{equation}
with the complex-valued function $\mathrm{f}$ given by
\begin{equation}\label{eq:def of f}
\mathrm{f}(\rf^0,\vb*{\rf}) = f(\rf^0)\e^{i\theta_f(\rf^0)}\abs{\eta_\delta(\abs{\vb*{\rf}-\vb*{x}})}\e^{2i\pip\rf^0}\,,
\end{equation}
where $f$ and $\theta_f$ are arbitrary real functions of $\rf^0$ and $\eta_\delta$ is a Gaussian peaking function with width $\delta$. Moreover, from now on, we consider correlations that belong to the same relationally localized $4$-simplex and are $\pi_\phi$-conserving, i.e.
\begin{equation}\label{eq:locality condition}
\delta\Psi = \delta\Psi(\rf^\mu_v,\mm^v)\delta^{(4)}(\rf^\mu_v-\rf^\mu_w)\delta(\mm^v-\mm^w)\,.
\end{equation}
Assuming specific relations of the peaking parameters, see Appendix~\ref{app:Derivation condensate dynamics}, the perturbation equations can be simplified further. In particular, the kernel $\delta\Psi$ entangling timelike and spacelike sectors, satisfies an equation of the form
\begin{equation}
0 = \delta\Psi''+t_1\delta\Psi'+t_0\delta\Psi+s_2\nabla_{\vb*{x}}^2\delta\Psi\,,
\end{equation}
where primes denote relational time derivatives and $t_i[f,\theta_f](x^0,\pmm)$ and $s_2[f,\theta_f](x^0,\pmm)$ are complex quantities depending on the functions $f$ and $\theta_f$. This form of the equations, containing zeroth, first and second order time derivatives as well as a second order spatial derivative, already bears resemblance with the perturbations equations of GR, summarized in Appendix~\ref{app:Classical perturbation theory}.

\section{Effective dynamics of cosmological scalar perturbations}\label{sec:dynamics}

At late times, it is demonstrated in~\cite{Marchetti:2020qsq} that quantum fluctuations of extensive operators acting on the Fock space $\mathcal{F}$ are generically small. In this regime, which we consider from here on, one can
associate classical cosmological quantities $\mathcal{O}$ with expectation values of appropriate one-body GFT operators $\hat{\mathcal{O}}$ on the states proposed in Eq.~\eqref{eq:linearized states}, defined as
\begin{equation}\label{eqn:observablesgen}
\mathcal{O}_{\Delta}(x^0,\vb*{x})\defeq \expval{\hat{\mathcal{O}}}{\Delta;x^0,\vb*{x}}= \bar{\mathcal{O}}(x^0)+\delta\mathcal{O}(x^0,\vb*{x})\,.
\end{equation}
The above expectation value is effectively localized in relational spacetime, and thus should be compared to a corresponding classical relational \cite{Goeller:2022rsx} (or, equivalently, harmonic gauge-fixed) observable \cite{Gielen:2018fqv}.\footnote{Defining a state-independent relational localization in GFT has been tackled recently in~\cite{Marchetti:2024nnk}.} By construction, \eqref{eqn:observablesgen} splits into a background, $\bar{\mathcal{O}}$, and a perturbation, $\delta\mathcal{O}$.\footnote{Expectation values of the observables considered in this chapter are real. Thus, $\bar{\mathcal{O}}$ unambiguously denotes the background part and \emph{not} complex conjugation.} Below we compute equation \eqref{eqn:observablesgen} for some important geometric and matter observables, and study their dynamics in detail. 

A crucial example of such observables is the spatial $3$-volume $\hat{V}$ \cite{Oriti:2016qtz}. Its expectation value $V_\Delta$ and the matching with classical dynamics is treated in detail in Appendix~\ref{app:volume matching}. $V_\Delta$ splits into a background contribution $\bar{V}(x^0,\pmm)= \mathrm{v}\abs{\slrcw(x^0,\pmm)}^2$ and a perturbation $\delta V(x^0,\pmm)=2\mathrm{v}\Re{F[\delta\Psi](x^\mu,\pmm)}$, where $\mathrm{v}$ is a volume eigenvalue~\cite{Jercher:2021bie} scaling as $\mathrm{v}\sim \rho_o^{3/2}$ and $F$ is a functional depending on background wavefunctions as well as the functions $f$ and $\theta_f$. The background volume satisfies
\begin{equation}
\frac{\bar{V}'}{3\bar{V}} = \frac{2}{3}\mu_\texttt{+}(\pmm)\,,\qquad \left(\frac{\bar{V}'}{3\bar{V}}\right)' = 0\,,
\end{equation}
which successfully matches classical flat Friedmann dynamics in harmonic gauge if $\mu_\texttt{+}(\pmm)=3\bar{\pi}_\phi/(8M_{\Pl}^2)$~\cite{Oriti:2016qtz,deCesare:2016axk,Marchetti:2021gcv}. This intermediate result has been obtained first in~\cite{Jercher:2021bie}, extracting the homogeneous Friedmann dynamics from the Lorentzian BC model restricted to spacelike tetrahedra. Having exactly the same form as the equations derived from an \enquote{EPRL-like} GFT model~\cite{Oriti:2016qtz}, it is a tentative hint at the universality of the BC and the EPRL model. Note that since the spacelike and timelike sector decouple at background level, contributions of timelike tetrahedra drop out for spacelike observables. 

A similar matching can be performed for the perturbations, which completely fixes the functions $f$ and $\theta_f$ (see Appendix~\ref{app:volume matching}). In this way, the dynamics of $\delta V$ take the simple form
\begin{equation}\label{eqn:pertvolume}
\left(\frac{\delta V}{\bar{V}}\right)'' +a^4 k^2\left(\frac{\delta V}{\bar{V}}\right) = -3\mathcal{H}\left(\frac{\delta V}{\bar{V}}\right)',
\end{equation}
where $\mathcal{H}=\bar{V}'/(3\bar{V})$ is the Hubble parameter and $k$ is the Fourier mode relative to $\vb*{x}$. The harmonic term entering with $a^4$ constitutes an essential improvement compared to previous work~\cite{Marchetti:2021gcv} and is a combined consequence of the Lorentzian reference frame and the use of entangled CPS. Importantly, both of these constructions are facilitated by the extended set of causal building blocks the complete BC model of Chapter~\ref{chapter:cBC} offers. The right-hand side of Eq.~\eqref{eqn:pertvolume} is reminiscent of a friction term, which may be associated with a macroscopic dissipation phenomenon into the quantum gravitational microstructure (as suggested e.g.\ in~\cite{Perez:2017krv}).

The spacelike number operator satisfies $\hat{V} = \mathrm{v}\hat{N}_\texttt{+}$ due to assumption of a single spin condensate. Thus, the dynamics of $N_\texttt{+}^{\Delta}$ follow immediately from those of $V_{\Delta}$. In contrast, the expectation value of the timelike number operator $\hat{N}_\texttt{-}$ satisfies at the background level $\bar{N}_{\texttt{-}}'/\bar{N}_{\texttt{-}} = 2\mu_{\texttt{-}} = \mathrm{const.}$ There are a priori no matching conditions for the parameter $\mu_\texttt{-}$ with respect to an observable of classical GR due to a lack of GFT-observables that characterize the geometry of timelike slices. Further research might reveal additional constraints on {$\mu_\texttt{-}$}. The matching conditions for the perturbed volume imply that $\delta\Xi$ is only time-dependent. This in turn leads to $\delta N_{\texttt{-}}$ only being time-dependent which can thus be absorbed in the background $\bar{N}_{\texttt{-}}$.

To identify observables related to the matter content of the effective cosmology, one can study expectation values of the scalar field operators $\hat{\upphi}_\pm$ and their conjugate momenta $\hat{\upvarpi}^\pm_\mf$, defined on each of the two sectors (see Appendix~\ref{app:Derivation of matter dynamics} for a derivation of the following). Since classically the matter field is an intensive quantity, we combine the expectation values $\upphi_\pm$ of the scalar field operators $\hat{\upphi}_\pm$ through the following weighted sum\footnote{This is analogous to how intensive quantities such as chemical potentials are combined in statistical physics~\cite{Callen}.}
\begin{equation}\label{eq:mf as weighted sum}
\mf_\Delta = \upphi_+\frac{N_+}{N}+\upphi_-\frac{N_-}{N}\,,
\end{equation}
where $N$ is the total (average) number of quanta, and $N_\pm$ are the (average) number of quanta in each sector. The above quantity can then be split in a background, $\bar{\mf}$, and perturbed part, $\delta\mf$.

At the background level, one can show that by requiring $\bar{\upphi}_\pm$ to be intensive quantities and assuming $\mu_+>\mu_-$ (which enter the background condensate solutions in Eqs.~\eqref{eq:bkgcondensates}), $\bar{\mf}=\bar{\upphi}_+$ is completely captured by spacelike data at late times and satisfies the classical equation of motion $\bar{\mf}''=0$. Moreover, the background matter analysis unambiguously identifies the peaking momentum value $p_\phi$ with the classical background momentum of the scalar field, $\bar{\pi}_\mf$ \cite{Marchetti:2021gcv}. 

At first order in perturbations, and under the same assumptions as above, one can write $\delta\phi=\left(\frac{\delta N_+}{\bar{N}_+}\right)\bar{\phi}=\left(\frac{\delta V}{\bar{V}}\right)\bar{\mf}$, so that, using Eq.~\eqref{eqn:pertvolume} and $\bar{\phi}''=0$, we obtain
\begin{equation}\label{eqn:pertmf}
\delta\mf''+a^4 k^2\delta\mf = J_{\mf}\,,
\end{equation}
with the source term $J_{\mf}$ given by $J_{\mf} = \left(-3\mathcal{H}\bar{\phi}+2\bar{\phi}'\right)\left(\frac{\delta V}{\bar{V}}\right)'$. For the expectation value of $\hat{\upvarpi}^\pm_\mf$ at perturbed level, $\delta\upvarpi^\pm_\mf$, a matching with the classical perturbed momentum variable $\delta\pi_\mf^0$, defined in Eq.~\eqref{eq:classical perturbed mm 0}, cannot be established. The main difficulty in matching these two quantities is that the classical equation~\eqref{eq:classical perturbed mm 0} depends on the $(00)$-component of the perturbed metric. To recover this quantity from the fundamental QG theory, one would need additional geometric operators other than the volume, see also the discussion at the end of this chapter.

A crucial quantity in classical cosmology is the comoving curvature perturbation $\mathcal{R}$~\cite{lyth}, proportional to the so-called Mukhanov-Sasaki variable~\cite{Mukhanov:1988jd,mukhanov2,sasaki}, see also Appendix~\ref{app:Classical perturbation theory}. This can be obtained by combining matter data with isotropic and anisotropic geometric information in a gauge-invariant way. Restricting to isotropic volume data, one can define an analogous \emph{curvature-like} variable
\begin{equation}
\tilde{\mathcal{R}}\equiv  -\frac{\delta V}{3\bar{V}}+\mathcal{H}\frac{\delta\mf}{\bar{\mf}'}\,,
\end{equation}
which is perturbatively gauge-invariant only for $k\ll 1$. The curvature-like variable $\tilde{\mathcal{R}}$ can be constructed within our framework by combining Eqs.~\eqref{eqn:pertvolume} and~\eqref{eqn:pertmf}\footnote{We note that in this case $\tilde{\mathcal{R}}$ is constructed out of relational observables, and is thus gauge-invariant by construction.} yielding
\begin{equation}\label{eq:GFT perturbed R}
\tilde{\mathcal{R}}'' + a^4 k^2\tilde{\mathcal{R}} = J_{\tilde{\mathcal{R}}}\,,
\end{equation}
with source term $J_{\tilde{\mathcal{R}}} = \left[3\mathcal{H}-\frac{1}{4 M_{\Pl}^2}\left(\bar{\mf}^2\right)'\right]\left(\frac{\delta V}{\bar{V}}\right)'$.

%
%
%

\begin{figure}
    \centering
    \begin{subfigure}[b]{0.53\textwidth}
    \includegraphics[width=\linewidth]{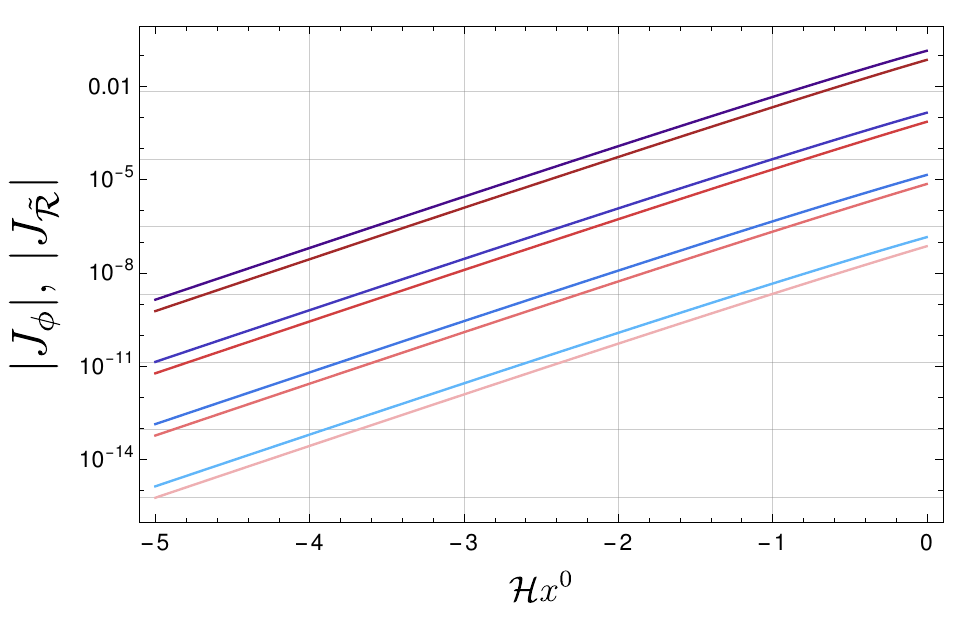}
    \end{subfigure}%
    \begin{subfigure}[b]{0.45\textwidth}
    \includegraphics[clip, trim = 20pt 200pt 100pt 100pt, width=\linewidth]{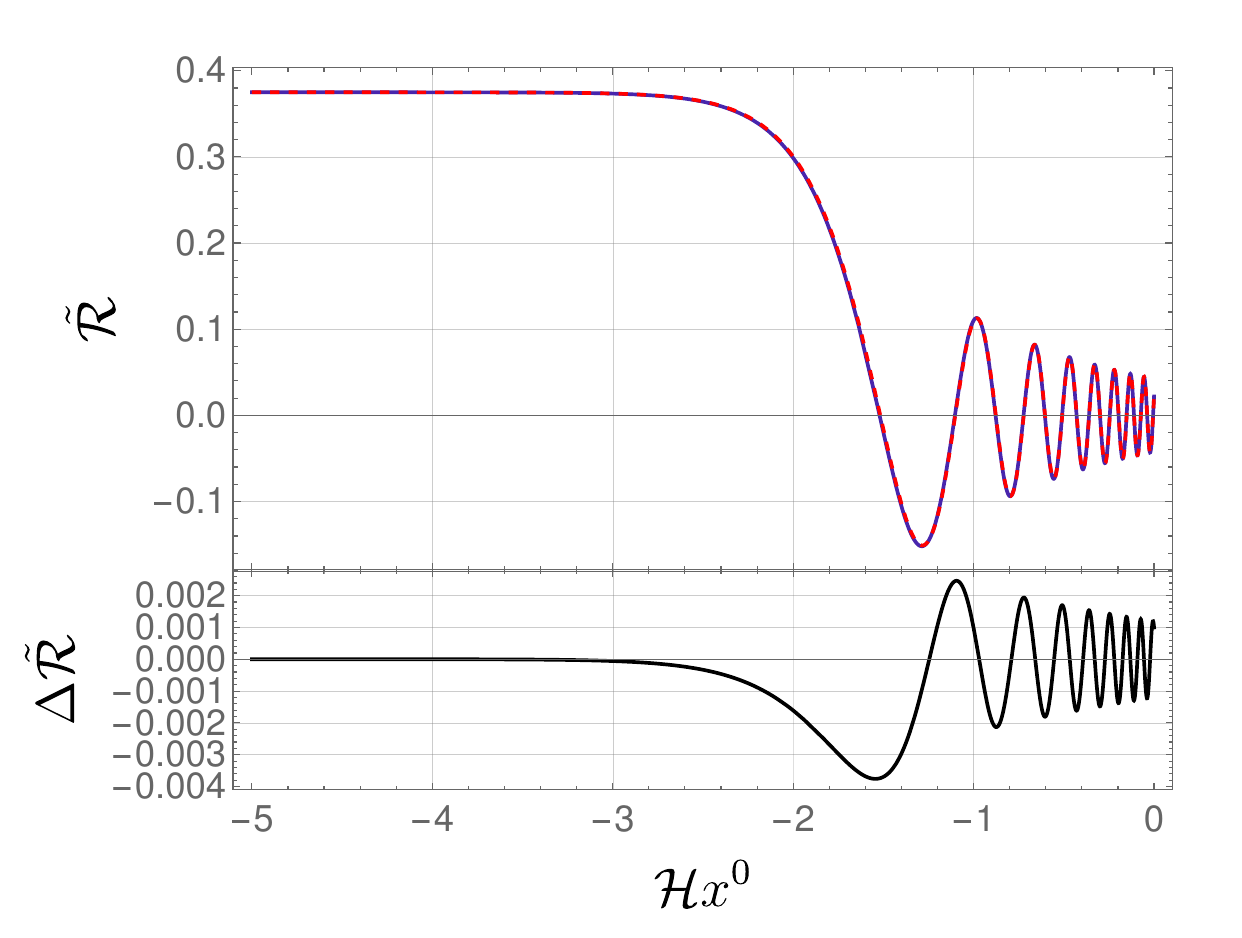}
    \end{subfigure}
    \caption{Left: absolute value of source terms $J_\mf$ (blue) and $J_{\tilde{\mathcal{R}}}$ (red) with $c = 0.1$ for modes $k/M_{\Pl}\in\{10^{-3},\: 10^{-2},\: 10^{-1},\: 10^0\}$ in the vicinity of the trans-Planckian regime where darker colors correspond to larger modes. In the trans-Planckian regime $k/M_{\Pl} \sim 1$, $J_\mf$ and $J_{\tilde{\mathcal{R}}}$ become of order one at late times, while they remain small at all times in the sub-Planckian regime. Right top: curvature-like perturbation $\tilde{\mathcal{R}}_\GFT$ (blue) and $\tilde{\mathcal{R}}_\GR$ (dashed red) for a fixed mode $k/M_\Pl = 10^2$ in the trans-Planckian regime. Right bottom: difference of the blue and red dashed curves above, $\Delta\tilde{\mathcal{R}}\equiv \tilde{\mathcal{R}}_\GFT-\tilde{\mathcal{R}}_\GR$. For both plots, the ratio of $\delta V/\bar{V}$ and $\delta\mf$ at initial time has been set to $c=0.1$.}
    \label{fig:sourceterms}
\end{figure}

Comparing Eqs.~\eqref{eqn:pertmf} and \eqref{eq:GFT perturbed R} with their classical GR counterparts \eqref{eq:classical mf perturbation equation} and~\eqref{eq:classical R perturbation equation}, we notice that they contain an additional source term, $J_\mf$ and $J_{\tilde{\mathcal{R}}}$, respectively. The intrinsically quantum gravitational nature of these terms can be made manifest by solving first Eq.~\eqref{eqn:pertvolume} with initial conditions chosen such that solutions match the GR ones in the super-horizon limit.\footnote{Note that for different initial conditions, the perturbed volumes $\delta V/\bar{V}$ in GFT and GR converge to different constants at early times. This deviation is thus not a dynamical deviation buy simply corresponds to a different choice in integration constant. The dynamical agreement between perturbations from GFT and GR is not affected by this choice.} Indeed, in this case we can write
\begin{equation}\label{eqn:sourceterms}
    J_{\mf}=c\left(\frac{a^2k}{M_{\Pl}}\right)j_{\mf}[\bar{\mf}]\,,\qquad J_{\tilde{\mathcal{R}}}=c\left(\frac{a^2k}{M_{\Pl}}\right)j_{\tilde{\mathcal{R}}}[\bar{\mf}]\,,
\end{equation}
where $c$ is the initial ratio of perturbed and background volume (and thus required to be small) and $j_{\mf}$ and $j_{\tilde{\mathcal{R}}}$ are functions depending on the background field $\bar{\mf}$ and the mode $k$. Visualized in the left panel of Fig.~\ref{fig:sourceterms}, effects of the source terms remain small at all times in the sub-Planckian regime $k/M_{\Pl}\ll 1$ such that classical dynamics are recovered. Effects of the source terms become important in the trans-Planckian regime, $k/M_{\Pl}\gtrsim 1$. This shows the corrections to be genuine quantum gravity effects.

An exemplary solution of Eq.~\eqref{eq:GFT perturbed R}, together with a comparison to the classical GR counterpart is depicted in the right panel of Fig.~\ref{fig:sourceterms}. As it is clear from the plots, QG corrections are only relevant for highly trans-Planckian modes, which are many orders of magnitude larger than the typical momentum scales of interest in cosmology ($\sim \text{Mpc}^{-1}$). 
It would be particularly interesting to investigate whether the above corrections can also be obtained from a continuum modified gravitational action, potentially establishing an effective field theory framework for the above dynamics. We leave this intriguing avenue to future work.\\

\begin{center}
    \textit{Summary.}
\end{center}

\noindent The main achievement of this chapter is the extraction of scalar cosmological perturbations from entangled GFT coherent states which agree with classical perturbations on sub-Planckian scales and are subject to corrections in the trans-Planckian regime. Consequently, these modifications constitute genuine quantum effects arising from the microscopic spacetime structure. This represents a significant advancement compared to previous attempts~\cite{Marchetti:2021gcv} where agreement of effective perturbations with classical GR has only been achieved in the super-horizon limit. Two principles guided the analysis of this chapter: 1) The causal properties of frame fields should be faithfully transferred to the quantum theory. This has been realized by restricting the GFT kinetic kernels such that clock and rods only propagate along timelike, respectively spacelike dual edges. 2) Inhomogeneities of cosmological observables emerge from relational nearest-neighbor two-body correlations between GFT quanta, encoded in the entangled GFT coherent states. Implementing both principles was made possible by the extended causal structure of the complete BC model introduced in Chapter~\ref{chapter:cBC}. This puts again emphasis on the importance of a complete set of causal building blocks for recovering Lorentzian continuum geometries from QG. Our results represent a crucial intermediate step towards connecting QG effects to observations, elaborated below, and provide a concrete example of the emergence of non-trivial geometries from QG entanglement~\cite{Giddings:2005id,Ryu:2006bv,VanRaamsdonk:2010pw,Bianchi:2012ev,Maldacena:2013xja,Giddings:2015lla,Cao:2016mst,Donnelly:2016auv,swingle2018spacetime,colafrancheschi2022emergent,Bianchi:2023avf}.

\begin{center}
    \textit{Closing remarks.}
\end{center}

\paragraph{Fixing the dynamical freedom.} In the absence of interactions, the mean-field equations leave one of the entangling functions, $\delta\Phi$, dynamically undetermined. This freedom has been fixed in Sec.~\ref{sec:dynamics} by requiring a low-energy agreement with GR. While this is a sound strategy, one could also derive dynamical equations for $\delta\Phi$, in principle. As shown in~\cite{Jercher:2023nxa}, going to higher-order Schwinger-Dyson equations does not yield independent equations as long as interactions are neglected. Thus, a crucial step in this direction is to first include interactions and then go to equations beyond mean-field. Whether the such obtained dynamical equations of $\delta\Phi$ are compatible with the low-energy requirement given above is unclear. Note however, that except the spatial derivative part, the coefficients entering the equations for $\delta V$, $\delta \phi$ and $\tilde{R}$ are universal in that they do not depend on the relation between $\delta\Phi$ and $\delta\Psi$.

\paragraph{Geometric operators.} For further development of this work in future research, the construction of additional geometric operators is crucial. This would allow extending our analysis to observables commonly used in theoretical and observational cosmology, such as the comoving curvature $\mathcal{R}$. As mentioned above, $\mathcal{R}$ cannot be constructed from the volume operator alone as it involves anisotropic data. Capturing such data at the GFT level requires relaxing isotropy and introducing anisotropic observables (see also~\cite{deCesare:2017ynn,Calcinari:2022iss}), e.g. the areas of orthogonal two-surfaces. Introducing anisotropic observables would also be important to reconstruct the full effective metric, including not only scalar perturbations, but also vector and, most importantly, tensor perturbations. The extrinsic curvature is another observable we expect to be highly informative. In particular, it may provide information on the temporal components of metric perturbations, improving our understanding of the timelike sector and the parameter $\mu_{\texttt{-}}$.


\paragraph{Matter content.}  To move towards a matter content more realistic than massless free scalar fields, one would have to include cosmic fluids. Considerable effort has been devoted to the study of dust in classical~\cite{Giesel:2012rb,Domagala:2010bm} and quantum~\cite{Giesel:2007wk,Giesel:2020bht,Giesel:2020xnb,Husain:2020uac} cosmology, since it constitutes a key component of the Universe and serves as a natural physical reference frame. Coupling dust to GFT models would therefore allow setting up more realistic cosmological models in which the dynamics of perturbations can be studied.

\paragraph{Early time.} Generalizing the current analysis to earlier times would be important to understand the imprint of the quantum gravity bounce on the perturbations. Moreover, for the perturbation theory developed here to be self-consistent also in this regime it is important to check if the energy density of the perturbations remains bounded and small compared to the background quantum geometry. This concerns in particular perturbations of trans-Planckian wavelength and was dubbed the \enquote{real trans-Planckian issue} in another setting~\cite{Agullo:2012sh,Agullo:2023rqq}. In general, however, the mean-field equations, though still being valid, become considerably more complicated as the density of the background condensate decreases~\cite{Marchetti:2021gcv}. 

\paragraph{Connecting to observations.} There are two different levels at which one could try to make contact with observations. First, one could phenomenologically incorporate the GFT modified perturbation dynamics into the standard cosmological model with a single inflaton Fock quantized on the GFT background at early times. Due to the quantum bounce at early times, the inflaton power spectrum receives corrections which could be compared to those of loop quantum cosmology~\cite{Ashtekar:2020gec}. Additionally, observable consequences may be produced by the quantum corrections found here as inflation can be sensitive to trans-Planckian physics~\cite{martin2001, Danielsson:2002kx, Brandenberger:2012aj}. Alternatively, one could \emph{derive} the physics of cosmological perturbations (including the properties of their power spectra) from the full QG theory. However, this would require the construction of additional \emph{relational} operators (see~\cite{Marchetti:2024nnk} for recent advancements), the inclusion of a more realistic matter content, and a generalization of the analysis performed here to early times, as discussed above. Also, an inflationary mechanism would have to be included. For a more extensive discussion of these points, see~\cite{Jercher:2023nxa}.


\clearpage
\fancyhead[CE]{\leftmark}
\fancyhead[LE]{\scshape\MakeLowercase\chaptertitlename\hspace{2pt}\large\thechapter}
\fancyhead[RO]{\scshape\MakeLowercase\chaptertitlename\hspace{2pt}\large\thechapter}
\fancyhead[CO]{\leftmark}
\fancyfoot[C]{\thepage}
\bookmarksetup{startatroot}
\addtocontents{toc}{\bigskip}
\chapter[\textsc{Summary and Outlook}]{Summary and Outlook}

One of the most pressing challenges of background independent QG is the recovery of classical continuum geometries from very different microscopic quantum degrees of freedom. In this monograph, we took up this challenge and studied the emergence of Lorentzian geometries from the two closely related non-perturbative and background independent QG approaches of spin-foams (Part~\hyperref[Part I]{I}) and group field theories (Part~\hyperref[Part II]{II}). 

Chapter~\ref{chapter:specdim} demonstrated that $\mathcal{N}$-periodic Euclidean spin-foam frusta satisfy the important and non-trivial consistency check of a large-scale spectral dimension of $4$, therefore connecting to the observed spacetime dimension. At lower scales, the spectral dimension exhibits a non-trivial flow controlled by the face amplitude parameter $\alpha$. These results mark a significant improvement to previous studies~\cite{Steinhaus:2018aav} as effects from quantum amplitudes and curvature-induced oscillations were included. In the continuum limit, $\mathcal{N}\rightarrow\infty$, the existence of a critical surface in the parameter space $(\alpha,\GN,\bi,\Lambda)$ was conjectured where the large-scale spectral dimension changes discontinuously from $0$ to $4$. Comparing the spectral dimension flow to that of other approaches such as CDT~\cite{Ambjorn:2005qt,Coumbe:2014noa} will require tuning to this critical surface. Another important open task is to transfer and generalize the results to the Lorentzian signature case. This work marks an important step in these directions.

Transferring the 4-frusta geometries of Chapter~\ref{chapter:specdim} to Lorentzian signature, it was shown in Chapter~\ref{chapter:LRC} that their classical dynamics coupled to a massless scalar field captures discrete spatially flat cosmology. The relational Friedmann equations emerge in a continuum limit which exists only in the causally regular sector when trapezoids and 3-frusta are timelike. Our results therefore present serious evidence that other than spacelike building blocks are required to recover the Lorentzian geometries of GR. This is a central theme of this dissertation.

Motivated by the previous results, the effective cosmological spin-foam path integral coupled to a massive scalar field, derived from the (2+1)-dimensional coherent state model of~\cite{Simao:2024don}, was investigated in Chapter~\ref{chapter:3d cosmology}. Discrete Lorentzian geometries can be recovered from expectation values, rendering the model viable for quantum cosmology from QG. Causality violations and the path integral measure, however, can potentially obstruct semi-classicality. These are important insights, motivating future studies on causality violations in full spin-foams, and indicating that considering merely the critical points of the action is insufficient for the non-perturbative evaluation of the gravity path integral. 

Part~\hyperref[Part I]{I} presents a multifaceted analysis of frusta geometries which allow for explicit computations while still being physically relevant with their direct connection to spatially flat cosmology. A connection to semi-classical geometries was established by studying the spectral dimension and the cosmological path integral, revealing intriguing insights on the role of timelike building blocks, causality violations and the path integral measure.

Chapter~\ref{chapter:cBC} commenced Part~\hyperref[Part II]{II} of this work, focussing on the GFT approach to QG. A causal completion of the BC model was developed, incorporating spacelike, lightlike and timelike tetrahedra. It is the first GFT and spin-foam model that captures the full set of causal building blocks, forming the foundation of the ensuing investigations on Lorentzian quantum geometries in Chapters~\ref{chapter:LG} and~\ref{chapter:perturbations}. An important open task is to study its semi-classical properties at spin-foam level, thereby connecting to the semi-classical analyses of the EPRL-CH model~\cite{Liu:2018gfc,Kaminski:2017eew,Simao:2021qno,Barrett:2009mw}.  

The impact of a complete set of causal building blocks on the phase structure of the BC model was studied in Chapter~\ref{chapter:LG} via a Landau-Ginzburg analysis, making use of the field-theoretic description GFTs offer. Crucially, the mean-field approximation is generically self-consistent, ensured by the exponential suppression of the Ginzburg-$Q$ induced by the hyperbolic part of the Lorentz group. The critical behavior is entirely driven by the representation labels of spacelike faces, while timelike faces do not contribute, as they characterize the compact part of $\SL$. These results lend support for the existence of a continuum gravitational phase characterized by the mean-field vacuum, motivating the extraction of cosmological perturbations from GFT coherent states in the subsequent chapter. 

Dynamics of scalar cosmological perturbations were extracted in Chapter~\ref{chapter:perturbations} from GFT coherent states encoding the entanglement within and between the spacelike and timelike sectors of the complete BC GFT model. The causal properties of a physical Lorentzian reference were faithfully encoded into the quantum theory, bearing relevance for a wider class of QG approaches. Cosmological perturbations were extracted from the expectation values of GFT observables with respect to the entangled coherent states governed by the GFT mean-field dynamics. As the main result, the effective equations agree with the classical perturbations equations of GR up to trans-Planckian, and thus quantum corrections.
 
The overarching insight of Part~\hyperref[Part II]{II} is that Lorentzian geometries beyond spatially homogeneous cosmology can be extracted from the complete BC GFT model developed here. The underlying assumption of a phase excited by many GFT quanta is corroborated by a self-consistent mean-field approximation within LG theory. The extraction of cosmological perturbations demonstrates the importance of a causally complete set of building blocks for bridging the gap between QG and GR.\\

While the findings of this work constitute a significant step towards understanding the emergence of Lorentzian geometries from spin-foams and GFTs, many more challenges remain and new questions arise. We close with the following reflections.

Connecting back to the discussion of the introduction, spin-foams and GFTs can take very different perspectives on how a semi-classical continuum limit of QG should be taken. This is strikingly demonstrated by the two parts of this monograph which are seemingly independent in their underlying assumptions, the utilized methods and the results obtained. However, as advocated for in the following, bringing spin-foams and GFTs closer again is desirable. First and foremost, the original connection persists: spin-foams prescribe the Feynman amplitudes of a field theory which is a GFT. Furthermore, taking the perspective of a quantum field theorist, to understand a field theory one benefits from exploring both, the properties of its amplitudes and its non-perturbative properties via the action and the full partition function. This duality suggests that deepening the connection between spin-foams and GFTs could be highly fruitful as insights and recent advancements from both approaches could be amalgamated. Such a mutual transfer of results between communities is very important, especially given the small number of practitioners worldwide. A particularly promising avenue lies in establishing a connection between the perturbative renormalization~\cite{Carrozza:2024gnh,BenGeloun:2011rc,Carrozza:2012uv,Carrozza:2013wda,BenGeloun:2013vwi} of quantum geometric GFTs with the refinement limit~\cite{Asante:2022dnj} of the corresponding spin-foam amplitudes. First steps in this direction have been taken in~\cite{Dona:2022vyh,Riello:2013bzw,Frisoni:2021uwx,Frisoni:2021dlk,Dona:2018pxq} studying radiative corrections to the edge amplitudes of the Lorentzian EPRL spin-foam model. Connecting these results to the perturbative renormalization of tensor-invariant GFTs will require evaluating spin-foam amplitudes on other than simplicial building blocks. 

Another important step in bringing spin-foams and GFTs closer lies in model building. Current developments in spin-foams predominantly involve the EPRL model, with its causal completion given by the CH-extension. In contrast, the results of Chapters~\ref{chapter:cBC}--\ref{chapter:perturbations} are obtained exclusively for the completion of the BC GFT model which, as argued in the introduction of Chapter~\ref{chapter:cBC}, also constitutes a viable QG model. Therefore, setting up and investigating an explicit formulation of the EPRL-CH GFT model is indispensable for linking the two approaches and for comparing results. This would furthermore allow investigating whether these two seemingly distinct models could yield the same continuum physics under a renormalization procedure, placing them in the same universality class. \emph{Tentative} indications supporting this perspective come from area Regge calculus~\cite{Dittrich:2021kzs} and GFT condensate cosmology~\cite{Jercher:2021bie}. This conjecture would be further strengthened by a LG analysis of a yet to be developed EPRL GFT model that could reveal the same critical exponents. If the BC and EPRL model indeed belong to the same universality class, the simpler BC model could be employed to study the emergent continuum physics.

Lastly, a crucial point of synthesis that was encountered already in the main body of this thesis is the importance of an extended causal structure, common to both approaches. This includes in particular timelike building blocks which were a necessary ingredient for: I) causal regularity, the continuum limit and semi-classicality in Part~\hyperref[Part I]{I}, and II) the construction of a physical Lorentzian reference frame and the extraction of cosmological perturbations in Part~\hyperref[Part II]{II}. This insight contrasts much of the current research in spin-foams and GFTs, which is prevalently limited to models that involve exclusively spacelike tetrahedra. While the proximity to LQG and the computationally more feasible representation theory of $\SUT$ (rather than $\SL$ or $\SUO$) justify this restriction, going beyond is crucial to connect to the physics of Lorentzian spacetimes. Given the insights of Lorentzian effective spin-foams~\cite{Asante:2021zzh,Asante:2021phx} on causality violations, the question arises how full spin-foam models such as the EPRL-CH model deal with such configurations. In particular, it remains unclear whether these configurations persist in a suitably defined semi-classical and continuum limit. Tackling these questions constitutes an intriguing research direction that could profit from recent advancements in numerical methods, see~\cite{Dona:2019dkf,Gozzini:2021kbt,Dona:2022dxs,Dona:2023myv,Steinhaus:2024qov,Asante:2024eft}. Also, from the GFT perspective, it remains as an intriguing question to determine the role of causality violations and how restrictions on the generated complexes, e.g. through a dual-weighting~\cite{Kazakov:1995ae,Benedetti:2011nn}, could enforce causal regularity. 
\clearpage
\pagenumbering{numstyle}
\thispagestyle{empty}
\renewcommand*{\dictumauthorformat}[1]{#1}
\renewcommand*{\dictumwidth}{.5\textwidth}

\vspace*{\fill}
\dictum[\normalfont{\textsc{Hermann Hesse -- Stufen (1941)}}]{\textit{
Wie jede Blüte welkt und jede Jugend\\
Dem Alter weicht, blüht jede Lebensstufe,\\
Blüht jede Weisheit auch und jede Tugend\\
Zu ihrer Zeit und darf nicht ewig dauern.\\
Es muß das Herz bei jedem Lebensrufe\\
Bereit zum Abschied sein und Neubeginne,\\
Um sich in Tapferkeit und ohne Trauern\\
In andre, neue Bindungen zu geben.\\
Und jedem Anfang wohnt ein Zauber inne,\\
Der uns beschützt und der uns hilft, zu leben.\\
{$\phantom{0}$}\\
Wir sollen heiter Raum um Raum durchschreiten,\\
An keinem wie an einer Heimat hängen,\\
Der Weltgeist will nicht fesseln uns und engen,\\
Er will uns Stuf' um Stufe heben, weiten.\\
Kaum sind wir heimisch einem Lebenskreise\\
Und traulich eingewohnt, so droht Erschlaffen;\\
Nur wer bereit zu Aufbruch ist und Reise,\\
Mag lähmender Gewöhnung sich entraffen.\\
{$\phantom{0}$}\\
Es wird vielleicht auch noch die Todesstunde\\
Uns neuen Räumen jung entgegen senden,\\
Des Lebens Ruf an uns wird niemals enden,\\
Wohlan denn, Herz, nimm Abschied und gesunde!
}}

\clearpage
\renewcommand*{\raggeddictum}{\raggedright}
\renewcommand*{\raggeddictumauthor}{\raggedright}
\renewcommand*{\raggeddictumtext}{\raggedright}
\thispagestyle{empty}
\dictum[\normalfont{\textsc{Werner Herzog -- Encounters at the end of the world (2008)}}]{\textit{\dots, but why?}}

\clearpage
\appendix
\phantomsection
\addcontentsline{toc}{part}{Part III: Appendices}
\addtocontents{toc}{\protect\setcounter{tocdepth}{0}}
\pagenumbering{Roman}
\setcounter{page}{1}
\fancyhead[CE]{\leftmark}
\fancyhead[LE]{\scshape\MakeLowercase\chaptertitlename\hspace{2pt}\large\thechapter}
\fancyhead[RO]{\large\thesection}
\fancyhead[CO]{\rightmark}
\fancyfoot[C]{\thepage}
\renewcommand{\headrulewidth}{0.5pt}
\renewcommand{\footrulewidth}{0pt}
\setcounter{page}{3}
\chapter[\textsc{The Lorentz Group in (2+1) and (3+1) Dimensions}]{Representation Theory of the \\Lorentz Group in (2+1) and (3+1) Dimensions}

\section{Facts and conventions on SU(1,1)}\label{app:su11}

\subsection{Unitary irreducible representations}

The unitary irreducible representations of $\mathrm{SU}(1,1)$ were classified for the first time by Bargmann in \cite{Bargmann:1946me}, and the analysis of generalized eigenstates was carried out by Lindblad in \cite{Lindblad:1969zz}.  In the following collection of facts based on these works, we follow the conventions of the latter. 

The $\mathfrak{su}(1,1)$ algebra is spanned by the generators $F^i=(L^3, K^1, K^2)$, defined in the fundamental representation with the standard Pauli matrices as $\varsigma^i/2=(\sigma_3/2, i\sigma_2/2, -i\sigma_1/2)$. The respective subgroups read
\begin{equation}
  e^{i \alpha L^3}=\begin{pmatrix}
    e^{i \frac{\alpha}{2}} & 0 \\
    0 & e^{-i \frac{\alpha}{2}}
  \end{pmatrix}\,, \quad   e^{i t K^1}=\begin{pmatrix}
    \cosh \frac{t}{2} & i\sinh\frac{t}{2} \\
    -i\sinh\frac{t}{2} & \cosh \frac{t}{2}
  \end{pmatrix}\,, \quad
    e^{i u K^2}=\begin{pmatrix}
     \cosh \frac{u}{2} & \sinh\frac{u}{2} \\
    \sinh\frac{u}{2} & \cosh \frac{u}{2}
  \end{pmatrix}\,,
\end{equation}
and the Casimir element is given by $Q=(L^3)^2-(K^1)^2-(K^2)^2$. There are two families of unitary irreducible representations characterized by the eigenvalues of $Q$ and $L^3$, called the \textit{discrete} and \textit{continuous series}. Regarding the first, the Hilbert space $\mathcal{D}^q_k$ is spanned by the orthonormal states
\begin{equation}
  \begin{gathered}
  Q\ket{k,m}=k(k+1)\ket{k,m}\,, \quad  k \in -\frac{\mathbb{N}}{2}\,,\\[7pt]
  L^3 \ket{k,m}=m\ket{k,m}\,, \quad m\in q(- k  + \mathbb{N}^0)\,, \quad q=\pm\,.
  \end{gathered}
\end{equation}
For the continuous series, an orthonormal basis for the Hilbert space $\mathcal{C}^\delta_j$ is given by 
\begin{equation}
  \begin{gathered}
  Q\ket{j,m}=j(j+1)\ket{j,m}\,, \quad j=-\frac{1}{2}+is\,,\quad s\in\mathbb{R}^+\,,\\[7pt]
  L^3 \ket{j,m}=k\ket{j,m}\,, \quad m\in \delta + \mathbb{Z}\,, \quad \delta\in\{0,\frac{1}{2}\}\,. 
  \end{gathered}
\end{equation}
An alternative orthonormal basis of $\mathcal{C}^\delta_j$ can be obtained from generalized eigenstates of the non-compact operator $K^2$. The eigenstates satisfy
\begin{equation}
  \begin{gathered}
  Q\ket{j,\lambda,\sigma}=j(j+1)\ket{j,\lambda,\sigma}\,, \quad j=-\frac{1}{2}+is\,,\quad s\in\mathbb{R}^+\,,\\[7pt]
  K^2 \ket{j,\lambda,\sigma}=\lambda \ket{j,m}\,, \quad \lambda \in \mathbb{C}\,, \\[7pt]
  P \ket{j,\lambda,\sigma}=(-1)^\sigma \ket{j,\lambda,\sigma}\,, \quad \sigma\in\{0,1\}\,,
  \end{gathered}
\end{equation} 
where $P$ is an outer automorphism of the Lie algebra taking $(L^3, K^1, K^2)\mapsto (-L^3, -K^1, K^2)$. They are complete and orthonormal in the sense that
\begin{equation}
  \begin{gathered}
  \sum_\sigma \int_{\mathbb{R}+i\alpha} \dd{\lambda} \,  \braket{j,m}{j, \lambda, \sigma} \braket*{j, \overline{\lambda}, \sigma}{j,n}=\delta_{m,n}\,, \quad \alpha \in \mathbb{R}\,, \\[7pt]
  \sum_m  \braket*{j, \overline{\lambda'}, \sigma'}{j,m} \braket{j,m}{j, \lambda, \sigma} = \delta(\lambda-\lambda')\,, \quad \Im \lambda =\Im \lambda'\,,
  \end{gathered}
\end{equation}
and indeed there is a family of bases $\left\{\, \ket{j, \lambda+i \alpha, \sigma} \, | \, \lambda \in \mathbb{R}\,, \sigma \in \{0,1\}\right\}_{\alpha}$ indexed by $\alpha \in \mathbb{R}$.

\subsection{Harmonic analysis}

According to \cite{Pukanszky:1963,Takahashi:1961,Ruehl1970}, given any function $f\in L^2(\SUO)$,
\begin{equation}\label{idexp}
\begin{aligned}
  f(\one)&=\sum_\delta \int_{-\infty}^\infty \dd{s}\, s\tanh(\pi s)^{1-4\delta} \mathrm{Tr}_{\mathcal{C}^\delta_j}\left[\int_{\mathrm{SU}(1,1)}\dd{g}\, f(g) \vb*{D}^{j}(g)  \right] \\[7pt]
  &+\sum_{q} \sum_{2k=-1}^{-\infty} (-2k-1) \mathrm{Tr}_{\mathcal{D}^q_k}\left[\int_{\mathrm{SU}(1,1)}\dd{g}\, f(g) \vb*{D}^{k}(g)  \right]\,,
\end{aligned}
\end{equation}
with $\dd{g}$ the Haar measure and $\vb*{D}$ the $\SUO$ Wigner matrices. Observe that the lowest $k=-\frac{1}{2}$ discrete representation is absent from the decomposition of $f$. Setting $f(g)\defeq\int\dd{h} \overline{f_1(h)} f_2(gh)$ it follows from Eq.~\eqref{idexp} that
\begin{equation}
\begin{aligned}
  \int \dd{g} \, \overline{f_1(g)} f_2(g) &= \sum_\delta \int_{-\infty}^\infty \dd{s}\, s\tanh(\pi s)^{1-4\delta} (\overline{c_1})^{j(\delta)}_{\lambda\sigma,\lambda'\sigma'} (c_2)^{j(\delta)}_{\lambda\sigma,\lambda'\sigma'}  \\[7pt]
  &+\sum_{q} \sum_{2k=-1}^{-\infty} (-2k-1) (\overline{c_1})^{k(q)}_{m m'} (c_2)^{k(q)}_{m m'}\,, 
\end{aligned}
\end{equation}
where all lower repeated indices are appropriately contracted, and the Fourier coefficients read
\begin{equation}
\begin{aligned}
  (c_i)^{k(q)}_{m m'}&= \int\dd{g} \, f_i(g) D^{k(q)}_{mm'}(g)\,, \\[7pt]
  (c_i)^{j(\delta)}_{\lambda\sigma,\lambda'\sigma'}&= \int\dd{g} \, f_i(g) D^{j(\delta)}_{\lambda\sigma, \lambda' \sigma'}(g)\,.
\end{aligned}
\end{equation}
The coefficients for the continuous series could of course also be written with respect to the $L^3$ eigenbasis. 
 Yet another consequence of Eq.~\eqref{idexp} are the orthogonality relations
\begin{align}
  \int \dd{g} \, \overline{D^{k(q)}_{m n}(g)} D^{k'(q')}_{m' n'}(g) &= \frac{\delta_{q,q'} \delta_{k,k'} }{-2k-1}\;   \delta_{m,m'} \delta_{n,n'}\,, \\[7pt]
    \int \dd{g} \, \overline{D^{j(\delta)}_{m,n}(g)} D^{j'(\delta')}_{m' n'}(g) &= \frac{\delta_{\delta,\delta'} \delta(j-j')}{s\tanh(\pi s)^{1-4\delta}}\;  \delta_{m,m'} \delta_{n,n'}\,,\\[7pt]
  \int \dd{g} \, \overline{D^{j(\delta)}_{\lambda\sigma,\mu \epsilon}(g)} D^{j'(\delta')}_{\lambda' \sigma', \mu' \epsilon '}(g) &= \frac{ \delta_{\delta,\delta'} \delta(j-j')}{s\tanh(\pi s)^{1-4\delta}}\; \delta(\lambda-\lambda') \delta(\mu-\mu') \,,
  \label{coefforto}
\end{align}
which hold for the matrix coefficients of the discrete and continuous series in the $L^3$ eigenbasis, and the coefficients of the continuous series in the $K^2$ eigenbasis, respectively. 

\section{Facts and conventions on SL(2,C)}\label{sec:Aspects of SL2C and its Representation Theory}

In this appendix, we provide a summary of necessary formulas for computations involving $\SL$ representation theory based on~\cite{Ruehl1970}.

\subsection{Representation theory of $\SL$}\label{app:Representation Theory of SL2C}

Unitary irreducible representations of $\SL$ are realized as the space of homogeneous functions on $\C^2$ with degree $(\lambda,\mu)\in\C^2$, $\lambda-\mu\in\mathbb{Z}$, parametrized  as
\begin{equation}\label{eq:(lambda,mu)}
(\lambda,\mu) = (i\rho+\nu-1,i\rho-\nu-1)\,,
\end{equation}
with $\nu\in\mathbb{Z}/2$. For the remainder, we restrict to the principal series with $\rho\in\R$. Denoting the representation space by $\mathcal{D}^{(\rho,\nu)}$, the group action of $\SL$ on this space is defined as~\cite{Ruehl1970}
\begin{equation}\label{eq:group action}
\left(\vb*{D}^{(\rho,\nu)}(g)F\right)(\vb*{z}) \defeq F(g^T\vb*{z})\,,
\end{equation}
where $g\in\SL$, $\vb*{z}\in\C^2$ and $\vb*{D}^{(\rho,\nu)}$ is the $\SL$ Wigner matrix. In the canonical basis, the Wigner matrices have components $D^{(\rho,\nu)}_{jmln}$ with \textit{magnetic indices} $(j,m,l,n)$ in the range 
\begin{equation}
j,l\in\{\abs{\nu},\abs{\nu}+1,...\}\,, \quad m\in\{-j,...,j\}\,,\quad n\in\{-l,...,l\}\,.
\end{equation}
They form an orthogonal basis of $L^2\left(\SL\right)$, satisfying 
\begin{equation}\label{eq:orthogonality relation of SL2C wigner matrices}
\int\limits_{\SL}\dd{h}\overline{D^{(\rho_1,\nu_1)}_{j_1 m_1 l_1 n_1}(h)}D^{(\rho_2,\nu_2)}_{j_2 m_2 l_2 n_2}(h)
=
\frac{\delta(\rho_1-\rho_2)\delta_{\nu_1, \nu_2}\delta_{j_1, j_2}\delta_{l_1, l_2}\delta_{m_1, m_2}\delta_{n_1, n_2}}{\rho_1^2+\nu_1^2}\,,
\end{equation}
and the complex conjugation property~\cite{Speziale:2016axj}
\begin{equation}\label{eq:complex conjugate of Wigner matrix}
\overline{D^{(\rho,\nu)}_{jmln}(g)}
=
(-1)^{j-l+m-n}D^{(\rho,\nu)}_{j-ml-n}(g)\,.
\end{equation}
The Cartan decomposition of group elements $g\in\SL$ is given by~\cite{Ruehl1970} 
\begin{equation}\label{eq:Cartan decomp g}
g = u\e^{\frac{\eta}{2}\sigma_3}v\,,\qquad u,v\in\SUT\,,\qquad \eta\in\R_+\,.
\end{equation}
Therein we have $\eta$ as the rapidity parameter of a boost along the $z$ axis while $u$ and $v$ are arbitrary $\SUT$ rotations. This induces a decomposition of the Haar measure on $\SL$,
\begin{equation}\label{eq:Cartan decomp measure}
\dd{g} = \frac{1}{\pi}\dd{u}\dd{v}\dd{\eta}\sinh^2(\eta)\,,
\end{equation}
as well as a decomposition of $\SL$ Wigner matrices
\begin{equation}\label{eq:Cartan decomp D}
D^{(\rho,\nu)}_{jmln}(g) = \sum_{q=-\mathrm{min}(j,l)}^{\mathrm{min}(j,l)}D^{j}_{mq}(u)d^{(\rho,\nu)}_{jlq}(\eta)D^{l}_{qn}(v)\,,
\end{equation}
with $d^{(\rho,\nu)}$ being the reduced $\SL$ Wigner matrix~\cite{Ruehl1970}.\footnote{As explained below in Appendix~\ref{app:Empty integrals}, the skirt radius $a$ of the hyperbolic part of $\SL$ is included by the substitution $\tilde{\eta} = a\eta$.}

Square-integrable functions $f\in L^2(\SL)$ exhibit a Plancherel decomposition, or \textit{spin representation}, which is explicitly given by
\begin{equation}
f(g) = \int\dd{\rho}\sum_\nu (\rho^2+\nu^2)\sum_{j,m,l,n}f^{\rho,\nu}_{jmln}D^{(\rho,\nu)}_{jmln}(g)\,,
\end{equation}
where $(\rho^2+\nu^2)$ is the Plancherel measure on $\SL$. Functions $f\in L^2(\SL)$ which are class functions, i.e. satisfying $f(g) = f(hgh^{-1})$ for all $h\in\SL$, are expanded in terms of traces
\begin{equation}\label{eq:class function}
f(g) = \int\dd{\rho}\sum_\nu(\rho^2+\nu^2)f^{\rho,\nu}\Tr\left(\vb*{D}^{(\rho,\nu)}(g)\right)\,,
\end{equation}
where here the trace in the representation $(\rho,\nu)$ is also referred to as \textit{character}. A particularly important example of a class function is the $\delta$-distribution on $\SL$ which is given for $f^{\rho,\nu} = 1$, thus being written as
\begin{equation}\label{eq:delta on SL2C}
\delta(g) = \int\dd{\rho}\sum_{\nu}(\rho^2+\nu^2)\Tr\left(\vb*{D}^{(\rho,\nu)}(g)\right)\,.
\end{equation}

The Casimir operators of $\SL$ act on states in the canonical basis $\ket{(\rho,\nu);jm}\in\mathcal{D}^{(\rho,\nu)}$ as\footnote{If the skirt radius $a$ is included, the two Casimirs come with a pre-factor of $1/a^2$.}
\begin{align}
\mathrm{Cas}_1\ket{(\rho,\nu);jm} & =  (-\rho^2+\nu^2-1)\ket{(\rho,\nu);jm}\label{eq:cas1}\,,\\[7pt]
\mathrm{Cas}_2\ket{(\rho,\nu);jm} & = \rho\nu\ket{(\rho,\nu);jm}\label{eq:cas2}\,.
\end{align}
%

\paragraph{A note on conventions.} There is a variety of textbooks and articles on the unitary irreducible representation theory of $\SL$ using different conventions. There are three choices of conventions one has to make\footnote{I would like to thank Jos\'{e} Sim\~{a}o for clarifying discussions on this matter and refer to~\cite{Simao:2024chp} for further reading.}:
\begin{enumerate}
    \item[1)] The parametrization of $(\lambda,\mu)$ in terms of $\rho$ and $\nu$: In our case this is given by Eq.~\eqref{eq:(lambda,mu)}. Different conventions are for example used in~\cite{Ruehl1970} with labels $(\rho_R,\nu_R)$ that are related to our choice by $(\rho_R,\nu_R) = (2\rho,2\nu)$.
    \item[2)] Haar measure: Since $\SL$ is non-compact, the Haar measure is determined up to a multiplicative factor. In Eq.~\eqref{eq:Cartan decomp measure}, this factor is given by $1/\pi$, which is the same choice as in~\cite{Dona:2021ldn}. In contrast,~\cite{Ruehl1970} uses a pre-factor of $1/4\pi$.
    \item[3)] Group action: The group action defined in Eq.~\eqref{eq:group action} has been chosen as the left action of the transpose of $g$ on the argument of the function of $F\in\mathcal{D}^{(\rho,\nu)}$, corresponding to the conventions of~\cite{Ruehl1970}. Other choices, such as a right action or the action by the inverse group element are conceivable.
\end{enumerate}
Conventions 1) and 2) determine the orthogonality relation of $\SL$ Wigner matrices and the multiplicative factor of the Plancherel measure. Also, empty $\SL$ integrals and the projection onto the trivial representation discussed below depend on choices 1) and 2). Convention 3) determines the precise form of the Wigner matrices important for the formula of the reduced $\SL$ Wigner matrix coefficients in terms of hypergeometric functions, used in Appendix~\ref{sec:asymptotics of d}.

\subsection{Constant function and the trivial representation}\label{sec:trivial represenation}

Since $\SL$ is non-compact, the constant function is not part of the $L^2$-space. Thus, to be able to project onto constant field configurations, one has to extend the space of functions to that of so-called hyperfunctions, following~\cite{Ruehl1970}. This allows us to define a pseudo-projector\footnote{\enquote{Pseudo} because it is a projection up to a divergent volume factor of $\SL$.} onto the trivial representation, given by the integral expression
\begin{equation}\label{eq:zero mode projection pre integration}
\int\limits_\SL\dd{g}D^{(\rho,\nu)}_{jmln}(g)\,.
\end{equation}
Exploiting the Cartan decomposition in Eqs.~\eqref{eq:Cartan decomp g}, \eqref{eq:Cartan decomp measure} and~\eqref{eq:Cartan decomp D}, we find
\begin{equation}
\int\limits_\SL\dd{g}D^{(\rho,\nu)}_{jmln}(g) = \sum_q\int\limits_{\SUT}\dd{u}D^j_{mq}(u)\int\limits_{\SUT}\dd{v}D^l_{qn}(v^{-1})\frac{1}{\pi}\int\limits_{\R^+}\dd{\eta}\sinh^2(\eta)d^{(\rho,\nu)}_{jlq}(\eta)\,.
\end{equation}
One can perform the $\SUT$ integrals explicitly, yielding Kronecker-$\delta$'s on the magnetic indices $(j,m,l,n)$. The condition that $j\geq\abs{\nu}$ (see above) together with $j=0$ forces the discrete representation label $\nu = 0$.

Following~\cite{Ruehl1970} with the conventions as in Eqs.~\eqref{eq:(lambda,mu)} and~\eqref{eq:Cartan decomp measure}, the reduced Wigner matrix $d^{(\rho,0)}_{000}(\eta)$, also referred to as character, is given by
\begin{equation}\label{eq:d0}
d^{(\rho,0)}_{000}(\eta) = \frac{\sin(\rho\eta)}{\rho\sinh(\eta)}\,.
\end{equation}
Then, we can compute the above integral 
\begin{equation}
\begin{aligned}
 \frac{1}{\pi}\int\limits_{\R^+}\dd{\eta}\sinh^2(\eta)\frac{\sin(\rho\eta)}{\rho\sinh(\eta)} =& \frac{1}{\pi}\frac{1}{4i\rho}\int\limits_{\R}\dd{\eta}\left[\e^{i\eta(\rho-i)}-\e^{i\eta(\rho+i)}\right]\\[7pt]
=&
\frac{1}{2i\rho}\left[\delta(\rho-i)-\delta(\rho+i)\right]\longrightarrow -\delta(\rho-i)\,.
\end{aligned}
\end{equation}
In the last step, we used the unitary equivalence of $(\rho,0)$ and the $(-\rho,0)$ representations, yielding a factor of $2$ and that $\frac{1}{i\rho}\delta(\rho-i)$ acts as a distribution as $-\delta(\rho-i)$, yielding a minus sign.\footnote{We emphasize that the explicit evaluation of Eq.~\eqref{eq:zero mode projection pre integration} depends on the conventions used in Eqs.~\eqref{eq:(lambda,mu)} and~\eqref{eq:Cartan decomp measure}. Both of these conventions differ from Ref.~\cite{Ruehl1970}, where the same computation would yield in total $-\frac{1}{4}\delta(\rho_R-2i)$.} 

In total, the integral of Eq.~\eqref{eq:zero mode projection pre integration} yields
\begin{equation}\label{eq:zero mode projection}
\int\dd{g}D^{(\rho,\nu)}_{jmln}(g) = -\delta(\rho-i)\delta_{\nu,0}\delta_{j,0}\delta_{m,0}\delta_{l,0}\delta_{n,0}\,.
\end{equation}
This makes manifest that the identity in the space of distributions on $L^2(\SL)$ is written as
\begin{equation}\label{eq:trivial rep}
\one = D^{(i,0)}_{0000}(g)\,,
\end{equation}
such that any constant function $f$ can be written as $f\one$. Notice furthermore that Eq.~\eqref{eq:trivial rep} is consistent with the orthogonality relation in Eq.~\eqref{eq:orthogonality relation of SL2C wigner matrices}, i.e.
\begin{equation}
\int\dd{g}D^{(\rho,\nu)}_{jmln}(g)\one = \int\dd{g}D^{(\rho,\nu)}_{jmln}(g)D^{(i,0)}_{0000}(g) = \frac{\delta(\rho-i)\delta_{\nu,0}}{\rho^2+\nu^2}\delta_{j,0}\delta_{m,0}\delta_{l,0}\delta_{n,0}\,.
\end{equation}

\begin{figure}
    \centering
    \includegraphics[width=0.6\linewidth]{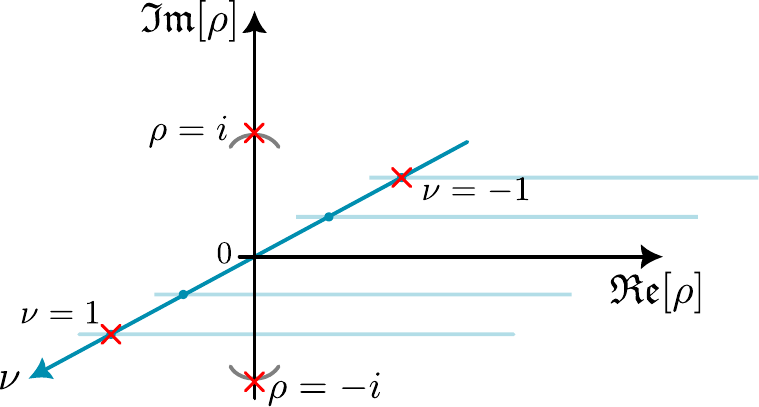}
    \caption{A visualization of the space of all irreducible unitary $\SL$ representations. The principal series is parametrized by $(\rho,\nu)\in\R\times\mathbb{Z}/2$ and the complementary series is given by $\nu = 0$ and $\rho\in (-i,i)$. Values of representations for which both Casimir operators vanish are marked by red crosses. These are also the points which are obtained from the complementary series in a limit $\rho\rightarrow\pm i$~\cite{Naimark1964}. Since the limits converge to two points, respectively, the space of irreducibles is equipped with a non-Hausdorff topology.}
    \label{fig:SL2C-irreps}
\end{figure}

In order to appreciate this result further, one needs to consider the full space of irreducible unitary representations of $\SL$, as presented in~\cite{Naimark1964} and depicted in Fig.~\ref{fig:SL2C-irreps}. Detailed therein, all the irreducible unitary representations are captured by the principal series, the complementary series and the trivial representation. The pseudo-projector defined above maps only to the trivial representation $(\pm i,0)$ and not to $(0,\pm 1)$. Consequently, constant field configurations will only involve $\rho = \pm i$ representations and vanishing magnetic indices $j=0,m=0$.

\subsection{Regularization of Dirac delta function}\label{sec:Regularization of Dirac delta function}

In Chapter~\ref{chapter:LG}, one faces projections onto the trivial representation which lead to products of $\delta(\rho-i)$, normalized by volume factors of $\SL.$ We show in this subsection that these expressions are indeed regularized in the sense that they are objects similar to Kronecker-deltas.

Consider Eq.~\eqref{eq:zero mode projection} evaluated on vanishing magnetic indices,
\begin{equation}
-\delta(\rho-i) = \int\dd{g}D^{(\rho,0)}_{0000}(g)\,,
\end{equation}
and evaluate both sides on $\rho = i$, yielding
\begin{equation}
\eval{-\delta(\rho-i)}_{\rho = i} = \int\dd{g}D^{(i,0)}_{0000}(g) = \int\dd{g} = V_{\texttt{+}}\, ,
\end{equation}
where we used that $D^{(i,0)}_{0000}(g)$ is the identity for all $g\in\SL$ and where $V_{\texttt{+}}$ is the divergent $\SL$ volume factor. Then,
\begin{equation}
\eval{-\frac{\delta(\rho-i)}{V_\texttt{+}}}_{\rho=i} = 1\,,
\end{equation}
and we have in total
\begin{equation}\label{eq:regularized delta}
\delta_{\rho,i}\defeq -\frac{\delta(\rho-i)}{V_\texttt{+}} =
\begin{cases}
1,&\quad\text{for }\rho = i\,,\\[7pt]
0,&\quad\text{for }\rho\neq i\,.
\end{cases}
\end{equation}
Importantly, the symbol $\delta_{\rho,i}$ is obtained independent of conventions, i.e. choosing a different Haar measure or a different convention for Eq.~\eqref{eq:(lambda,mu)} changes the expressions for the volume factors and the pre-factors in front of $\delta(\rho-i)$ but not their ratio entering $\delta_{\rho,i}$.

\subsection{Asymptotics of reduced Wigner matrix}\label{sec:asymptotics of d}

Relevant for the non-local correlation function in Chapter~\ref{chapter:LG}, we derive here the asymptotics of the reduced Wigner matrix $d^{(0,\nu)}_{jlq}(\eta)$ in the limit of large $\eta$. To that end, the reduced Wigner matrix is expressed through the hypergeometric function ${}_2F_1$,
\begin{equation}\label{eq:reduced d with 2F1}
\begin{aligned}
& d^{(0,\nu)}_{jlq}(\eta) = c(j,l,q)\sqrt{\prod_{+,-}(j\pm\nu)!(j\pm q)!(l\pm \nu)!(l\pm q)!} \\[7pt]
&\times
\sum_{s,t}\frac{(-1)^{s+t}(\nu+q+s+t)!(j+l-\nu-q-s-t)!}{s!(j-\nu-s)!(j-q-s)!(\nu+q+s)!t!(l-\nu-t)!(l-q-t)!(\nu+q+t)!}\\[7pt]
&\times
\e^{-\eta(\nu+q+1+2t)} {}_2F_1(l+1, \nu+q+1+s+t, j+l+2,1-\e^{-2\eta})\,.
\end{aligned}
\end{equation}
The integers $s$ and $t$ are constrained such that the factorials in the above formula are defined, yielding the inequalities
\begin{equation}
s,t\geq 0\,,\quad j-s\geq\nu\,,\quad j-s\geq q\,,\quad \nu+q+s\geq 0\,,\quad \nu+q+t\geq 0\,.
\end{equation}

The hypergeometric function is diverging for $\eta\rightarrow\infty$ if $j-\nu-q-s-t \leq 0$, which is generically true in the magnetic index space spanned by $j,\nu,q,s$ and $t$. This can be seen more explicitly from the following expansion of the hypergeometric function in the case where $a,b,c\in\mathbb{Z}$,
\begin{equation}
{}_2F_1(a,b,c;x)
=
\frac{\log(1-x)}{x^{c-1}}\sum_{p}\alpha_p x^p + \frac{1}{x^{c-1}}\sum_{q}\beta_q x^q + (1-x)^{c-a-b}\sum_{n}\gamma_n x^n\,.
\end{equation}
Since we are interested in the behavior near $x=1$, we consider only the logarithmic and the $(1-x)^{c-a-b}$ term as relevant. Importantly, ${}_2F_1$ enters the reduced Wigner matrix in Eq.~\eqref{eq:reduced d with 2F1} not isolated but as a product with $\e^{-\eta(\nu+q+1+2t)}$, which is written as $(1-x)^{\frac{1}{2}(\nu+q+1+2t)}$ for $x = 1-\e^{-2\eta}$. Combining these two factors, we observe that the logarithmic term is suppressed, while only 
\begin{equation}
\sim (1-x)^{\frac{1}{2}(2j+1-\nu-q-2s)}= \e^{-\eta(2j+1-\nu-q-2s)}
\end{equation}
remains. Clearly, as $x\rightarrow 1$, the term for which the exponent is minimal will dominate. As a result, we find numerically as well as analytically that the minima are given by the configuration
\begin{equation}
q = \nu\,,\quad  s = j-\nu\,,\quad t = 0\,. 
\end{equation}
This result is supported by the numerical findings where the dominating terms of the rescaled hypergeometric function have been obtained. Inserting these labels into Eq.~\eqref{eq:reduced d with 2F1}, we find 
\begin{equation}
\begin{aligned}
& \sum_q\sum_{s,t}f(q,s,t)(1-x)^{\frac{1}{2}(2j+1-\nu-q-2s)}{}_2F_1(l+1, \nu+q+1+s+t, j+l+2, x)\\[7pt]
&\overset{x\rightarrow 1}{\longrightarrow}f(\nu,j-\nu,0)(1-x)^{\frac{1}{2}}\sim \e^{-\eta}\,.
\end{aligned}
\end{equation}
Remarkably, the dependence of $d^{(0,\nu)}(\eta)$ on $\eta$ and $\nu$ decouples in the limit $\eta\rightarrow \infty$. This behavior has crucial consequences for the Landau-Ginzburg analysis as it renders timelike faces irrelevant for the critical behavior.

\section{Homogeneous spaces}\label{app:Homogeneous spaces}

The $\SL$ subgroups $\SUT$, $\ISO$ and $\SUO$, denoted as $\mathrm{U}^{(\texttt{+})}, \mathrm{U}^{(0)}$ and $\mathrm{U}^{(\texttt{-})}$, respectively, are of particular interest for us throughout Chapters~\ref{chapter:cBC} -- \ref{chapter:LG}. Explicitly, these are defined as~\cite{Ruehl1970}
\begin{equation}
\begin{gathered}
\SUT \defeq \left\{g\in\SL\;\middle\vert\; gg^{\dagger} = e\right\}\,,\quad 
\SUO \defeq \left\{g\in\SL\;\middle\vert\; g\sigma_3 g^{\dagger} = \sigma_3\right\}\,,\\[7pt]
\ISO \defeq \left\{g\in\SL\;\middle\vert\; g(e+\sigma_3)g^{\dagger} = e+\sigma_3\right\}\,,
\end{gathered}
\end{equation}
respectively stabilizing the normal vectors
\begin{equation}\label{eq:reference normal vectors}
X_\texttt{+} = (\pm 1,0,0,0)\,,\quad
X_0 = \frac{1}{\sqrt{2}}(\pm 1,0,0,1)\,,\quad
X_\texttt{-} = (0,0,0,1)\,.
\end{equation}

Forming the quotient space $\SL/\mathrm{U}^{(\alpha)}$ with respect to these groups yields homogeneous spaces $\mathrm{H}_\alpha$ which can be understood as embedded manifolds in $\R^{1,3}$
\begin{subequations}\label{eq:hom spaces in R13}
\begin{align}
\mathrm{H}_{\texttt{+}} &\defeq \left\{(t,x,y,z)\in\R^{1,3}\;\middle\vert\; t^2-x^2-y^2-z^2 = 1,\; t\gtrless 0\right\}\,,\label{eq:defH3}\\[7pt]
\mathrm{H}_0 &\defeq \left\{(t,x,y,z)\in\R^{1,3}\;\middle\vert\; t^2-x^2-y^2-z^2 = 0,\; t\gtrless 0\right\}\,,\label{eq:defC}\\[7pt]
\mathrm{H}_{\texttt{-}} &\defeq \left\{(t,x,y,z)\in\R^{1,3}\;\middle\vert\; t^2-x^2-y^2-z^2 = -1\right\}\,,\label{eq:defH21}
\end{align}
\end{subequations}
with line elements
\begin{equation}\label{eq:line elements}
\dd{\mathrm{H}_{\texttt{+}}^2} = \dd{\eta}^2+\sinh^2(\eta)\dd{\Omega^2}\,,\qquad
\dd{\mathrm{H}_0^2} = r^2\dd{\Omega^2}\,,\qquad
\dd{\mathrm{H}_{\texttt{-}}^2} = \dd{\eta}^2-\cosh^2(\eta)\dd{\Omega^2}\,,
\end{equation}
where $\eta\in\R, r\in\R_+$ and $\dd{\Omega}$ the normalized measure on the $2$-sphere $S^2$. Integration measures on $\mathrm{H}_{\texttt{+}}$ and $\mathrm{H}_{\texttt{-}}$ are given by
\begin{equation}
\dd{X_{\texttt{+}}} = \frac{1}{2\pi}\dd{\Omega}\dd{\eta}\sinh^2(\eta)\,,\,,\qquad
\dd{X_{\texttt{-}}} = \frac{1}{2\pi}\dd{\Omega}\dd{\eta}\cosh^2(\eta)\,.
\end{equation}
Since $\mathrm{H}_{\texttt{+}}$ and $\mathrm{H}_{\texttt{-}}$ are obtained by forming the quotient space of $\SL$ with respect to the groups $\mathrm{U}^{(\alpha)}$, the choice of Haar measure on $\SL$, given in Eq.~\eqref{eq:Cartan decomp measure}, induces a measure on $\mathrm{H}_{\texttt{+}}$ and $\mathrm{H}_{\texttt{-}}$. To account for that we added a factor of $1/(2\pi)$.

The light cone, equipped with topology $\R\times S^2$, exhibits a degenerate line element, characteristic for a null surface. Correspondingly, the integration measure obtained by naively taking the determinant of the induced metric is independent of the degenerate non-compact direction. However, by parametrizing null vectors as $X_0 = (\lambda,\lambda\hat{\vb*{r}})$, with $\lambda\in\R_+,\hat{\vb*{r}}\in S^2$, the measure on the light cone is in fact given by
\begin{equation}\label{eq:measure on H0}
\dd{X_0} = \frac{1}{2\pi}\dd{\Omega}\dd{\lambda}\lambda\,.
\end{equation}
For further details on null hypersurfaces and their geometric treatment, see~\cite{Ciambelli:2019lap}.

\subsection{Action of SL(2,C) on homogeneous spaces}

Since the homogeneous spaces $\SL/\mathrm{U}^{(\alpha)}$ arise as quotient spaces, the respective elements are given in terms of equivalence classes. So for $a\in\SL$, $[a]_{\alpha}\in\SL/\mathrm{U}^{(\alpha)}$ denotes an equivalence class which satisfies $
[a]_{\alpha} = [au]_{\alpha}$ for all  $u\in \mathrm{U}^{(\alpha)}$. On $\SL/\mathrm{U}^{(\alpha)}$, $\SL$ acts in a canonical way, defined by
\begin{equation}\label{eq:SL2C-action on equivalence classes}
g\cdot [a]_{\alpha} \defeq [ga]_{\alpha}\,,
\end{equation}
from which it follows that the stabilizer subgroup $\mathrm{U}_{[a]_{\alpha}}$ of $[a]_{\alpha}$ is given by
\begin{equation}\label{eq:definition of stabilizer subgroup}
\mathrm{U}_{[a]_{\alpha}}
\defeq
\left\{aua^{-1}\;\middle\vert\; u\in \mathrm{U}^{(\alpha)}\right\} = a\mathrm{U}^{(\alpha)}a^{-1}\,.
\end{equation}
Since conjugation is a group isomorphism the stabilizer subgroup $U_{[a]_{\alpha}}$ is isomorphic to $\mathrm{U}^{(\alpha)}$.

A different way of defining the $\SL$ action on homogeneous spaces is by exploiting the isomorphism of Minkowski space $\R^{1,3}$ and the space of $2\times 2$ Hermitian matrices
\begin{align}
\Phi: \R^{1,3} &\overset{\cong}{\longrightarrow} \mathrm{H}_2(\C)\\[7pt]
X &\longmapsto \Phi(X)\,.
\end{align}
If the Minkowski inner product of $X$ is either $+1,0$ or $-1$, then $X$ can be represented as an equivalence class in the respective $\SL$ quotient space. In this representation, the equivalence of $\SL$ quotient spaces and submanifolds of Minkowski space is made transparent with the cost of the action, defined by
\begin{equation}\label{eq:SL2C action on normal using H2}
g\cdot X \defeq  \Phi^{-1}(g\Phi(X)g^{\dagger})\,,
\end{equation}
being less straightforward compared to Eq.~\eqref{eq:SL2C-action on equivalence classes}.

\subsection{Empty integrals}\label{app:Empty integrals}

In Chapter~\ref{chapter:LG}, the projection onto constant field configurations frequently yields empty integrals over $\SL$ and the homogeneous spaces $\mathrm{H}_{\alpha}$, yielding diverging volume factors $V_{\alpha}$. Consequently, a regularization of these terms is required. When removing the regulator from the final result, it is therefore important to understand to which degree these volume factors diverge.

From the Cartan decomposition of the Haar measure on $\SL$ in Eq.~\eqref{eq:Cartan decomp measure} one concludes that the volume factor of $\SL$ and $\mathrm{H}_\texttt{+}$ diverge in the same way. This can also be seen by observing that $\SL$ is of topology $\mathrm{H}_{\texttt{+}}\times S^3$, where the measure on $S^3$ is normalized, such that the non-compact direction is in fact the hyperboloid $\mathrm{H}_{\texttt{+}}$. Introducing a cutoff $L$, the regularized volume factor $V_{\texttt{+}}$ is therefore defined as
\begin{equation}
V_{\texttt{+}} = \frac{1}{\pi}\int\limits_{0}^{L}\dd{\eta}\sinh^2(\eta)\:\underset{L\gg 1}{\longrightarrow} \:\frac{1}{4\pi}\e^{2L}\,.
\end{equation}
Similarly, we find for the empty integral on the one-sheeted hyperboloid $\mathrm{H}_{\texttt{-}}$ the asymptotic behavior $V_{\texttt{-}}\:\longrightarrow \:\frac{1}{4\pi}\e^{2L}$. 

On the light cone with measure given in Eq.~\eqref{eq:measure on H0}, we perform a coordinate transformation $\lambda\rightarrow\e^\eta$ and introduce an upper cutoff $L$, such that $V_0$ is given by
\begin{equation}
V_0 = \frac{1}{2\pi}\int\limits_{\R_+}\dd{\lambda}\int\dd{\Omega}\lambda = \frac{1}{2\pi}\int\limits_{0}^L\dd{\eta}\int\dd{\Omega}\e^{2\eta} \:\underset{L\gg 1}{\longrightarrow} \:\frac{1}{4\pi}\e^{2L}\,.  
\end{equation}

From the scaling behavior of the volume factors in terms of the upper cutoff $L$, we find that the $V_\alpha$ all diverge to the same degree. In particular, expressions like $V_\alpha/V_\beta = 1$ after regularizing.  

Notice that here, we parametrized the non-compact direction of the $\mathrm{H}_\alpha$ with the dimensionless variable $\eta$. When extracting the scaling behavior of the Ginzburg-$Q_{\alpha\beta}$ in the main body of this work, one would like to work with a dimensionful variable for the non-compact direction. This can be straightforwardly implemented by the substitution $\tilde{\eta} = a\eta$. Notice that assume the same scale $a$ for all three cases. For the hyperboloids $\mathrm{H}_\texttt{+}$ and $\mathrm{H}_\texttt{-}$, $a$ has the interpretation of a skirt radius. On the light cone $\mathrm{H}_0$, $a$ acts as a multiplicative factor of vectors along the degenerate direction without affecting its geometry. Then, the scaling of the volume factors for large cutoffs is given by
\begin{equation}
    V_\alpha \sim\frac{a^3}{4\pi}\e^{2L/a}\,,
\end{equation}
frequently employed in Chapter~\ref{chapter:LG}.

\subsection{Projection onto U\textsuperscript{($\pmb{\alpha}$)}-invariant subspaces}\label{app:Projection onto invariant subspaces}

The simplicity constraints of the complete BC model in Chapter~\ref{chapter:cBC} imply that the group field $\varphi_\alpha$ is effectively defined on the homogeneous space $\mathrm{H}_\alpha$. Imposing these constraints is achieved via a (pseudo-)projection\footnote{Again, \enquote{pseudo} refers to the fact that for the non-compact groups $\ISO$ and $\SUO$, applying the projector twice yields a divergent volume factor.} onto $\mathrm{U}^{(\alpha)}$-invariant subspaces. In this appendix we derive the spin representation of $\varphi_\alpha$ after such a projection and the imposition of closure. To that end,  the derivation in the appendix of~\cite{Jercher:2021bie} is generalized to the case of the normal vector being either timelike, lightlike or spacelike.

Define a (pseudo-)projector $P_{\alpha}^{(\rho,\nu)}$ from the $\SL$ representation space $\mathcal{D}^{(\rho,\nu)}$ onto the $\mathrm{U}^{(\alpha)}$-invariant subspace as
\begin{equation}
P_{\alpha}^{(\rho,\nu)} \defeq \int\limits_{\mathrm{U}^{(\alpha)}}\dd{u}\vb*{D}^{(\rho,\nu)}(u)\,.
\end{equation}
Let $\vert\mathcal{I}^{(\rho,\nu),\alpha}\rangle$ be an $\mathrm{U}^{(\alpha)}$-invariant vector in $\mathcal{D}^{(\rho,\nu)}$,
\begin{equation}
\vb*{D}^{(\rho,\nu)}(u)\vert\mathcal{I}^{(\rho,\nu),\alpha}\rangle = \vert\mathcal{I}^{(\rho,\nu),\alpha}\rangle,\quad \forall u\in \mathrm{U}^{(\alpha)}\,. 
\end{equation}
Then, the projector is conveniently rewritten as
\begin{equation}
P^{(\rho,\nu)}_{\alpha} \eqdef \vert\mathcal{I}^{(\rho,\nu),\alpha}\rangle\langle\mathcal{I}^{(\rho,\nu),\alpha}\vert\,,
\end{equation}
with matrix coefficients in the canonical basis given by\footnote{The symbols $\mathcal{I}^{(\rho,\nu),\alpha}_{jm}$ are equal to the $\mathcal{W}$-symbols of Ref.~\cite{Perez:2000ep} for the case of $\alpha = +,-$ and they represent an extension of the work done in~\cite{Perez:2000ep} since also lightlike normal vectors are taken into account for $\alpha = 0$.}\textsuperscript{,}\footnote{In this article, we work in the canonical basis of $\SL$ representations which are denoted in bra-ket notation as $\ket{(\rho,\nu);jm}$. The Wigner matrices are best under control in this case, e.g. with respect to the orthogonality relation \eqref{eq:orthogonality relation of SL2C wigner matrices} or the behavior under complex conjugation in Eq.~\eqref{eq:complex conjugate of Wigner matrix}. Furthermore, the interaction term of mixed type will contain the convolution of Wigner matrices arising from different normal vectors. In different bases, for instance in the pseudo-basis for functions with $\alpha = -$, the coefficients for basis change would be required explicitly, and they are not available. The price we pay for using the canonical basis is that the evaluation of $\SUO$ and $\ISO$ elements on the Wigner matrices does not yield an immediate simplification as it is for the $\SUT$ case, where $D^{(\rho,\nu)}_{jmln}(u) = \delta^{jl}D^j_{mn}(u)$ holds.}
\begin{equation}
\bra{(\rho,\nu);jm}P^{(\rho,\nu)}_{\alpha}\ket{(\rho,\nu);ln} 
=
P_{jmln}^{(\rho,\nu),\alpha}
=
\bar{\mathcal{I}}^{(\rho,\nu),\alpha}_{jm}\mathcal{I}^{(\rho,\nu),\alpha}_{ln}\,.
\end{equation}
The results of the subsequent appendix show that the invariant coefficients project onto simple representations. That is,
\begin{align}
    P_{jmln}^{(\rho,\nu),+} &= \delta_{\nu,0}P^{(\rho,0),\texttt{+}}_{jmln}\,,\label{eq:I+}\\[7pt]
    P^{(\rho,\nu),0}_{jmln} &= \delta_{\nu,0}P^{(\rho,0),0}_{jmln}\,,\label{eq:I0}\\[7pt]
    P^{(\rho,\nu),-}_{jmln} &= \delta_{\nu,0}P^{(\rho,0),-}_{jmln}+\delta(\rho)\chi_{\nu\in 2\mathbb{N}^+}P^{(0,\nu),-}_{jmln}\,.\label{eq:I-}
\end{align}
%
%
To summarize the spin representation of all three group fields $\varphi_\alpha$ neatly in one formula, it is advantageous to single out the part of the projector which enforces either $\nu=0$ or $\rho = 0$, 
\begin{equation}
\varpi_\alpha^{(\rho,\nu)} \defeq
\begin{cases}
\delta_{\nu,0}\,,\qquad&\text{for }\alpha=\texttt{+},0\,,\\
\delta_{\nu,0}+\delta(\rho)\chi_\nu\,,\qquad&\text{for }\alpha=\texttt{-}\,.
\end{cases}
\end{equation}

A basis for functions on $\SL^4\times\mathrm{H}_\alpha$ which satisfy closure and simplicity is then  given by
\begin{equation}\label{eq:basis functions}
\Psi^{\vbr\vbn,\alpha}_{\vb*{j}\vb*{m}}(\vbg,X_{\alpha})
=
\prod_{c=1}^4 \varpi_\alpha^{(\rho_c,\nu_c)}\sum_{l_c n_c} D^{(\rho_c,\nu_c)}_{j_c m_c l_c n_c}(g_c g_{X_{\alpha}})\bar{\mathcal{I}}^{(\rho_c,\nu_c),\alpha}_{l_c n_c}\,,
\end{equation}
where $g_{X_{\alpha}}\in\SL$ is a representative of the equivalence class of $X_{\alpha} = [g_X]_{\alpha}\in\SL/\mathrm{U}^{(\alpha)}$. These basis functions satisfy closure, simplicity and invariance under change of representative $g_{X_{\alpha}}\rightarrow g_{X_{\alpha}}u$. Consequently, a function $\varphi_\alpha\in L^2\left(\SL^4\times\mathrm{H}_\alpha \middle/ \sim\right)$, where $\sim$ encodes the quotient structure due to geometricity, is expanded in terms of $\SL$ representation labels as
\begin{equation}
\varphi(\vbg,X_{\alpha})
=
\left[\prod_{c=1}^4 \sum_{\nu_c}\int\dd{\rho_c}(\rho_c^2+\nu_c^2)\varpi_\alpha^{(\rho_c,\nu_c)}\sum_{j_c m_c l_c n_c}\right]\varphi^{\vbr\vbn,\alpha}_{\vb*{j} \vb*{m}}\prod_c D^{(\rho_c,\nu_c)}_{j_c m_c l_c n_c}(g_c g_{X_{\alpha}})\bar{\mathcal{I}}^{(\rho_c,\nu_c),\alpha}_{l_c n_c},
\end{equation}
where the factor $\left(\rho_c^2+\nu_c^2\right)$ stems from the Plancherel measure of functions on $\SL$. If in addition the normal vector is integrated over, the expansion is given by 
\begin{equation}\label{eq:spin representation of group field after normal integration}
 \int\dd{X_{\alpha}}\varphi(\vbg,X_{\alpha}) 
= 
\left[\prod_{c=1}^4 \sum_{\nu_c}\int\dd{\rho_c}(\rho_c^2+\nu_c^2)\varpi_\alpha^{(\rho_c,\nu_c)}\sum_{j_c m_c l_c n_c}\right]\varphi^{\vbr\vbn,\alpha}_{\vb*{j} \vb*{m}}B^{\vbr\vbn,\alpha}_{\vb*{l} \vb*{n}}\prod_c D^{(\rho_c,\nu_c)}_{j_c m_c l_c n_c}(g_c)\,,
\end{equation}
with generalized BC intertwiners~\cite{Jercher:2022mky}
\begin{equation}\label{eq:definition of generalized BC intertwiners}
B^{\vbr\vbn,\alpha}_{\vb*{j} \vb*{m}} \defeq\int\limits_{\mathrm{H}_\alpha}\dd{X}_{\alpha}\prod_{c=1}^4\sum_{l_c n_c}D^{(\rho_c,\nu_c)}_{j_c m_c l_c n_c}(g_{X_\alpha})\bar{\mathcal{I}}^{(\rho_c,\nu_c),\alpha}_{l_c n_c}\,.
\end{equation}

\subsection{Harmonic analysis on homogeneous spaces}\label{app:Harmonic analysis on homogeneous spaces}

Here, the mathematical basis is given for the explicit computations of the kernels $K_{\alpha\beta}$ in Sec.~\ref{sec:Explicit Expressions for the Vertex Amplitudes}, which define the vertex amplitude of the complete BC model. To begin with it is instructive to consider the representation expansion of the $\delta$-function on $\SL$,
\begin{equation}\label{eq:spin rep of delta on SL2C}
\delta(g_1^{-1}g_2)
=
\sum_{\nu}\int\dd{\rho}(\rho^2+\nu^2)\sum_{jm}\overline{D^{(\rho,\nu)}_{jmjm}(g_1^{-1}g_2)}\,.
\end{equation}
Imposing simplicity with respect to the group $\mathrm{U}^{(\alpha)}$ effectively yields the $\delta$-function on the homogeneous space $\SL/\mathrm{U}^{(\alpha)}\ni X,Y$,
\begin{equation}\label{eq:spin rep of delta on general homogeneous space}
\delta(X,Y)
=
\sum_{\nu}\int\dd{\rho}(\rho^2+\nu^2)\varpi_\alpha^{(\rho,\nu)}\overline{K_{\alpha\alpha}^{(\rho,\nu)}(X,Y)}\,,
\end{equation}
where $K_{\alpha\alpha}$ is defined in Eq.~\eqref{eq:definition of K} and $\varpi$ singles out the simple representations. With the expansions of the $\delta$-functions on $\SL/\mathrm{U}^{(\alpha)}$ derived in~\cite{VilenkinBook} and presented in the following, one can relate the kernels $K_{\alpha\alpha}$ with integral geometric expressions that can be explicitly computed.

\paragraph{Harmonic analysis on $\pmb{\mathrm{H}_\texttt{+}}$.}

Following~\cite{VilenkinBook}, the Gel'fand transform of functions $f\in L^2\left(\mathrm{H}_{\texttt{+}}\right)$ is given by
\begin{equation}\label{eq:Gel'fand transform on 3-hyperboloid}
F(\xi;\rho) = \int\limits_{\mathrm{H}_{\texttt{+}}}\dd{X}f(X)\left( X^{\mu}\xi_{\mu}\right)^{i\rho-1}\,,
\end{equation}
with inverse defined as
\begin{equation}\label{eq:inverse Gel'fand transform on 3-hyperboloid}
f(X) = \int\limits_{0}^{\infty}\dd{\rho}\rho^2\int\limits_{S^2}\dd{\Omega}F(\xi;\rho)\left(\pm X^{\mu}\xi_{\mu}\right)^{-i\rho-1}\,,
\end{equation}
where we absorbed the prefactor of $(4\pi)^3$ appearing in~\cite{VilenkinBook} into the measure. Here, the null vector $\xi\in\mathrm{H}_{0}$ is parametrized as $\xi = (1,\hat{\vb*{\xi}}(\phi,\theta))$, where $\hat{\vb*{\xi}}(\phi,\theta)\in S^2$, $\phi$ and $\theta$ are the angles on the sphere. Inserting Eq.~\eqref{eq:Gel'fand transform on 3-hyperboloid} into Eq.~\eqref{eq:inverse Gel'fand transform on 3-hyperboloid} and imposing that this reproduces the original function $f(X)$, we conclude that the $\delta$-function on $\mathrm{H}_{\texttt{+}}$ is written as
\begin{equation}\label{eq:delta on H3}
\delta(X,Y) = \int\dd{\rho}\rho^2\int\dd{\Omega}\left(Y^{\mu}\xi_{\mu}\right)^{i\rho-1}\left(X^{\mu}\xi_{\mu}\right)^{-i\rho-1}.
\end{equation}
Notice that the same definition holds on the lower sheet of the two-sheeted hyperboloid. Given Eq.~\eqref{eq:spin rep of delta on general homogeneous space} and comparing with Eq.~\eqref{eq:delta on H3}, we obtain the expression for $K_{\texttt{++}}$, 
\begin{equation}\label{eq:D_++ before evaluation}
K_{++}^{(\rho,0)}(X,Y)
=
\frac{1}{2}\int\dd{\Omega}\left(X^\mu\xi_{\mu}\right)^{i\rho-1}\left(Y^{\mu}\xi_{\mu}\right)^{-i\rho-1}\,.
\end{equation}
The factor of $\frac{1}{2}$ appears as a consequence of change of integration range $\rho\in(-\infty,\infty)\rightarrow \rho\in[0,\infty)$, using the unitary equivalence of representations. 

\paragraph{Harmonic analysis on $\pmb{\mathrm{H}_{\texttt{-}}}$}

A decomposition of functions on imaginary Lobachevskian space $\mathrm{H}_{\texttt{-}}/\mathbb{Z}_2$ has been derived in~\cite{VilenkinBook}. Importantly, this space differs from the one-sheeted hyperboloid $\mathrm{H}_{\texttt{-}}$ by the fact that opposite points are identified, $X = -X$. The expansion of functions on $\mathrm{H}_{\texttt{-}}/\mathbb{Z}_2$ contains components with both, discrete and continuous labels $\rho$ and $\nu$. Explicitly, for $f\in L^2(\mathrm{H}_{\texttt{-}})$, it is given by~\cite{VilenkinBook,Perez:2000ep}
\begin{equation}\label{eq:inverse Gel'fand transform on H21}
f(X) = \int\limits_0^{\infty}\dd{\rho}\rho^2\int\dd{\Omega}F(\xi;\rho)\abs{X^{\mu}\xi_{\mu}}^{-i\rho-1}
+
128\pi\sum_{k = 1}^{\infty} k^2\int\dd{\Omega}F(\xi,X;2k)\delta(X^{\mu}\xi_{\mu})\,,
\end{equation}
with inverses~\cite{VilenkinBook}
\begin{align}
F(\xi;\rho) &= \int\limits_{\mathrm{H}_{\texttt{-}}}\dd{X}f(X)\abs{X^{\mu}\xi_{\mu}}^{i\rho-1}\,,\label{eq:Gel'fand transform on H21 with rho}\\[7pt]
F(\xi,X;2k) &= \frac{1}{k}\int\limits_{\mathrm{H}_{\texttt{-}}}\dd{Y}f(Y)e^{-i2k\Theta(X,Y)}\delta(Y^{\mu}\xi_{\mu})\,,\label{eq:Gel'fand transform on H21 with nu}
\end{align}
where $k\in\mathbb{N}^+$ and $\cos(\Theta) \defeq \abs{X\cdot Y}$. Importantly, note that the discrete representation parameter $\nu\in\mathbb{Z}/2$ is restricted to positive and even integers, $\nu\in 2\mathbb{N}^+$. Similar to the previous section, we insert Eqs.~\eqref{eq:Gel'fand transform on H21 with rho} and~\eqref{eq:Gel'fand transform on H21 with nu} into Eq.~\eqref{eq:inverse Gel'fand transform on H21} and impose that this gives back the original function, yielding the form of the $\delta$-function on $\mathrm{H}_{\texttt{-}}$
\begin{equation}\label{eq:delta on H21}
\delta(X,Y)
=
\int\limits_{0}^{\infty}\dd{\rho}\rho^2\int\dd{\Omega}\abs{Y^{\mu}\xi_{\mu}}^{i\rho-1}\abs{X^{\mu}\xi_{\mu}}^{-i\rho-1}
+
128\pi\sum_{k=1}^{\infty}k^2\int\dd{\Omega}\frac{1}{k}e^{-i2k\Theta(X,Y)}\delta(Y^{\mu}\xi_{\mu})\delta(X^{\mu}\xi_{\mu})\,.
\end{equation}
Notice the symmetry of the $\delta$-distribution under $(X,Y)\rightarrow (-X,-Y)$ which assures that it is effectively defined on $\mathrm{H}_{\texttt{-}}/\mathbb{Z}_2$. Comparing Eqs.~\eqref{eq:delta on H21} and~\eqref{eq:spin rep of delta on general homogeneous space}, we identify the two components of the kernel $K_{\texttt{-}\texttt{-}}$, 
\begin{align}
K_{\texttt{-}\texttt{-}}^{(\rho,0)}(X,Y)
& =
\frac{1}{2}\int\dd{\Omega}\abs{X^{\mu}\xi_{\mu}}^{i\rho-1}\abs{Y^{\mu}\xi_{\mu}}^{-i\rho-1}\,,\label{eq:D-- rho}\\[7pt]
K_{\texttt{-}\texttt{-}}^{(0,\nu)}(X,Y)
& =
\frac{128\pi}{\abs{\nu}}\int\dd{\Omega}e^{i2\nu\Theta(X,Y)}\delta(Y^{\mu}\xi_{\mu})\delta(X^{\mu}\xi_{\mu})\,,\label{eq:D-- nu}
\end{align}
where $\cos(\Theta) \defeq \abs{X\cdot Y}$. Notice that the formula for $K_{--}^{\nu}$ contains the absolute value of $\nu$, guaranteeing unitary equivalence of the kernel. 

\paragraph{Harmonic analysis on the light cone.}

Following~\cite{VilenkinBook}, the Gel'fand transform of $f\in L^2(\mathrm{H}_0)$ and its inverse are given by~\cite{VilenkinBook}
\begin{align}\label{eq:Gel'fand transform on cone}
F(X;\rho) &= \int\limits_{0}^{\infty}\dd{t}f(tX)t^{-i\rho}\,,\\[7pt]
f(X) &= \int\limits_{\R}\dd{\rho}\rho^2F(X;\rho)\label{eq:inverse Gel'fand transform on cone}\,,
\end{align}
where, in comparison to~\cite{VilenkinBook}, we absorbed a factor of $\rho^2/4\pi$ into the measure for convenience. Physically, the fact that $\nu = 0$ in the expansion reflects that tetrahedra with a lightlike normal vector cannot have timelike faces. Remarkably, the condition of $\nu = 0$ further implies that the faces need to be spacelike, since the first Casimir operator in Eq.~\eqref{eq:cas1} is strictly negative.

Parametrizing lightlike vectors $X\in\mathrm{H}_0$ as $X = \lambda\xi$, where $\xi$ is as defined above, the measure on $\mathrm{H}_0$ was shown above to take the form
\begin{equation}\label{eq:induced measure on C+}
\dd{X} = \lambda\dd{\lambda}\dd{\Omega}\,.
\end{equation}
Inserting Eq.~\eqref{eq:Gel'fand transform on cone} into Eq.~\eqref{eq:inverse Gel'fand transform on cone} yields an expression for the $\delta$-function on $\mathrm{H}_0$
\begin{equation}\label{eq:delta function on cone}
\delta(\lambda\xi,\lambda'\xi') 
=
\frac{\delta(\theta-\theta')\delta(\phi-\phi')}{\sin(\theta)}\int\dd{\rho}\lambda^{i\rho-1}(\lambda')^{-i\rho-1}\,,
\end{equation}
where $(\theta,\phi)$ and $(\theta',\phi')$ are the angles parametrizing $\xi$ and $\xi'$, respectively. Then, comparing this equation with Eq.~\eqref{eq:spin rep of delta on general homogeneous space}, we identify the kernel
\begin{equation}\label{eq:D00}
K_{00}^{(\rho,0)}(\lambda'\xi',\lambda\xi)
=
\frac{\delta(\theta'-\theta)\delta(\phi'-\phi)}{\sin(\theta)}(\lambda')^{-i\rho-1}\lambda^{i\rho-1}\,.
\end{equation}
The term $\frac{\delta(\theta-\theta')\delta(\phi-\phi')}{\sin(\theta)}$ is interpreted as a $\delta$-function on the two-sphere, which acts regularly upon integration.

\chapter[\textsc{Euclidean Frusta, the Laplacian and the Spectral Dimension}]{Euclidean Frusta, the Laplacian and the Spectral Dimension}

\section{Regge action, Hessian determinant and geometry of Euclidean frusta}\label{app:Euclidean frusta}

The semi-classical approximation of the spin-foam frusta vertex amplitude in the Euclidean EPRL-FK model, employed in Chapter~\ref{chapter:specdim}, is defined via the Regge action of a 4-frustum, a cosmological constant term and the Hessian determinant. Moreover, the Laplacian on frusta configurations is defined in terms of the 3- and 4-volume as well as the dual edge lengths. In this appendix, we provide the necessary ingredients to define $\mathcal{A}_v$ and the Laplace operator $\Delta$.

For the reader's convenience, the expression of the semi-classical vertex amplitude is repeated,
\begin{equation}
\mathcal{A}_v 
    = 
    \frac{1}{\pi^7(1-\bi^2)^{21/2}}\left(\frac{\e^{ \frac{i}{\GN}\Sreg }}{-\det H}+\frac{\e^{-\frac{i}{\GN}\Sreg }}{-\det H^{*}}+2\frac{\cos(\frac{ \bi} {\GN} \Sreg-\frac{\Lambda}{\GN}V^{(4)})}{\sqrt{\det H \det H^{*}}}\right)\,.
\end{equation}
Here, $\det H$ is the Hessian determinant given by~\cite{Bahr:2017eyi}
\begin{equation}\label{eq:determH}
\begin{aligned}
\det H &= 16 j_n^3 j_{n+1}^3 k_n^{15} K\left(K-i K^2+iQ\right)^3 \\[7pt]
& \times \left(1+K^2-2 Q\right)^3(K+i)^6 (K-3 i)^2 \left(1+3 K^2-2Q-2 i K (Q-1)\right)^3\,,
\end{aligned}
\end{equation}
with
\begin{equation}\label{eq:Q theta K}
Q \defeq 2+\frac{j_n+j_{n+1}}{2k_n}\,,\qquad \theta \defeq \arccos \frac{1}{\tan\phi}\,,\qquad K \defeq \sqrt{-\cos 2\phi}\,,
\end{equation}
and $\phi = \frac{j_n-j_{n+1}}{4k_n} $ the slope angle.

The Eucliudean Regge action~\cite{Regge:1961ct} of a single 4-frustum is given by 
\begin{equation}\label{eq:Regge action}
    \Sreg = 6(j_n - j_{n+1})\left(\frac{\pi}{2}-\theta\right) +12k_n\left(\frac{\pi}{2}-\arccos(\cos^2(\theta))\right)\,.
\end{equation}
Identifying the spins with areas, the action $\Sreg$ is strictly speaking the area Regge action~\cite{Asante:2018wqy,Barrett:1997tx}. However, in the highly symmetric restriction to 4-frusta, the transition between area and length variables is one-to-one~\cite{Dittrich:2023rcr}. For the dihedral angles to be well-defined, the inequality
\begin{equation}\label{eq:spin ineq}
    -\frac{1}{\sqrt{2}} \leq \frac{j_n - j_{n+1}}{4 k_n} \leq \frac{1}{\sqrt{2}}
\end{equation}
is required to hold, posing a stronger condition than the definition of $\phi$.

Volume, $V_n^{(4)}$, and height, $H_n$ (carrying an index \enquote{$n$}, not to be confused with the Hessian determinant above), of a $4$-frustum in the $n$th slice are respectively given by
\begin{equation}\label{eq:4Volume}
V^{(4)}_{n} = \frac{1}{2} k_n (j_n + j_{n+1}) \sqrt{1 - \frac{(j_n - j_{n+1})^2}{8 k_n^2}}\,,
\end{equation}
and
\begin{equation}\label{eq:4Height}
H_{n} =\frac{2 k_n}{\sqrt{j_n} + \sqrt{j_{n+1}}} \sqrt{1 - \frac{(j_n - j_{n+1})^2}{8 k_n^2}}\,.
\end{equation}
For the 4-frustum to be embeddable in 4-dimensional Euclidean space, the height $H_n$ must be real. Readily, this condition is equivalent to Eq.~\eqref{eq:spin ineq}, following from the condition of dihedral angles being well-defined.

The action of the Laplace operator on a test scalar field is given by
\begin{equation}\label{eq:defLaplace app}
    -(\Delta \phi)_{\vn} = -\sum_{\vm\sim\vn}\Delta_{\vn\vm}\left(\phi_{\vn} - \phi_{\vm}\right) = \frac{1}{V^{(4)}_{\vn}}\sum_{\vm\sim\vn}\frac{V^{(3)}_{\vn\vm}}{l^*_{\vn\vm}}\left(\phi_{\vn}-\phi_{\vm}\right)\,,
\end{equation}
where $\vn,\vm$ denote vertices of the dual 2-complex $\Gamma$. This 2-complex is constructed as follows: $4$-frusta are aligned along the $t$-axis with dual vertices being obtained as the average of the corner points of the $4$-frustum. Consequently, these points lie on the $t$-axis on half of the $4$-frustum height. Dual edges, ``spacelike'' and ``timelike'', are chosen orthogonal to the 3-cells such that their lengths are minimal. Based on this construction, the length of the two types of dual edges is summarized in the following.

\paragraph{``Timelike'' dual edges.}

$3$-dimensional volumes between the vertices $\vm$ and $\vn$, which connect the $(n-1)$th and $n$th slice, are simply given by the volume of cubes with area $j_n$
\begin{equation}\label{eq:timelike 3-volume}
V_{\vm \vn}^{(3)} = j_n^{\frac{3}{2}}\,.
\end{equation}
Following the construction of a dual lattice outlined above, the length of dual edges is given as the half of the sum of heights of ``past'', $H_{n-1}$, and ``future'', $H_n$, hyperfrusta
\begin{equation}\label{eq:timelike dual edge length}
\begin{aligned}
l_\star^{\vm \vn} &= \frac{1}{2} (H_{n-1} + H_{n})\\[7pt] 
&=
\frac{ k_{n-1}}{\sqrt{j_{n-1}} + \sqrt{j_{n}}} \sqrt{1 - \frac{(j_{n-1} - j_{n})^2}{8 k_{n-1}^2}} + \frac{ k_n}{\sqrt{j_n} + \sqrt{j_{n+1}}} \sqrt{1 - \frac{(j_n - j_{n+1})^2}{8 k_n^2}}\,.
\end{aligned}
\end{equation}
%


This defines all the ingredients of the Laplace operator $\Delta_{\vm\vn}$ for neighbouring vertices $\vm,\vn$ which have a ``timelike'' separation, and we use the notation $\Delta_{n-1 n}$  below.

\paragraph{``Spacelike'' dual edges.}

In ``spacelike'' direction, so within a thick slice, $4$-frusta are connected with each other via boundary $3$-frusta. The corresponding $3$-volumes are given by
\begin{equation}\label{eq:spacelike 3-volume}
V^{(3)}_{\vm\vn}  = \frac{2k_n(j_n+\sqrt{j_nj_{n+1}}+j_{n+1})}{3(\sqrt{j_{n+1}}+\sqrt{j_n})}\sqrt{1-\frac{(j_{n+1}-j_n)^2}{16k_n^2}}\,.
\end{equation}
To obtain the length of ``spacelike'' dual edges
it suffices to project the geometry onto the plane spanned by the $t$-axis and the dual edge. In this picture, the dual edge connects two glued trapezoids and is orthogonal with respect to the connecting face. For $\Theta$ the dihedral angle between the $(n+1)$-cube and the boundary $3$-frustum, defined as~\cite{Bahr:2017eyi}
\begin{equation}
\Theta_n = \arccos\left(\frac{1}{\tan(\phi_n)}\right)\,,
\end{equation}
the dual edge length is then given by
\begin{equation}\label{eq:spacelike dual edge length}
l_\star^{\vm \vn} = \frac{1}{2} \left(\sqrt{j_n} + \sqrt{j_{n+1}}\right) \cos\left(\frac{\pi}{2} - \Theta_n\right)\,.
\end{equation}
Below, we denote the components of the Laplacian $\Delta_{\vn\vm}$ with vertices $\vn$ and $\vm$ having a ``spacelike'' separation by $\Delta_{nn}$.


\section{Spectrum of Laplacian on $\mathcal{N}$-periodic frusta}\label{app:Laplacian spectrum}

The first step in deriving the spectrum of the discrete Laplacian is to introduce the Fourier transform of the test field and to notice that the homogeneity of the geometry effectively reduces the Laplace coefficients $\Delta_{\vn\vm}$ to an $L\times L$ matrix. To make this explicit, let us first introduce some notation: We indicate the ``spatial'' component of $\vn\in\mathbb{Z}_L^4$ as $\mathbf{n}\in\mathbb{Z}_L^3$. 
Taking $\mathcal{N}$-periodicity into account, we indicate a slice 
as $n_0 + z\mathcal{N}$ where $n_0\in\mathbb{Z}_{\mathcal{N}}$ and $z\in\mathbb{Z}_{L/\mathcal{N}}$. Thus, the variable $z$ labels the $\mathcal{N}$-cell in which the slice is located and $n_0$ denotes the $n_0$th slice within a given $\mathcal{N}$-cell. Using this notation, the scalar test field is written as $\phi_{n_0,\mathbf{n}}^{(z)}$ with $\phi_{n_0+\mathcal{N},\mathbf{n}}^{(z)}\equiv \phi^{(z+1)}_{n_0,\mathbf{n}}$.

To write the Laplace operator in Fourier space, we consider a similar ansatz to the proposal of~\cite{Sahlmann:2009rk}, given by
\begin{equation}\label{eq:test field ansatz}
\phi_{n_0,\mathbf{n}}^{(z)} = c_{n_0}\e^{ip_0z}\e^{i\mathbf{n}\cdot\mathbf{p}},
\end{equation}
where $c_{n_0}$ is an $\mathcal{N}$-dimensional vector. A phase with spatial momentum $p_i$ is picked whenever changing  one lattice site in spatial direction $i$. In contrast, a phase with temporal momentum $p_0$ is only picked up when changing to another $\mathcal{N}$-cell. This pattern of picking up phases will be reflected in Eqs.~(\ref{eq:Deltac1}),~(\ref{eq:Deltac2}) and~(\ref{eq:Deltac3}). For a finite lattice of size $L^4$ with periodic boundary conditions, the momenta $p_{\mu}$ take values $p_{\mu} = \frac{2\pi}{L} k_{\mu}$ with $k_{\mu}\in\mathbb{Z}_L$. In the limit $L\rightarrow\infty$, the momenta lie in the Brillouin zone $p_\mu\in [-\pi,\pi]$. Inserting the ansatz of Eq.~\eqref{eq:test field ansatz} into the action of the Laplace operator in Eq.~\eqref{eq:defLaplace app}, we obtain for $n \notin \{0,\mathcal{N}-1\}$
\begin{equation}\label{eq:Deltac1}
-(\Delta c)_n = -\left[\Delta_{nn+1}(c_n-c_{n+1}) + \Delta_{nn-1}(c_n - c_{n-1}) + 2\Delta_{nn} c_n\sum_{i = 1}^3(1-\cos(p_i)) \right].
\end{equation}
$\Delta_{nn+1}$ are the components of the Laplace operator on dual edges connecting slices $n$ and $n+1$, defined by Eqs.~(\ref{eq:4Volume}),~(\ref{eq:timelike 3-volume}) and~(\ref{eq:timelike dual edge length}). $\Delta_{nn}$ are the components of the Laplace operator within a slice, defined by Eqs.~(\ref{eq:4Volume}),~(\ref{eq:spacelike 3-volume}) and ~(\ref{eq:spacelike dual edge length}), associated to ``spacelike'' separated vertices. 
Due to spatial homogeneity, $\Delta_{nn}$ is independent of the spatial direction and therefore factorizes from the spatial momenta. 
For the slices $n = 0, \mathcal{N}-1$ connecting neighboring $\mathcal{N}$-cells, exponential factors of $\e^{\pm i p_0}$ are picked up
\begin{align}
-(\Delta c)_0               &= -\left[x_{0}(c_0-c_{1})+x_{\mathcal{N}-1}(c_0-c_{\mathcal{N}-1}e^{-ip_0})+X_0c_0\right],\label{eq:Deltac2}\\
-(\Delta c)_{\mathcal{N}-1} &= -\left[x_{\mathcal{N}-1}(c_{\mathcal{N}-1}-c_{0}e^{ip_0})+x_{\mathcal{N}-2}(c_{\mathcal{N}-1}-c_{\mathcal{N}-2})+X_{\mathcal{N}-1}c_{\mathcal{N}-1}\right],\label{eq:Deltac3}
\end{align}
where for brevity, we introduced the notation 
\begin{equation}\label{eq:w and W}
x_n \defeq \Delta_{nn+1}
\quad , \quad 
X_n \defeq 2\Delta_{nn}\sum_i (1-\cos(p_i)) \, .
\end{equation}
Exploiting $\mathcal{N}$-periodicity and spatial homogeneity, we observe that the action of the Laplace operator reduces to a vector equation in $\mathcal{N}$ dimensions. Altogether, Eqs.~(\ref{eq:Deltac1}), (\ref{eq:Deltac2}) and~(\ref{eq:Deltac3}) are captured by
\begin{equation}
-(\Delta c)_m = -\sum_{n = 0}^{\mathcal{N}-1} M_{mn}c_n\,,
\end{equation}
with the matrix $M$ being defined as
\begin{equation}\label{eq:Laplace_momentum}
M \defeq 
\begin{pmatrix}x_{\mathcal{N}-1}+x_0 +X_0 & -x_0 & &\dots &  -x_{\mathcal{N}-1}e^{-ip_0} \\
-x_0 & x_0 + x_1 + X_1  &  &&\\
\vdots & & &  \ddots && \\
 -x_{\mathcal{N}-1}e^{ip_0}  & \dots & &-x_{\mathcal{N}-2} & x_{\mathcal{N}-2} + x_{\mathcal{N}-1} + X_{\mathcal{N}-1}
\end{pmatrix}\,.
\end{equation}
For a given spin configuration, the spectrum in momentum space is then given by the eigenvalues of the matrix $M(p_0,p_1,p_2,p_3)$, which must be computed for every combination of momenta. 

In the special case of $\mathcal{N}=1$, considered in Secs.~\ref{subsec:1-periodic spectral dimension} and~\ref{subsec:Cosmological constant}, the matrix $M$ reduces to the scalar,
\begin{equation}
M = 2x_0(1-\cos(p_0)) + X_0(p_1,p_2,p_3) = \sum_{\mu=0}^3\omega^{(\mu)}(p_{\mu})\,,
\end{equation}
decomposing into components $\omega^{(\mu)}(p_\mu)$. This form of the eigenvalues is particularly advantageous to compute the return probability, since on an infinite lattice, $P_1(\tau)$ can be written as a product of integrals
\begin{equation}\label{eq:P1}
P_1(\tau) = \prod_{\mu = 0}^3\int_{[-\pi,\pi]}\dd{p_{\mu}}\e^{-\tau\omega^{(\mu)}(p_{\mu})} = \left(\int\dd{p_0}\e^{-\tau\omega^{(0)}(p_0)}\right)\left(\int\dd{p_3}\e^{-\tau\omega^{(3)}(p_3)}\right)^3\,,
\end{equation}
where we have exploited spatial homogeneity.

The Laplace operator at $\mathcal{N} = 2$  becomes a $2\times 2$ matrix in momentum space
\begin{equation}
M = 
\begin{pmatrix}
x_0+x_1+X_0 & -x_0-x_1 e^{-ip_0} \\
-x_0-x_1 e^{ip_0} & x_0+x_1+X_1
\end{pmatrix}\,,
\end{equation}
with the corresponding eigenvalues given by
\begin{equation}
\omega_{\pm}(p_\mu) = x_0+x_1+\frac{X_0+X_1}{2}\pm\sqrt{x_0^2+x_1^2+2x_0x_1\cos(p_0)+\left(\frac{X_0-X_1}{2}\right)^2}\,.
\end{equation}
Compared to the $1$-periodic case in Eq.~\eqref{eq:P1}, this expression is more involved due to the intermingling of the $p_0$ and $p_i$ terms. As a consequence, the return probability from momentum integration,
\begin{equation}
P_2(\tau) = \sum_{\epsilon = \pm}\int\prod_{\mu = 0}^3\dd{p_\mu}\e^{-\tau\omega_{\epsilon}(\{p_\nu\})}\,,
\end{equation}
cannot be written as the product of $1$-dimensional integrals. Instead, full $4$-dimensional integration is required to compute $P_2$, leading to larger numerical computation times.

\section{Derivation of analytical estimate of the spectral dimension}\label{app:Derivation of analytical estimate}

Building up on the ideas of~\cite{Steinhaus:2018aav}, we present in this appendix a derivation of the analytical estimate of $\Ds$ used in Sec.~\ref{subsec:Analytical estimate of the spectral dimension}. Tackling first the spectral dimension of general $\mathcal{N}$-periodic spin-foam frusta, we obtain a qualitative expression for the spectral dimension which is, however, still too intricate to compute explicitly. Nevertheless, it serves as a support for the numerical results as well as a guidance for the limit $\mathcal{N}\rightarrow\infty$.

For the analysis of the spectral dimension, it is advantageous to introduce the average spin variable $r^2\defeq\frac{1}{n}\sum j_f^2$, where $n$ is the total number of degrees of freedom, being $n = 2\mathcal{N}$ in the case of $\mathcal{N}$-periodic spin-foam frusta. Technically, the variable $r$ is a radial coordinate in the space of configurations $j_f$. Likewise, the remaining variables can be seen as an angular part, and we therefore denote them by $\Omega$ in the following. Following the arguments of~\cite{Sahlmann:2009rk} and~\cite{Calcagni:2014cza,Steinhaus:2018aav}, let us assume for the moment that under the spin-foam measure
\begin{equation}\label{eq:Laplace scaling}
\Delta(j_f)\sim\frac{1}{r}\Delta\,,
\end{equation}
where $\Delta$ is the Laplace matrix on the equilateral hypercubical lattice. 

Within this assumption, let us have a closer look on the expectation value of the return probability with respect to semi-classical amplitudes. First, the summation over configurations can be approximated by an integral for $\frac{j_{\mathrm{max}}}{j_{\mathrm{min}}}\gg 1$. Performing in addition a change to spherical coordinates as described above, we obtain for the return probability expectation value
\begin{equation}
\langle P(\tau)\rangle = \frac{1}{Z}\int\dd{\Omega}\int\limits_{j_{\mathrm{min}}}^{j_{\mathrm{max}}}\dd{r}\;r^{n-1}\prod_v\mathcal{A}_v(r,\Omega)\Tr\left(\e^{\frac{\tau\Delta}{r}}\right)\,.
\end{equation}
Forming the logarithmic derivative of this expression yields
\begin{equation}
\frac{\tau}{\langle P\rangle}\frac{\partial \langle P\rangle}{\partial \tau} = \frac{1}{\langle P\rangle Z}\int\dd{r}\dd{\Omega}\;r^{n-1}\prod_v\mathcal{A}_v\Tr\left(\frac{\tau\Delta}{r}\e^{\frac{\tau\Delta}{r}}\right)\,.
\end{equation}
Since $\Tr\left(\e^{\frac{\tau\Delta}{r}}\right)$ is in fact a function of the ratio $\tau/r$, we can trade the derivative with respect to $\tau$ with an $r$ derivative,
\begin{equation}\label{eq:tdt = -rdr}
\Tr\left(\frac{\tau\Delta}{r}\e^{\frac{\tau\Delta}{r}}\right) = -r\frac{\partial}{\partial r}\Tr\left(\e^{\frac{\tau\Delta}{r}}\right)\,.
\end{equation}
Using the $r$ derivative, we can integrate by parts,
\begin{equation}
\frac{\tau}{\langle P\rangle}\frac{\partial \langle P\rangle}{\partial \tau} = \frac{1}{\langle P\rangle Z}\left[\partial I(\tau) + \int\dd{r}\dd{\Omega}r^{n-1}\prod_v\mathcal{A}_v\left(n + \sum_v\frac{r}{\mathcal{A}_v}\frac{\partial\mathcal{A}_v}{\partial r}\right)\Tr\left(\e^{\frac{\tau\Delta}{r}}\right)\right]\,.
\end{equation}
Here, $\partial I$ denotes the boundary term in the partial integration, explicitly given by
\begin{equation}\label{eq:bdr term}
\partial I(\tau) = -\int\dd{\Omega} \;r^n \prod_v\mathcal{A}_v\Tr\left(\e^{\frac{\tau\Delta}{r}}\right)\bigg|_{r = j_{\mathrm{min}}}^{j_{\mathrm{max}}}\,.
\end{equation}
The other terms inside the brackets stem from the $r$-derivative acting first on $r^n$ and then on the product of amplitudes. Notice that $-\frac{r}{\mathcal{A}_v}\frac{\partial\mathcal{A}_v}{\partial r}$ is exactly the effective scaling $\gamma$ of $\mathcal{A}_v$ introduced in Eq.~\eqref{eq:effective scaling}. Finally, writing general vertex amplitudes as
\begin{equation}
\mathcal{A}_v 
= r^{-\sfsc}\mathcal{C}_v(r,\Omega)\,,
\end{equation}
where $\mathcal{C}_v(r,\Omega)$ can be understood as a \emph{correction} term to the scaling part with constant~$\sfsc$, the spectral dimension can be expressed as
\begin{equation}\label{eq:approximation D app}
    \Ds = 2(\sfsc V - n)-2\sum_v\frac{\int\dd{\Omega}\dd{r}\;r^{n-1}\frac{r}{\mathcal{C}_v}\frac{\partial\mathcal{C}_v}{\partial r}\prod_{v'}\mathcal{A}_{v'}\Tr\left(\e^{\frac{\tau\Delta}{r}}\right)}{\int\dd{\Omega}\dd{r}\;r^{n-1}\prod_{v'}\mathcal{A}_{v'}\Tr\left(\e^{\frac{\tau\Delta}{r}}\right)}-2\frac{\partial I}{\langle P\rangle Z}\,,
\end{equation}
with $V = \mathcal{N}^4$ the number of dual vertices. For oscillating correction terms that attain many zeros, and therefore lead to divergences of the effective scaling, the integration domain of the above needs to be restricted accordingly. Aspects of well-definedness and convergence need to be addressed for each given $\mathcal{C}_v$ individually. Put into this form,~\eqref{eq:approximation D app} suggests that the pure scaling value of $\Ds^\alpha = 2((9-12\alpha)V-n)$ is corrected by a term arising from the effective scaling $-\frac{r}{\mathcal{C}_v}\frac{\partial \mathcal{C}_v}{\partial r}$ of the correction term $\mathcal{C}_v$ as well as the boundary term. If the value of $\Ds^\alpha$ lies outside the interval $[0,4]$, the boundary term counteracts to yield either zero or four.
\chapter[\textsc{Lorentzian 4-Frusta and Dihedral Angles}]{Lorentzian 4-Frusta and Dihedral Angles}

\section{Geometry of Lorentzian 4-frusta}\label{app:Lorentzian frustum}

In a (3+1)-dimensional Lorentzian setting, the edges, trapezoids and $3$-frusta connecting slices $n$ and $n+1$ can be either spacelike, timelike or null (null configurations are excluded in Chapter~\ref{chapter:LRC} by the choice of boundary data), reflected in the sign of the squared length, area and $3$-volume, respectively.
In the following, we express these geometric quantities as functions of $(l_n,l_{n+1},H_n)$ and show under which conditions the building blocks are spacelike or timelike.


\paragraph{Edges.} Struts contained in trapezoids which thus connect cubes that lie in distinct slices $n$ and $n+1$ have a squared edge length given by
\begin{equation}\label{eq:edge length}
m_n^2 = H_n^2 - 3\left(\frac{l_n-l_{n+1}}{2}\right)^2\,,
\end{equation}
and therefore we have that:
\begin{center}
\begin{tabular}{c c c c}
  edge is timelike if   & $m_n^2 > 0$ & $\Leftrightarrow$ & $H_n^2 > \frac{3}{4}(l_n-l_{n+1})^2$\,, \\[7pt]
  edge is spacelike if  & $m_n^2 < 0$ & $\Leftrightarrow$ & $H_n^2 < \frac{3}{4}(l_n-l_{n+1})^2$\,.
\end{tabular}
\end{center}

\paragraph{Trapezoids.} The squared area $k_n^2$ of a trapezoid  is given by
\begin{equation}
k_n^2 = \left(\frac{l_n+l_{n+1}}{2}\right)^2\left[H_n^2 - \left(\frac{l_n-l_{n+1}}{\sqrt{2}}\right)^2\right],
\end{equation}
from which we extract that:
\begin{center}
\begin{tabular}{c c c c}
  trapezoid is timelike if   & $k_n^2 > 0$ & $\Leftrightarrow$ & $H_n^2 > \frac{1}{2}(l_n-l_{n+1})^2$\,, \\[7pt]
  trapezoid is spacelike if  & $k_n^2 < 0$ & $\Leftrightarrow$ & $H_n^2 < \frac{1}{2}(l_n-l_{n+1})^2$\,.
\end{tabular}
\end{center}

\paragraph{$\mathbf{3}$-frusta.} To determine the signature and the $3$-volume of $3$-frusta, we consider first the squared height of the $3$-frustum,
\begin{equation}
h_n^2 = H_n^2 - \left(\frac{l_n-l_{n+1}}{2}\right)^2.
\end{equation}
Then, the signature of the $3$-frustum is determined by the signature of its height:
\begin{center}
\begin{tabular}{c c c c}
    $3$-frustum is timelike if & $h_n^2 > 0$ & $\Leftrightarrow$ & $H_n^2 > \frac{1}{4}(l_n-l_{n+1})^2$\,,\\[7pt]
    $3$-frustum is spacelike if & $h_n^2 < 0$ & $\Leftrightarrow$ & $H_n^2 < \frac{1}{4}(l_n-l_{n+1})^2$\,.
\end{tabular}
\end{center}
Using the squared height of the $3$-frustum, $h_n^2$, one can express the squared $3$-volume as
\begin{equation}
v_n^2 = \frac{(l_n^2+l_nl_{n+1}+l_{n+1}^2)^2}{9}h_n^2 =  \frac{(l_n^2+l_nl_{n+1}+l_{n+1}^2)^2}{9}\left[H_n^2 - \left(\frac{l_n-l_{n+1}}{2}\right)^2\right]\,. 
\end{equation}

\paragraph{$\mathbf{4}$-frusta.} Lastly, the squared $4$-volume of a $4$-frustum is given by
\begin{equation}
V_n^2 = \frac{(l_n^2+l_{n+1}^2)^2(l_n+l_{n+1})^2}{16}H_n^2\,.
\end{equation}
As a building block of top dimension, the Lorentzian $4$-frustum is constrained to have a positive $4$-volume, yielding the condition of a positive squared $4$-height, i.e. $H_n^2 > 0$. If $H_n^2 < 0$, the $4$-frustum can only be embedded in $4$-dimensional Euclidean space.

\section{Lorentzian angles}\label{app:Lorentzian angles}

In this appendix we derive the dihedral angles for the Lorentzian $4$-frustum following the conventions of~\cite{Sorkin:2019llw,Asante:2021zzh}. In the boundary of a $4$-frustum there are three distinct $3$-dimensional building blocks, being the initial and final cube as well as the six boundary $3$-frusta. Due to homogeneity and isotropy, there are three distinct dihedral angles per slab $(n,n+1)$, being $\varphi_{nn+1}$, $\varphi_{n+1n}$ and $\theta_n$ located at the initial and final square and the trapezoids, respectively. 

We compute these angles by embedding a single $4$-frustum in flat Minkowski space $\R^{1,3}$ and determining the outward-pointing normal vectors of the $3$-dimensional building blocks. For initial and final cube, these are respectively given by $N_{s,\mp} = (\mp 1 ,0 ,0, 0)$. To obtain normal vectors of the $3$-frusta, we form the wedge product of three spanning vectors and act with the Hodge-star operator $*$. As a result, the six normal vectors are given by
\begin{equation}
N_{\pm,i} = \frac{1}{\sqrt{\abs{H_n^2-\left(\frac{l_n-l_{n+1}}{2}\right)^2}}}\left(\frac{l_n-l_{n+1}}{2}, \pm H_n\vb*{e}_i\right)\,.
\end{equation}
The signature of $N_{\pm,i}$ is opposite of the signature of the $3$-frustum, i.e. $\eta(N_{\pm,i},N_{\pm,i}) = +1$ for a spacelike $3$-frustum, and $\eta(N_{\pm,i},N_{\pm,i}) = -1$ for a timelike $3$-frustum. Notice that the time orientation for timelike $N_{\pm,i}$ depends on $\mathrm{sgn}(l_n - l_{n+1})$. 

Lorentzian dihedral angles are defined in terms of the Minkowski product of the normal vectors associated to the two polyhedra. For the three cases we consider here, these are given by
\begin{align}
    \eta(N_{\pm i},N_{\pm j}) &= \frac{(l_n-l_{n+1})^2}{\abs{4H_n^2-(l_n-l_{n+1})^2}}\,,\\[7pt]
    \eta(N_{\pm,i},N_{s,\mp}) &= \mp\frac{l_n-l_{n+1}}{\sqrt{\abs{4H_n^2-(l_n-l_{n+1})^2}}}\,.
\end{align}
The precise form of the Lorentzian angles $\varphi_{nn+1},\varphi_{n+1n}$ and $\theta_n$ as a function of the Minkowski products depend then on the signature of the $3$-polyhedra. Moreover, the signature of the trapezoid is decisive for the associated angle to be either Lorentzian or Euclidean. With the conventions of~\cite{Sorkin:2019llw,Asante:2021zzh}, the angles in Eqs.~\eqref{eq:phi nn+1}--\eqref{eq:theta E} follow below. 

In the restricted setting considered in Chapter~\ref{chapter:LRC}, there are two types of dihedral angles, associated either to spacelike squares or to trapezoids connecting slices. In the following, we provide a list of the dihedral angles for the various cases as a function of the variables $(l_n,l_{n+1},H_n)$.

\paragraph{Dihedral angles at squares.} Cubes lie in spacelike hypersurfaces such that the squares contained in it must also be spacelike. As a consequence, the space orthogonal to a square is two-dimensional Minkowski space $\R^{1,1}$. In that space, the dihedral angle between the $3$-cube and the $3$-frustum meeting at that square is given by the Lorentzian angle between the respective projected normal vectors. While the normal vector to a $3$-cube is always timelike the signature of the vector normal to the $3$-frustum can be either spacelike or timelike, opposite to the signature of the $3$-frustum. Within a $4$-frustum, we refer to the dihedral angle located at the (past) $n$-th, respectively the (future) $(n+1)$-th slice as $\varphi_{nn+1}$ and $\varphi_{n+1n}$. Their explicit definitions in terms of the variables $(l_n,l_{n+1},H_n)$ are given by
\begin{equation}\label{eq:phi nn+1}
\varphi_{nn+1} = 
\begin{cases}
-\cosh^{-1}\left(\frac{l_{n+1}-l_n}{\sqrt{(l_n-l_{n+1})^2-4H_n^2}}\right) &\mathrm{if}\quad H_n^2 <\frac{1}{4}(l_n-l_{n+1})^2\quad\mathrm{and}\quad l_n<l_{n+1}\,,\\[10pt]
\cosh^{-1}\left(\frac{l_n - l_{n+1}}{\sqrt{(l_n-l_{n+1})^2-4H_n^2}}\right) \mp i\pi &\mathrm{if}\quad H_n^2 <\frac{1}{4}(l_n-l_{n+1})^2\quad\mathrm{and}\quad l_n>l_{n+1}\,,\\[10pt]
\sinh^{-1}\left(\frac{l_n-l_{n+1}}{\sqrt{4H_n^2-(l_n-l_{n+1})^2}}\right)\mp i\frac{\pi}{2} &\mathrm{if}\quad H_n^2 > \frac{1}{4}(l_n-l_{n+1})^2\,,
\end{cases}
\end{equation}
and
\begin{equation}\label{eq:phi n+1n}
\varphi_{n+1n} = 
\begin{cases}
-\cosh^{-1}\left(\frac{l_n-l_{n+1}}{\sqrt{(l_n-l_{n+1})^2-4H_n^2}}\right) &\mathrm{if}\quad H_n^2 <\frac{1}{4}(l_n-l_{n+1})^2\quad\mathrm{and}\quad l_n > l_{n+1}\,,\\[10pt]
\cosh^{-1}\left(\frac{l_{n+1}-l_n}{\sqrt{(l_n-l_{n+1})^2-4H_n^2}}\right) \mp i\pi &\mathrm{if}\quad H_n^2 <\frac{1}{4}(l_n-l_{n+1})^2\quad\mathrm{and}\quad l_n < l_{n+1}\,,\\[10pt]
\sinh^{-1}\left(\frac{l_{n+1}-l_n}{\sqrt{4H_n^2-(l_n-l_{n+1})^2}}\right) \mp i\frac{\pi}{2} &\mathrm{if}\quad H_n^2 > \frac{1}{4}(l_n-l_{n+1})^2\,.
\end{cases}
\end{equation}
We observe that the real parts of the angles $\varphi_{nn+1}$ and $\varphi_{n+1n}$ are related by a minus sign. Furthermore, we notice that for a timelike frustum, the dihedral angles $\varphi$ do not depend on the sign of $l_n-l_{n+1}$. That is because the associated normal vector is spacelike and therefore insensitive to the time orientation.

\paragraph{Dihedral angles at trapezoids.} The dihedral between two $3$-frusta is located at a trapezoid, which can be either spacelike or timelike.

To a spacelike trapezoid, implying the inequality $H_n^2<\frac{1}{2}(l_n-l_{n+1})^2$, we associate the Lorentzian dihedral angle
\begin{align}\label{eq:theta L}
\theta_n^{\mathrm{L}} = 
\begin{cases}
-\cosh^{-1}\left(\frac{(l_n-l_{n+1})^2}{(l_n-l_{n+1})^2-4H_n^2}\right)&\mathrm{if}\quad H_n^2<\frac{1}{4}(l_n-l_{n+1})^2\,,\\[10pt]
-\cosh^{-1}\left(\frac{(l_n-l_{n+1})^2}{4H_n^2-(l_n-l_{n+1})^2}\right) \mp i\pi &\mathrm{if}\quad H_n^2 > \frac{1}{4}(l_n-l_{n+1})^2\,.
\end{cases}
\end{align}

To a timelike trapezoid, i.e. $H_n^2 > \frac{1}{2}(l_n-l_{n+1})^2$, we associate the Euclidean angle 
\begin{equation}\label{eq:theta E}
\theta_n^{\mathrm{E}} = \cos^{-1}\left(\frac{(l_n-l_{n+1})^2}{4H_n^2-(l_n-l_{n+1})^2}\right)\,.
\end{equation}
\chapter[\textsc{Derivation of Effective Cosmological Amplitude}]{Derivation of Effective Cosmological Amplitude and Wynn's algorithm}

\section{A derivation of the effective cosmological amplitude}\label{app:derivation of the effective model}

In this section, we set up the effective cosmological amplitude employed in Chapter~\ref{chapter:3d cosmology}, which has been derived in~\cite{Jercher:2024hlr}. Starting point is the introduction of the underlying fundamental (2+1) spin-foam model in Sec.~\ref{app:A cuboidal spin-foam amplitude for Lorentzian gravity}. We then perform a semi-classical analysis in Sec.~\ref{app:The semi-classical limit of the vertex} followed by a sequence of modifications in Sec.~\ref{app:Spin-foam amplitude simplifications}. 

\subsection{A cuboidal spin-foam amplitude for Lorentzian gravity}\label{app:A cuboidal spin-foam amplitude for Lorentzian gravity}

The underlying spin-foam model of interest in Chapter~\ref{chapter:3d cosmology} is the Ponzano-Regge model for (2+1)-dimensional Lorentzian quantum gravity~\cite{Barrett:1993db,Freidel:2000uq,Freidel:2002hx} formulated in a coherent state representation developed recently in~\cite{Simao:2024don}. Although the model has originally been put forth in a simplicial setting, its generalization to other spacetime discretizations is straightforward. The model is based upon the theory of unitary irreducible representations of $\mathrm{SU}(1,1)$, introduced in Appendix~\ref{app:su11}.

Given a cuboidal lattice $\mathcal{X}^{(3)}$, where each 2-cell $\Box$ is identified by a single integer $a$, and each 1-cell $\diagup$ by a pair $ab$ of adjacent 2-cells, a \textit{history} $\uppsi$ is an assignment of data to $\mathcal{X}^{(3)}$, as follows:
\begin{enumerate}
    \item To each 1-cell $\diagup_{ab}$ one assigns a spin $s_{ab}\in \mathbb{R}^+$ or $k_{ab}\in -\frac{\mathbb{N}}{2}$ of the continuous $\mathcal{C}^0_s$ or discrete $\mathcal{D}^q_k$ series of unitary representations of $\mathrm{SU}(1,1)$, respectively. The discrete spin is moreover complemented by a sign $\tau_{ab}:=-q_{ab}$, selecting between the positive or negative family of the series. An edge with discrete spin is termed timelike, and spacelike otherwise. Semi-classically, the spin is in correspondence with the length of the edges of $\mathcal{X}^{(3)}$.
    \item To each ordered pair $(\Box_a, \Box_b)$ of adjacent boundary 2-cells one assigns an ordered pair $(n_{ab},n_{ba})$ of $\mathrm{SU}(1,1)$ group elements, together with an orientation $\mathfrak{o}_{ab}=\pm$.
\end{enumerate}
The spin-foam model prescribes an \emph{amplitude map} $\mathcal{A}: (\mathcal{X}^{(3)},\uppsi) \rightarrow \mathbb{C}$, i.e. it assigns a complex number to each spin-foam history $\uppsi$ on a given lattice $\mathcal{X}^{(3)}$. Its fundamental building block is the \textit{vertex amplitude} $\mathcal{A}_v: \uppsi \rightarrow \mathbb{C}$, obtained by evaluating the amplitude map on a single 3-cell lattice $\cube$. The vertex amplitude is most clearly expressed in a diagrammatic representation, 
\begin{equation}
\label{cube_vertex}
    \mathcal{A}_v(\uppsi)= \scalebox{0.4}{\cubevertex}\,,
\end{equation}
the meaning of which we discuss next.

The diagram is composed of boxes $\raisebox{2pt}{\scalebox{0.5}{\boxtikz}}_a$ and links $\raisebox{2pt}{\scalebox{0.5}{\link}}_{ab}$, the latter indexed by the two boxes which intersect each link. A history $\uppsi$ induces a coloring of the diagram, according to the rules:
\begin{enumerate}
    \item Each link is in correspondence with an edge $\diagup_{ab}$. If the associated spin is of the discrete (continuous) series, the endpoints are colored white $\raisebox{2pt}{\scalebox{0.5}{\link}}$ (black $\raisebox{2pt}{\scalebox{0.5}{\linkblack}}$), and the link is said to be timelike (spacelike) as it corresponds to a timelike (spacelike) edge.
    \item Each box $\raisebox{2pt}{\scalebox{0.5}{\boxtikz}}_a$ is associated with a 2-cell $\Box_a$. For a single 3-cell lattice all adjacent 2-cells $(\Box_a, \Box_b)$ are boundary, so each link $\raisebox{2pt}{\scalebox{0.5}{\link}}_{ab}$ inherits a pair $(n_{ab},n_{ba})$ and an orientation $\mathfrak{o}_{ab}$. Accordingly, to the end-point $\raisebox{2pt}{\scalebox{0.5}{\ept}}_{ab}$ close to the box $\raisebox{2pt}{\scalebox{0.5}{\boxtikz}}_a$ there corresponds the element $n_{ab}\in\mathrm{SU}(1,1)$, and the orientation prescribes an ordering to the link $\overset{\rightarrow}{\raisebox{2pt}{\scalebox{0.5}{\link}}}_{ab}$. Semi-classically, the $n_{ab}$ are in correspondence with the edge vectors of the cuboid.
\end{enumerate}
Note that in Eq.~\eqref{cube_vertex} we have omitted most of the $\uppsi$ data for simplicity. The diagram is now evaluated as follows: for a timelike link, colored with a spin $k_{ab}$ of the discrete series, we set
\begin{equation}
\begin{gathered}
\label{brakettl}
\!{}^{n_{ab}}\overset{\rightarrow}{\scalebox{0.65}{\braketblack}}{}^{n_{ba}}= \bra{\tau_{ab}}g^\dagger_a \sigma_3 g_b \ket{\tau_{ba}}^{2k_{ab}}\,, \\
\ket{+_{ab}}:=n_{ab}\left(\begin{smallmatrix} 1 \\ 0 \end{smallmatrix}\right)\,, \quad \ket{-_{ab}}:=n_{ab}\left(\begin{smallmatrix} 0 \\ 1 \end{smallmatrix}\right)\,,
\end{gathered}
\end{equation}
where $g_a,g_b \in \mathrm{SU}(1,1)$ are associated to the two boxes, and remain unspecified. For a spacelike link, colored with a spin $s_{ab}$ of the continuous series, we take
\begin{equation}
\begin{gathered}
\label{braketsl}
\!{}^{n_{ab}}\overset{\rightarrow}{\scalebox{0.65}{\brakettikz}}{}^{n_{ba}}=   \mathcal{C}_{n_{ab},n_{ba}} \bra{l^+_{ab}}g_a^\dagger \sigma_3 g_b\ket{l^-_{ba}}^{-1+2is_{ab}}\,,\\
 \mathcal{C}_{n_{ab},n_{ba}} := e^{s_{ab} \bra{l^+_{ab}}g_a^\dagger \sigma_3 g_b \ket{l^+_{ba}}^2}\,, \quad \ket{l^\pm_{ab}}:=\frac{n_{ab}}{\sqrt{2}}  \left(\begin{smallmatrix} 1 \\ \pm 1 \end{smallmatrix}\right)\,.
\end{gathered}
\end{equation}
Here $\gamma$ denotes the Euler-Mascheroni constant. The states $\ket{\tau_{ab}}$ and $\ket{l^\pm_{ab}}$ are $\mathrm{SU}(1,1)$ Perelomov coherent states~\cite{Perelomov:1986tf} in the defining representation. The term $ \mathcal{C}_{n_{ab},n_{ba}}$ corresponds to a Gaussian constraint that has been introduced by hand, ensuring a well-behaved semi-classical limit by implementing the otherwise absent gluing between edges $n_{ab}$ and $n_{ba}$. The vertex amplitude is now obtained from Eq.~\eqref{cube_vertex} by taking the product of all links, and integrating over the boxes. Explicitly, for a certain choice of orientation,
\begin{equation}\label{eq:SU11 general vertex}
     \mathcal{A}_{v}(\uppsi)=\int\limits_{\mathrm{SU}(1,1)^5} \prod_a \mathrm{d} g_a\,  \prod_{ab\; \tikz\draw[black,fill=white] (0,0) circle (.3ex); }  \bra{\tau_{ab}}g^\dagger_a \sigma_3 g_b  \ket{\tau_{ba}}^{2k_{ab}}\prod_{ab \; \tikz\draw[black,fill=black] (0,0) circle (.3ex);}     \mathcal{C}_{n_{ab},n_{ba}} \bra{l^+_{ab}}g_a^\dagger \sigma_3 g_b\ket{l^-_{ba}}^{-1+2is_{ab}}\,,
\end{equation}
where we have gauge-fixed one of the Haar integrals  to regularize the amplitude~\cite{Engle:2008ev}.

The full amplitude for extended lattices $\mathcal{X}^{(3)}$ follows straightforwardly from gluing the diagrams in Eq.~\eqref{cube_vertex}. If the gluing is such that a closed loop arises, the link is assigned a value
\begin{equation}
\label{eq:looptl}
 \scalebox{0.5}{\looptikz}_k   = \sum_{q=\pm} (-2k-1)\mathrm{Tr}\left[\vb*{D}^{k(q)}(g)\right]\,,
\end{equation}
or 
\begin{equation}
\label{eq:loopsl}
  \scalebox{0.5}{\looptikz}_s  =\sum_{\delta=0,\frac{1}{2}} s\tanh^{1-4\delta} (\pi s) \, \mathrm{Tr}\left[\vb*{D}^{j(\delta)}(g)\right]\,,
\end{equation}
depending on the history $\uppsi$. Note that the unitary representations of $\mathrm{SU}(1,1)$ are infinite-dimensional, so that the previous two equations must be understood in the sense of their regularization. An example configuration $\mathcal{X}^{(3)}$ inducing such a loop is given by four cuboids sharing an edge, for which a corresponding diagram would take the form
\begin{equation}
     \mathcal{A}(\cube^4, \uppsi)= \scalebox{0.4}{\cubefour}\,,
\end{equation}
where the loop associated to the internal edge is dashed for clarity. On such an extended diagram one must still gauge-fix all redundant group integrations.

Finally, the spin-foam \emph{partition function} is given by a sum over histories $\uppsi$ in agreement with a fixed choice of boundary data. By boundary data $\partial \uppsi$ we mean the data assignments of $\uppsi$ restricted to the boundary edges $\diagup_{ab}\in \partial \mathcal{X}^{(3)}$. Thus, for given $\mathcal{X}^{(3)}$ and boundary data $\Phi$,
\begin{equation}\label{eq:ptfunc}
    Z(\mathcal{X}^{(3)}, \Phi):= \sumint{\uppsi\,|\, \partial \uppsi=\Phi} \mathcal{A}(\mathcal{X},\uppsi)\,.
\end{equation}
Recovering the previous example of the $\cube^4$ lattice, the partition function would read
\begin{equation}
    Z(\cube^4, \Phi):= \sum_{k\,  \mapsto  \, \raisebox{1pt}{\scalebox{0.3}{\looptikz}}} \int_{s\,  \mapsto  \, \raisebox{1pt}{\scalebox{0.3}{\looptikz}}}\mathrm{d}s  \;\scalebox{0.4}{\cubefour}\,,
\end{equation}
and include a sum and integral over assignments of spins $k$ and $s$ to the bulk looped link. Notice that the causal character of bulk edges is being summed over. This can be traced back to the fact that the Plancherel decomposition of functions on $\SUO$ contains contributions of both the continuous and the discrete series.

\subsection{The semi-classical limit of the vertex}\label{app:The semi-classical limit of the vertex}

Let us return to the vertex amplitude $\mathcal{A}_{v}$, and consider a history with particular color (i.e. causal character) assignments
\begin{equation}
\label{vertex_color}
    \mathcal{A}_{v}(\uppsi)= \scalebox{0.4}{\cubevertex}\,,
\end{equation}
being here two opposing 2-cells with spacelike edges connected by timelike edges. 
The general strategy to take the semi-classical limit is to write the link functions in terms of complex exponentials weighted by the spins, such that stationary phase methods can be applied \cite[Th. 7.7.5]{Hormander2003}. Defining then
\begin{equation}
    \!{}^{n_{ab}}\overset{\rightarrow}{\scalebox{0.65}{\braketblack}}{}^{n_{ba}}= e^{S_{ab}^{\mathrm{tl}}}\,, \quad \!{}^{n_{ab}}\overset{\rightarrow}{\scalebox{0.65}{\brakettikz}}{}^{n_{ba}}=   \bra{l^+_{ab}}g_a^\dagger \sigma_3 g_b\ket{l^-_{ba}}^{-1}  e^{S_{ab}^{\mathrm{sl}}}\,,
\end{equation}
the model assigns the actions
\begin{equation}
\begin{gathered}
S_{ab}^{\mathrm{tl}}= 2k_{ab} \ln  \bra{\tau_{ab}}g_a^\dagger \sigma_3 g_b \ket{\tau_{ba}}\,, \\
S_{ab}^{\mathrm{sl}}=2is_{ab} \ln \bra{l^+_{ab}}g_a^\dagger \sigma_3 g_b \ket{l^-_{ab}} + s_{ab} \bra{l^+_{ab}}g_a^\dagger \sigma_3 g_b \ket{ l^+_{ab}}^2\,,
\end{gathered}
\end{equation}
to the spin-foam amplitude. The critical point equations $\sum_{ab} \delta_{g_a} S_{ab}=0$ and $\mathfrak{Re}\{ S_{ab}\}=0$ (the value at which it is maximal) can be shown to imply, for spacelike $ab$, 
\begin{equation}
\begin{gathered}
\label{gluesl}
g_b \ket{l^+_{ba}}=\vartheta_{ab} g_a \ket{l^+_{ab}} \\
g_b \ket{l^-_{ba}}=\vartheta_{ab}^{-1} g_a \ket{l^-_{ab}}
\end{gathered}\,, \quad \vartheta_{ab}\in \mathbb{R}^+\,,
\end{equation}
while for timelike $ab$
\begin{equation}
\begin{gathered}
\label{gluetl}
g_b \ket{\tau_{ba}}=\varrho_{ab} g_a \ket{\tau_{ab}} \\
g_b \ket{-\tau_{ba}}=\overline{\varrho}_{ab} g_a \ket{-\tau_{ab}}
\end{gathered}\,, \quad \varrho_{ab} \in e^{i \mathbb{R}}\,,
\end{equation}
together with a closure relation 
\begin{equation}
\label{eq:closure}
   \forall a\,, \quad  \sum_b^{ab\, \mathrm{tl}} - \mathfrak{o}_{ab} k_{ab}  v_{ab} +  \sum_b^{ab\, \mathrm{sl}} \mathfrak{o}_{ab} s_{ab} v_{ab} =0\,.
\end{equation}
In the equation above $v_{ab}$ stands for a 3-vector, whose definition depends on the coloring of the link $ab$. For a spacelike link,
\begin{equation}
\label{geo_vec_sl}
    v_{ab}:= \pi(n_{ab}) \hat{e}_2\in H^{\mathrm{sl}}\subset \mathbb{R}^{1,2} \,, \quad \bra{l^+_{ab}}\sigma_3 \varsigma^I \ket{l^-_{ab}}=i v_{ab}^I\,,
\end{equation}
while for a timelike link 
\begin{equation}
\label{geo_vec_tl}
    v_{ab}:= \tau_{ab} \pi(n_{ab}) \hat{e}_0\in H^\tau \subset \mathbb{R}^{1,2} \,, \quad \tau_{ab}\bra{\tau_{ab}}\sigma_3 \varsigma^I \ket{\tau_{ab}}=v_{ab}^I\,.
\end{equation}
The symbol $\varsigma^I=(\sigma_3, -i\sigma_2, i\sigma_1)^I$ denotes a tuple of Pauli matrices, and $\mathfrak{o}_{ab}$ is a sign which depends on the choice of orientation of the link $ab$: it is positive when the orientation is incoming at $n_{ab}$ and negative otherwise. The geometrical vectors are obtained from $\mathrm{SU}(1,1)$ group elements via the spin homomorphism
\begin{equation}
    \begin{gathered}
        \pi: \; \mathrm{SU}(1,1)\; \rightarrow \; \mathrm{SO}_0(1,2) \\
        g \sigma_\mu g^\dagger = \pi(g)^\nu_{\;\;\mu} \sigma_\nu\,, \quad\mu,\nu=0,1,2\,,
    \end{gathered}
\end{equation}
projecting down to the connected identity component of the Lorentz group. Note that Eq.~\eqref{eq:closure} implies that the critical points are characterized by (possibly skewed) 2-polygons at each of the six boxes, whose lengths are determined by the spin assignments. The data $n_{ab}\in \mathrm{SU}(1,1)$, in turn, are in correspondence with the edge vectors.

Further clarity can be achieved by following the algorithm of \cite{Dona:2017dvf, Dona:2020yao}, which allows for determining critical points of spin-foam amplitudes based on non-simplicial polytopes. The idea is to consider sets of three quadrilaterals $\Box_a,\Box_b,\Box_c$ which are pair-wise adjacent, so as to determine $g_a,g_b,g_c$ at criticality; one then finds two other quadrilaterals adjacent to one of the former, and reiterates the method until all critical points are identified. There are only two types of such sets in Eq.~\eqref{vertex_color}: either all quadrilaterals meet at spacelike edges (a case which reduces to what was studied in \cite{Simao:2024don}), or two quadrilaterals share a timelike edge. Consider then the latter, and assume the timelike edge is shared between  $\Box_a$ and $\Box_c$. The sets of equations \eqref{gluesl} and \eqref{gluetl} imply
\begin{equation}
\label{system}
    \begin{cases}
        g_a^{-1} g_b n_{ba} n_{ab}^{-1} = e^{-i \theta_{ab} v_{ab} \cdot \varsigma^\dagger} \\
        g_b^{-1} g_c n_{cb} n_{bc}^{-1} = e^{-i \theta_{bc} v_{bc} \cdot \varsigma^\dagger} \\
        g_a^{-1} g_c n_{ca} n_{ac}^{-1} = e^{-i \rho_{ac} v_{ac} \cdot \varsigma^\dagger} 
    \end{cases} \,,
\end{equation}
where we have introduced $\theta_{ab}:=\ln \vartheta_{ab}$ and $\rho_{ac}:=i\ln \varrho_{ac}$.\footnote{Recall that $|v_{ab}|^2=|v_{bc}|^2=-1$ and $|v_{ac}^2|=1$.} As in \cite{Dona:2017dvf, Dona:2020yao}, we make the explicit gauge choice of setting $n_{ab}=n_{ba}$ for all boundary data to simplify one set of solutions of Eqs.~\eqref{gluesl}--\eqref{eq:closure}. In complete analogy to the fully spacelike case of \cite{Simao:2024don}, straightforward - if tedious - algebra then yields the angle formulas
\begin{equation}
\label{eq:angle1}
  \rho_{ac}=0 \quad \vee \quad  \tan \rho_{ac} = \frac{v_{cb} \cdot v_{ab}\times v_{ac}}{(v_{ac}\times v_{cb})\cdot (v_{ab}\times v_{ac})}\,,
\end{equation}
\begin{equation}
\label{eq:angle2}
  \theta_{ab}=0 \quad \vee \quad  \tanh \theta_{ab} = \frac{v_{cb} \cdot v_{ac}\times v_{ab}}{(v_{ab}\times v_{cb})\cdot (v_{ac}\times v_{ab})}\,,
\end{equation}
\begin{equation}
\label{eq:angle3}
  \theta_{cb}=0 \quad \vee \quad  \tanh \theta_{cb} = \frac{v_{ac} \cdot v_{ab}\times v_{cb}}{(v_{ac}\times v_{cb})\cdot (v_{ab}\times v_{cb})}\,,
\end{equation}
involving Minkowski vector $\times$ and scalar $\cdot$ products. The angle associated to a spacelike edge lies in the corresponding orthogonal plane which is isomorphic to $\R^{1,1}$. As a result, the angle $\theta_{ab}$ is Lorentzian and thus defined via a tangent hyperbolic. In contrast, the angle associated to a timelike edge lies in the corresponding orthogonal plane which is isomorphic to $\R^2$. Thus, the angle formula for $\rho_{ac}$ contains a trigonometric tangent. Note moreover that the equations yield two sectors of solutions: if some angle is zero then all remaining ones must vanish as well as per Eq.~\eqref{system}; this propagates to every other set of three quadrilaterals in the cuboid, as can be seen from equations analogous to Eq.~\eqref{system} for all other sets, and all critical group elements are identified with the identity $g_a = \one$. As to the second sector, once all angles $\theta,\rho$ are determined (provided the relevant equations admit solutions, see \cite{Jercher:2024kig}), one may resort to equations of the type \eqref{system} to determine all critical $g_a$.

Putting everything together, the asymptotic amplitude for an arbitrary assignment of causal characters reads
\begin{equation}
\label{as1}
    \mathcal{A}_{v} = e^{\frac{7i\pi}{4}}\frac{2^{\Delta_{\mathrm{tl}}-10}}{(2\pi)^{5/2}} \left(\frac{1}{\sqrt{\det H_{\one}}} + \Theta \frac{e^{2 i\sum_{ab}^{\mathrm{sl}} s_{ab} \theta_{ab}  + 2i \sum_{ab}^{\mathrm{tl}} (-k_{ab}) \rho_{ab}}}{\sqrt{\det H_{\vartheta}} \prod_{ab}^{\mathrm{sl}} \vartheta_{ab} } \right) + \mathcal{O}\left(j^{\frac{11}{2}}\right) \,,
\end{equation}
where $\Theta=0,1$ is a binary toggle for the second sector of solutions: if the boundary data is such that the quadrilaterals are either all timelike or all spacelike, then $\Theta=1$. The numerical factors heading the equation are obtained from 1) Hörmander's theorem \cite[Th. 7.7.5]{Hormander2003} and 2) a spin redundancy ($g_a \mapsto -g_a$) of factor $2$ depending on the number $\Delta_{\mathrm{tl}}$ of non-gauge-fixed squares with entirely timelike edges. The matrices appearing in the asymptotic formula are the Hessian matrices of the total spin-foam action $S=\sum_{ab}^\mathrm{tl} S_{ab}^{\mathrm{tl}}+\sum_{ab}^\mathrm{sl} S_{ab}^{\mathrm{sl}}$ evaluated at the two critical points; we denote by $H_{\one}$ the Hessian  at all $g_a=\pm \one$, and by $H_\vartheta$ the Hessian at the non-trivial critical point. The Hessian matrices are block matrices where each component $H^{ab}$ is a $3\times 3$ matrix, the form of which depends on the causal character assigned to the edge $\diagup_{ab}$, and where $g_6=\one$ has been gauge-fixed. The diagonal blocks read
\begin{equation}\label{eq:diag Hess}
H_{IJ}^{aa}=-\sum_b^{ab\, \mathrm{tl}} \frac{k_{ab}}{2}\left[\eta_{IJ} -v_{ab,I}^{(a)} v_{ab,J}^{(a)} \right]  -\sum_b^{ab\, \mathrm{sl}}  \frac{i s_{ab}}{2}\left[\eta_{IJ} +v_{ab,I}^{(a)} v_{ab,J}^{(a)}-i\vartheta_{ab}^2 m_{ab,I}^{(a)} m_{ab,J}^{(a)} \right]\,,
\end{equation}
while for the off-diagonal $a\neq b$ blocks we have
\begin{equation}
    {}^{\mathrm{tl}}H_{IJ}^{ab}=\frac{k_{ab}}{2}\left[\eta_{IJ} -v_{ab,I}^{(a)} v_{ab,J}^{(a)} -i \epsilon_{IJK} \mathfrak{o}_{ab} \eta^{KL} v_{ab,L}^{(a)}\right]\,,
\end{equation}
\begin{equation}\label{eq:off-diag sl Hess}
       {}^{\mathrm{sl}}H_{IJ}^{ab}=\frac{i s_{ab}}{2}\left[\eta_{IJ} +v_{ab,I}^{(a)} v_{ab,J}^{(a)} + \epsilon_{IJK}  \mathfrak{o}_{ab} \eta^{KL} v_{ab,L}^{(a)}-i\vartheta_{ab}^2 m_{ab,I}^{(a)} m_{ab,J}^{(a)} \right]\,.
\end{equation}
In the equations above we have denoted $v_{ab,I}^{(a)}:= \left[\pi(g_a) v_{ab}\right]_I$ for simplicity, and introduced the future null vectors
\begin{equation}
m_{ab}=\pi(n_{ab})(\hat{e}_0-\hat{e}_1) \in C^+ \subset \mathbb{R}^{1,2}\,, \quad \bra{l^+_{ab}}\sigma_3 \varsigma^I\ket{l^+_{ab}}=m_{ab}^I\,, \quad ab \;\text{sl}\,,
\end{equation}
constructed from the boundary data associated to spacelike edges.

The exponential function of Eq.~\eqref{as1} is a global phase. On the other hand, note from Eqs.~\eqref{geo_vec_sl} and~\eqref{geo_vec_tl} that the geometrical vectors determined by the boundary data $\Upphi$ are oblivious to a phase change at each coherent state. A precedent has therefore appeared in the literature \cite{Barrett:2009gg, Barrett:2009mw, Kaminski:2017eew, Dona:2017dvf} where the phase of each coherent state is fixed (given the global information of the vertex amplitude) in order to bring the asymptotic amplitude into a more symmetric form. Proceeding as such, under a concrete choice of phase 
\begin{equation}
    \begin{cases}
        \ket{\tau_{ab}} \mapsto e^{i k_{ab} \rho_{ab}} \ket{\tau_{ab}} \\
        \ket{l^-_{ab}} \mapsto e^{-i s_{ab} \theta_{ab}} \ket{l^-_{ab}}\,,
    \end{cases}
\end{equation}
the first term of Eq.~\eqref{as1} becomes
\begin{equation}\label{eq:vertex semi-classics}
    \mathcal{A}_{v}^{\mathrm{asy}} = e^{\frac{7i\pi}{4}}\frac{2^{\Delta_{\mathrm{tl}}-10}}{(2\pi)^{5/2}}\Biggl(\frac{e^{ -i\sum_{ab}^{\mathrm{sl}} s_{ab} \theta_{ab}  - i \sum_{ab}^{\mathrm{tl}} (-k_{ab}) \rho_{ab}}}{\sqrt{\det H_{\one}}}  + \Theta \frac{e^{ i\sum_{ab}^{\mathrm{sl}} s_{ab} \theta_{ab}  + i \sum_{ab}^{\mathrm{tl}} (-k_{ab}) \rho_{ab}}}{\sqrt{\det H_{\vartheta}} \prod_{ab}^{\mathrm{sl}} \vartheta_{ab} } \Biggr) \,.
\end{equation}
The above expression captures the well-known result that spin-foam asymptotics tend to reproduce the cosine of the boundary Regge action \cite{Barrett:2009gg, Barrett:2009mw, Kaminski:2017eew, Liu:2018gfc, Simao:2021qno, Dona:2017dvf, Dona:2020yao}. Here, we note the additional property of the current model that the second term of the cosine may be absent depending on the causal structure, which has been shown in~\cite{Jercher:2024kig}. 

\subsection{Spin-foam amplitude simplifications}\label{app:Spin-foam amplitude simplifications}

Given the cuboidal lattice $\mathcal{X}^{(3)}$ and some a priori unrestricted history $\uppsi$, the (2+1) coherent model prescribes a spin-foam amplitude of the form
\begin{equation}
     \mathcal{A}(\mathcal{X}^{(3)}, \uppsi)= \scalebox{0.4}{\cubethree}\,,
\end{equation}
where all open lines are taken to be joined to adjacent diagrams. There is one hexagonal diagram per 3-cell in $\mathcal{X}^{(3)}$. An equivalent form of the amplitude can be obtained by making use of a completeness relation of coherent states on the Hilbert spaces of the relevant representations,
\begin{equation}
    \label{eq:restl}
       \one_{k(q)} = (-2k-1) \int \mathrm{d} g \; \vb*{D}^{k (q)} (g) \ket{k, -qk}\bra{k, -qk} \vb*{D}^{ k (q) \dagger}(g) =: d_k\int \mathrm{d}g \raisebox{1pt}{\scalebox{0.7}{\restime}}\,,
\end{equation}
\begin{equation}
\label{eq:ressl}
   \one_{j(\delta)} = s \tanh (\pi s)^{1-4\delta}\int \mathrm{d} g \; \vb*{D}^{j (\delta)} (g) \ket{j, ij, 0}\bra{j, \overline{ij}, 0} \vb*{D}^{ j (\delta) \dagger}(g) =:  d_j\int \mathrm{d}g \raisebox{1pt}{\scalebox{0.7}{\resspace}}\,,
\end{equation}
which follow from the orthonormality of $\mathrm{SU}(1,1)$ Wigner matrices in $L^2(SU(1,1))$, discussed in Appendix~\ref{app:su11}. Consequently, any adjacent diagrams can be rewritten as 
\begin{equation}
\label{glue}
     \scalebox{0.4}{\cubetwo} = \prod_{i=1}^4 d_{j_i}\int\mathrm{d}g_i \scalebox{0.4}{\cubetwores}\,,
\end{equation}
having made in this example an explicit assumption on the causal character of the joined lines - namely that they are timelike, and that the appropriate resolution is $d_k \int \mathrm{d}g \raisebox{1pt}{\scalebox{0.7}{\restime}}$. Furthermore, recall that each coherent state appearing in Eqs.~\eqref{eq:restl} and~\eqref{eq:ressl} admits a geometrical interpretation: note that \cite{Simao:2024don}
\begin{equation}
    \bra{j, \overline{ij}, 0} \vb*{D}^{ j (\delta) \dagger}(g)\, \vb*{D}^{ j (\delta)}( \varsigma^I)\,  \vb*{D}^{ j (\delta)}( g)\ket{j, ij, 0 } = - \frac{\gamma}{\pi} \bra{l^+}g^\dagger \sigma_3 \varsigma^I g\ket{ l^-}^{2j}\,, 
\end{equation}
\begin{equation}
    \bra{k, -qk} \vb*{D}^{ k (q) \dagger}(g)\, \vb*{D}^{ k (q)}( \varsigma^I)\,  \vb*{D}^{ k (q)}( g) \ket{ k, -qk } =  \bra{-q}g^\dagger \sigma_3 \varsigma^I g\ket{-q}^{2k}\,, 
\end{equation}
and Eqs.~\eqref{geo_vec_sl} and~\eqref{geo_vec_tl} show that the right-hand side above relates to vector components of 3-vectors in either the spacelike or timelike hyperboloids, respectively. One is thus justified in thinking of Eq.~\eqref{glue} as a gluing identity between vertices, where the integrations range over all possible geometric vectors assigned to the boundary of each diagram. 

The first step in our simplification leverages the geometrical interpretation of the coherent states, and enforces a particular boundary geometry. The process consists in performing a symmetry reduction, by fixing the integration domain at each gluing to a particular group element, and consequently a particular coherent state, in correspondence with the geometry of a 3-frustum. In other words, we insist on complementing each history $\uppsi \mapsto \uppsi'$ with coherent states at the boundary of each diagram, and operate the reduction
\begin{equation}
    \mathcal{A}(\mathcal{X}^{(3)},\uppsi) =  \prod_{i=1}^4 d_{k_i}\int \mathrm{d}g_i \scalebox{0.4}{\cubetwores}\; \mapsto\;  \hat{\mathcal{A}}_1(\mathcal{X}^{(3)},\uppsi')
    :=  \scalebox{0.4}{\cubetwosplit}\,,
\end{equation}
arriving at a first modified amplitude $\hat{\mathcal{A}}_1$.\footnote{Notice that we also dropped the Plancherel factors $d_j$ and $d_k$, corresponding to a modification of the face amplitude of the model. In~\cite{Bahr:2015gxa,Bahr:2017bn} and Chapter~\ref{chapter:specdim}, different choices of face amplitudes were parametrized by a parameter $\alpha$ and cuboid and frustum intertwiners have been introduced with a normalization factor modifying the edge amplitude $\mathcal{A}_e$. Here, we effectively set $\mathcal{A}_e = 1$ and introduce Plancherel factors only for closed loops. For highly oscillating amplitudes in one variable as considered in Secs.~\ref{sec:Numerical evaluation: Strut in the bulk}, modified edge and face amplitudes are expected to have a negligible influence on the qualitative behavior of expectation values. They can however influence the numerical stability of series accelerations and numerical integration. The choices made here are going to prove numerically feasible.} The structure of this new amplitude is such that it factorizes into a product of frusta amplitudes, each of which is of a similar structure to that of the original vertex amplitude $\mathcal{A}_v$ of Eq.~\eqref{cube_vertex}. The amplitudes $\hat{\mathcal{A}}_1$ and $\mathcal{A}_v$ are however not strictly equivalent. The timelike coherent states of the identity resolution satisfy
\begin{equation}
 \!{}^{n_{ab}}\overset{\rightarrow}{\scalebox{0.65}{\braketsquare}}{}^{n_{ba}} = \!{}^{n_{ab}}\overset{\rightarrow}{\scalebox{0.65}{\braketblack}}{}^{n_{ba}}\,,
\end{equation}
by the definitions in Eqs.~\eqref{brakettl} and~\eqref{eq:restl}. In particular, the pairings of Eq.~\eqref{brakettl} do not come with an additional constraint $\mathcal{C}$, as gluing is implicitly ensured~\cite{Simao:2024don}. The same is not true for spacelike pairings: the closure constraint $\mathcal{C}_{n_{ab},n_{ba}}$ of Eq.~\eqref{braketsl} is missing, and it is only that
\begin{equation}
 \!{}^{n_{ab}}\overset{\rightarrow}{\scalebox{0.65}{\braketsquareblack}}{}^{n_{ba}} = \!{}^{n_{ab}}\overset{\rightarrow}{\scalebox{0.65}{\brakettikz}}{}^{n_{ba}}  \,.
\end{equation}
Following~\cite{Simao:2024don}, the constraint $\mathcal{C}_{n_{ab},n_{ba}}$ must be added by hand in order for the vertex amplitude to have a well-behaved asymptotic formula. Hence, we perform another modification to the amplitude by including the constraint on every vertex, 
\begin{equation}
     \hat{\mathcal{A}}_1(\mathcal{X}^{(3)},\uppsi') =  \scalebox{0.4}{\cubetwosplit}\; \mapsto\;   \hat{\mathcal{A}}_2 ( \mathcal{X}^{(3)},\uppsi')
    :=  \scalebox{0.4}{\cubetwosplitorig}\,,
\end{equation}
such that the amplitude $ \hat{\mathcal{A}}_2 (\mathcal{X}^{(3)},\uppsi')$ amounts to a simple product of the vertex amplitude $\mathcal{A}_v$ over every frustum in $\mathcal{X}$, i.e. 
\begin{equation}
     \hat{\mathcal{A}}_2 (\mathcal{X}^{(3)},\uppsi') = \prod_{{\cube} \in \mathcal{X}}  \mathcal{A}_v(\uppsi'|_{\cube}) \prod_{\mathrm{bulk }\diagup \in \mathcal{X}}  \mathcal{A}_f(\uppsi'|_\diagup)\,,
\end{equation}
where $ \mathcal{A}_f$ is the face amplitude associated to bulk edges $\diagup \in \mathcal{X}^{(3)}$. It corresponds to the Plancherel measure appearing in Eqs.~\eqref{brakettl} and~\eqref{braketsl}, depending on the history, as
\begin{equation}\label{eq:face amplitudes}
     \mathcal{A}_f = \begin{cases}
        -2k-1\,, \quad \diagup \text{ tl} \\
        s \tanh(\pi s )\,, \quad \diagup \text{ sl}
    \end{cases}\,.
\end{equation}

As it stands, the amplitude $ \hat{\mathcal{A}}_2$ does not yet single out a frustum geometry. The interpretation of the spin-foam vertex amplitude in terms of a geometrical polyhedron follows from the semi-classical limit, where only histories derived from convex polyhedra dominate. Yet another simplification step is therefore to replace the amplitude $ \hat{\mathcal{A}}_2$ with the semi-classical formula at each frustum, evaluating it on the geometrical data of a 3-frustum. That is,
\begin{equation}
\label{eq:a3}
     \hat{\mathcal{A}}_2(\uppsi', \mathcal{X}^{(3)}) \; \mapsto \;  \hat{\mathcal{A}}^{\mathrm{grav}} (\uppsi', \mathcal{X}^{(3)}) = \prod_{{\cube} \in \mathcal{X}} \mathcal{A}^{\mathrm{asy}}_{v}(\uppsi'|_{\cube}) \prod_{\mathrm{bulk }\diagup \in \mathcal{X}}  \mathcal{A}_f(\uppsi'|_\diagup)\,,
\end{equation}
with $\mathcal{A}^{\mathrm{asy}}_{v}$ the semi-classical amplitude of a 3-frustum, characterized in the next section. The amplitude $\mathcal{A}^{\mathrm{grav}}$ is the gravitational part of the full effective amplitude utilized in Chapter~\ref{chapter:3d cosmology}, anticipating an additional scalar field coupling, described in Sec.~\ref{sec:Minimally coupled massive scalar field}.  

\begin{figure}
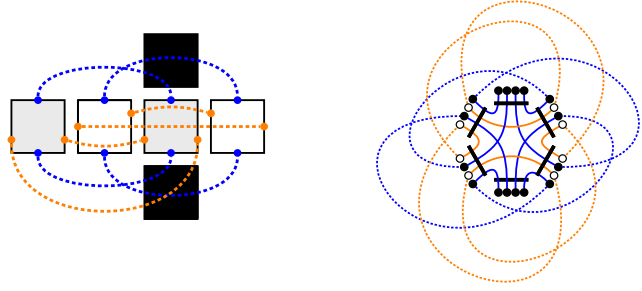

    \centering
     \scalebox{0.7}{\unfold} \scalebox{0.45}{\eldrich}
    \vspace{-1cm} \caption{Left: \enquote{unfolded} Lorentzian cuboid with spacelike top and bottom faces (black) and four faces of arbitrary signature (gray and white). Edge identifications necessary to fold the boundary back into a cuboid are not explicitly drawn and understood implicitly. To obtain a spatial $T^2$ toroidal topology, similarly-colored white and gray faces are identified, as are their respective edges. The orange-colored edge identifications lead to a single bulk edge. Blue-colored edge identifications readily induce a toroidal topology of the black faces. Right: representation of edge identifications of the left diagram in the amplitude diagram. One can verify that the orange line consists of a single loop. There are four open blue lines, two at the top and two at the bottom spacelike face. These correspond to the initial and final tori each of which is characterized by two radii.} 
    \label{fig:torus}
\end{figure}

For the remainder, we additionally assume a toroidal spatial topology. Due to spatial homogeneity, multiple building blocks in spatial direction would simply enter as additional powers of the amplitude $\mathcal{A}_v^{\mathrm{asy}}$, as discussed in Chapters~\ref{chapter:specdim} and~\ref{chapter:LRC}. The toroidal topology is accomplished by appropriately identifying faces and edges of a 3-frustm, described and depicted in detail in Fig.~\ref{fig:torus}. As a result, the polyhedral complexes $\mathcal{X}^{(3)}_{\mathcal{V}}$ considered are given by $\mathcal{V}$ frusta-like 3-cells, organized in a linear chain along the temporal direction.

\subsection{Identifying boundary data with Lorentzian 3-frusta}\label{app:Identifying boundary data with Lorentzian 3-frusta}

In order to specify the spin-foam amplitude $\hat{\mathcal{A}}_3$ to the setting of 3-dimensional spatially flat Lorentzian cosmology with toroidal topology, we prescribe here the necessary choice of boundary data. 

One first needs to appropriately identify the geometric data of a 3-frustum $(l_0,l_1,m)$ with the boundary data $\Phi = \{t_{ab},n_{ab}\}$, consisting of spins $t_{ab}$ (standing collectively for spins of both continuous $j$ and discrete $k$ types) and group elements $n_{ab}\in\SUO$. The spins $t_{ab}$, which lie in the discrete (continuous) series, and are associated to the timelike (spacelike) 
edges $(ab)$, can be semi-classically identified with the geometrical edge length.\footnote{The Casimir spectrum of the continuous series of $\mathrm{SU}(1,1)$ representations is given by $s^2+\frac{1}{4}$ with $s\in\R$ and thus exhibits a length gap. Different identifications of spacelike geometrical edge length and $s$ are possible, and we choose here $l = s+\frac{1}{2}$ with a real off-set to map the length gap. As we are going to detail in Sec.~\ref{sec:Evaluation of the partition function}, this choice yields an effective amplitude finite in the bulk spatial edge length.} Thus, for edges $(ab)$ of squares, we identify the spins of the continuous series $s_{ab}$ with the length $l$ as $l = s_{ab}+\frac{1}{2}$. For timelike struts, we identify $-k_{ab}$ with the strut length $m$, while for spacelike struts, we set $m=s_{ab}+\frac{1}{2}$. 

As a consequence of the symmetry reduction, the normalized geometric edge vectors $v_{ab}^{(a)}$, which determine the group elements $n_{ab}$ up to a phase, are a function of the edge lengths $(l_0,l_1,m)$. Choosing an embedding of the 3-frustum into $\R^{1,2}\ni(t,x,y)$ where the squares lie in constant-$t$ planes with edges parallel to the $x$ and $y$ directions, the edge vectors of squares are given by $(0,\pm 1, 0)$ or $(0,0,\pm 1)$. Strut edge vectors take the form
\begin{equation}\label{eq:strut vec}
v_{ab} = \frac{1}{m}\left(\epsilon^0_{ab}\sqrt{\frac{(l_0-l_1)^2}{2}+\vb*{m}^2},\epsilon^1_{ab}\frac{l_0-l_1}{2},\epsilon^2_{ab}\frac{l_0 - l_1}{2}\right).
\end{equation}
Here, the signs $\epsilon^I_{ab} =\pm$ are chosen such that the vectors $v_{ab}$ and $v_{ba}$ are parallel and closure holds as in Eq.~\eqref{eq:closure}. The $0$-component of the strut vectors correspond to the height of the 3-frustum which follows immediately from the chosen embedding.

For such chosen boundary data, let us summarize the squared volumes of the (sub-)cells of a (2+1) frustum.

\paragraph{Height.} The squared height of $3$-frusta is given by
\begin{equation}
    \vb*{H}^2 = \vb*{m}^2+\frac{(l_0-l_1)^2}{2}\,.
\end{equation}
Embeddability of a $3$-frustum in $\R^{1,2}$ requires $\vb*{H}^2 > 0$ and thus poses the condition $\vb*{m}^2 > -\frac{(l_0-l_1)^2}{2}$. Configurations that violate this bound correspond to Euclidean $3$-frusta.

\paragraph{Trapezoids.} The squared height of trapezoids is given by
\begin{equation}
\vb*{h}^2 = \vb*{H}^2-\frac{(l_0-l_1)^2}{4} = \vb*{m}^2 +\frac{(l_0-l_1)^2}{4}\,,
\end{equation}
from which the squared area of trapezoids follows as
\begin{equation}
\vb*{v}^2 = \left(\frac{l_0+l_1}{2}\right)^2 \vb*{h}^2 = \left(\frac{l_0+l_1}{2}\right)^2\left[\vb*{m}^2 +\frac{(l_0-l_1)^2}{4}\right]\,.
\end{equation}
A trapezoid is therefore spacelike if $-\frac{(l_0-l_1)^2}{2} < \vb*{m}^2 < -\frac{(l_0-l_1)^2}{4}$, and timelike otherwise.

\paragraph{3-volume.} The squared volume of a 3-frustum is given by
\begin{equation}\label{eq:3-volume}
\vb*{V}^2 =\left(\frac{l_0^2+l_0 l_1+l_1^2}{3}\right)^2\vb*{H}^2 = \left(\frac{l_0^2+l_0 l_1+l_1^2}{3}\right)^2\left[\vb*{m}^2+\frac{(l_0-l_1)^2}{2}\right]\,,
\end{equation}
which is positive if the 3-frustum is Lorentzian, i.e. if it can be embedded into $\R^{1,2}$. 

As detailed in Sec.~\ref{sec:Lorentzian 3-frusta}, the presence of different causal characters of subcells depending on the ratio $(l_0-l_1)^2/m_0^2$ suggests dividing the theory into different sectors, similar to Chapter~\ref{chapter:LRC}. Sectors I, II and III are defined by the inequalities $-(l_0-l_1)^2/2 < \vb*{m}^2 < -(l_0-l_1)^2/4$, $-(l_0-l_1)^2/4 < \vb*{m}^2 < 0$, and $\vb*{m}^2 > 0$, respectively.

\subsection{Asymptotic vertex amplitude and measure factors}\label{sec:Asymptotic vertex amplitude and measure factors}

The semi-classical amplitude evaluated on 3-frusta geometry amounts to
%
\begin{equation}\label{eq:vertex semi-classics 2}
 \mathcal{A}_v^{\mathrm{asy}} = \left(\upmu_{\one}(l_0,l_1,m)\e^{-i\,\mathfrak{Re}\{S_{\mathrm{R}}\}}+\Theta\,\upmu_{\vartheta}(l_0,l_1,m)\e^{i\,\mathfrak{Re}\{S_{\mathrm{R}}\}}\right) \,,
\end{equation}
where $\Theta$ is one in Sector I and vanishes in II and III. The phases $\mathfrak{Re} \{S_{\mathrm{R}}\}$ contain the Lorentzian Regge action, taking over the three sectors the values\footnote{Strictly speaking, the identification of spatial edge length $l = s+\frac{1}{2}$ introduces a relative phase of the exponents. The effects of such a shift are however negligigble, in particular since the edge length are assumed to be large for the asymptotic formula to hold.}
\begin{equation}\label{eq:S}
\begin{aligned}
S_{\textsc{i}} &= 4\abs{l_0-l_1}\left[i\frac{\pi}{2}-\cosh^{-1}\left(\frac{l_0-l_1}{\sqrt{4m^2-(l_0-l_1)^2}}\right)\right]-4m\left[i\frac{\pi}{2}-\cosh^{-1}\left(\frac{(l_0-l_1)^2}{4m^2-(l_0-l_1)^2}\right)\right]\,,\\
S_{\textsc{ii}} &= 4(l_0-l_1)\sinh^{-1}\left(\frac{l_1-l_0}{\sqrt{4m^2-(l_0-l_1)^2}}\right)+4m\left[i\frac{\pi}{2}+\cosh^{-1}\left(\frac{(l_0-l_1)^2}{4m^2-(l_0-l_1)^2}\right)\right]\,,\\
S_{\textsc{iii}} &= 4(l_0-l_1)\sinh^{-1}\left(\frac{l_1-l_0}{\sqrt{4m^2-(l_0-l_1)^2}}\right)+4m\left[\frac{\pi}{2}-\cos^{-1}\left(\frac{(l_0-l_1)^2}{4m^2-(l_0-l_1)^2}\right)\right]\,,
\end{aligned}
\end{equation}
according to Eq.~\eqref{eq:vertex semi-classics} and the angle formulas in Eqs.~\eqref{eq:angle1}--\eqref{eq:angle3}. Importantly, only the real part of the Lorentzian Regge action enters $\mathcal{A}_v^{\mathrm{asy}}$ which is a result of the stationary phase approximation of the full spin-foam quantum amplitude. 


The functions $\upmu_{\one,\vartheta}$ constitute measure factors arising from the spin-foam asymptotics and consist of the inverse square root of the Hessian determinant and factors of $\vartheta$, given as exponentials of dihedrals angles, at the non-identity solution of the critical points,
\begin{equation}
\upmu_{\one} = e^{\frac{7i\pi}{4}}\frac{2^{-10}}{(2\pi)^{5/2}} 
\frac{1}{\sqrt{\det H_{\one}}}\,,\qquad \upmu_{\vartheta} = e^{\frac{7i\pi}{4}}\frac{2^{-10}}{(2\pi)^{5/2}}
\frac{1}{\vartheta_m^4 \sqrt{\det H_{\vartheta}} }\,,
\end{equation}
where $\vartheta_m$ is associated to the struts. 
Since $\vartheta = 1$ at the identity solution $g_a = \one$, the $15\times 15$-matrix $H_\one$ simplifies significantly with many entries being zero. As a consequence, the determinant $\det H_{\one}$ can be computed explicitly, with its functional form given by
\begin{equation}\label{eq:det H 1}
\det H_{\one} = \frac{1}{m^4}(l_0l_1)^3\sum_{\substack{n_0,n_1,n_m,n_s \\ n_0+n_1+n_m+n_s = 13}}c^{\one}_{n_0,n_1,n_m,n_s}l_0^{n_0} l_1^{n_1} m^{n_m} \left(\sqrt{\frac{(l_0-l_1)^2}{2}+\vb*{m}^2}\right)^{n_s}\,,
\end{equation}
where $c^{\one}_{n_0,n_1,n_m,n_s}$ are constant complex coefficients. At the non-identity solution $\vartheta\neq 1$, the Hessian $H_\vartheta$ has less zero entries compared to $H_\one$ which furthermore take a more involved form. As a result, we have not been able to find an analytical formula for $\det H_{\vartheta}$. Still, one can give the functional form also for $\det H_{\vartheta}$, which is now
\begin{equation}\label{eq:det H vartheta}
\det H_\vartheta = \frac{1}{m^4}(l_0l_1)^3\sum_{\substack{n_0,n_1,n_m,n_s \\ n_0+n_1+n_m+n_s = 13}}c^\vartheta_{n_0,n_1,n_m,n_s}l_0^{n_0} l_1^{n_1} m^{n_m} \left(\sqrt{\frac{(l_0-l_1)^2}{2}+\vb*{m}^2}\right)^{n_s}\,.
\end{equation}
In contrast to the case before, the $c^\vartheta_{n_0,n_1,n_m,n_s}$ are now not constants anymore but complex-valued scale-invariant functions of the variables $l_0,l_1$ and $m$.
%
%

This concludes the characterization of the gravitational part of the effective amplitude $\hat{\mathcal{A}}^{\mathrm{grav}}$. In Sec.~\ref{sec:Minimally coupled massive scalar field}, an additional massless scalar field is minimally coupled which completes the effective model employed in Secs.~\ref{sec:Numerical evaluation: Strut in the bulk} and~\ref{sec:1slice}.

\section{Wynn's algorithm for convergence acceleration}\label{app:Wynn}

Let us introduce here the Shanks transform and Wynn's $\epsilon$-algorithm which have been applied to effective spin-foams\footnote{A closely related series convergence acceleration, known as Aitken's $\Delta^2$-process~\cite{Weniger2003} has been applied to infinite bulk variable summations in~\cite{Dona:2023myv}.}  in~\cite{Dittrich:2023rcr}. In Secs.~\ref{sec:Numerical evaluation: Strut in the bulk} and~\ref{sec:1slice} of the main body, we will frequently encounter unbounded sums of the form
\begin{equation}
\mathfrak{S} = \lim\limits_{N\rightarrow\infty}\mathfrak{S}_N = \lim\limits_{N\rightarrow\infty}\sum_{j = 1}^{N}a_j\,,
\end{equation}
for some complex sequence $\{a_j\}$. Assuming the partial sum $\mathfrak{S}_N$ to be known and of the form~\cite{Weniger2003}
\begin{equation}
\mathfrak{S}_N = \mathfrak{S} + \sum_{j=0}^{k-1}c_j\lambda_j^N\,,
\end{equation}
with $c_j$ coefficients and the $1 > \abs{\lambda_0} > \dots > \abs{\lambda_{k-1}}$ referred to as transients, there are $2k+1$ unknowns given by the limiting value $\mathfrak{S}$, the $c_j$ and the $\lambda_j$. In order to solve for $\mathfrak{S}$, one utilizes $2k+1$ consecutive sequence values $\mathfrak{S}_N,\mathfrak{S}_{N+1},\dots,\mathfrak{S}_{N+2k}$ to obtain the $k$th Shanks transform~\cite{Schmidt1941,Shanks1955} $e_k(\mathfrak{S}_N)$ as a ratio of determinants. For sufficiently large $N$, the Shanks transform $e_k(\mathfrak{S}_N)$ approximates $\mathfrak{S}$ faster than taking the limit $N\rightarrow \infty$ of the partial sums $\mathfrak{S}_N$.

In~\cite{Wynn1956}, Wynn introduced a non-linear recursive relation to efficiently compute the Shanks transform, commonly referred to as $\epsilon$-algorithm. Given the partial sums $\mathfrak{S}_1,\dots,\mathfrak{S}_N$, define~\cite{Weniger2003}
\begin{equation}
\epsilon_{-1}^{(N)} = 0\,,\qquad \epsilon_0^{(N)} = \mathfrak{S}_N\,.
\end{equation}
Then, $\epsilon$'s of higher $k$ are obtained via the relation
\begin{equation}
\epsilon_{k+1}^{(N)} = \epsilon_{k-1}^{(N+1)} + \frac{1}{\epsilon_k^{(N+1)}-\epsilon_k^{(N)}}\,.
\end{equation}
Wynn has shown~\cite{Wynn1956} that $\epsilon^{(N)}_{2k} = e_k(\mathfrak{S}_N)$ thus providing a fast algorithm for computing the $k$th Shanks transform of a sequence $\mathfrak{S}_N$.\footnote{A Mathematica algorithm for convergence acceleration via Wynn's method can be found in the repository \href{https://resources.wolframcloud.com/FunctionRepository/resources/SequenceLimit/}{SequenceLimit}. An implementation in \textsc{Julia} can be found in \href{https://github.com/Jercheal/3d-cosmology/blob/master/wynn.jl}{https://github.com/Jercheal/3d-cosmology}.}


The algorithm can in principle also be applied to sequences of more than one variable. Let $\{a_{j_1\dots j_m}\}$ be such a sequence for which we would like to compute the sum
\begin{equation}
\mathfrak{S} = \sum_{j_1=1}^\infty\dots\sum_{j_m=1}^\infty a_{j_1\dots j_m}.
\end{equation}
To apply the algorithm above to this sum choose a common cutoff of the indices $j_1,\dots,j_m$, defining a subset $J_N\subset \mathbb{N}^m$. There are different schemes to implement this cutoff, for instance the restriction that every $j_i\leq N$ or $j_1+\dots j_m\leq N$. While the choice of cutoff scheme does not matter in the limit of $N\rightarrow\infty$, it can be expected to be relevant for finite $N$. Given one choice of such $J_N$, define the sum
\begin{equation}
\mathfrak{S}_N = \sum_{(j_1,\dots,j_m)\in J_N}a_{j_1\dots j_m}\,,
\end{equation}
to which Wynn's algorithm can be applied. In Sec.~\ref{sec:1slice}, we successfully use this algorithm to accelerate the convergence of a  sequences with two indices. 
\chapter[\textsc{Correlation Functions and Explicit Expressions of} $\chi$]{Correlation Functions with Geometricity Constraints and Explicit Expressions of $\pmb{\chi}$}\label{app:Correlation functions with geometricity constraints}

\section{Derivation of the correlation function}\label{app:A derivation of the correlation function}

In this appendix, we derive a computable expression for the correlation function studied in Chapter~\ref{chapter:LG} of the main body. Challenges for this derivation, arising from the structure of the complete BC GFT model (introduced in Chapter~\ref{chapter:cBC}), are given by 1) the presence of multiple fields $\varphi_\alpha$, 2) the presence of spacelike and timelike terms in the spin representation, and 3) the geometricity constraints.

The fluctuations $\delta\varphi_\alpha$ with the normal integrated out satisfy the equation
\begin{equation}
\sum_\beta\int\dd{\vbg'}\dd{\vbf'}G_{\alpha\beta}(\vb*{g},\vb*{\phi};\vb*{g}',\vb*{\phi}')\delta\varphi_\beta(\vb*{g}',\vb*{\phi}')=0\,,
\end{equation}
with $G_{\alpha\beta}$ being the modified propagator composed of the kinetic kernel and the Hessian of the linearized interaction. It is important for further analysis to study the symmetries of the effective kernel as a bi-local function of the geometric and matter variables.

The effective kinetic kernel reduces to a function of the absolute value of scalar field differences, $\abs{\vbf-\vbf'}$. Concerning the geometric variables, notice that the matrix $\chi_{\alpha\beta}(\vbg,\vbg')$ entering the Hessian is either constant or contains $\delta$-functions on $g$ and $g'$ as detailed in the next section. Similarly, the Laplace operator $\Delta_{\SL}$ is invariant under left and right translation. Since we assume in addition that the functions $Z^\phi_\alpha(\vbg,\vbg')$ only depend on the trace of $g_c^{-1}g_c$, the effective kinetic kernel $G_{\alpha\beta}$ is invariant under the simultaneous left and right group action of $\SL$ acting on both arguments, i.e. for all $\vb*{a}\in\SL^4$ and $\vb*{b}\in\SL^4$, we have
\begin{equation}
G_{\alpha\beta}(\vbg,\vbf;\vbg',\vbf') = G_{\alpha\beta}(\vb*{a}\vbg\vb*{b},\vbf;\vb*{a}\vbg'\vb*{b},\vbf')\,.
\end{equation}
Consequently, $G_{\alpha\beta}$ only depends on the trace of $g_c^{-1}g_c'$, such that one can write
\begin{equation}\label{eq:2point G}
G_{\alpha\beta}(\vbg,\vbf;\vbg',\vbf') = G_{\alpha\beta}(\vbg^{-1}\vbg',\abs{\vbf-\vbf'})\,.
\end{equation}
This turns $G_{\alpha\beta}$ into a class function which has important consequences for the spin representation, discussed below. Note that these symmetries are exactly those of a $2$-point function on a domain with local and non-local variables. 

To derive the correlation function, going to spin representation is a necessary intermediate step. In comparison to previous work~\cite{Marchetti:2022nrf,Marchetti:2022igl}, the extended causal structure of the cBC model introduces also timelike faces, rendering this step more involved. 

We consider first those components of $G_{\alpha\beta}$, which only contain spacelike labels $\rho$ in their spin representation, i.e. $(\alpha\beta)\neq(\texttt{-}\texttt{-})$. The $(\texttt{-}\texttt{-})$-component, which contains also timelike labels $\nu$, is summarized thereafter. $G_{\alpha\beta}$ satisfies the symmetries of a two-point function, as detailed above, and consequently, its spin representation is given in terms of traces,
\begin{equation}
G_{\alpha\beta}(\vbg,\vbf;\vbg',\vbf') = \int\dd{\mu(\vbr,\vbk)}\e^{i\vbk (\vbf-\vbf')}\prod_c D^{(\rho_c,0)}_{j_c m_c j_c m_c}(g_c^{-1}g_c')G_{\alpha\beta}^{\vbr}(\vbk)\,,
\end{equation}
where $\dd{\mu}$ is short-hand notation containing the Plancherel measure and $2\pi$-factors of the $\vbk$-integrations. Summation over repeated magnetic indices is understood. Using the expansion of $G_{\alpha\beta}$, the equations of motion in spin representation are given by
\begin{equation}
\sum_\beta G_{\alpha\beta}^{\vbr}(\vbk)\delta\varphi^{\vbr,\beta}_{\vb*{j}\vb*{m}}(\vbk)B^{\vbr,\beta}_{\vb*{l}\vb*{n}} = 0\,,
\end{equation}
where the magnetic indices in this equation are uncontracted. An important detail here is that the generalized Barrett-Crane intertwiner (defined in Appendix~\ref{app:Projection onto invariant subspaces}) cannot be erased from the equation as it enters the sum over the signature label $\beta$. 

Starting from this equation, our goal is to obtain the correlation function first in spin representation and ultimately in group representation. To that end, we set up an \textit{effective action} for the field $\delta\varphi$,
\begin{equation}
S_{\mathrm{eff}}[\delta\varphi_\alpha]=\frac{1}{2}\sum_{\alpha,\beta}\int\dd{\mu(\vbr,\vbk)}\delta\varphi^{\vbr,\alpha}_{\vb*{j}\vb*{m}}(\vbk)B^{\vbr,\alpha}_{\vb*{l}\vb*{n}}G_{\alpha\beta}^{\vbr}(\vbk)\varphi^{\vbr,\beta}_{\vb*{j}\vb*{m}}(\vbk)B^{\vbr,\beta}_{\vb*{l}\vb*{n}}\,,
\end{equation}
which, upon variation, yields the equations of motion above. The generating functional for $n$-point functions is given by
\begin{equation}\label{eq:Z[J] before int}
Z[J] = \int\mathcal{D}[\delta\varphi_\alpha]\,\e^{-S_{\mathrm{eff}}[\delta\varphi_\alpha]}\,\exp\left(\sum_\alpha\int\delta\varphi^{\vbr,\alpha}_{\vb*{j}\vb*{m}}(\vbk)B^{\vbr,\alpha}_{\vb*{l}\vb*{n}}J^{\vbr,\alpha}_{\vb*{j}\vb*{m}\vb*{l}\vb*{n}}(\vbk)\right)\,.
\end{equation}
The fact that the source $J$ couples to both the field $\delta\varphi$ and the Barrett-Crane intertwiner is necessary to ensure the correct equations of motion that explicitly incorporate the Barrett-Crane intertwiner. Due to the Gaussian form of $Z[J]$, one can perform a completion of the square by introducing new variables
\begin{equation}
V^{\vbr,\alpha}_{\vb*{j}\vb*{m}\vb*{l}\vb*{n}}(\vbk)\defeq \delta\varphi^{\vbr,\alpha}_{\vb*{j}\vb*{m}}(\vbk)B^{\vbr,\alpha}_{\vb*{l}\vb*{n}}-\sum_\beta \left(G^{\vbr}(\vbk)^{-1}\right)_{\alpha\beta}J^{\vbr,\beta}_{\vb*{j}\vb*{m}\vb*{l}\vb*{n}}(\vbk)\,,
\end{equation}
and explicitly perform the integration. As a result, $Z[J]$ takes the form
\begin{equation}\label{eq:Z[J] after int}
Z[J] = \frac{1}{\mathrm{D}}\exp\left(-\frac{1}{2}\sum_{\alpha,\beta}\int J^{\vbr,\alpha}_{\vb*{j}\vb*{m}\vb*{l}\vb*{n}}(\vbk)\left(G^{\vbr}(\vbk)^{-1}\right)_{\alpha\beta}J^{\vbr,\alpha}_{\vb*{j}\vb*{m}\vb*{l}\vb*{n}}(\vbk)\right)\,,
\end{equation}
where $\mathrm{D}$ is the determinant factor resulting from the Gaussian integration. This factor drops out when computing expectation values and is therefore irrelevant for the correlation function. Taking the second functional derivative of $\ln Z[J]$ as defined by Eq.~\eqref{eq:Z[J] before int}, one finds
\begin{equation}\label{eq:2nd der before int}
    \eval{\frac{\delta^2\ln Z[J]}{\delta J^{\vbr,\alpha}_{\vb*{j}\vb*{m}\vb*{l}\vb*{n}}(\vbk)\delta J^{\vbr',\beta}_{\vb*{j}'\vb*{m}'\vb*{l}'\vb*{n}'}(\vbk')}}_{J = 0} = \expval{\delta\varphi^{\vbr,\alpha}_{\vb*{j}\vb*{m}}(\vbk)B^{\vbr,\alpha}_{\vb*{l}\vb*{n}}\delta\varphi^{\vbr',\beta}_{\vb*{j}'\vb*{m}'}(\vbk')B^{\vbr',\beta}_{\vb*{l}'\vb*{n}'}}\,.
    \end{equation}
On the other hand, if we use the expression of Eq.~\eqref{eq:Z[J] after int} for $Z[J]$, then same derivative yields
\begin{equation}\label{eq:2nd der after int}
    \eval{\frac{\delta^2\ln Z[J]}{\delta J^{\vbr,\alpha}_{\vb*{j}\vb*{m}\vb*{l}\vb*{n}}(\vbk)\delta J^{\vbr',\beta}_{\vb*{j}'\vb*{m}'\vb*{l}'\vb*{n}'}(\vbk')}}_{J = 0} = \left(G^{\vbr}(\vbk)^{-1}\right)_{\alpha\beta}\delta(\vbk+\vbk')\prod_{c}\delta(\rho_c-\rho_c')\delta_{j_c,j_c'}\delta_{m_c,m_c'}\delta_{l_c,l_c'}\delta_{n_c,n_c'}\,.
\end{equation}
In Eq.~\eqref{eq:2nd der before int}, one can identify the correlation function in spin representation, which in this setting enters with two additional Barrett-Crane intertwiners. To finally obtain the correlation function $C_{\alpha\beta}(\vbg,\vbf;\vbg',\vbf')$, we start with the defining expression
\begin{equation}
C_{\alpha\beta}(\vbg,\vbf;\vbg',\vbf') = \expval{\delta\varphi_\alpha(\vbg,\vbf)\delta\varphi_\beta(\vbg',\vbf')}\,.
\end{equation}
Going to spin representation of the right-hand side of the equation, we find as the integrand Eq.~\eqref{eq:2nd der before int}. Using Eq.~\eqref{eq:2nd der after int}, the correlation function is therefore finally given by
\begin{equation}\label{eq:spinrepCab}
C_{\alpha\beta}(\vbg,\vbf;\vbg',\vbf') = \int\dd{\mu(\vbr,\vbk)}\e^{i\vbk (\vbf-\vbf')}\left(G^{\vbr}(\vbk)^{-1}\right)_{\alpha\beta}\prod_c D^{(\rho_c,0)}_{j_c m_c j_c m_c}(g_c^{-1}g_c')\,.
\end{equation}
Now, written in this form, it is apparent that the correlation function satisfies the same symmetries as $G_{\alpha\beta}$. Consequently, the form of $C_{\alpha\beta}$ is given by
\begin{equation}
C_{\alpha\beta}(\vbg,\vbf;\vbg',\vbf') = C_{\alpha\beta}(\vbg^{-1}\vbg',\abs{\vbf-\vbf'})\,.
\end{equation}
This allows to replace $(\vb*{g}',\vb*{\phi}')$ with $(\vb*{e},\vb*{0})$, finally yielding  the function $C_{\alpha\beta}(\vb*{g},\vb*{\phi})$.

The $(\texttt{-}\texttt{-})$-component of the correlation function contains additional contributions from timelike faces. As explained in Sec.~\ref{sec:Correlation functions in spin representation}, these components are simply obtained as the scalar inverse of the effective kinetic kernel $G_{\texttt{-}\texttt{-}}^{(\vbr\vbn)_t}$ with $t>0$ timelike labels $(0,\nu)$. Thus, $C_{\texttt{-}\texttt{-}}$ in position space is given by 
\begin{equation}\label{eq:spinrepC--}
\begin{aligned}
    &C_{\minus\minus}(\vb*{g},\vb*{\phi})
    =
    \int\dd{\mu}(\vbr,\vbk)\e^{i\vb*{k}\vbf}\prod_c D^{(\rho_c,0)}_{j_cm_cj_cm_c}(g_c)\left(G^{\vbr}(\vbk)^{-1}\right)_{\minus\minus}\\[7pt]
    +&
    \sum_{t=1}^4\sum_{(c_1,...,c_t)}\sumint\dd{\mu((\vbr\vbn)_t,\vbk)}\e^{i\vb*{k}\vbf}\prod_{c=c_1}^{c_t} D^{(0,\nu_c)}_{j_c m_c j_c m_c}(g_c)\prod_{c'=c_{t+1}}^{c_4}D^{(\rho_{c'},0)}_{j_{c'}m_{c'}j_{c'}m_{c'}}(g_{c'})\frac{1}{G_{\minus\minus}^{(\vbr\vbn)_t}(\vbk)}\,,
\end{aligned}
\end{equation}
with $(\vbr\vbn)_t \equiv (\nu_{c_1},...,\nu_{c_t},\rho_{c_{t+1}},...\rho_{c_4})$ and where the sum-integral symbol denotes integration over the $\rho$'s and summation over the $\nu$'s.

\section{Explicit expressions for the matrix $\chi$}\label{sec:Explicit expressions for chi}

For the type of interactions considered in this work, the form of the matrix $\chi_{\alpha\beta}$ crucially depends on the interplay of combinatorics and the details of the causal structure. This is summarized with the notion of a causal vertex graph. This appendix gives an exhaustive list of all the possible causal vertex graphs if the underlying vertex graph is double-trace melonic, quartic melonic, necklace and non-colored simplicial.

In order to keep the notation clean, we present the fields in the interaction $\mathfrak{V}$ as 
\begin{equation}
\varphi_{1234}^\alpha\equiv \int\dd{X_\alpha}\varphi(g_1,g_2,g_3,g_4,\vbf,X_\alpha)\,,
\end{equation}
where repeated numbers imply a contraction via group integration. Also, we suppress the dependence on the scalar fields $\vbf$, as this is not of relevance for what is about to follow. The resulting $\chi_{\alpha\beta}$ matrices contain regularized symbols $\delta_{\rho,i}$ and we write for convenience
\begin{equation}
\delta_{\rho_1,i}\equiv\delta_1\,,\qquad  \delta_{\rho_1,i}\delta_{\rho_2,i}\equiv \delta^2_{12}\,,\qquad  \prod_{c=1}^3\delta_{\rho_c,i}\equiv \delta^3_{123}\,,\qquad  \prod_{c=1}^4\delta_{\rho_c,i} \equiv\delta^4\,.
\end{equation}

Notice also that it suffices to perform the computation of a certain combination of signatures and combinatorics for one exemplary set of signatures, e.g. the simplicial case with $(n_\plus,n_0,n_\minus) = (2,2,1)$ will be similar to $(1,2,2)$, but with rows exchanged. The form of the determinant of the effective kernel $G_{\alpha\beta}$ does not change and thus, the pole structure of the correlator remains unaffected. 

Interactions with a single type of signature (taken here as an example to be spacelike), their pictorial representation and the resulting $\chi$, which is in this case a scalar. Notice that the $\chi$'s agree with the functions $\mathcal{X}$ of~\cite{Marchetti:2020xvf,Marchetti:2022igl}.
\begin{align}
    &\mathfrak{V} = \varphi^{\plus}_{1234}\varphi^{\plus}_{1234}\varphi^{\plus}_{5678}\varphi^{\plus}_{5678}\,,  &&{\scriptsize \cvftpppp}\,, &&&\chi = 4\left(2\delta^4+1\right)\,,\\
    &\mathfrak{V} = \varphi_{1234}^\plus\varphi_{5674}^\plus\varphi_{5678}^\plus\varphi_{1238}^\plus\,,  &&{\scriptsize \cvfpppp}\,,    &&&\chi = 4(\delta^4+\delta^3_{123}+2\delta_4)\,,\\
    &\mathfrak{V} = \varphi^\plus_{1234}\varphi^\plus_{3456}\varphi^\plus_{5678}\varphi^\plus_{7812}\,,  && {\scriptsize \cvfnpppp}\,,   &&&\chi = \chi = 4(\delta^4+\delta^2_{12}+\delta^2_{34})\,,
\end{align}
\begin{align}
    &\mathfrak{V} = \varphi^\plus_{1234}\varphi^\plus_{4567}\varphi^\plus_{7389}\varphi^\plus_{9620}\varphi^\plus_{0851}\,,  && {\scriptsize \cvfsppppp}\,,  &&& \chi = 5(\delta^3_{123}+\delta^3_{234}+\delta^3_{341}+\delta^3_{412})\,.
 \end{align}

Interactions with all but one tetrahedron of the same signature. As an example, all tetrahedra are spacelike except one timelike tetrahedron.
\begin{align}
    &\mathfrak{V} = \varphi^{\plus}_{1234}\varphi^{\plus}_{1234}\varphi^{\plus}_{5678}\varphi^{\minus}_{5678}\,, && {\scriptsize \cvftpppm}\,,\chi=
    \begin{pmatrix}
    2(\delta^4+\delta^3_{123}+\delta_4)\ & \delta^4+\delta^3_{123}+\delta_4\\[7pt]
    \delta^4+\delta^3_{123}+\delta_4& 0
    \end{pmatrix}\,,\\
    &\mathfrak{V} = \varphi_{1234}^\plus\varphi_{5674}^\plus\varphi_{5678}^\plus\varphi_{1238}^\minus\,,  && {\scriptsize \cvfpppm}\,,   \chi = 
    \begin{pmatrix}
    2(\delta^4+\delta^3_{123}+\delta_4)\ & \delta^4+\delta^3_{123}+\delta_4\\[7pt]
    \delta^4+\delta^3_{123}+\delta_4& 0
    \end{pmatrix}\,,\\
    &\mathfrak{V} = \varphi^\plus_{1234}\varphi^\plus_{5634}\varphi^\minus_{5678}\varphi^\plus_{1278}\,,  && {\scriptsize \cvfnpppm}\,,  \chi = 
    \begin{pmatrix}
    2(\delta^4+\delta^2_{12}+\delta^2_{34})\ & \delta^4+\delta^2_{12} +\delta^2_{34}\\[7pt]
    \delta^4+\delta^2_{12}+\delta^2_{34}\ & 0
    \end{pmatrix}\,,
\end{align}
\begin{equation}
\begin{aligned}
    \mathfrak{V} = \varphi^\plus_{1234}\varphi^\plus_{4567}\varphi^\plus_{7389}\varphi^\plus_{9620}\varphi^\minus_{0851}\,, \qquad {\scriptsize \cvfsppppm}\,, \\ \chi = 
    \begin{pmatrix}
    3\sum_c\prod_{c'\neq c}\delta_{\rho_{c'},i}\ & \sum_c\prod_{c'\neq c}\delta_{\rho_{c'},i}\\[7pt]
    \sum_c\prod_{c'\neq c}\delta_{\rho_{c'},i}\ & 0
    \end{pmatrix}\,.
\end{aligned}
\end{equation}

Interactions with two types of signature, here $\np,\nm$, with $\np,\nm\geq 2$.
\begin{align}
    &\mathfrak{V} = \varphi^{\plus}_{1234}\varphi^{\plus}_{1234}\varphi^{\minus}_{5678}\varphi^{\minus}_{5678}\,, && {\scriptsize \cvftppmma}\,, &&&\chi = 
    \begin{pmatrix}
     2\ & 4\delta^4\\[7pt]
     4\delta^4\ & 2    
    \end{pmatrix}\,,\\
    &\mathfrak{V} = \varphi^{\plus}_{1234}\varphi^{\minus}_{1234}\varphi^{\plus}_{5678}\varphi^{\minus}_{5678}\,, && {\scriptsize \cvftppmmb}\,, &&&\chi =    
    \begin{pmatrix}
    2\delta^4\ & 2\delta^4+2\\[7pt]
    2\delta^4+2\ & 2\delta^4    
    \end{pmatrix}\,,\\
    &\mathfrak{V} = \varphi_{1234}^\plus\varphi_{5674}^\minus\varphi_{5678}^\minus\varphi_{1238}^\plus\,, && {\scriptsize \cvfppmma}\,, &&&\chi = 
    \begin{pmatrix}
    2\delta_4\ & 2(\delta^4+\delta^3_{123})\\[7pt]
    2(\delta^4+\delta^3_{123})\ & 2\delta_4
    \end{pmatrix}\,,
\end{align}
\begin{align}
    &\mathfrak{V} = \varphi_{1234}^\plus\varphi_{5674}^\plus\varphi_{5678}^\minus\varphi_{1238}^\minus\,, && {\scriptsize \cvfppmmb}\,, &&&\chi = 
    \begin{pmatrix}
    2\delta^3_{123}\ & 2(\delta^4+\delta_4)\\[7pt]
    2(\delta^4+\delta_4)\ & 2\delta^3_{123}
    \end{pmatrix}\,,\\
    &\mathfrak{V} = \varphi_{1234}^\plus\varphi_{5674}^\minus\varphi_{5678}^\plus\varphi_{1238}^\minus\,, && {\scriptsize \cvfppmmc}\,, &&&\chi = 
    \begin{pmatrix}
    2\delta^4\ & 2(\delta^3_{123}+\delta_4)\\[7pt]
    2(\delta^3_{123}+\delta_4)\ & 2\delta^4
    \end{pmatrix}\,,\\  
    &\mathfrak{V} = \varphi^\plus_{1234}\varphi^\minus_{5634}\varphi^\minus_{5678}\varphi^\plus_{1278}\,, &&  {\scriptsize \cvfnppmma}\,, &&& \chi =  
    \begin{pmatrix}
    2\delta^2_{34}\ & 2(\delta^4+\delta^2_{12})\\[7pt]
    (\delta^4+\delta^2_{12})\ & 2\delta^2_{34}
    \end{pmatrix}\,,\\
    &\mathfrak{V} = \varphi^\plus_{1234}\varphi^\minus_{5634}\varphi^\plus_{5678}\varphi^\minus_{1278}\,, && {\scriptsize \cvfnppmmb}\,, &&&    \chi = 
    \begin{pmatrix}
    2\delta^4\ & 2(\delta^2_{12} +\delta^2_{34})\\[7pt]
    2(\delta^2_{12} +\delta^2_{34})\ & 2\delta^4
    \end{pmatrix}\,,
\end{align}
\begin{equation}
\begin{aligned}
    \mathfrak{V} = \varphi^\plus_{1234}\varphi^\plus_{4567}\varphi^\plus_{7389}\varphi^\minus_{9620}\varphi^\minus_{0851}\,,\qquad {\scriptsize \cvfspppmm}\,,\\
    \chi =  \begin{pmatrix}
        2(\delta^3_{123}+\delta^3_{234})+\delta^3_{341}+\delta^3_{412}\ & \delta^3_{123}+\delta^3_{234}+2(\delta^3_{341}+\delta^3_{412})\\[7pt]
        \delta^3_{123}+\delta^3_{234}+2(\delta^3_{341}+\delta^3_{412})\ & \delta^3_{123}+\delta^3_{234}
        \end{pmatrix}\,.
\end{aligned}
\end{equation}

Quartic interactions with three different signatures.
\begin{align}
    &\mathfrak{V} = \varphi^{\plus}_{1234}\varphi^{\plus}_{1234}\varphi^{0}_{5678}\varphi^{\minus}_{5678}\,, && {\scriptsize \cvftppmza}\,, &&& \chi = 
    \begin{pmatrix}
     2\ & 2\delta^4\ & 2\delta^4 \\[7pt]
     2\delta^4\ & 0\ & 1 \\[7pt]
     2\delta^4\ & 1\ & 0
    \end{pmatrix}\,,\\
    &\mathfrak{V} = \varphi^{\plus}_{1234}\varphi^{0}_{1234}\varphi^{\plus}_{5678}\varphi^{\minus}_{5678}\,, &&  {\scriptsize \cvftppmzb}\,, &&& \chi = 
    \begin{pmatrix}
     2\delta^4\ & \delta^4+1\ & \delta^4+1 \\[7pt]
     \delta^4+1\ & 0\ & \delta^4 \\[7pt]
     \delta^4+1\ & \delta^4\ & 0
    \end{pmatrix}\,,\\
    &\mathfrak{V} = \varphi^\plus_{1234}\varphi_{5674}^0\varphi_{5678}^\minus\varphi_{1238}^\plus\,, && {\scriptsize \cvfppmza}\,, &&& \chi = 
    \begin{pmatrix}
    2\delta_4\ & \delta^4+\delta^3_{123}\ & \delta^4+\delta^3_{123}\\[7pt]
    \delta^4+\delta^3_{123}\ & 0\ & \delta_4\\[7pt]
    \delta^4+\delta^3_{123}\ & \delta_4\ & 0
    \end{pmatrix}\,,
\end{align}
\begin{align}
    &\mathfrak{V} = \varphi_{1234}^\plus\varphi_{5674}^\plus\varphi_{5678}^0\varphi_{1238}^\minus\,, &&  {\scriptsize \cvfppmzb}\,, &&&\chi = 
    \begin{pmatrix}
    2\delta^3_{123}\ & \delta^4+\delta_4\ & \delta^4+\delta_4\\[7pt]
    \delta^4+\delta_4\ & 0\ & \delta^3_{123}\\[7pt]
    \delta^4+\delta_4\ & \delta^3_{123}\ & 0
    \end{pmatrix}\,,\\
    &\mathfrak{V} = \varphi_{1234}^\plus\varphi_{5674}^0\varphi_{5678}^\plus\varphi_{1238}^\minus\,, && {\scriptsize \cvfppmzc}\,, &&& \chi = 
    \begin{pmatrix}
    2\delta^4\ & \delta^3_{123}+\delta_4\ & \delta^3_{123}+\delta_4\\[7pt]
    \delta^3_{123}+\delta_4\ & 0\ & \delta^4\\[7pt]
    \delta^3_{123}+\delta_4\ & \delta^4\ & 0
    \end{pmatrix}\,,\\
    &\mathfrak{V} = \varphi^\plus_{1234}\varphi^0_{5634}\varphi^\minus_{5678}\varphi^\plus_{1278}\,, && {\scriptsize \cvfnppmza}\,, &&& \chi = 
    \begin{pmatrix}
    2\delta^2_{34}\ & \delta^4+\delta^2_{12}\ & \delta^4+\delta^2_{12}\\[7pt]
    \delta^4+\delta^2_{12}\ & 0\ & \delta^2_{34}\\[7pt]
    \delta^4 + \delta^2_{12}\ & \delta^2_{34}\ & 0
    \end{pmatrix}\,,\\
    &\mathfrak{V} = \varphi^\plus_{1234}\varphi^0_{5634}\varphi^\plus_{5678}\varphi^\minus_{1278}\,, &&  {\scriptsize \cvfnppmzb}\,, &&& \chi = 
    \begin{pmatrix}
    2\delta^4\ & \delta^2_{12}+\delta^2_{34}\ & \delta^2_{12}+\delta^2_{34}\\[7pt]
    \delta^2_{12}+\delta^2_{34}\ & 0\ &\delta^4\\[7pt]
    \delta^2_{12}+\delta^2_{34}\ & \delta^4\ & 0
    \end{pmatrix}\,.
\end{align}
Simplicial interactions with all three signatures.
\begin{equation}
\begin{aligned}
    \mathfrak{V} = \varphi^\plus_{1234}\varphi^\plus_{4567}\varphi^\plus_{7389}\varphi^0_{9620}\varphi^\minus_{0851}\,, \qquad{\tiny \cvfspppmz}\,,\\
    \chi = \begin{pmatrix}
    2(\delta^3_{123}+\delta^3_{234})+\delta^3_{341}+\delta^3_{412}\ & \delta^3_{123}+\delta^3_{341}+\delta^3_{412}\ &\delta^3_{234}+\delta^3_{341}+\delta^3_{412}\\[7pt]
    \delta^3_{123}+\delta^3_{341}+\delta^3_{412}\ & 0\ & \delta^3_{123}\\[7pt]
    \delta^3_{234}+\delta^3_{341}+\delta^3_{412}\ & \delta^3_{234}\ & 0
    \end{pmatrix}\,,
\end{aligned}
\end{equation}
\begin{equation}
\begin{aligned}
    \mathfrak{V} = \varphi^\plus_{1234}\varphi^\plus_{4567}\varphi^0_{7389}\varphi^0_{9620}\varphi^\minus_{0851}\,, \qquad{\tiny \cvfsppmzz}\,,\\
    \chi = \begin{pmatrix}
    \delta^3_{123}+\delta^3_{234}\ & 2\delta^3_{412}+\delta^3_{123}+\delta^3_{341}\ &\delta^3_{234}+\delta^3_{341}\\[7pt]
    2\delta^3_{341}+\delta^3_{234}+\delta^3_{412}\ & \delta^3_{123}+\delta^3_{234}\ & \delta^3_{123}+\delta^3_{412}\\[7pt]
    \delta^3_{123}+\delta^3_{412}\ & \delta^3_{234}+\delta^3_{341}\ & 0
    \end{pmatrix}\,.
\end{aligned}
\end{equation}
\chapter[\textsc{Condensate Dynamics and Classical Perturbations}]{Condensate Dynamics and Classical Perturbations}

\section{Derivation of condensate dynamics}\label{app:Derivation condensate dynamics}

Here, we provide the detailed derivations of the dynamical equations for the perturbed condensate introduced in Chapter~\ref{chapter:perturbations}. To that end, consider an expansion of the kinetic kernels
\begin{align}
\mathcal{K}_\texttt{+}((\rf^{0})^2,\mm^2) &= \sum_{n = 0}^\infty\frac{\mathcal{K}_\texttt{+}^{(2n)}(\mm^2)}{(2n)!}(\rf^0)^{2n}\,,\label{eq:K+ expansion}\\[7pt]
\mathcal{K}_\texttt{-}(\abs{\vb*{\rf}}^2,\mm^2) &= \sum_{n = 0}^\infty\frac{\mathcal{K}_\texttt{-}^{(2n)}(\mm^2)}{(2n)!}\abs{\vb*{\rf}}^{2n}\label{eq:K- expansion}\,,
\end{align}
the existence of which is supported by the studies of~\cite{Li:2017uao}. Notice that the reference fields are coupled to the GFT model via Eq.~\eqref{eq:kineticrestriction}, such that their expansion differs slightly from that discussed in~\cite{Marchetti:2021gcv}. The reduced condensate wavefunctions $\slrcw$ and $\tlrcw$ are expanded in derivatives
\begin{align}
\slrcw(\rf^0+x^0,\mm) &= \sum_{n = 0}^\infty \frac{\slrcw^{(n)}(x^0,\mm)}{n!}(\rf^0)^n\,,\label{eq:slrcw expansion}\\[7pt]
\tlrcw(\rf^0+x^0,\mm) &= \sum_{n=0}^\infty\frac{\tlrcw^{(n)}(x^0,\mm)}{n!}(\rf^0)^n\,,\label{eq:tlrcw expansion}
\end{align}
where $\slrcw^{(n)}$ denotes the $n$-th derivative with respect to the clock, applying similarly to $\tlrcw$. These expansions will be employed for background and perturbed part of the equations of motion.

\subsection{Background equations}\label{sec:Derivation of background equations}

At background level, the two equations of motion in spin-representation are given by
\begin{align}
0 &= \int\dd{\rf^0}\dd{\mf'}\mathcal{K}_\texttt{+}\left(\rf^0,(\mf-\mf')^2\right)\sigma(\rf^0+x^0,\mf')\,,\label{eq:spacelike bkg eom}\\[7pt]
0 &= \int\dd[4]{\rf}\dd{\mf'}\mathcal{K}_\texttt{-}\left(\vb*{\rf},(\mf-\mf')^2\right)\tau(\rf^0+x^0,\vb*{\rf}+\vb*{x},\mf')\,,\label{eq:timelike bkg eom}
\end{align}
where here and in the remainder, empty rod-integrations are regularized. For a further analysis, a Fourier transform of the matter field variables $\mf\rightarrow\mm$ is performed. Following~\cite{Marchetti:2021gcv}, we assume a peaking of both condensate wavefunctions on a fixed scalar field momentum $\pmm$, realized by a Gaussian peaking.
The dynamical equations of the reduced condensate wavefunctions are derived in the following separately for the spacelike, respectively timelike sector. 

\paragraph{Spacelike background dynamics.} Using the expansions of Eqs.~\eqref{eq:K+ expansion} and~\eqref{eq:slrcw expansion}, the (regularized) spacelike background equation~\eqref{eq:spacelike bkg eom} evaluates to
\begin{equation}
\begin{aligned}
0 &= \int\dd{\rf^0}\mathcal{K}_\texttt{+}((\rf^0)^2,\mm^2)\slrcw(\rf^0+x^0,\mm)\eta_{\epsp}(\rf^0,\pip)\\[7pt]
&= 
\sum_{m,n}\frac{\mathcal{K}_\texttt{+}^{(2m)}(\mm^2)\slrcw^{(n)}(x^0,\mm)}{(2m)!n!}\int\dd{\rf^0}\eta_{\epsp}(\rf^0,\pip)(\rf^0)^{2m+n}\\[7pt]
&\approx
\mathcal{K}_\texttt{+}^{(0)}\left[\left(I_0+I_2\frac{\mathcal{K}_\texttt{+}^{(2)}}{2\mathcal{K}_\texttt{+}^{(0)}}\right)\slrcw(x^0,\mm)+I_1\partial_0\slrcw(x^0,\mm)+\frac{1}{2}I_2\partial_0^2\slrcw(x^0,\mm)\right]\,.
\end{aligned}
\end{equation}
Following~\cite{Marchetti:2021gcv}, we introduced the function $I_{2m+n}(\epsp,\pip)$, defined as
\begin{equation}\label{eq:In}
I_{n}(\epsp,\pip) \defeq\mathcal{N}_{\epsp}\sqrt{2\pi\epsp}\left(i\sqrt{\frac{\epsp}{2}}\right)^{n}\e^{-z_\texttt{+}^2}H_n\left(\sqrt{\frac{\epsp}{2}}\pip\right)\,,
\end{equation}
where $H_n$ are the Hermite polynomials and $z_\texttt{+}^2 = \epsp(\pip)^2/2$. Truncating the expansion at order $\epsp$ leads to the condition that only terms with $2m+n\leq 2$ contribute. Introducing the quantities
\begin{equation}
E_\texttt{+}^2(\mm) \defeq \frac{2}{\epsp(2z_\texttt{+}^2-1)}-\frac{\mathcal{K}_\texttt{+}^{(2)}}{\mathcal{K}_\texttt{+}^{(0)}}\,,\qquad \tpip \defeq\frac{\pip}{2z_\texttt{+}^2-1}\,,
\end{equation}
one finally obtains
\begin{equation}
\partial_0^2\slrcw(x^0,\mm)-2i\tpip\partial_0\slrcw(x^0,\mm)-E_\texttt{+}^2(\mm)\slrcw(x^0,\mm) = 0\,.
\end{equation}

\paragraph{Timelike background dynamics.} On the timelike sector, the procedure to obtain the equations of motion differs slightly because of the different peaking properties of $\tau$ and the mere rod-dependence of the timelike kernel $\mathcal{K}_\texttt{-}$. Starting with Eq.~\eqref{eq:timelike bkg eom} and inserting the expansions of Eqs.~\eqref{eq:K- expansion} and~\eqref{eq:tlrcw expansion}, one obtains
\begin{equation}
\begin{aligned}
0 &= \int\dd[4]{\rf}\mathcal{K}_\texttt{-}(\abs{\vb*{\rf}}^2,\mm^2)\tlrcw(\rf^0+x^0,\mm)\eta_{\epsm}(\rf^0,\pim)\eta_{\delta}(\abs{\vb*{\rf}},\pi_x)\\[7pt]
&= \sum_n\frac{\tlrcw^{(n)}(x^0,\mm)}{n!}\int\dd{\rf^0}\eta_{\epsm}(\rf^0,\pim)(\rf^0)^{n}\int\dd[3]{\rf}\mathcal{K}_\texttt{-}(\abs{\vb*{\rf}}^2,\mm)\eta_{\delta}(\abs{\vb*{\rf}},\pi_x)\,.
\end{aligned}
\end{equation}
Assuming that the spatial integral is non-zero, the equations factorize. Truncating at linear order in $\epsm$ finally yields
\begin{equation}
I_0^-\tlrcw(x^0,\mm)+I_1^-\partial_0\tlrcw(x^0,\mm)+\frac{1}{2}I_2^-\partial_0^2\tlrcw(x^0,\mm)\approx 0\,,
\end{equation}
where $I_n^-$ is defined equivalently to Eq.~\eqref{eq:In} but evaluated on the timelike peaking parameters $\epsm$ and $\pim$. Introducing
\begin{equation}
E_\texttt{-}^2 \defeq \frac{2}{\epsm(2z_\texttt{-}^2-1)}\,,\qquad \tpim \defeq \frac{\pim}{2z_\texttt{-}^2-1}\,,
\end{equation}
the background equation for the timelike reduced condensate wavefunction reads
\begin{equation}
\partial_0^2\tlrcw(x^0,\mm)-2i\tpim\partial_0\tlrcw(x^0,\mm)-E_\texttt{-}^2\tlrcw(x^0,\mm) = 0\,.
\end{equation}
Notice that due to the interplay of peaking and kernel dependencies, the quantity $E_\texttt{-}$ does not carry a matter momentum dependence, in cotrast to $E_\texttt{+}(\mm)$.

\paragraph{Classical limit.} As elaborated previously~\cite{Oriti:2016qtz,Marchetti:2020umh,Marchetti:2020qsq}, the semi-classical limit of the condensate is obtained at late relational time scales where the moduli of the condensate wavefunctions are dominant but where interactions are still negligible. It has been furthermore shown in~\cite{Pithis:2016cxg}, that in this limit, expectation values of for instance the volume operator are sharply peaked. In this limit, background solutions are given by
\begin{align}
\slrcw(x^0,\pmm) = \slrcw_0\e^{(\mu_\texttt{+} + i\tpip)x^0}\,,\qquad \tlrcw(x^0,\pmm) = \tlrcw_0\e^{(\mu_\texttt{-} + i\tpim)x^0}\,,
\end{align}
where $\slrcw_0$ and $\tlrcw_0$ are determined by initial conditions. The parameters $\mu_\pm$ are defined as $\mu_\pm^2(\pmm) \defeq E_\pm^2(\pmm)-(\tilde{\pi}_0^\pm)^2$.

\subsection{Perturbation equations}\label{sec:Derivation of perturbation equations}

Continuing the analysis of the equations of motion, we derive in this section the perturbed equations of motion for the spacelike and then the timelike sector.

\paragraph{Spacelike perturbed dynamics.} Dynamics of the spacelike sector at first order of perturbations are governed by
\begin{equation}\label{eq:pert eom sl}
\begin{aligned}
0 &= \int\dd[4]{\rf}\dd{\mf'}\mathcal{K}_{+}((\rf^0)^2,(\mf-\mf')^2)\int\dd[4]{\rf'}\dd{\mf''}\Bigg{[}\delta\Psi(\rf^\mu+x^\mu,\mf',\rf^{\mu\prime},\mf'')\bar{\tau}(\rf^{\mu\prime},\mf'')\\[7pt]
&+\delta\Phi(\rf^\mu+x^\mu,\mf',\rf^{\mu\prime},\mf'')\bar{\sigma}(\rf^{0\prime},\mf'') \Bigg{]}\,.
\end{aligned}
\end{equation}
Let us repeat the set of assumptions that were posed in the main body. As a first simplification, the bi-local kernel $\delta\Psi$ is chosen to depend only on one copy of relational frame data, i.e.
\begin{align}\label{eq:locality condition app}
\delta\Psi(\rf^\mu,\mm,\rf^{\mu\prime},\mm') &= \delta\Psi(\rf^\mu,\mm)\delta^{(4)}(\rf^\mu-\rf^{\prime\mu})\delta(\mm-\mm')\,.
\end{align}
From a simplicial gravity perspective, locality with respect to the reference fields $\rf^\mu$ corresponds to correlations only within the same $4$-simplex,  which can be compared to nearest-neighbor interactions in statistical spin systems. For the momenta of the matter field $\mm$, the condition is interpreted as momentum conservation across tetrahedra of the same $4$-simplex. Next, the ansatz
\begin{equation}\label{eq:relation of dPsi and dPhi app}
\delta\Phi(\rf^\mu,\mm) = \mathrm{f}(\rf^\mu)\delta\Psi(\rf^\mu,\mm)\,,  
\end{equation}
with the complex valued function $\mathrm{f}$ defined as
\begin{equation}\label{eq:def of f app}
\mathrm{f}(\rf^0,\vb*{\rf}) = f(\rf^0)\e^{i\theta_f(\rf^0)}\abs{\eta_\delta(\abs{\vb*{\rf}-\vb*{x}})}\e^{2i\pip\rf^0}\,.
\end{equation}
Here, $f$ and $\theta_f$ are real functions that only depend on the reference clock $\rf^0$. In addition, the following relations between the peaking parameters $\epsilon^\pm$ and $\pi_0^\pm$ of the different sectors are considered,
\begin{equation}\label{eq:peaking parameter relations}
\epsp = \epsm\,,\qquad  \pip = -\pim\,.
\end{equation}
Employing all of these assumptions, the expansions of $\mathcal{K}_\texttt{+},\slrcw$ and $\tlrcw$ according to Eqs.~\eqref{eq:K+ expansion},~\eqref{eq:slrcw expansion} and~\eqref{eq:tlrcw expansion}, and truncating at linear order in $\epsp$ and $\delta$, one obtains
\begin{equation}
\begin{aligned}
0 &= \mathcal{K}_\texttt{+}^{(0)}(\pmm^2)\Bigg{[}\left(I_0+I_2\frac{\mathcal{K}_\texttt{+}^{(2)}}{2\mathcal{K}_\texttt{+}^{(0)}}\right)\delta\Psi\left(J_{0,\vb*{0}}\bar{\tlrcw}+f\e^{i\theta_f}\bar{\slrcw}\right)+I_1\partial_0\left(\delta\Psi(J_{0,\vb*{0}}\bar{\tlrcw}+f\e^{i\theta_f}\bar{\slrcw})\right)\\[7pt]
&+\frac{I_2}{2}\partial_0^2\left(\delta\Psi(J_{0,\vb*{0}}\bar{\tlrcw}+f\e^{i\theta_f}\bar{\slrcw})\right) + \bar{\tlrcw}I_0\frac{J_{0,(0,0,2)}}{2}\nabla_{\vb*{x}}^2\delta\Psi\Bigg{]}\,.
\end{aligned}
\end{equation}
All fields, $\slrcw$, $\tlrcw$ and $\delta\Psi$ are evaluated at $x^0$, respectively $x^i$ and the peaked matter momentum $\pmm$. 
Due to the spatial peaking of the timelike condensate $\tau$, coefficients $J_{m,(n_1,n_2,n_3)}$ appear in the expression above, defined as
\begin{equation}
J_{m,(n_1,n_2,n_3)} \defeq \int\dd[3]{\rf}\eta_\delta(\abs{\vb*{\rf}},\pi_x)\abs{\vb*{\rf}}^{2m}\prod_{i=1}^3(\rf^i)^{n_i}\,.
\end{equation}
The relevant coefficients for the derivation of the equations of motion are $J_{0,\vb*{0}}$, $J_{2,\vb*{0}}$ and $J_{0,(0,0,2)}$, explicitly defined as~\cite{Marchetti:2021gcv}
\begin{equation}
\begin{gathered}
    J_{0,\vb*{0}} = -2\mathcal{N}_\delta\sqrt{2\pi}\pi^2\delta^{3/2}z^2\e^{-z^2}\,,\qquad  J_{2,\vb*{0}} = 4\mathcal{N}_\delta\sqrt{2\pi}\pi^2\delta^{5/2}z^4\e^{-z^2}\,,\\
     J_{0,(0,0,2)} = \frac{16}{3}\mathcal{N}_\delta\sqrt{2\pi}\pi\delta^{5/2}z^4\e^{-z^2}\,,
\end{gathered}
\end{equation}
keeping only first-order contributions in the peaking parameter $\delta$, where $z^2 = \delta\pi_x^2/2$. Factorizing $I_2/2$ from the spacelike perturbed equations of motion above, we finally obtain
\begin{equation}\label{eq:dPsi}
0 = \partial_0^2\left(\delta\Psi(J_{0,\vb*{0}}\bar{\tlrcw}+f\e^{i\theta_f}\bar{\slrcw})\right)-2i\tpip\partial_0\left(\delta\Psi(J_{0,\vb*{0}}\bar{\tlrcw} +f\e^{i\theta_f}\bar{\slrcw})\right)+
- E_\texttt{+}^2\delta\Psi\left(J_{0,\vb*{0}}\bar{\tlrcw}+f\e^{i\theta_f}\bar{\slrcw}\right) + \alpha\tlrcw\nabla_{\vb*{x}}^2\delta\Psi\,,
\end{equation}
with $\alpha$ defined as $\alpha\defeq \frac{I_0 J_{0,(0,0,2)}}{I_2}$. 

\paragraph{Timelike perturbed dynamics.} The perturbed equations of motion on the timelike sector are given by
\begin{equation}\label{eq:pert eom tl}
\begin{aligned}
0 &= \int\dd[4]{\rf}\dd{\mf'}\mathcal{K}_{-}(\abs{\vb*{\rf}}^2,(\mf-\mf')^2)\int\dd[4]{\rf'}\dd{\mf''}\Bigg{[}\delta\Psi(\rf^{0\prime},\vb*{\rf}',\mf',\rf^\mu+x^\mu,\mf'')\bar{\sigma}(\rf^{0\prime},\mf'')\\[7pt]
&+\delta\Xi(\rf^\mu+x^\mu,\mf',\rf^{0\prime},\vb*{\rf}',\mf'')\bar{\tau}(\rf^{\mu\prime},\mf'') \Bigg{]}\,.
\end{aligned}
\end{equation}
Using the expansions of Eqs.~\eqref{eq:K- expansion},~\eqref{eq:slrcw expansion} and~\eqref{eq:tlrcw expansion} as well as the relations of Eqs.~\eqref{eq:relation of dPsi and dPhi app} and~\eqref{eq:peaking parameter relations}, one arrives at
\begin{equation}
\begin{aligned}
0 &= \int\dd[3]{\rf}\mathcal{K}_\texttt{-}(\abs{\rf}^2,\pmm^2)\left[I_0\delta\Psi\bar{\slrcw}+\bar{I}_1\partial(\delta\Psi\bar{\slrcw})+\frac{I_2}{2}\partial_0^2\left(\delta\Psi\bar{\slrcw}\right)\right]\\[7pt]
  &+ \mathcal{K}_\texttt{-}^{(0)}(\pmm^2)\Bigg{[}\left(I_0J_{0,\vb*{0}}+I_0J_{2,\vb*{0}}\frac{\mathcal{K}_\texttt{-}^{(2)}}{\mathcal{K}_\texttt{-}^{(0)}}\right)\delta\Xi\bar{\tlrcw}+J_{0 ,\vb*{0}}I_1\partial\left(\delta\Xi\bar{\tlrcw}\right)\\[7pt]
  &+ J_{0,\vb*{0}}\frac{I_2}{2}\partial_0^2\left(\delta\Xi\bar{\tlrcw}\right)+I_0\frac{J_{0,(0,0,2)}}{2}\bar{\tlrcw}\nabla_{\vb*{x}}^2\delta\Xi\Bigg{]}\,,
\end{aligned}
\end{equation}
where the coefficients $I_0,I_2,J_{0,\vb{0}},J_{2,\vb*{0}}$ and $J_{0,(0,0,2)}$ are defined as above. Inserting the background solutions of $\slrcw$ and $\tlrcw$ in the classical, this can be further written as
\begin{equation}
\begin{aligned}
0 &= \bar{\slrcw}\int\dd[3]{\rf}\mathcal{K}_\texttt{-}(\abs{\vb*{\rf}},\pmm^2)\left[\left(\frac{2I_0}{I_2}+(\pip)^2+\mu_\texttt{+}^2\right)\delta\Psi+2\mu_\texttt{+}\partial_0\delta\Psi+\partial_0^2\delta\Psi\right]\\[7pt]
&+\mathcal{K}_\texttt{-}^{(0)}J_{0,\vb*{0}}\bar{\tlrcw}\Bigg{[}\left(\frac{2I_0}{I_2}+\frac{I_0J_{2,\vb*{0}}}{I_2J_{0,\vb*{0}}}\frac{\mathcal{K}_\texttt{-}^{(2)}}{\mathcal{K}_\texttt{-}^{(0)}}+(\pip)^2+\mu_\texttt{-}^2\right)+2\mu_\texttt{-}\partial_0\delta\Xi+\partial_0^2\delta\Xi+\frac{\alpha}{J_{0,\vb*{0}}}\nabla_{\vb*{x}}^2\delta\Xi\Bigg{]}\,.
\end{aligned}
\end{equation}
Using the definition of $\mu_\texttt{-}^2$ and introducing
\begin{align}
\beta\defeq  -\frac{I_0J_{2,\vb*{0}}}{I_2J_{0,\vb*{0}}}\frac{\mathcal{K}_\texttt{-}^{(2)}}{\mathcal{K}_\texttt{-}^{(0)}}\,,\qquad \gamma\defeq \frac{\alpha}{J_{0,\vb*{0}}}\,,
\end{align}
the perturbed equation of motion on the timelike sector is finally given by
\begin{equation}
\begin{aligned}
0 &=  \bar{\slrcw}\int\dd[3]{\rf}\mathcal{K}_\texttt{-}(\abs{\vb*{\rf}},\pmm^2)\left[\partial_0^2\delta\Psi+2\mu_\texttt{+}\partial_0\delta\Psi-\frac{\mathcal{K}_\texttt{+}^{(2)}}{\mathcal{K}_\texttt{+}^{(0)}}\delta\Psi\right]\\[7pt]
&+ 
\bar{\tlrcw}\mathcal{K}_\texttt{-}^{(0)}J_{0,\vb*{0}}\Bigg{[}\partial_0^2\delta\Xi+2\mu_\texttt{-}\partial_0\delta\Xi-\beta\delta\Xi+\gamma\nabla_{\vb*{x}}^2\delta\Xi\Bigg{]}\,.
\end{aligned}
\end{equation}
Since the space-dependence of the first term is integrated out, solutions $\delta\Xi$ need to be space-independent, i.e. $\delta\Xi(x^\mu,\pmm) \equiv \delta\Xi(x^0,\pmm)$. Hence, the space-derivative acting on $\delta\Xi$ vanishes and the equation reduces to a second-order inhomogeneous ordinary differential equation.

\section{Matching classical volume perturbations}\label{app:volume matching}

We present here a derivation of the dynamics of $\delta V$, given in Eq.~\eqref{eqn:pertvolume}. To that end, one computes the expectation value of $\hat{V}$ with respect to $\ket{\Delta;x^0,\vb*{x}}$, 
\begin{equation}\label{eq:V expval}
\begin{aligned}
     \expval{\hat{V}}{\Delta;x^0,\vb*{x}} &= \mathrm{v}\int\dd[4]{\rf}\dd{\mm}\bar{\sigma}(\rf^0,\mm)\sigma(\rf^0,\mm)\\[7pt]
    &\quad+ 2\mathrm{v}\mathfrak{Re}\left\{\int\dd[4]{\rf}\dd{\mm}\delta\Psi(\rf^\mu,\mm)\bar{\sigma}(\rf^0,\mm)\bar{\tau}(\rf^{\mu},\mm)\right\}\\[7pt]
    &\quad+2\mathrm{v}\mathfrak{Re}\left\{\int\dd[4]{\rf}\dd{\mm}\mathrm{f}(\rf^\mu)\delta\Psi(\rf^\mu,\mm)\bar{\sigma}(\rf^0,\mm)\bar{\sigma}(\rf^{0},\mm)\right\}\,.
\end{aligned}
\end{equation}
The first term defines the background volume $\bar{V} = \mathrm{v}\abs{\slrcw(x^0,\pmm)}^2$ which readily satisfies
\begin{equation}
    \frac{\bar{V}'}{3\bar{V}} = \frac{2}{3}\mu_\texttt{+}(\pmm)\,,\qquad \left(\frac{\bar{V}'}{3\bar{V}}\right)' = 0\,.
\end{equation}
The matching to the classical background volume dynamics is prescribed in Sec.~\ref{sec:dynamics}.

The remaining two terms in Eq.~\eqref{eq:V expval} make up the perturbations of the volume
\begin{equation}
\begin{aligned}\label{eq:pert volume expval}
&\delta V(x^\mu,\pmm) = 2\mathrm{v}\mathfrak{Re}\left\{\delta\Psi(x^\mu,\pmm)f(x^0)\e^{i\theta_f(x^0)}\bar{\slrcw}(x^0,\pmm)\bar{\slrcw}(x^0,\pmm)\right\}\\[7pt]
&\quad+2\mathrm{v}\mathfrak{Re}\left\{J_0\delta\Psi(x^\mu,\pmm)\bar{\slrcw}(x^0,\pmm)\bar{\tlrcw}(x^0,\pmm)+\frac{J_2}{2}\bar{\slrcw}(x^0,\pmm)\bar{\tlrcw}(x^0,\pmm)\nabla_{\vb*{x}}^2\delta\Psi(x^\mu,\pmm)\right\}\,.
\end{aligned}
\end{equation}
To study the dynamics of $\delta V$ it is convenient to perform a split of the complex-valued function $\delta\Psi$ into its modulus and phase, $\delta\Psi = R(x^\mu,\pmm)\e^{i\Theta}$. We pose the condition that this phase is in fact constant.\footnote{Assuming instead a time-dependent phase and splitting the equation into real and imaginary part, one finds $\Theta' = c/R^2$ with some time-dependent factor $c$. Since $R$ is however space-dependent and we require $\Theta$ to be only time-dependent, the function $c$ must vanish, and we conclude that $\Theta$ is in fact constant.} As a result, the overall phases of the first and second term inside the real parts of $\delta V$ are respectively given by
\begin{equation}
\theta_1 = \Theta + \theta_f(x^0) -2\tpip x^0\,,\qquad \theta_2 = \Theta\,.
\end{equation}
Exploiting once more the dynamical freedom on $\delta\Phi$, and thus on the function $\mathrm{f}(\rf^0,\vb*{\rf})$ entering Eq.~\eqref{eq:relation of dPsi and dPhi app}, we set $\theta_f = \frac{\pi}{2}+2\tpip x^0$. In momentum space of the rod variable, the resulting form of $\delta V$ is given by
\begin{equation}\label{eq:dV=AR}
\frac{\delta V(x^0,k)}{2\mathrm{v}\slrcw_0\tlrcw_0} = \left[\cos(\Theta)\e^{(\mu_\texttt{+}+\mu_\texttt{-})x^0}\left(J_0-\frac{J_2}{2}k^2\right)+\sin(\Theta)\e^{2\mu_\texttt{+} x^0}\frac{\slrcw_0}{\tlrcw_0}f\right]R\,. 
\end{equation}
Put in this form, the perturbed volume $\delta V$ is directly related to the modulus $R$ by a time- and momentum-dependent factor $A$, 
\begin{equation}
\frac{\delta V(x^0,k)}{2\mathrm{v}\slrcw_0\tlrcw_0} \eqdef A(x^0,k)R\,.
\end{equation}
Therefore, the dynamics of $\delta V$ are essentially governed by the dynamics of $R$. Introducing the function
\begin{equation}
 g_f\defeq \left(\slrcw_0 f\e^{\mu_\texttt{+}x^0}+J_0\tlrcw_0\e^{\mu_\texttt{-} x^0}\right)\,,
\end{equation}
and employing the dynamical equation of $\delta\Psi = R\e^{i\Theta}$ in Eq.~\eqref{eq:dPsi} for a constant phase, the dynamics of $R$ are given by
\begin{equation}\label{eq:R equation}
  0 = R'' +2\frac{g_f'}{g_f}R'+\left(\frac{g_f''}{g_f}-\mu_\texttt{+}^2\right)R -\frac{\alpha\tlrcw_0\e^{\mu_\texttt{-}x^0}}{g_f}k^2 R\,,  
\end{equation}
which straightforwardly follows from the dynamics of $\delta\Psi$. Combining Eqs.~\eqref{eq:dV=AR} and~\eqref{eq:R equation}, the dynamical equation for the perturbed volume $\delta V$ is given by
\begin{equation}\label{eq:dV with A}
    \delta V'' +\left[2\frac{g_f'}{g_f}-2\frac{A'}{A}\right]\delta V'+\left[\frac{g_f''}{g_f}-\mu_\texttt{+}^2+2\left(\frac{A'}{A}\right)^2-\frac{A''}{A}-2\frac{g_f'}{g_f}\frac{A'}{A}\right]\delta V-\frac{\alpha\tlrcw_0\e^{\mu_\texttt{-}x^0}}{g_f} k^2\delta V = 0\,.
\end{equation}
The above equation, and thus any solution of it, clearly depends on the function $g_f$ encoding the aforementioned mean-field dynamical freedom. Remarkably, however, this freedom can be fixed entirely by requiring the above equation to take the same functional form (at least in the late time, classical regime) of the corresponding GR one, given in Eq.~\eqref{eq:classical perturbed volume equation}.

To see this explicitly, we start from the spatial derivative term, whose pre-factor $a^4$, as mentioned in the introduction of Chapter~\ref{chapter:perturbations}, could not be recovered by considering a perturbed condensate of only spacelike tetrahedra~\cite{Marchetti:2021gcv}. Exactly because of the additional timelike degrees of freedom, and thus of the above dynamical freedom, here we can easily recover the appropriate pre-factor, by simply requiring the function $g_f$ to satisfy
\begin{equation}
-\frac{\alpha\tlrcw_0\e^{\mu_\texttt{-} x^0}}{g_f} = a^4 = \slrcw_0^{8/3}\e^{8\mu_\texttt{+} x^0/3}\,,
\end{equation}
where $a$ is the scale factor. The above condition corresponds to the following choice of $f$:
\begin{equation}
f = -\frac{\tlrcw_0}{\slrcw_0}\e^{(\mu_\texttt{-} -\mu_\texttt{+})x^0}\left(J_0+\alpha a^{-4}\right)\,,
\end{equation}
fixing the aforementioned dynamical freedom completely.\footnote{Note that the initial conditions for scale factor are chosen such that the present day value at time $x^0_*$ is normalized, i.e. $a(x^0_*) = 1$. Therefore, $a < 1$ for all times $x < x^0_*$ and therefore, the volume factor $a^{-4}$ in the equation above is not negligible.} As a result of this fixing, the function $g_f$ satisfies the following derivative properties
\begin{equation}\label{eq:g_f derivatives}
\frac{g_f'}{g_f} = \mu_\texttt{-} -\frac{8}{3}\mu_\texttt{+},\qquad \frac{g_f''}{g_f} = \left(\mu_\texttt{-} -\frac{8}{3}\mu_\texttt{+}\right)^2\,.
\end{equation}
Inserting the expression of $f$ into the function $A(x^0,k)$, one obtains
\begin{equation}
\frac{A'}{A} = \mu_\texttt{+}+\mu_\texttt{-} +\frac{8}{3}\mu_\texttt{+}\frac{\alpha\frac{\tlrcw_0}{\slrcw_0}\sin(\Theta)a^{-4}}{\cos(\Theta)\left(J_0-\frac{J_2}{2}k^2\right)-\frac{\tlrcw_0}{\slrcw_0}\sin(\Theta)\left(J_0+\alpha a^{-4}\right)}\,.
\end{equation}
As we see from the above equation, in general $A$ is a complicated function of the momenta $k$. As a consequence, the same holds for  the factors in front of $\delta V'$ and $\delta V$ in Eq.~\eqref{eq:dV with A}. This is in sharp contrast to what happens in GR, see again Eq.~\eqref{eq:classical perturbed volume equation}. However, this undesired $k$-dependence can be easily removed by choosing $\Theta = n\frac{\pi}{2}$ with odd integer $n$ and assuming that $J_0$ is negligible with respect to $\alpha a^{-4}$. {Notice that this is equivalent to $(\delta\pi_x/\epsilon\pi_0^+)^2 a^{-4}\gg 1$ which is ensured by the condition $\pi_x\gg\pi_0^+$.} Under these assumptions, the derivatives of $A$ take the form
\begin{equation}\label{eq:A derivatives}
\frac{A'}{A} = -\frac{5}{3}\mu_\texttt{+}+\mu_\texttt{-},\qquad \frac{A''}{A} = \left(-\frac{5}{3}\mu_\texttt{+}+\mu_\texttt{-}\right)^2\,.
\end{equation}
Combining Eqs.~\eqref{eq:g_f derivatives} and~\eqref{eq:A derivatives}, the perturbed volume equation attains the form
\begin{equation}\label{eq:GFT perturbed volume equation}
\delta V'' - 3\mathcal{H}\delta V' +a^4 k^2\delta V = 0\,,
\end{equation}
where we identified $\mathcal{H} = \frac{2}{3}\mu_\texttt{+}$ from the background equations. Expressed instead in terms of the ratio $\delta V/\bar{V}$, the relative perturbed volume equation is given by
\begin{equation}\label{eq:GFT perturbed relative volume equation}
\left(\frac{\delta V}{\bar{V}}\right)''+3\mathcal{H}\left(\frac{\delta V}{\bar{V}}\right)' +a^4 k^2\left(\frac{\delta V}{\bar{V}}\right) = 0\,.
\end{equation}
Remarkably, the two coefficients in front of the zeroth and first derivative term in Eq.~\eqref{eq:GFT perturbed volume equation} are both completely fixed by the background parameter $\mu_\texttt{+}$.\footnote{The values of these two coefficients is a direct consequence of matching the spatial derivative term. If the exponent of $a$ is chosen to be $\lambda\in\R$ instead of $4$, the first derivative coefficient is given by $-2\mu_\texttt{+}(2\lambda+1)$. Since the $a^4$-factor is crucial for obtaining the appropriate behavior of perturbations, we fix $\lambda = 4$.} In fact, the parameter $\mu_\texttt{-}$, characterizing the behavior of the timelike condensate, does not enter the perturbed volume equation at all. 

\paragraph{Remark on number of quanta.} As indicated in the main body, the dynamics of spacelike GFT quanta, $\bar{N}_\texttt{+}$ and $\delta N_{\texttt{+}}$, is fully determined by $\bar{V}$ and $\delta V$ due to the single spin assumption. The matching conditions presented in this appendix have however a non-trivial effect on the dynamics of the number of timelike quanta, $\bar{N}_\texttt{-}$ and $\delta N_{\texttt{-}}$. At background level, 
%
\begin{equation}
    \frac{\bar{N}_\texttt{-}'}{\bar{N}_\texttt{-}} = 2\mu_\texttt{-}\,,\qquad \left(\frac{\bar{N}_\texttt{-}'}{\bar{N}_\texttt{-}}\right)' = 0\,.
\end{equation}
The matching conditions do not involve $\mu_\texttt{-}$ except the assumption $\mu_\texttt{+}>\mu_\texttt{-}$ and its characterization is left as an intriguing task to future research. With the matching conditions above, the perturbation $\delta N_{\texttt{-}}$ satisfies
\begin{equation}
    \delta N_\texttt{-} = 2\mathfrak{Re}\left\{\int\dd{\rf^0}\delta\Xi(\rf^0,\mm)\bar{\tlrcw}^2(\rf^0,\mm)\eta^2_{\epsp}(\rf^0-x^0;\pip)\right\}\,.
\end{equation}
Since $\delta\Xi$ is only time-dependent, as we have shown in Appendix~\ref{app:Derivation condensate dynamics}, it follows that $\delta N_\texttt{-}(x^\mu,\pmm)\equiv\delta N_\texttt{-}(x^0,\pmm)$ only depends on the relational time. Thus, from a relational perspective, the perturbation of the timelike tetrahedra number can be absorbed into the background.

\section{Derivation of matter dynamics}\label{app:Derivation of matter dynamics}

We derive here  the dynamics of the matter scalar field $\phi$. Its classical relational dynamics is captured by
the expectation values of suitably defined matter and momentum operators
\begin{align}
\hat{\upphi}_\pm &= \frac{1}{i}\int\dd{\vbg}\dd{\rf^\mu}\dd{\mm}\dd{X_\pm}\hat{\varphi}^\dagger(\vbg,\rf^\mu,\mm,X_\pm)\pdv{}{\mm}\hat{\varphi}(\vbg,\rf^\mu,\mm,X_\pm)\,,\label{eq:mf operator}\\
\hat{\upvarpi}^\pm_\mf &= \int\dd{\vbg}\dd{\rf^\mu}\dd{\mm}\dd{X_\pm}\hat{\varphi}^\dagger(\vbg,\rf^\mu,\mm,X_\pm)\:\mm\:\hat{\varphi}(\vbg,\rf^\mu,\mm,X_\pm)\label{eq:mm operator}\,.
\end{align}
Note again that, in contrast to the reference fields $\rf^\mu$, we do not assume a priori that the scalar field propagates only along dual edges of a certain causal character. In analogy to above, we separate the expectation value of the above operators on the condensate states $\ket{\Delta;x^0,\vb*{x}}$ in background and perturbations. Expectation values of $\hat{\Phi}_\pm$ at the background and perturbed level evaluate to
\begin{equation}
    \bar{\upphi}_\texttt{+} = \frac{1}{i}\bar{\slrcw}(x^0,\mm)\eval{\pdv{}{\mm}\slrcw(x^0,\mm)}_{\mm=\pmm}\,,\qquad     \bar{\upphi}_\texttt{-} = \frac{1}{i}\bar{\tlrcw}(x^0,\mm)\eval{\pdv{}{\mm}\tlrcw(x^0,\mm)}_{\mm=\pmm}\,,
\end{equation}
and 
\begin{align}
\delta\upphi_\texttt{+} &= \frac{1}{i}\int\dd[4]{\rf}\dd{\mm}\left[\bar{\sigma}\partial_{\mm}(\delta\Phi\bar{\sigma})+\bar{\delta\Phi}\sigma\partial_{\mm}\sigma+\bar{\sigma}\partial_{\mm}(\delta\Psi\bar{\tau})+\bar{\delta\Psi}\tau\partial_{\mm}\sigma\right]\,,\label{eq:dphi+ pre}\\
\delta\upphi_\texttt{-} &= \frac{1}{i}\int\dd[4]{\rf}\dd{\mm}\left[\bar{\tau}\partial_{\mm}(\delta\Psi\bar{\sigma})+\bar{\delta\Psi}\sigma\partial_{\mm}\tau+\bar{\tau}\partial_{\mm}(\delta\Xi\bar{\tau})+\bar{\delta\Xi}\tau\partial_{\mm}\tau\right]\,,\label{eq:dphi- pre}
\end{align}
respectively. Since we work in momentum space while peaking on momentum, the operators $\hat{\upvarpi}_\mf^\pm$ and  $\hat{N}_\pm$ are closely defined and thus, their expectation values are related by
\begin{equation}
\bar{\upvarpi}_\mf^\pm = \pmm\bar{N}_\pm(x^0,\pmm)\,,\qquad \delta\upvarpi_\phi^\pm = \pmm\delta N_\pm\,.
\end{equation}
In the following two paragraphs, we analyze the dynamics of these expectation values and suggest a matching to the quantities $\mf$ and $\mm$ of general relativity.

\paragraph{Background part.} To compute $\bar{\upphi}_\pm$, we recall the decomposition of the condensate wavefunctions into radial and angular part, $r_{\pm}(x^0,\mm)$ and $\theta_\pm(x^0,\mm)$, respectively. Keeping only dominant contributions in $r_\pm$, one obtains
\begin{equation}
\bar{\upphi}_\pm = \bar{N}_\pm\eval{\partial_{\mm}\theta_\pm}_{\mm = \pmm}\,.
\end{equation}
Solutions of the background phases $\theta_\pm$ are given by
\begin{equation}
\theta_\pm = \tilde{\pi}^\pm x^0 - \frac{Q_\pm}{\mu_\pm r_\pm^2} + C_\pm\,,
\end{equation}
where $Q_\pm$ and $C_\pm$ are integration constants. Then, the zeroth order expectation value of $\hat{\upphi}_\pm$ is given by
\begin{equation}
\bar{\upphi}_\pm = \eval{-\partial_{\mm}\left(\frac{Q_\pm}{\mu_\pm}\right)+2\frac{Q_\pm}{\mu_\pm r_\pm^2}(\partial_{\mm}{\mu_\pm})x^0+\bar{N}_\pm\partial_{\mm}{C_\pm}}_{\mm=\pmm}\,.
\end{equation}
As a consequence of the peaking properties of $\sigma$ and $\tau$, the timelike condensate parameter $\mu_\texttt{-}$ is independent of $\mm$, i.e. $\partial_{\mm}\mu_\texttt{-} = 0$. If we choose in addition $C_\pm$ to be independent of $\mm$, $\bar{\upphi}_\pm$ is an intensive quantity for both $\pm$, as one would expect for a scalar field:
\begin{equation}\label{eqn:phibar+}
\bar{\upphi}_\texttt{+} = \eval{-\partial_{\mm}\left(\frac{Q_\texttt{+}}{\mu_\texttt{+}}\right)+2\frac{Q_\texttt{+}}{\mu_\texttt{+}}(\partial_{\mm}{\mu_\texttt{+}})x^0}_{\mm=\pmm}\,,\qquad \bar{\upphi}_\texttt{-} = \eval{-\frac{1}{\mu_\texttt{-}}\partial_{\mm}Q_\texttt{-}}_{\mm=\pmm}\,.
\end{equation}

In order to connect these expectation values to the scalar field variable $\mf$ of GR, one needs to define a way to combine the expectation values $\upphi_\pm$. To that end, we notice that the scalar field is intensive and canonically conjugate to the extensive quantity $\hat{\upvarpi}_\phi$. In analogy to the chemical potential in statistical physics, one possible way to combine $\upphi_\texttt{+}$ and $\upphi_\texttt{-}$ is to consider the weighted sum
\begin{equation}\label{eq:mf as weighted sum app}
\mf = \upphi_\texttt{+}\frac{N_\texttt{+}}{N}+\upphi_\texttt{-}\frac{N_\texttt{-}}{N}\,,
\end{equation}
where all the quantities appearing are the full expectation values, containing zeroth- and first-order terms. $N$ denotes the expectation value of the total number of GFT particles, i.e. $N = N_\texttt{+} + N_\texttt{-}$. Expanding all the quantities to linear order, we identify the background scalar field as
\begin{equation}
\bar{\phi} = \bar{\upphi}_\texttt{+}\frac{\bar{N}_\texttt{+}}{\bar{N}} + \bar{\upphi}_\texttt{-}\frac{\bar{N}_\texttt{-}}{\bar{N}}\,.
\end{equation}
{Assuming that $\bar{N}_\texttt{+}\gg\bar{N}_\texttt{-}$ at late times, corresponding to $\mu_\texttt{+} > \mu_\texttt{-}$ and reflecting that the background is predominantly characterized by the spatial geometry}, the matter field can be approximated as
\begin{equation}
\bar{\mf} \approx \bar{\upphi}_\texttt{+}\,.
\end{equation}
{Using Eq.~\eqref{eqn:phibar+}, we see that the scalar field is linear in relational time, as expected classically.
Thus, we can easily match the classical GR background equations for $\bar{\mf}$:  imposing $Q_\texttt{+} = \mm^2$, yields}
\begin{equation}\label{eq:mf background}
\bar{\mf}' = \pmm,\qquad \bar{\mf}'' = 0\,,
\end{equation}
{as required}. Besides the relation $\mu_\texttt{+} > \mu_\texttt{-}$, the background matching does not impose any further conditions on $Q_\texttt{-}$ and the precise form of $\mu_\texttt{-}$. 

For $\upvarpi_\mf^\pm$, we notice that this quantity grows with the system size, given by the respective number of tetrahedra $\bar{N}_\pm$. At lowest order, we therefore identify the classical quantity $\bar{\pi}_\mf$ as%
\begin{equation}
\bar{\pi}_\mf = \frac{\bar{\upvarpi}_\mf^\texttt{+} + \bar{\upvarpi}_\mf^\texttt{-}}{\bar{N}} = \frac{\bar{N}_\texttt{+} + \bar{N}_\texttt{-}}{\bar{N}_\texttt{+} + \bar{N}_\texttt{-}} \pmm= \pmm\,,
\end{equation}
which corresponds to the peaked matter momentum $\pmm$. With this identification, the GFT parameter $\mu_\texttt{+}$ can be expressed by the peaked matter momentum as
\begin{equation}
M_{\Pl}^2\:\mu_\texttt{+}^2(\pmm) = \frac{8}{3}\bar{\pi}_\mf^2 = \frac{8}{3}\pmm^2\,,
\end{equation}
where again a factor of Planck mass has been added to ensure the correct energy dimensions.

\paragraph{First-order perturbations.} Given the expectation values $\delta\upphi_\pm$ in Eqs.~\eqref{eq:dphi+ pre} and~\eqref{eq:dphi- pre}, we perform a partial integration in $\mm$ and only keep dominating terms, yielding
\begin{align}
\delta\upphi_\texttt{+} &= 2\mathfrak{Re}\left\{\int\dd[4]{\rf}\dd{\mm}\left[\delta\Phi\bar{\sigma}^2\partial_{\mm}\theta_\texttt{+}+\delta\Psi\bar{\tau}^2\partial_{\mm}\theta_\texttt{+}\right]\right\}\,,\\
\delta\upphi_\texttt{-} &= 2\mathfrak{Re}\left\{\int\dd[4]{\rf}\dd{\mm}\left[\delta\Psi\bar{\sigma}^2\partial_{\mm}\theta_\texttt{-}+\delta\Xi\bar{\tau}^2\partial_{\mm}\theta_\texttt{-}\right]\right\}\,.
\end{align}
Using the relation of $\delta\Phi$ and $\delta\Psi$ in Eq.~\eqref{eq:relation of dPsi and dPhi app}, as well as the assumptions on the peaking parameters of $\sigma$ and $\tau$ in Eq.~\eqref{eq:peaking parameter relations}, the first-order expectation value $\delta\upphi_\texttt{+}$ evaluates to
\begin{equation}\label{eq:deltaUpphi_+}
\delta\upphi_\texttt{+} = \eval{\delta N_\texttt{+}(x^\mu,\mm)\partial_{\mm}\theta_\texttt{+}}_{\mm=\pmm} = \frac{\delta N_\texttt{+}}{\bar{N}_\texttt{+}}\bar{\phi}.
\end{equation}
In contrast to $\delta\upphi_\texttt{+}$, the evaluation of $\delta\upphi_\texttt{-}$ is more intricate since the peaking properties of $\bar{\tau}^2$ yield a time derivative expansion when integrating over the reference field. However, as we show next, the perturbed scalar field $\delta\mf$ does not explicitly depend on $\delta\upphi_\texttt{-}$ under the assumption that $\mu_\texttt{+}>\mu_\texttt{-}$. Following the definition of $\mf$ in Eq.~\eqref{eq:mf as weighted sum app}, at linear order in perturbations, one obtains
\begin{equation}
\delta\phi \approx \bar{\phi}\left(\frac{\delta N_\texttt{+}-\delta N_\texttt{-}}{\bar{N}_\texttt{+}}\right)+\bar{\upphi}_\texttt{-}\frac{\delta N_\texttt{-}}{\bar{N}_\texttt{+}}.
\end{equation}
Since the timelike number perturbation $\delta N_\texttt{-}$ is only time-dependent, and therefore part of the background, the factors of $\delta N_\texttt{-}/\bar{N}_\texttt{+}$ are negligible and one is left with $\delta\mf = \left(\delta V/\bar{V}\right)\bar{\mf}$. Applying Eqs.~\eqref{eq:GFT perturbed relative volume equation} and~\eqref{eq:mf background} for $\delta V/\bar{V}$ and $\bar{\mf}$, respectively, the dynamical equation for $\delta\mf$ from GFT is given by
\begin{equation}\label{eq:GFT perturbed mf equation}
\delta\mf''+a^4 k^2\delta\mf = \left(-3\mathcal{H}\bar{\phi}+2\bar{\phi}'\right)\left(\frac{\delta V}{\bar{V}}\right)'.
\end{equation}
Notice that the right-hand side of this partial differential equation constitutes a source term that is absent in the classical equation of $\mf_\GR$, given in Eq.~\eqref{eq:classical mf perturbation equation}, formulated in harmonic gauge.

Let us consider now the first-order matter momentum variable $\delta\upvarpi_\mf^\pm$ which, as for the background variable, scales with the system size. In order to connect this quantity to the intrinsic quantity $\delta\mm$ of GR, dividing $\delta\upvarpi_\mf^\pm$ by the particle number is required. In principle, there are two different ways to do so, both of which we present in the following. 

First, one can define $\delta\mm$ as the first-order term of
\begin{equation}
\delta\mm\overset{(1)}{=}\frac{\upvarpi_\mf^\texttt{+}+\upvarpi_\mf^\texttt{-}}{N_\texttt{+}+N_\texttt{-}} = 0,
\end{equation}
where all the quantities entering this expression contain both, zeroth- and first-order perturbations. However, in this case $\delta\mm = 0$. Operatively, this could be interpreted as a perturbation of the background momentum $\bar{\pi}_\mf$. Since this is a constant of motion, any such perturbation would vanish by construction. Alternatively, one could perturb only the momenta and keep the particle numbers at zeroth order. In this case, $\delta\mm$ is given by
\begin{equation}
\delta\mm \overset{(2)}{=} \frac{\delta\upvarpi_{\mf}^+ + \delta\upvarpi_{\mf}^-}{\bar{N}_\texttt{+}+\bar{N}_\texttt{-}} \approx \pmm\frac{\delta N_\texttt{+}}{\bar{N}_\texttt{+}} = \pmm\frac{\delta V}{\bar{V}}.
\end{equation}
None of the options above offer a matching to the classical perturbed momentum variable $\delta\pi_\mf^0$, defined in Eq.~\eqref{eq:classical perturbed mm 0} as the $0$-component of the conjugate momentum of $\mf$ at linear order. The main difficulty in matching these two quantities is that the classical equation~\eqref{eq:classical perturbed mm 0} depends on the perturbation of the lapse function, $A$. To recover this quantity from the fundamental QG theory, one would need additional (relational) geometric operators other than the volume.

\section{Classical perturbation theory}\label{app:Classical perturbation theory}

In this appendix, we provide an overview of the perturbation equations for geometry and matter in classical GR. To allow for a comparison with relational GFT results, we mostly use harmonic coordinates $\{x^\mu\}$ which are adapted to the reference field $\{\rf^\mu\}$ via the relation $\rf^\mu = \kappa^\mu x^\mu$ (no summation over $\mu$), where $\kappa^\mu$ are some dimensionful proportionality factors \cite{Gielen:2018fqv}. Assuming that the reference fields satisfy the Klein-Gordon equation at all orders, one finds
\begin{equation}\label{eq:harmonic gauge}
\Gamma^{\lambda}_{\mu\nu}\mathrm{g}^{\mu\nu} = 0\,,
\end{equation}
which poses a condition on the metric.

\subsection{Geometry}\label{sec:Classical perturbation theory - geometry}

At zeroth order, the line element of a spatially flat Friedmann-Lema\^{i}tre-Robertson-Walker (FLRW) spacetime with signature of $(+,-,-,-)$ is given by
\begin{equation}
\dd{s}^2 = \bar{\mathrm{g}{}}_{\mu\nu}\dd{x}^\mu\dd{x}^\nu = N^2\dd{t}^2 - a^2\dd{\vb*{x}}^2\,,
\end{equation}
where $N$ is the lapse function, $a$ is the scale factor and $\dd{\vb*{x}}^2$ the line element of $3$-dimensional Euclidean flat space. Imposing harmonic gauge on the background yields $a^3/N = c_H$, where $c_H$ is an integration constant. For the remainder, we set $c_H = 1$, and we assume that the matter content is dominated by the matter field $\mf$ with conjugate momentum $\mm$. Within these assumptions, the dynamics of the geometry at background level are captured by
\begin{equation}\label{eq:classical 3H^2 eq}
3\mathcal{H}^2 = \frac{1}{2 M_{\Pl}^2}\bar{\pi}_\mf^2\,,\qquad \mathcal{H}' = 0\,,
\end{equation}
where $\mathcal{H} = a'/a$ is the Hubble parameter {in harmonic coordinates} and $\bar{\pi}_\mf$ is the background contribution of the canonical conjugate of the scalar field, defined in Eq.~\eqref{eq:clasical mm def}. Introducing the background volume $\bar{V} = a^3$, the geometric equations can be recast to
\begin{equation}
3\left(\frac{\bar{V}'}{3\bar{V}}\right)^2 = \frac{1}{2 M_{\Pl}^2}\bar{\pi}_\mf^2\,,\qquad \left(\frac{\bar{V}'}{3\bar{V}}\right)' = 0\,.
\end{equation}

To derive perturbed volume equations, consider first-order scalar perturbations of the FLRW metric with line element
\begin{equation}\label{eq:perturbed line element}
\dd{s}^2 = a^6(1+2A)\dd{t}^2 - a^4\partial_iB\dd{t}\dd{x^i} - a^2\left((1-2\psi)\delta_{ij}+2\partial_i\partial_j E\right)\dd{x}^i\dd{x}^j\,,
\end{equation}
with scalar perturbation functions $A,B,\psi$ and $E$. Einstein's equations at linear order yield~\cite{Marchetti:2021gcv,Battarra:2014tga}
\begin{align}
 \frac{1}{2 M_{\Pl}^2}\bar{\mf}'\delta\mf'+3\mathcal{H}\psi'-a^4\nabla^2\psi-\mathcal{H}\nabla^2\left(E'-a^2 B\right) &= 0\,,\label{eq:pEFE1}\\
\mathcal{H} A + \psi'-\frac{1}{2 M_{\Pl}^2}\bar{\mf}'\delta\mf &= 0\,,\label{eq:pEFE2}\\
 E''-a^4\nabla^2 E &= 0\label{eq:pEFE3}\,,
\end{align}
where $\delta\mf$ is the scalar field perturbation. Combining Eq.~\eqref{eq:pEFE1} and the time-derivative of Eq.~\eqref{eq:pEFE2}, we obtain
\begin{equation}\label{eq:psi''}
\psi'' = -\mathcal{H} A'-3\mathcal{H}\psi'+a^4\nabla^2\psi+\mathcal{H}\nabla^2\left(E'-a^2 B\right)\,.
\end{equation}
To obtain an equation for the perturbed volume, which is an observable accessible also from the GFT side, consider on a slice of constant time the  local volume element
\begin{equation}
\sqrt{-\mathrm{g}_{(3)}} = \bar{V} + \delta V = a^3\left(1-3\psi+\nabla^2E\right)\,.
\end{equation}
Thus, we identify the perturbed spatial volume as
\begin{equation}
\frac{\delta V}{\bar{V}} = -3\psi+\nabla^2 E\,.
\end{equation}
Taking the second derivative of $\delta V/\bar{V}$ and using Eqs.~\eqref{eq:pEFE3} and~\eqref{eq:psi''}, one obtains
\begin{equation}\label{eqn:classicalpertvolumegeneral}
\left(\frac{\delta V}{\bar{V}}\right)'' + 3\mathcal{H}\left(\frac{\delta V}{\bar{V}}\right)'-a^4\nabla^2\left(\frac{\delta V}{\bar{V}}\right) = 3\mathcal{H}\left(A'+a^2\nabla^2 B\right)\,.
\end{equation}

At first order in perturbations, the harmonic gauge conditions are given by~\cite{Battarra:2014tga}
\begin{equation}\label{eq:pert HGC}
\begin{aligned}
0 &= A'+3\psi'-\nabla^2(E'-a^2B)\,,\\
0 &= (a^2B)'+a^4(A-\psi-\nabla^2 E)\,,
\end{aligned}
\end{equation}
which, imposed on Einstein's equations~\eqref{eq:pEFE1} - \eqref{eq:pEFE3}, yield~\cite{Battarra:2014tga,Marchetti:2021gcv}
\begin{align}
\psi''-a^4\nabla^2\psi &=0\,, &&A'' -a^4\nabla^2 A + 4a^4\nabla^2\psi =0\,, \\
E'' -a^4\nabla^2 E&=0\,, &&(a^2B)''-a^4\nabla^2(a^2B)-8a^2(a^2\psi)' =0\,.
\end{align}
Expressed in terms of the volume, the first harmonic gauge condition is expressed as
\begin{equation}
    A'+a^2\nabla^2 B = \left(\frac{\delta V}{\bar{V}}\right)'\,,
\end{equation}
such that the volume equation becomes
\begin{equation}\label{eq:classical relative perturbed volume equation}
\left(\frac{\delta V}{\bar{V}}\right)''-a^4 \nabla^2\left(\frac{\delta V}{\bar{V}}\right) = 0\,,
\end{equation}
or equivalently
\begin{equation}\label{eq:classical perturbed volume equation}
\delta V'' - 6\mathcal{H}\delta V'+9\mathcal{H}^2\delta V-a^4 \nabla^2\delta V = 0\,.
\end{equation}
To change to Fourier space in the rod variable one can simply perform the substitution $\nabla^2 \rightarrow -k^2$ here and in the following.

Following~\cite{Battarra:2014tga}, there is a residual gauge freedom in performing a coordinate transformation $\xi^\mu \mapsto x^\mu+\xi^\mu$, with $\xi^\mu = (\xi^0,\partial^i\xi)$  satisfying
\begin{equation}
(\xi^0)''-a^4\nabla^2\xi^0 = \xi''-a^4\nabla^2\xi = 0\,,
\end{equation}
such that harmonicity is conserved. Under this transformation, the perturbation functions transform as~\cite{Battarra:2014tga,Marchetti:2020umh}
\begin{equation}\label{eq:pert gauge trafo}
\begin{gathered}
\psi \mapsto \psi +\mathcal{H}\xi^0\,,\qquad A \mapsto A -(\xi^0)' - 3\mathcal{H}\xi^0\,,\\
E \mapsto E -\xi\,,\qquad B \mapsto B +a^2\xi^0-a^{-2}\xi'\,.
\end{gathered}
\end{equation}
After introducing the matter equations in the following, we combine the geometric and matter quantities in a single fully gauge-invariant quantity, the so-called curvature perturbation $\mathcal{R}$. 

\subsection{Matter}\label{sec:Classical perturbation theory - matter}

The matter content of the classical theory consists of four reference scalar fields $\rf^\mu$ as well as one additional free minimally coupled real scalar field $\mf$, defined by the continuum action
\begin{equation}
S[\rf^\mu,\mf] = -\frac{1}{2 M_{\Pl}^2}\int\dd[4]{x}\sqrt{-\mathrm{g}}\mathrm{g}^{ab}\left(\partial_a\mf\partial_b\mf+\sum_{\mu=0}^3\partial_a\rf^\mu\partial_b\rf^\mu\right)\,.
\end{equation}
{In this form, the action poses a well-defined variational principle, yielding the Klein-Gordon equations for appropriate boundary conditions. One of such admissible conditions are von Neumann boundary conditions which assume vanishing variation of the gradients at the boundary. For reference fields in harmonic coordinates, as used in the remainder of this subsection, $\rf^\mu = \kappa^\mu x^\mu$, this clearly applies since $\partial_\mu\rf^\nu = \delta_\mu^\nu\kappa^\nu$ is constant and thus has vanishing variation.}

The energy momentum tensor in arbitrary coordinates is given by 
\begin{equation}\label{eq:classical EM-tensor}
  M_{\Pl}^2\; T_{ab}=\sum_{\lambda=0}^3\left(\partial_a\chi^\lambda\partial_b\chi^\lambda-\frac{\mathrm{g}_{ab}}{2}\mathrm{g}^{mn}\partial_m\chi^\lambda\partial_n\chi^\lambda\right)+\partial_a\phi\partial_b\phi-\frac{\mathrm{g}_{ab}}{2}\mathrm{g}^{mn}\partial_m\phi\partial_n\phi\,,
\end{equation}
{which we assume to be dominated by the matter field $\mf$.} The full equations of motion for $\mf$ are given by the massless Klein-Gordon equation
\begin{equation}
\partial_\mu\left(\sqrt{-\mathrm{g}}\mathrm{g}^{\mu\nu}\partial_\nu\mf\right) = 0\,.
\end{equation}
Linearizing in both, the scalar field and the metric, we obtain the zeroth order equation $\bar{\phi}'' = 0$, and the first-order perturbation equation
\begin{equation}
\delta\mf''-a^4\nabla^2\delta\mf = \left[A'+3\psi'-\nabla^2E'+a^2\nabla^2B\right]\bar{\mf}'\,,
\end{equation}
respectively. Supplementing the latter with the harmonic gauge condition in Eq.~\eqref{eq:pert HGC}, $\delta\mf$ satisfies
\begin{equation}\label{eq:classical mf perturbation equation}
\delta\mf''-a^4\nabla^2\delta\mf = 0\,.
\end{equation}

To define the GR counterpart of the GFT observables $\hat{\upvarpi}_\mf^\pm$, defined in Eq.~\eqref{eq:mm operator}, we introduce the momentum conjugate to the scalar field, commonly defined as
\begin{equation}\label{eq:clasical mm def}
\mm^\mu\defeq  \pdv{\tilde{\mathcal{L}}}{(\partial_\mu\mf)} = -\sqrt{-\mathrm{g}}\mathrm{g}^{\mu\nu}(\partial_\nu\phi)\,,
\end{equation}
where $\tilde{\mathcal{L}}$ is the Lagrangian density, defined by the matter field action above. Expanding up to linear order, the scalar field momentum $\pi_\phi^\mu$ is given by
\begin{equation}
\pi^\mu_\phi = -\overline{\sqrt{-\mathrm{g}}}\bar{\mathrm{g}}^{\mu 0}\partial_0\bar{\phi}-\delta\sqrt{-\mathrm{g}}\bar{\mathrm{g}}^{\mu 0}\partial_0\bar{\phi}-\overline{\sqrt{-\mathrm{g}}}\left(\delta \mathrm{g}^{\mu 0}\partial_0\bar{\phi}+\bar{\mathrm{g}}^{\mu\nu}\partial_\nu\delta\phi\right)\,,
\end{equation}
which can be split into background and perturbed part
\begin{equation}
\pi_\phi^\mu = \bar{\pi}_\phi^\mu+\delta\pi_\phi^\mu\,.
\end{equation}
At the background level and in harmonic gauge, $\bar{\pi}_\mf^\mu$ is given by
\begin{equation}
\bar{\pi}_\phi^0 = \partial_0\bar{\phi}\,,\qquad \bar{\pi}_\phi^i = 0\,.
\end{equation}
The perturbed part of $\mm^\mu$ is given by
\begin{align}
\delta\mm^0 &= \left(-A-3\psi+\nabla^2 E\right)\bar{\mf}'+\delta\mf'\,,\label{eq:classical perturbed mm 0}\\[7pt]
\delta\mm^i &= a^2\bar{\mf}'\partial^iB-a^4\partial^i\delta\mf\,.\label{eq:classical perturbed mm i}
\end{align}
Applying the zeroth- and first-order equations for $\mf$, the perturbed momentum satisfies the relativistic energy-momentum conservation equation
\begin{equation}
\partial_\mu\delta\mm^\mu = \left(\delta\mm^0\right)'+\partial_i\delta\mm^i = 0\,.
\end{equation}
Re-expressing this equation in terms of observables that are available in GFT, being $\bar{V},\delta V,\bar{\mf}$ and $\delta\mf$, we find
\begin{equation}\label{eq:classical perturbed mm equation}
\left(\delta\mm^0\right)'-\delta\mf''-\left(\frac{\delta V}{\bar{V}}\right)'\bar{\mf}' = -A'\bar{\mf}'\,.
\end{equation}
While the left-hand side is given in terms of variables available in GFT, the right-hand side contains the variable $A$, which is not accessible by the GFT observable that are available at the present state.

\paragraph{Classsical Mukhanov-Sasaki-like equation.}

As Eq.~\eqref{eq:pert gauge trafo} shows, the harmonic gauge conditions leaves a residual gauge freedom. Under these transformations, the perturbed scalar field $\delta\mf$ changes as
\begin{equation}
\delta\mf\mapsto \delta\mf -\bar{\mf}'\xi^0\,.
\end{equation}
Given this transformation behavior, one can combine $\psi$ and $\delta\mf$ to a fully gauge-invariant quantity, the so-called gauge-invariant curvature perturbation
\begin{equation}\label{eq:classical R}
\mathcal{R} \defeq \psi + \mathcal{H}\frac{\delta\mf}{\bar{\mf}'}\,.
\end{equation}
Since in harmonic gauge, $\psi$ and $\delta\mf$ satisfy the same equation, $\mathcal{R}$ satisfies~\cite{Battarra:2014tga}
\begin{equation}
\mathcal{R}''-a^4\nabla^2 \mathcal{R} = 0\,.
\end{equation}

In the context of GFT, one does not have direct access to the quantity $\psi$ but rather to the perturbed volume $\delta V$. For comparison of classical and GFT mechanics, we define the \textit{curvature-like perturbation} $\tilde{\mathcal{R}}$ as
\begin{equation}\label{eq:classical Rtilde}
\tilde{\mathcal{R}}\defeq -\frac{\delta V}{3\bar{V}}+\mathcal{H}\frac{\delta\mf}{\bar{\mf}'}\,.
\end{equation}
Again, since $\delta V/\bar{V}$ and $\delta\mf$ satisfy the same equation, $\tilde{\mathcal{R}}$ obeys
\begin{equation}\label{eq:classical R perturbation equation}
\tilde{\mathcal{R}}''-a^4\nabla^2\tilde{\mathcal{R}} = 0\,.
\end{equation}
Notice however, that $\tilde{\mathcal{R}}$ is not gauge-invariant but changes as
\begin{equation}
\tilde{\mathcal{R}}\mapsto \tilde{\mathcal{R}}-\nabla^2\xi\,.
\end{equation}
Still, since $\xi$ is assumed to satisfy the equation above, the equation for $\tilde{\mathcal{R}}$ does not change under gauge transformations. 

\clearpage
\bookmarksetup{startatroot}
\addcontentsline{toc}{chapter}{\textsc{Bibliography}}
\fancyhead[CE]{\scshape Bibliography}
\fancyhead[LE]{}
\fancyhead[RE]{}
\fancyhead[CO]{\scshape Bibliography}
\fancyhead[LO]{}
\fancyhead[RO]{}
\renewcommand{\headrulewidth}{0.5pt}
\renewcommand{\footrulewidth}{0pt}

\linespread{1.0}

\bibliography{references/specdim.bib,references/3d_cosmology.bib,references/perturbations.bib,references/LRC.bib,references/LG.bib,references/intro.bib}
\bibliographystyle{ieeetr}

\linespread{1.5}
\cleardoublepage
\bookmarksetup{startatroot}
\phantomsection
\addcontentsline{toc}{chapter}{\textsc{Ehrenwörtliche Erklärung}}
\chapter*{Ehrenwörtliche Erklärung}\label{Ehrenwörtliche Erklärung}
\thispagestyle{empty}
\pagestyle{empty}
Hiermit erkläre ich ehrenwörtlich,
\begin{enumerate}
  \item dass mir die geltende Promotionsordnung bekannt ist;
  \item dass ich die Dissertation selbst angefertigt habe, keine Textabschnitte eines Dritten oder eigener Prüfungsarbeiten ohne Kennzeichen übernommen und alle von mir benutzten Hilfmittel, persönlichen Mitteilungen und Quellen in meiner Arbeit angegeben habe; 
  \item dass meine Eigenständigkeitserklärung sich auch auf nicht zitierfähige, generierende KI-Anwendungen bezieht;
  \item dass bei der Auswahl und Auswertung des hier präsentierten Materials mich die nachstehen aufgeführten Personen in der jeweils beschriebenen Weise unterstützt
  haben:
  \begin{itemize}
    \item Sebastian Steinhaus bei der Betreuung über die gesamte Dauer der Promotion;
    \item Sebastian Steinhaus und Johannes Thürigen bei den in Kapitel~\ref{chapter:specdim} vorgestellten Ergebnissen, basierend auf~\cite{Jercher:2023rno};
    \item Sebastian Steinhaus bei den in Kapitel~\ref{chapter:LRC} vorgestellten Ergebnissen, basierend auf~\cite{Jercher:2023csk};
    \item Jos\'{e} Diogo Sim\~{a}o und Sebastian Steinhaus bei den in Kapitel~\ref{chapter:3d cosmology} vorgestellten Ergebnissen, basierend auf~\cite{Jercher:2024hlr,Jercher:2024kig};
    \item Daniele Oriti und Andreas Pithis bei den in Kapitel~\ref{chapter:cBC} vorgestellten Ergebnissen, basierend auf~\cite{Jercher:2022mky};
    \item Roukaya Dekhil, Daniele Oriti und Andreas Pithis bei den in Kapitel~\ref{chapter:LG} vorgestellten Ergebnissen, basierend auf~\cite{Dekhil:2024ssa,Dekhil:2024djp};
    \item Daniele Oriti, Luca Marchetti und Andreas Pithis bei den in Kapitel~\ref{chapter:perturbations} vorgestellten Ergebnissen, basierend auf~\cite{Jercher:2021bie,Jercher:2023kfr,Jercher:2023nxa}.
  \end{itemize}
  Die korrespondierenden Appendizes basieren ebenfalls auf diesen Zusammenarbeiten.  
  \item dass die Hilfe einer kommerziellen Promotionsvermittlerin/eines kommerziellen Promotionsvermittlers nicht in Anspruch genommen wurde und dass Dritte weder unmittelbar noch mittelbar geldwerte Leistungen für Arbeiten erhalten haben, die im Zusammenhang mit dem Inhalt der vorgelegten Dissertation stehen;
  \item dass die Dissertation noch nicht als Prüfungsarbeit für eine staatliche oder andere wissenschaftliche Prüfung eingereicht wurde;
  \item dass eine gleiche, eine in wesentlichen Teilen ähnliche oder eine andere Abhandlung nicht bei einer anderen Hochschule als Dissertation eingereicht wurde.
\end{enumerate}
\vspace*{\fill}
\hrule

\end{document}